\documentclass[useAMS,usenatbib]{mn2e}

\usepackage{tabularx}
\usepackage{graphicx}
\usepackage{txfonts}
\usepackage{longtable}
\usepackage{hhline}
\usepackage{arydshln}
\usepackage{multirow}
\usepackage{lscape}
\usepackage{array}

\newcommand{\hersc}{{\it Herschel}}
\newcommand{\lab}{LABOCA}
\newcommand{\pck}{{\it Planck}}

\newcommand{\spitz}{{\it  Spitzer}}

\newcommand{\lsun}{$L_\odot$}
\newcommand{\msun}{$M_\odot$}

\newcommand{\mic}{$\mu$m}

\newlength{\pointwidth}
\begin{document}

\title[Mapping the cold dust properties with \hersc]{Mapping the cold dust temperatures and masses of nearby Kingfish galaxies with Herschel}

\author[Galametz et al.]
{\parbox{\textwidth}{M. Galametz$^{1}$\thanks{e-mail: mgalamet@ast.cam.ac.uk}, 
R. C. Kennicutt$^{1}$, 
M. Albrecht$^{2}$,
G. Aniano$^{3}$,
L. Armus$^{4}$,
F. Bertoldi$^{2}$,
D. Calzetti$^{5}$,
A. F. Crocker$^{5}$,
K. V. Croxall$^{6}$,
D. A. Dale$^{7}$,
J. Donovan Meyer$^{8}$,
B. T. Draine$^{3}$,
C. W. Engelbracht$^{9}$,
J. L. Hinz$^{9}$,
H. Roussel$^{10}$,
R. A. Skibba$^{9}$,
F.S. Tabatabaei$^{11}$,
F. Walter$^{11}$,
A. Weiss$^{12}$,
C. D. Wilson$^{13}$,
M. G. Wolfire$^{14}$}\vspace{0.5cm}\\
\parbox{\textwidth}{$^{1}$Institute of Astronomy, University of Cambridge, Madingley Road, Cambridge CB3 0HA, UK\\
$^{2}$Argelander-Institut f\"ur Astronomie, Abteilung Radioastronomie, Auf dem H\"ugel, D-53121 Bonn, Germany\\
$^{3}$Department of Astrophysical Sciences, Princeton University, Princeton, NJ 08544, USA\\
$^{4}$Spitzer Science Center, California Institute of Technology, MC 314-6, Pasadena, CA 91125, USA\\
$^{5}$Department of Astronomy, University of Massachusetts, Amherst, MA 01003, USA\\
$^{6}$Department of Physics and Astronomy, University of Toledo, Toledo, OH 43606, USA\\
$^{7}$Department of Physics $\&$ Astronomy, University of Wyoming, Laramie, WY 82071, USA\\
$^{8}$Department of Physics and Astronomy, SUNY Stony Brook, Stony Brook, NY 11794-3800, USA\\
$^{9}$Steward Observatory, University of Arizona, Tucson, AZ 85721, USA 	     \\
$^{10}$Institut d'Astrophysique de Paris, Universit\'e Pierre et Marie Curie (UPMC), CNRS (UMR 7095), 75014 Paris, France\\
$^{11}$Max-Planck-Institut f\"ur Astronomie, K\"onigstuhl 17, D-69117 Heidelberg, Germany\\
$^{12}$Max-Planck-Institut f\"ur Radioastronomie, Auf dem H\"ugel 69, D-53121 Bonn, Germany\\
$^{13}$Department of Physics $\&$ Astronomy, McMaster University, Hamilton, Ontario L8S 4M1, Canada\\
$^{14}$Department of Astronomy, University of Maryland, College Park, MD 20742, USA}}

\maketitle{}


\begin{abstract}
 Taking advantage of the unprecedented combination of sensitivity and angular resolution afforded by the \hersc\ Space Observatory at far-infrared and submillimeter wavelengths, we aim to characterize the physical properties of cold dust within nearby galaxies, as well as the associated uncertainties, namely the robustness of the parameters we derive using different modified blackbody models. For a pilot subsample of the KINGFISH key program, we perform two-temperature fits of the \spitz\ and \hersc\ photometric data (from 24 to 500 \mic), with a warm and a cold component, both globally and in each resolution element. We compare the results obtained from different analysis strategies. At global scales, we observe a range of values of the modified blackbody fit parameter $\beta$$_c$ (0.8 to 2.5) and T$_c$ (19.1 to 25.1K).  
 We compute maps of our modelling parameters with $\beta$$_c$ fixed or treated as a free parameter to test the robustness of the temperature and dust surface density maps we deduce. When the emissivity is fixed, we observe steeper temperature gradients as a function of radius than when it is allowed to vary. When the emissivity is fitted as a free parameter, barred galaxies tend to have uniform fitted emissivities. Gathering the parameters obtained each resolution element in a T$_c$-$\beta$$_c$ diagram underlines an anti-correlation between the two parameters. It remains difficult to assess whether the dominant effect is the physics of dust grains, noise, 
or mixing along the line of sight and in the beam. We finally observe in both cases that the dust column density peaks in central regions of galaxies and bar ends (coinciding with molecular gas density enhancements usually found in these locations). We also quantify how the total dust mass varies with our assumptions about the emissivity index as well as the influence of the wavelength coverage used in the fits. We show that modified blackbody fits using a shallow emissivity ($\beta$ $<$ 2.0) lead to significantly lower dust masses compared to the $\beta$ $<$ 2.0 case, with dust masses lower by up to 50 $\%$ if $\beta$$_c$ =1.5 for instance. The working resolution affects our total dust mass estimates: masses increase from global fits to spatially-resolved fits. 
\end{abstract}
  


\section{Introduction}

The interstellar medium (ISM) plays a key role in the star formation history of galaxies. While stars evolve and die, they re-inject dust and gas in the ISM that will then be processed to form a new generation of stars. Dust is formed in the envelopes of late-evolved stars and is transformed in supernova shock waves or stellar winds by violent processes such as destruction, vaporization, sputtering or erosion \citep{Jones1994,Jones1996,Serra_Dias_Cano2008}. The size of dust grains also increases through processes like accretion or coagulation accretion \citep{Stepnik2003,Dominik1997,Dominik2008,Hirashita2009}. They participate in the gas thermal balance and screen parts of the ISM through the absorption of starlight and subsequent emission of infrared (IR) photons. 

Dust grains contribute to the chemistry of the ISM. Molecular hydrogen predominantly forms on grains, and molecules can freeze out on grain surfaces and undergo additional transformation \citep{Hasegawa1993,Vidali2004,Cazaux2005}. Atomic and molecular gas are commonly traced by H{\sc i} and CO observations. CO can be a poor tracer of molecular gas, especially in low-metallicity environments, due to the dependency of the CO-to-H$_2$ conversion factor with density and metallicity \citep{Wilson1995,Taylor1998,Israel2000,Leroy2005,Wolfire2010}. We know that dust and gas are closely tied in the ISM. This suggests that used with H{\sc i} and CO, dust could also be used as an additional tracer of the gas.

\begin{table*} 
\caption{Galaxy Data}
\label{Galaxy_data}
 \centering
 \begin{tabular}{ccccccccccc}
\hline
\hline
   &&&&&&&&&&\\
   Galaxy & Optical  & $\alpha$$_0$ & $\delta$$_0$ & Major diam. & Minor diam. &Distance & \multicolumn{2}{c}{12+log(O/H)} & SFR & L$_{TIR}$\\    
   & Morphology  & (J2000) & (J2000) & (arcmin) &  (arcmin)& (Mpc) & (PT) & (KK) & (\msun\ yr$^{-1}$) & ($\times$ 10$^{9}$ \lsun)  \\   
   (1) & (2) & (3) & (4) & (5) & (6) & (7) & \multicolumn{2}{c}{(8)} & (9) & (10) \\
      &&&&&&&&&&\\
     \hline
        &&&&&&&&&&\\
     NGC~337   & SBd    & 00h 59m 50.7s & - 07d 34' 44" & 2.9 & 1.8 & 19.3 & 8.18 & 8.84 & 1.30 & 12.0\\
     NGC~628   & SAc    & 01h 36m 41.8s & ~15d 47' 17" & 10.5 & 9.5 & 7.2 & 8.35 & 9.02 & 0.68 & 8.0 \\
     NGC~1097 & SBb    & 02h 46m 18.0s & - 30d 16' 42"  & 9.3 & 6.3 & 14.2 & 8.47 & 9.09 & 4.17 &45.0 \\
     NGC~1291 & SB0/a & 03h 17m 19.1s & - 41d 06' 32"  & 9.8 & 8.1 & 10.4 & 8.52 & 9.20 &  0.35 & 3.5\\
     NGC~1316 & SAB0  & 03h 22m 41.2s & - 37d 12' 10"  & 12.0 & 8.5 & 21.0 & 8.77 & 9.52 & -  & 8.0 \\
     NGC~1512 & SBab  & 04h 03m 55.0s & - 43d 20' 44"  & 8.9 & 5.6 & 11.6 & 8.56 & 9.11 & 0.36 & 3.8\\
     NGC~3351 & SBb    & 10h 43m 57.5s & ~11d 42' 19"  & 7.4 & 5.0 & 9.3 & 8.60 & 9.19 & 0.58 & 8.1\\
     NGC~3621 & SAd    & 11h 18m 18.3s & - 32d 48' 55"  & 12.3 & 7.1 & 6.5 & 8.27 & 8.80 &  0.51 & 7.9\\
     NGC~3627 & SABb & 11h 20m 13.4s & ~12d 59' 27" & 9.1 & 4.2 & 9.4 & 8.34 & 8.99 &  1.7 & 28.0\\
     NGC~4826 & SAab & 12h 56m 42.8s & ~21d 40' 50" & 10.0 & 5.4 & 5.3 & 8.54 & 9.20 & 0.26 & 4.2\\
     NGC~7793 & SAd    & 23h 57m 50.4s & - 32d 35' 30"  & 9.3  & 6.3 & 3.9 & 8.31 & 8.88 & 0.26 & 2.3\\
        &&&&&&&&&&\\
 \hline
\end{tabular}
\begin{list}{}{}
\item[$^{(1)}$] {\small Galaxy Name}
\item[$^{(2)}$] {\small Morphological type} 
\item[$^{(3)}$] J2000.0 Right ascension
\item[$^{(4)}$] J2000.0 Declination
\item[$^{(5)}$] Approximate major diameter in arcminutes
\item[$^{(6)}$] Approximate minor diameter in arcminutes
\item[$^{(7)}$] Distance in megaparsecs
\item[$^{(8)}$] Mean disk oxygen abundance from \citet{Moustakas2010}
\item[$^{(9)}$] Star formation rate in solar masses per year from Table 1 of \citet{Calzetti2010_2}
\item[$^{(10)}$] Total infrared luminosities in the 3-to-1100 \mic\ in units of 10$^{9}$ solar from \citet{Kennicutt2011}
\end{list}
 \end{table*} 

It is crucial to yield an exhaustive inventory on the range and distribution of dust temperatures as well as the radiative heating sources of dust grains. IRAS data combined with ground-based observations (SCUBA on JCMT for instance, observing at 450 and 850 \mic), have first led to the detection of the cool phases of dust in nearby galaxies \citep{Dunne2000,Dunne_Eales_2001,James2002}. The Infrared Space Observatory (ISO) and the \spitz\ {\it Space Telescope} (with a better sensitivity) then enabled us to resolve in detail the warm and cool dust reservoirs and more accurately model the Spectral Energy Distributions (SED) of nearby objects up to 160 \mic\ \citep{Draine2007}. 

Several issues still have to be investigated. Numerous studies thus highlighted the necessity of observing galaxies longward of 160 \mic\ at high resolution to properly sample the submillimeter (submm) regime of these SEDs and better characterise the total temperature range and dust masses \citep[e.g.][]{Gordon2010}. Moreover, IR emission is known to be a good tracer of the embedded star formation in galaxies \citep[][]{Kennicutt1998,Calzetti2007,Calzetti2010_2}. Nevertheless, the amount of IR emission not directly linked with the ongoing star formation is still poorly quantified. To address this issue, we need to understand the heating sources of the different grain populations: dust heating by the young stellar populations in H{\sc ii} and photodissociation regions (PDR), UV photons escaping from those regions or old stellar population.

The \hersc\ {\it Space Observatory}  \citep{Pilbratt2010} is currently observing the FIR/submm wavelength range to probe the coldest phases of dust that were poorly unexplored. Its onboard photometers PACS (from 70 to 160 \mic) and SPIRE (from 250 to 500 \mic) enable the imaging of nearby galaxies at a resolution and sensitivity never reached to date by a space telescope. The \pck\ satellite, launched by the same rocket, performs an all-sky survey at nine wavelengths between 350 \mic\ and 1 cm, complementing \hersc\ coverage, at a lower resolution. 
\citet{Planck_collabo_2011_NearbyGalaxies} studies have suggested that some nearby galaxies are both luminous and cool (presence of dust with T$<$20K), with properties comparable to those of galaxies observed with \hersc\ at higher redshifts \citep{Rowan-Robinson2010,Magnelli2012}. Using \hersc\ observations,  \citet{Engelbracht2010} compute separate SEDs for the center and the disks of nearby galaxies and find that the cool dust is on average 15$\%$ hotter in the center than in the disk and correlate this effect with morphological type and bar strength. 
Using a radial approach, \citet{Pohlen2010} show that SPIRE surface brightness ratios - used as a proxy for cold dust temperatures - seem to decrease with radius in spirals, and conclude that dust in the outer regions is colder than in the centre of the galaxies. \citet{Skibba2011} also investigate the relation between dust heating sources and the star formation activity with metallicity and the morphological types of their galaxies. On a more resolved scale, \citet{Bendo2011} derive PACS and SPIRE ratio maps of spiral galaxies. They find that those ratios seem to be preferentially correlated with the global starlight, including evolved stars, than with emission from star forming regions, results consistent with those obtained by \citet{Boquien2011} for the spiral local group galaxy M33 or by \citet{Auld2012} for extragalactic dust clouds of Centaurus A, among others. This supports the idea of a different source of heating for the cold (15-30 K) dust than that of the warm dust.  Using \hersc\ observations of the dwarf irregular galaxy NGC~6822, \citet{Galametz2010} show on the contrary that SPIRE surface brightness ratios are strongly correlated with the 24 \mic\ surface brightness, a commonly used tracer of the star formation activity. This might indicate that star forming regions are non negligible heating sources of cold dust in that object and suggest that the source of the dust heating probably differ with the environment and vary from normal spirals to low metallicity starbursts. 

Observations of galaxies with \hersc\ and from the ground, have also suggested that the properties of cold dust grains could differ from those often used in SED models such as blackbody modified with a beta-2 emissivity law or use of graphite grains to model carbon dust. In metal-poor objects in particular, using graphite grains can lead to unphysical gas-to-dust mass ratios (G/D) compared to those expected from the metallicity of the galaxy \citep{Meixner2010,Galametz2010}. Moreover, observations of dwarf galaxies in submm often lead to the detection of excess emission compared to extrapolations from \spitz-based fits \citep[]{Dumke2004, Galliano2003, Galliano2005, Bendo2006, Marleau2006, Galametz2009,OHalloran2010, Bot2010_2}. The origin of this excess is still an open issue and several explanations have been investigated so far: variations of the emissivity of the dust grains with temperature, ``spinning dust" emission, cold dust etc. Many investigations still need to be carried out to disentangle between those different hypotheses.

We have obtained photometric \hersc/PACS (70 to 160 \mic) and \hersc/SPIRE (250 to 500 \mic) observations of nearby galaxies as part of the open time key programme KINGFISH \citep[Key Insights on Nearby Galaxies: A Far-Infrared Survey with \hersc,][]{Kennicutt2011}. The KINGFISH sample comprises 61 galaxies probing a wide range of galaxy types (from elliptical to irregular galaxies), metallicity (7.3 $\le$ 12+log(O/H) $\le$ 9.3) and star-forming activity (10$^{-3}$ $\le$ SFR $\le$ 7 \msun\ yr$^{-1}$). The KINGFISH galaxies are located between 3 and 31 Mpc, leading to ISM resolution elements of 0.5 to 5.4 kpc at SPIRE 500 \mic\ resolution (36\arcsec). We refer to \citet{Kennicutt2011} for a detailed description of the sample and the observation strategy.

We aim to benefit from the good coverage and resolution of \hersc\ to sample the complete thermal dust emission spectrum and probe the spatially-resolved properties of the cold dust grain population. We refer to the work of \citet{Aniano2012a} (hereafter [A12a]) and Aniano et al. 2012b (in prep, full KINGFISH sample, hereafter [A12b]) for an analysis of other dust properties (PAH contribution to the total dust mass, distribution of the starlight intensity, dust mass surface density, among others) in the KINGFISH sample using the \citet{Draine_Li_2007} dust models. The two studies are complementary. The following paper particularly focuses on investigating the distribution of cold dust temperatures and dust mass surface densities in our objects. The SED model applied in this work (two modified-blackbody fit) allows for variations in the emissivity index of the dust grains and enables us to test the robustness of applying different assumptions of emissivity index to our extragalactic objects and quantify how this affects the temperatures and masses derived. 

This paper is structured as follows. In $\S$2, we describe the \hersc\ observations and the convolution techniques we apply to degrade images to the lowest resolution (that of SPIRE 500 \mic). In $\S$3, we discuss the integrated properties of the cold dust (temperature and emissivity). We also study how uncertainties in the flux measurements can lead to degeneracies between parameters. In $\S$4, we produce maps of the dust properties of our galaxies and discuss the choice of a fixed or free emissivity parameter on the temperature and emissivity maps obtained. We also study the distribution of the dust mass surface densities in our objects. We finally analyze how the total dust mass depends on the type of modified blackbody model we apply, the wavelength coverage and the working resolution.


\section{Observations and Data Processing}

\begin{figure*}
    \centering
    \begin{tabular}{cm{1cm}c} 
	 {\large NGC~337}  &&  {\large NGC~0628} \\
         	\includegraphics[height=5.2cm]{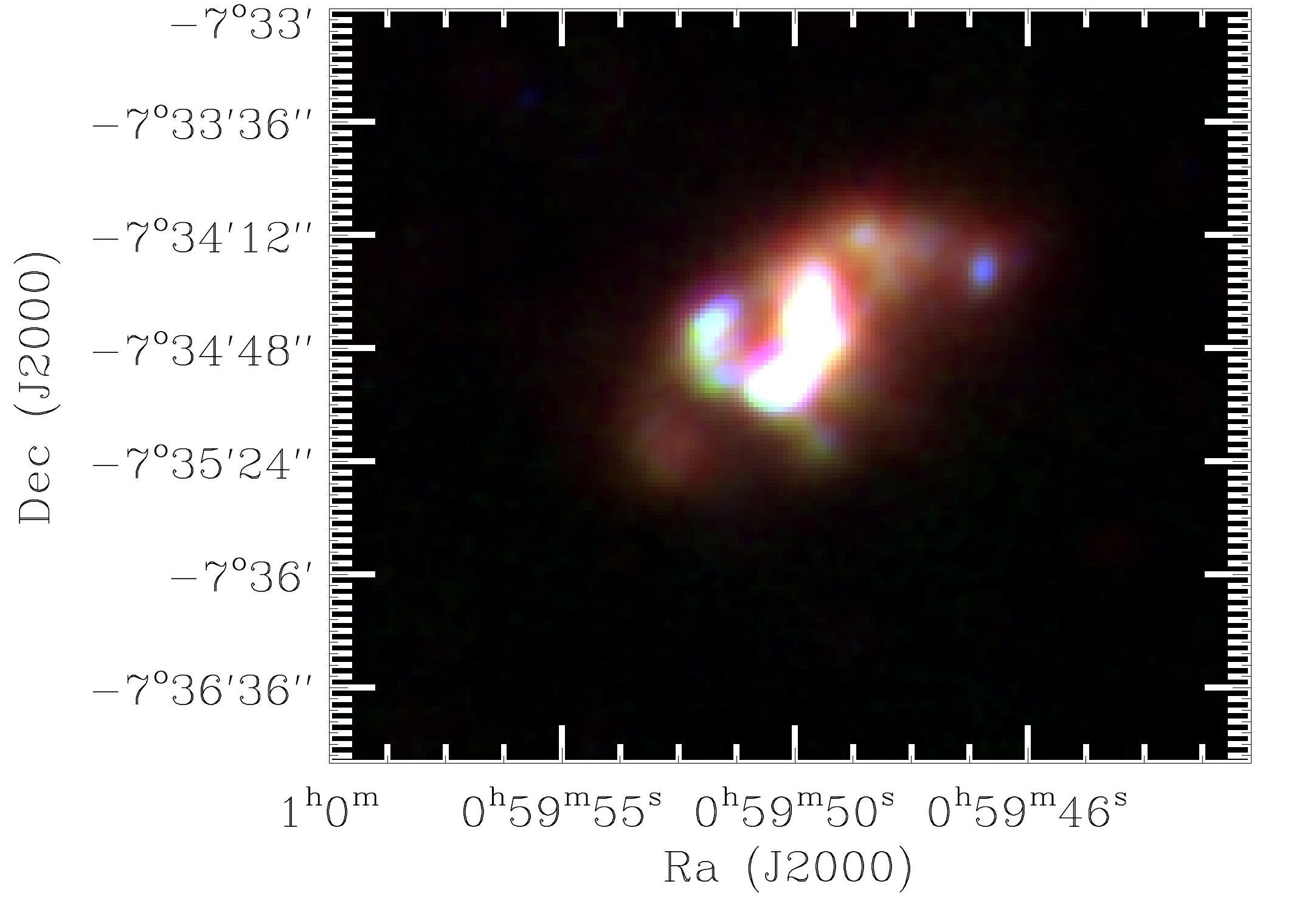}  && \includegraphics[height=5cm]{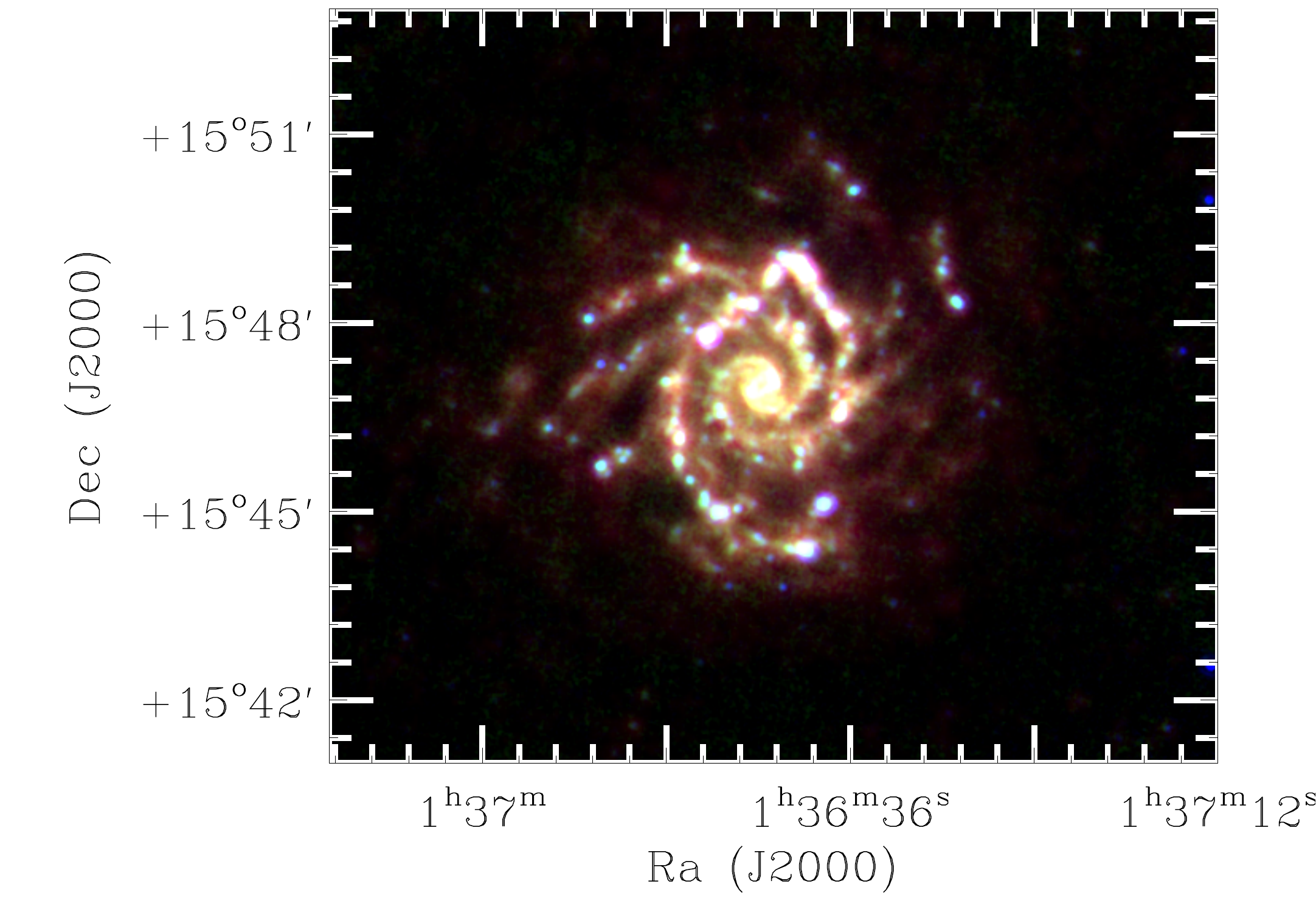}  \\
         \\
            {\large NGC~1097}  &&   {\large NGC~1291} \\
             \includegraphics[height=5.2cm]{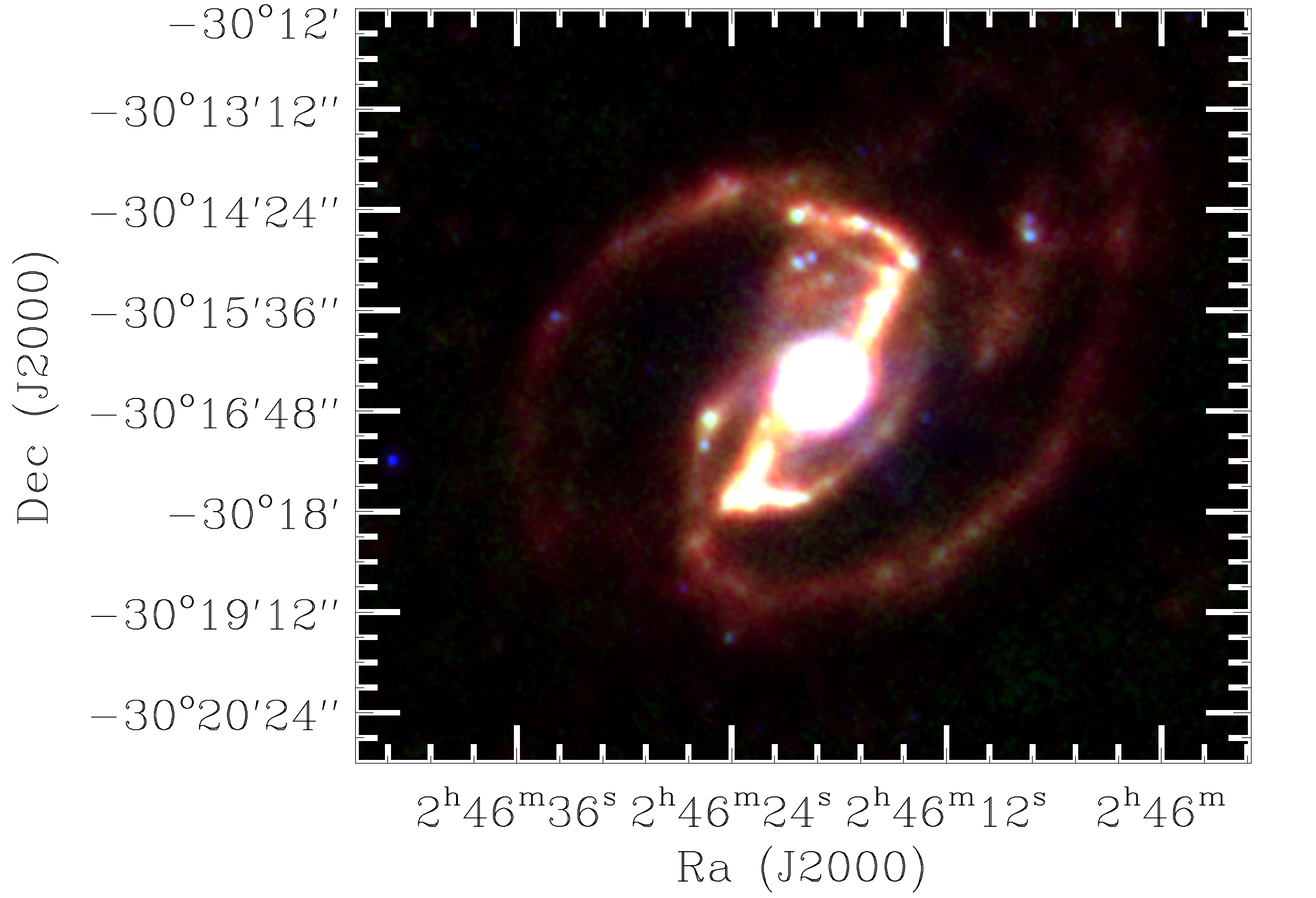} && \includegraphics[height=5cm]{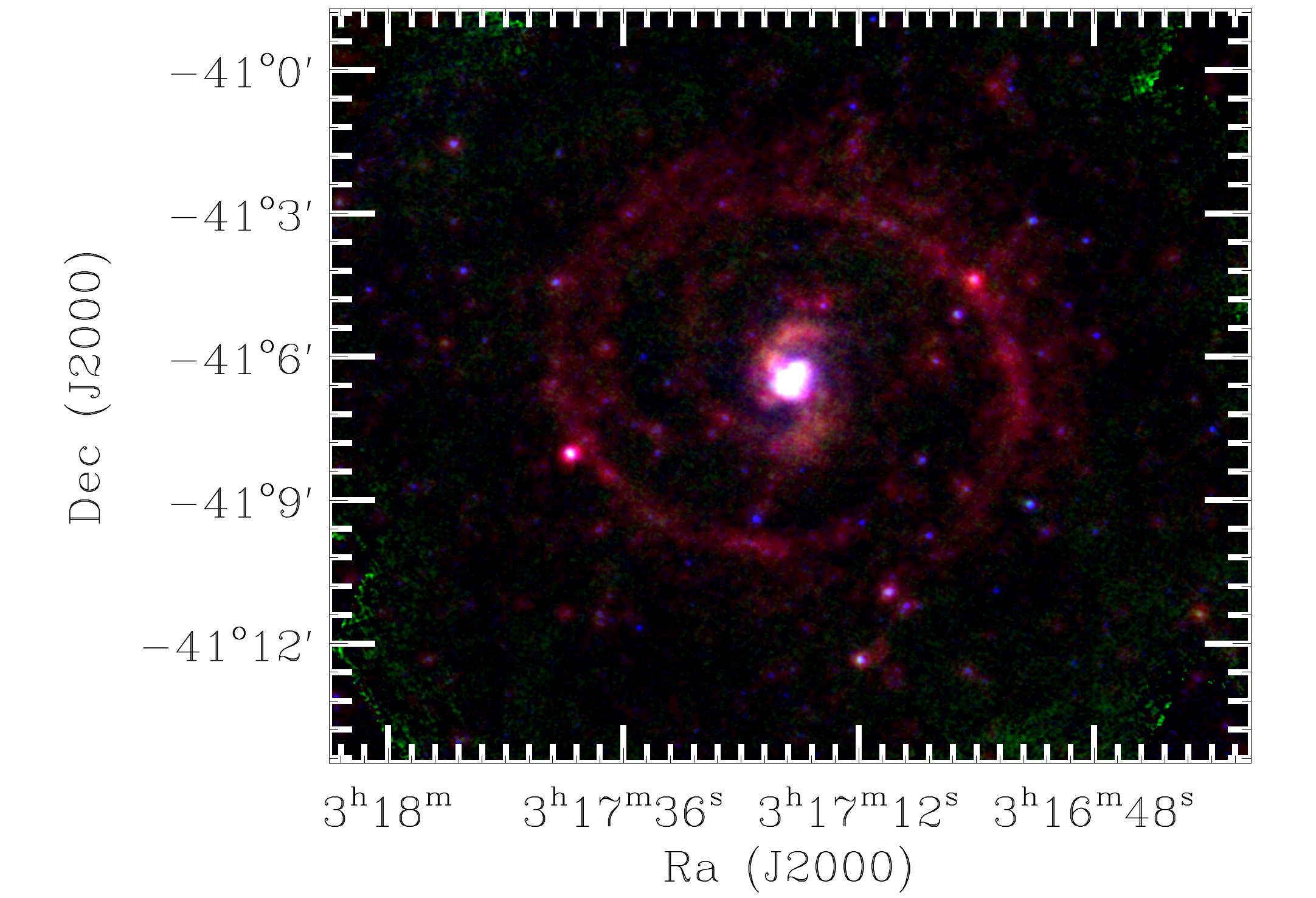} \\
         \\
	 {\large NGC~1316}  &&  {\large NGC~1512} \\
         	\includegraphics[height=5.2cm]{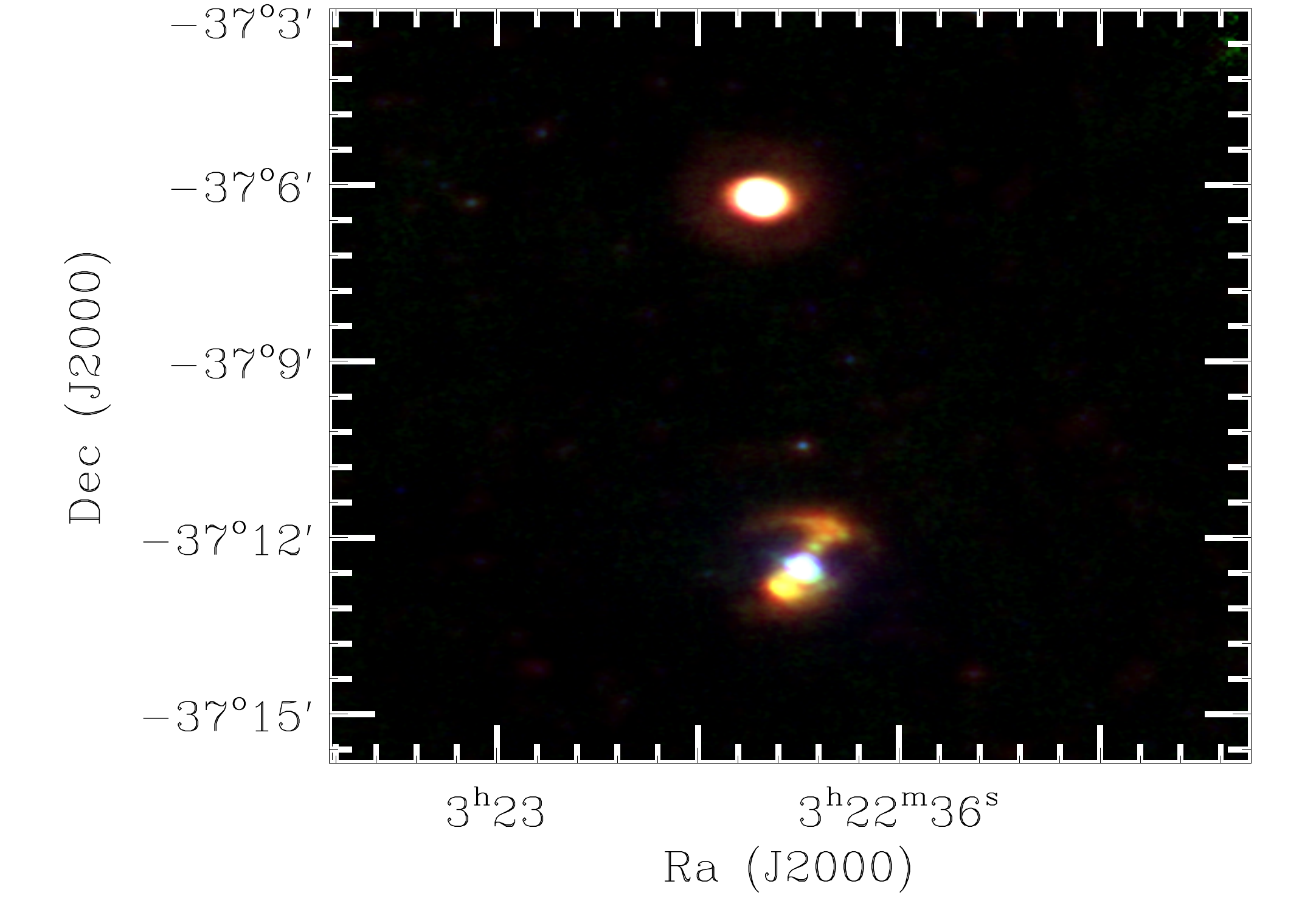}  &&  \includegraphics[height=5cm]{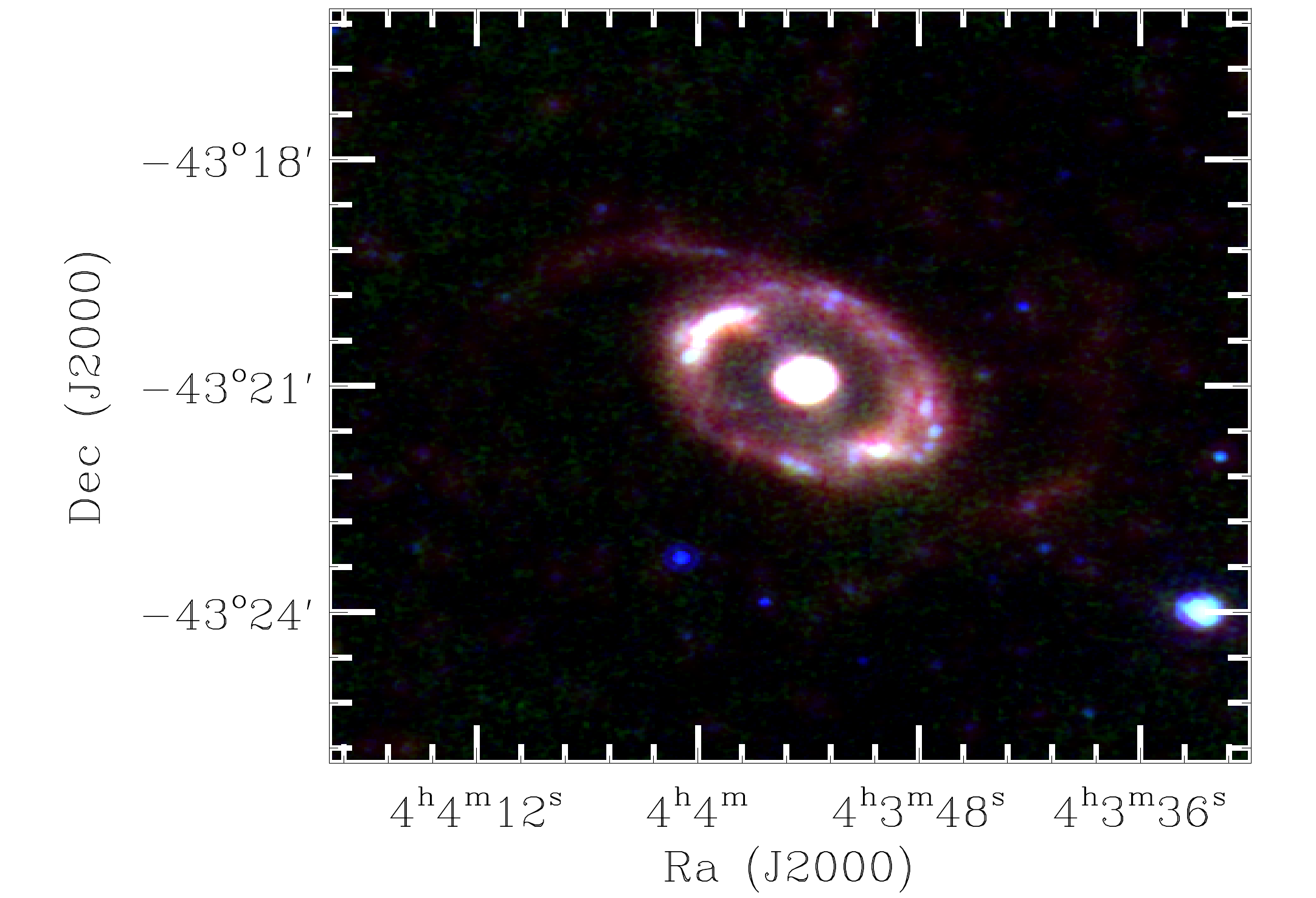}  \\
         \\
            {\large NGC~3351}  &&   {\large NGC~3621} \\
             \includegraphics[height=5.2cm]{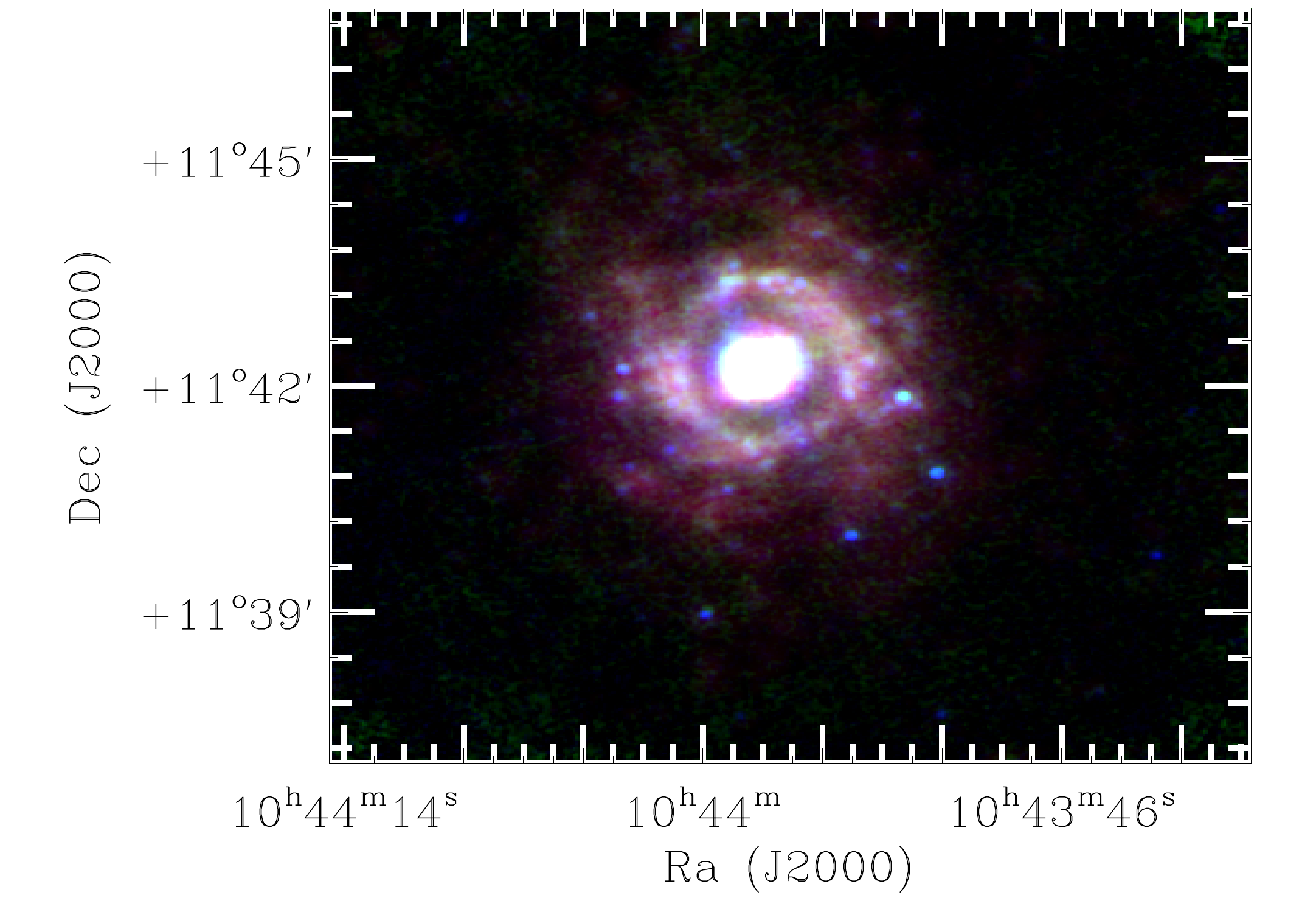} && \includegraphics[height=5cm]{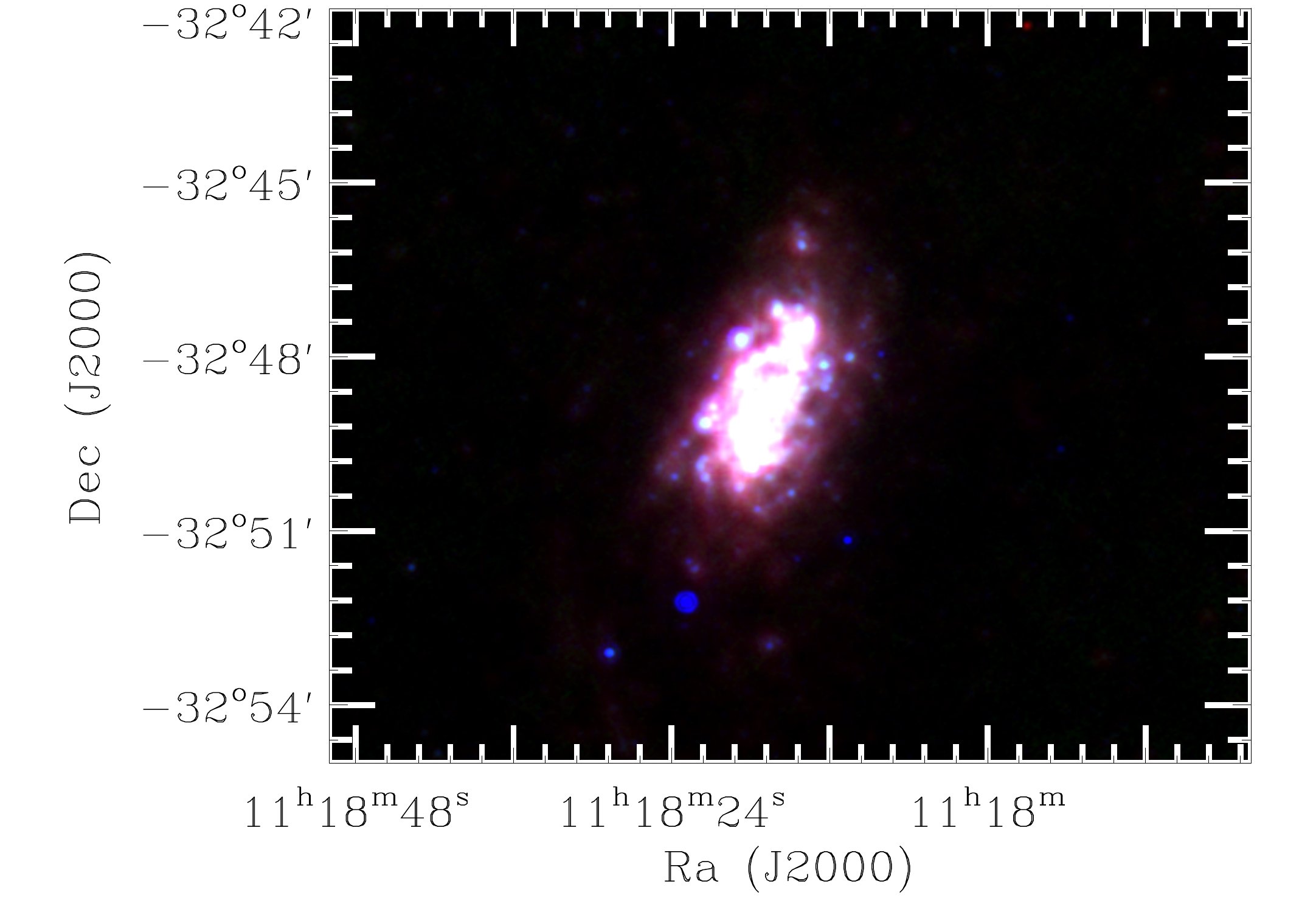} \\
             \\
          \end{tabular}
    \caption{3-color images of the sample. In each panel, north is up and east is left. Blue: \spitz/MIPS 24 \mic. Green: \hersc/PACS 100 \mic. Red: \hersc/SPIRE 250 \mic. We note that NGC~1316 and its companion NGC~1317 are seen in our \hersc\ field of view. NGC 1317 is north of NGC 1316.}
\end{figure*}
\addtocounter {figure}{-1}
\begin{figure}
    \centering
    \begin{tabular}{cc} 
	 {\large NGC~3627} \\ 
         \includegraphics[height=5.5cm]{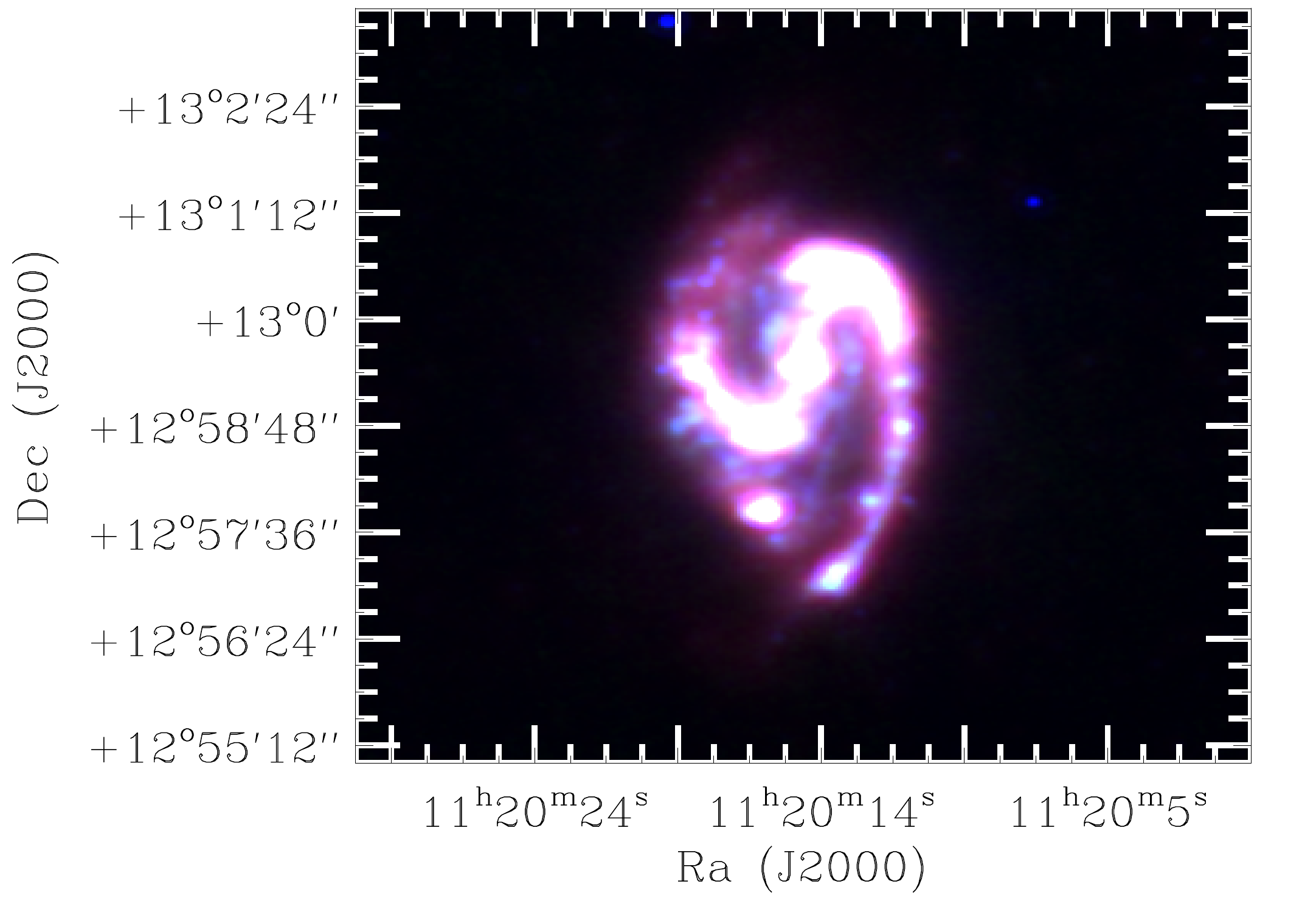}  \\
           & \\    
         	{\large NGC~4826} \\
	\includegraphics[height=5.5cm]{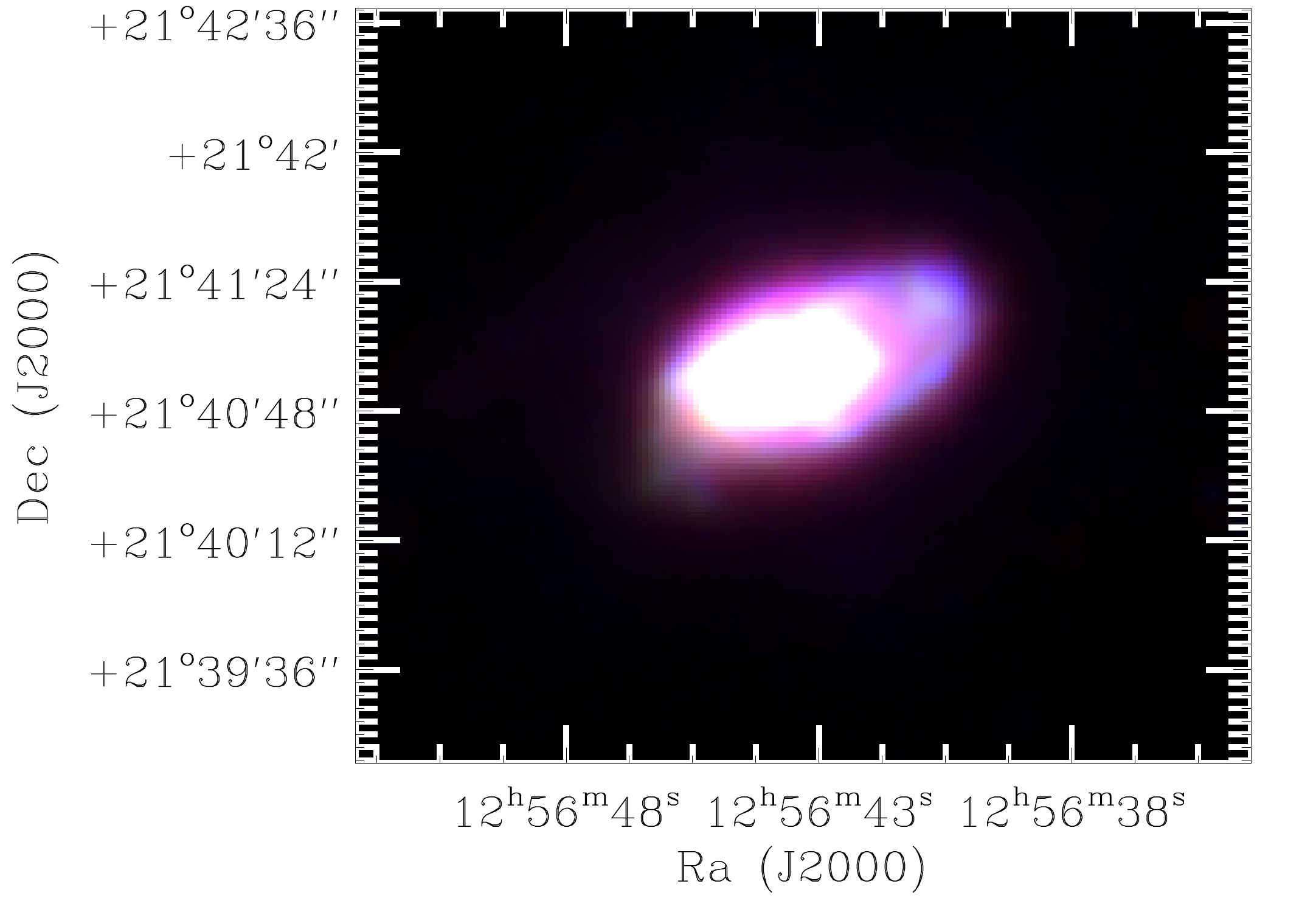} \\
		      &   \\
	{\large NGC~7793}    \\
         \includegraphics[height=5.5cm]{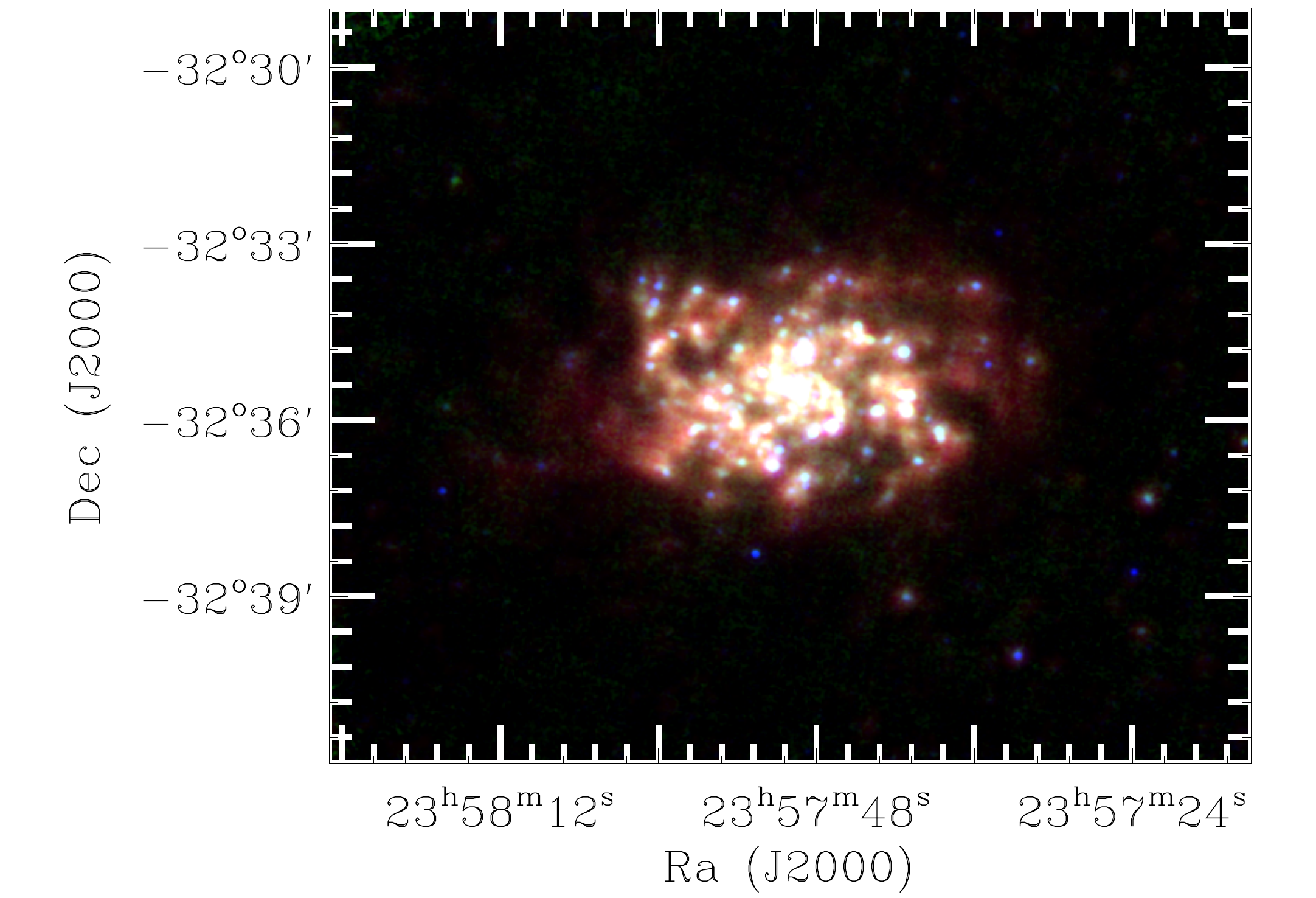}  \\              
           \end{tabular}
    \caption{Continued. }
    \label{Images}  
\end{figure}   

\subsection{The sample}

The data taken with the \hersc\ {\it Space Telescope} now enable us to properly constrain the cold dust properties in nearby galaxies. In this paper, we focus our study on 11 galaxies, 9 spirals and 2 lenticulars (NGC~1291 and NGC~1316) observed as part of the \hersc\ open time key programme KINGFISH. Those galaxies were primarily selected because they have also been mapped at 870 \mic\ with the \lab\ instrument located on APEX in Chile, data that will be analysed in a second paper in this series. Our galaxies {\it 1)} are observable from APEX ($\delta$$_o$~$<$+21$^{\circ}$) {\it 2)} are large enough to use the LABOCA mapping mode (the field of view of the complete array is 11\farcm4) {\it 3)} have diffuse emission bright enough to be detected with LABOCA. Our sample is thus biased toward relatively high surface brightness normal spirals. Albrecht et al. (in prep) describe the LABOCA observations and data reduction.We will combine \lab\ data with the results obtained in this paper to investigate possible 870 \mic\ submm excess in our objects (Galametz et al., in prep). The properties of our sample are summarized in Table~\ref{Galaxy_data}. Morphologies and sizes as well as references and methods used to derive distances, and the total infrared luminosity (L$_{TIR}$) are detailed in \citet{Kennicutt2011}. The SFRs are derived using a combination of H$\alpha$ and 24 \mic\ integrated luminosities and taken from \citet{Calzetti2010_2}. 
This formula is used in $\S$4.1 of this paper. Two values are given for the oxygen abundance, one coming from the empirical calibration of \citet{Pilyugin2005} (PT), the other from the calibration of \citet{Kobulnicky2004} (KK) as discussed in details in \citet{Moustakas2010}. The sample covers different morphologies and a wide range of bar strengths. It contains isolated targets as well as galaxies with disturbed outskirts resulting from past interactions such as NGC~3627. It also contains mergers such as NGC~1316 (Fornax A), an early-type radio galaxy that could be interacting with NGC~1317, also detected in our \hersc\ field of view. NGC~1316 possesses a core-jet lobe structure \citep{Ekers1983} and a low-luminosity X-ray active galactic nucleus \citep{KimD2003}. We discuss how this could affect our results later in the paper. The star formation rates (SFR) of our sample range from 0.3 to 8 \msun\ yr$^{-1}$ and the L$_{TIR}$ covers an order of magnitude, from 2.3 $\times$ 10$^9$ to 4.5 $\times$ 10$^{10}$ \lsun. Our sample thus spans a wide range of environments and physical conditions to which dust is exposed. One of them, the barred galaxy NGC~1097, was observed and studied during the Science Demonstration of \hersc. It hosts a prominent starbursting ring with a radius of $\sim$ 900 pc \citep{Hummel1987}. We refer to \citet{Sandstrom2010} and \citet{Beirao2010} for PACS photometric and spectroscopic studies of this circumnuclear ring.  We also refer to Hinz et al. (submitted to ApJ) for a radial study of temperatures in NGC~1291, another galaxy of this sample.
We benefit from the good spatial resolution of \hersc\ to resolve with SPIRE 500 \mic\ (FWHM = 36 arcsec) structures of $\sim$700 pc for the closest galaxy of our sample NGC~7793 and $\sim$4 kpc for the furthest galaxy NGC~1316. Three-color images (24 / 100 / 250 \mic) of the galaxies are provided in Fig.~\ref{Images}.

\subsection{The Herschel observations}

The KINGFISH programme observed their 61 nearby targets with the PACS and SPIRE instruments between December 2009 and June 2011. A description of the data processing steps are given in \citet{Engelbracht2010}, \citet{Sandstrom2010} and the KINGFISH overview paper of \citet{Kennicutt2011}. A few details are summarized in the next sections.

\subsubsection{PACS observations}
The PACS photometer observes at 70, 100 and 160 \mic\ with FWHMs of 5\farcs2, 7\farcs7 and 12\arcsec\ respectively \citep{Poglitsch2010}. Observations of KINGFISH galaxies with this instrument were obtained with 15'-long cross-scans (perpendicular scans) with a medium scan speed of 20"~s$^{-1}$.  
PACS calibration uncertainties are $\sim$5 $\%$ according to Version 4 of the PACS Observer's Manual\footnote {http://herschel.esac.esa.int/Docs/PACS/html/pacs$\_$om.html}. The data reduction is performed from raw data to Level~1 data with the Version 5 of the \hersc\ Interactive Processing Environment (HIPE) using the standard procedure with additional second-level deglitching steps. We use the Scanamorphos technique to process the data from Level~1. Scanamorphos is an IDL software that subtracts the brightness drifts caused by the low-frequency noise using the redundancy built in the observations. We refer to \citet{Roussel2012} for a complete description of the software. PACS data emerge from the pipeline with a pixel size of 1\farcs4, 1\farcs7 and 2\farcs85 at 70, 100 and 160 \mic\ respectively. We applied the needed correction factors to obtain the latest version (v6) of the responsivity calibration. The root-mean-square (rms) noise levels are [0.14-0.18], [0.24-0.35] and [0.74-1.9] mJy~arcsec$^{-2}$ at 70, 100 and 160 \mic\ respectively for our subsample.  
The Scanamorphos pipeline aims to preserve low surface brightness diffuse emission. Nevertheless, the PACS instrument is less sensitive than the MIPS instrument to low surface brightness emission. Our analysis of the resolved properties of our objects will be restricted to pixels for which we have a 3-sigma detection in all the bands.

\begin{table*}
\caption{\large Global IRAS, MIPS, PACS and SPIRE flux densities (in units of Jy)}
\label{Fluxes}
 \centering
 \begin{tabular}{c|ccccccccccc}
\hline
\hline
   &&&&&&&&&&&\\
  & NGC337 &    NGC628  & NGC1097 & NGC1291 & NGC1316 & NGC1512 & NGC3351 & NGC3621   & NGC3627 & NGC4826  & NGC7793  	\\ 
     &&&&&&&&&&&\\
     \hline
   &&&&&&&&&&&\\
2a$^{1}$ 	&253	&879 	& 758 	&884 	&864	&1001	&592	&791	&745	&716	&716 \\
 2b$^{1}$ 	&194	&808 	& 612 	&836 	&583	&928	&441	&555	&486	&427	&526 \\
 $\alpha$$^{1}$ 	&140	&90 		&130 	&90 		&50		&83		&11		&160	&167	&114	&98 \\
   &&&&&&&&&&&\\   
     \hline
     &&&&&&&&&&&\\
MIPS24$^{2}$ &0.7$\pm$0.03&3.2$\pm$0.1&6.6$\pm$0.3&0.6$\pm$0.02&0.4$\pm$0.02&0.5$\pm$0.02&2.6$\pm$0.1&3.7$\pm$0.2&7.4$\pm$0.3&2.7$\pm$0.2&2.1$\pm$0.1\\
IRAS25$^{3}$ &0.7$\pm$0.1&2.9$\pm$0.6&8.6$\pm$1.7&0.5$\pm$0.1&0.4$\pm$0.09&0.5$\pm$0.1&2.8$\pm$0.6&4.2$\pm$0.8&8.6$\pm$1.7&2.9$\pm$0.6&1.7$\pm$0.3\\
IRAS60$^{3}$ &9.4$\pm$1.9&21.6$\pm$4.3&54.4$\pm$10.9&2.8$\pm$0.6&3.3$\pm$0.7&3.8$\pm$0.8&19.7$\pm$3.9&32.2$\pm$6.4&66.3$\pm$13.3&36.7$\pm$7.3&18.1$\pm$3.6\\
MIPS70$^{2}$ &11.2$\pm$0.8&33.9$\pm$4.1&59.8$\pm$4.7&5.3$\pm$0.6&5.4$\pm$0.4&6.8$\pm$0.8&21.9$\pm$2.7&50.2$\pm$3.9&91.9$\pm$7.0&52.9$\pm$6.5&32.9$\pm$4.0\\
PACS70$^{4}$ &13.0$\pm$0.7&36.7$\pm$1.8&77.5$\pm$3.9&5.3$\pm$0.3&5.8$\pm$0.3&8.0$\pm$0.5&25.3$\pm$1.3&49.5$\pm$2.5&104$\pm$5.0&54.7$\pm$2.7&32.0$\pm$1.6\\
IRAS100$^{3}$  &20.2$\pm$4.0&54.5$\pm$10.9&115$\pm$23.0&11.5$\pm$2.3&9.2$\pm$1.8&12.6$\pm$2.5&41.1$\pm$8.2&82.1$\pm$16.4&137$\pm$27.3&81.7$\pm$16.3&54.1$\pm$10.7\\
PACS100$^{4}$ &19.5$\pm$1.0&74.0$\pm$3.7&116$\pm$6.0&12.8$\pm$0.7&9.3$\pm$0.5&13.8$\pm$0.7&46.1$\pm$2.3&94.4$\pm$4.7&179$\pm$9.0&95.7$\pm$4.8&65.8$\pm$3.3\\
MIPS160$^{2}$ &20.1$\pm$2.4&112$\pm$17&154$\pm$19&26.3$\pm$4.1&12.6$\pm$1.8&19.6$\pm$3.1&56.9$\pm$8.9&139$\pm$17&216$\pm$28&85.8$\pm$13&107$\pm$17\\
PACS160$^{4}$ &19.6$\pm$1.0&116$\pm$6.0&134$\pm$7.0&20.3$\pm$1.1&11.5$\pm$0.6&18.7$\pm$1.0&55.1$\pm$2.8&128$\pm$6.0&202$\pm$10&94.1$\pm$4.7&91.2$\pm$4.6\\
SPIRE250$^{4}$   &9.8$\pm$0.7&65.5$\pm$4.7&72.2$\pm$5.1&15.9$\pm$1.1&4.8$\pm$0.4&15.6$\pm$1.1&32.4$\pm$2.3&71.2$\pm$5.1&96.7$\pm$6.9&42.4$\pm$3.0&56.3$\pm$4.0\\
SPIRE350$^{4}$  &4.4$\pm$0.3&30.6$\pm$2.2&30.8$\pm$2.2&8.0$\pm$0.6&2.1$\pm$0.2&8.7$\pm$0.6&13.7$\pm$1.0&31.7$\pm$2.3&37.6$\pm$2.7&16.4$\pm$1.1&28.4$\pm$2.0\\
SPIRE500$^{4}$ &1.9$\pm$0.1&13.3$\pm$1.0&12.6$\pm$0.9&3.5$\pm$0.3&0.8$\pm$0.1&4.2$\pm$0.3&5.3$\pm$0.4&14.6$\pm$1.0&14.4$\pm$1.0&6.3$\pm$0.4&13.9$\pm$1.0\\
   &&&&&&&&&&&\\
\hline
\end{tabular}
\begin{list}{}{}
\item[$^{1}$] {\small Parameters of the photometric apertures used to derive \hersc\ flux densities. 2a and 2b : lengths of the major and minor axes; $\alpha$: position angle of the apertureÕs major axis measured east of north.}
\item[$^{2}$] {\small from \citet{Dale2007}.} 
\item[$^{3}$] {\small from the SCANPI tool and the HIRES atlas (see text).} 
\item[$^{4}$] {\small from \citet{Dale2012}.}
\end{list}
 \end{table*}

\subsubsection{SPIRE observations}
The SPIRE instrument produces maps at 250, 350 and 500 \mic, with FWHMs of 18\arcsec, 25\arcsec\ and 36\arcsec\ respectively \citep{Griffin2010}. Observations of KINGFISH galaxies with this instrument were obtained in scan mode. The data reduction was performed from raw data with the Version 5 of HIPE. Details on the data reduction are given in the KINGFISH Data Products Delivery User's Guide \footnote{http://herschel.esac.esa.int/UserReducedData.shtml}. The SPIRE maps were built with a nearest-neighbor projection on sky and averaging of the time ordered data. 
SPIRE maps emerge from the pipeline with a pixel size of 6\arcsec, 10\arcsec\ and 14\arcsec\ at 250, 350 and 500 \mic\ respectively. We use the SPIRE beam areas quoted in the SPIRE Observer's manual, namely 423, 751 and 1587 square arcseconds at 250, 350 and 500 \mic, with $\sim$1$\%$ uncertainty. The rms noise levels are [0.5-0.75], [0.24-0.34] and [0.14-0.21] MJy~sr$^{-1}$ at 250, 350 and 500 \mic\ respectively for our subsample. Calibration uncertainties are estimated to be $\sim$7$\%$ for the three wave bands\footnote{http://herschel.esac.esa.int/Docs/SPIRE/html/spire$\_$om.html}.

\subsection{Background galaxies and sky subtraction} 

The background galaxies subtraction and sky subtraction steps are similar to those applied in \citet{Dale2012} to derive the global photometry for the KINGFISH sample. We identify the bright background galaxies that could contaminate the photometry and ``remove" them by replacing the data with random values with similar noise characteristics. The removal of these background objects does not significantly affect the total flux estimates ($<$1$\%$). The companion of NGC~1316 (NGC~1317) is also removed from the \hersc\ fields of view to avoid strong contamination in the chosen photometric aperture.
A global sky subtraction is already applied on the PACS/SPIRE maps during the reduction process. However, this value was estimated on the entire map. We determine a refined local sky using apertures randomly distributed in the background. We choose the total ``sky" area to be greater than the area we choose to perform the photometry (see Table~\ref{Fluxes}). We do not expect Galactic cirrus to contaminate any of the galaxies of our sample. Indeed, the lowest Galactic latitude for the sample is +26$^{\circ}$ for NGC~3621.

 \subsection{Convolution to lower resolution maps}
   
To compare the fluxes at different wavelengths, we need to convolve every observation to a common resolution. We refer to \citet{Aniano2011} for details on how to construct convolution kernels between \spitz\ and \hersc\ bands. The main steps are similar to the technique described in \citet{Gordon2008}. We review the basic steps here. 

In-flight Point Spread Functions (PSFs) are available for \hersc. An ideal convolution kernel K(x,y) which transforms an original PSF (PSF$_{o}$) to a lower resolution final PSF (PSF$_{f}$) is defined as:

\begin{equation}
K(x,y) = FT^{-1}\left[\frac{FT[PSF_f(x,y)]}{ FT[PSF_o(x,y)]}\right] .
\label{kernel_equ}
\end{equation} 

To attenuate the high-frequency noise in the original PSF, a filter is first applied to the FT[PSF$_o$(x,y)]:

\begin{equation}
Filter (\omega) = \left \{
\begin{array}{ll}
1 &  for~\omega \le \omega_{low} \\
\frac{1}{2} \left[1+{\rm cos}\left(\pi \times \frac{\omega - \omega_{low}}{\omega_{high} - \omega_{low}}\right)\right] & for~\omega_{low} \le\omega \le \omega_{high} \\
0 & for~\omega_{high} \le \omega , \\
\end{array} \right .
\end{equation}

\noindent where {\it $\omega$$_{high}$} is the cutoff spatial frequency and  {\it $\omega$$_{low}$}= 0.7 {\it $\omega$$_{high}$}. {\it $\omega$$_{high}$} is appropriately chosen to reach a high accuracy but minimizes the noise linked with the process. The final kernels are circularized to produce rotationally symmetric kernels, thus independent of the relative orientation of the spacecraft. 

We convolve MIPS, PACS and SPIRE images to the SPIRE 500 \mic\ images using the previous kernels and the IDL procedure {\it conv$\_$image}. We finally project the maps to a common sampling grid (that of the SPIRE 500 \mic\ maps) with the IDL {\it Hastrom} procedure. 
\citet{Aniano2011} caution about convolutions from MIPS 160 to SPIRE 500 \mic. The deconvolution of an image to higher resolution can indeed be risky if the image contains noise and artefacts. We compare the resolved PACS 160 \mic\ flux densities (pixels about a 3-$\sigma$ in the band) for the whole sample with the resolved MIPS 160 \mic\ flux densities (maps convolved to SPIRE 500 resolution). We find a very good agreement between the two values and therefore decide to use both in the modelling. As mentioned in [A12a], among others, slight discrepancies can be observed between MIPS and PACS fluxes. We know for instance that the PACS instrument is less sensitive to diffuse emission than MIPS. Using both MIPS and PACS (at 70 and 160 \mic) data enables us to be more careful and conservative in our modelling. The model will indeed try to find a compromise between the two values.


 \section{Integrated properties}  
 
\subsection{Naming convention}

The SED peak emission of nearby galaxies spans a relatively narrow range of dust temperatures, from $\sim$15 to 40 K \citep[][for instance]{Bendo2003,Stevens2005,Dale2005,Boselli2010,Dale2012}. In this study, we decide to use an isothermal fit to describe the cold dust component and to call our parameter T$_c$ the ``cold dust temperature". It is nevertheless important to keep in mind that the temperature distribution of the cold dust phase is more complex than a delta function, with a range of cold grain temperatures along the line of sight. The ``cold dust temperatures" derived in this paper (and in other studies using similar fitting procedures) should be considered in reality as an approximation of the cold temperature distribution in our galaxies. Moreover, an emissivity index is, by definition, the power-law index describing the variation of dust opacity with wavelength, so an intrinsic property of the grains. Calling our parameter $\beta$$_c$ the ``cold dust emissivity index" is a commonly-used expression. Our parameter $\beta$$_c$ is more accurately the ``effective emissivity index'', namely a power-law index resulting from the sum of different grain populations, each of them having their own intrinsic emissivity index. The effective and intrinsic emissivity indices are identical in the isothermal case. 

\subsection{The model}

We first want to derive integrated dust properties of our sample to investigate the variations of the cold dust temperature and emissivity between objects. A combination of data from 24 to 500 \mic\ provides a complete sampling of the warm and cold dust thermal emission of our galaxies from which global dust properties can be deduced. The submm data are, in particular, crucial to constrain the coldest phases of dust that could account for a large amount of the dust mass. They also enlighten us as to how the properties of cold dust vary from galaxy to galaxy.

\begin{figure} 
    \centering
        \begin{tabular}{p{7cm} }
            \includegraphics[width=7.5cm ,height=5.8cm]{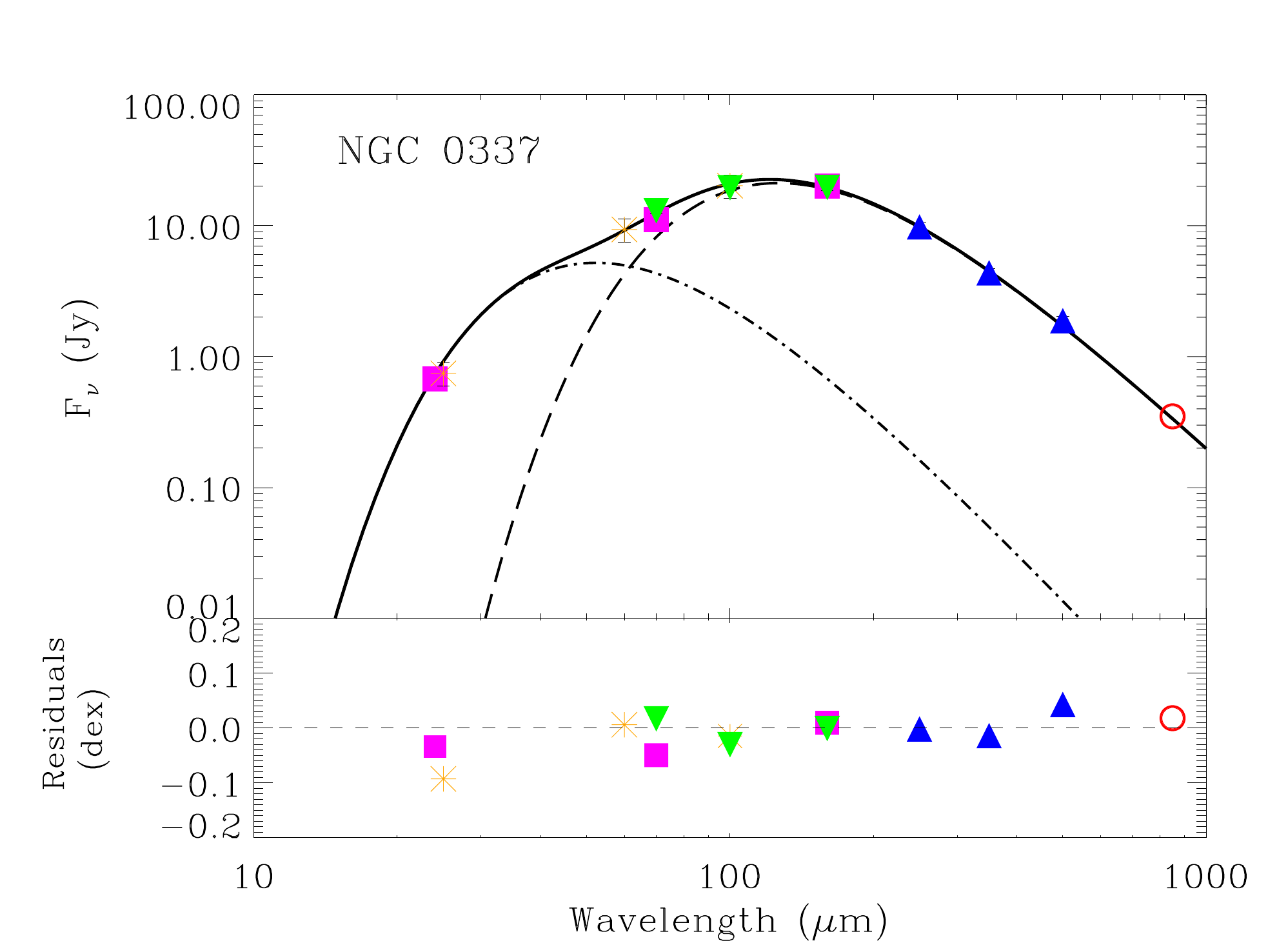}\\
             \includegraphics[width=7.5cm ,height=5.8cm]{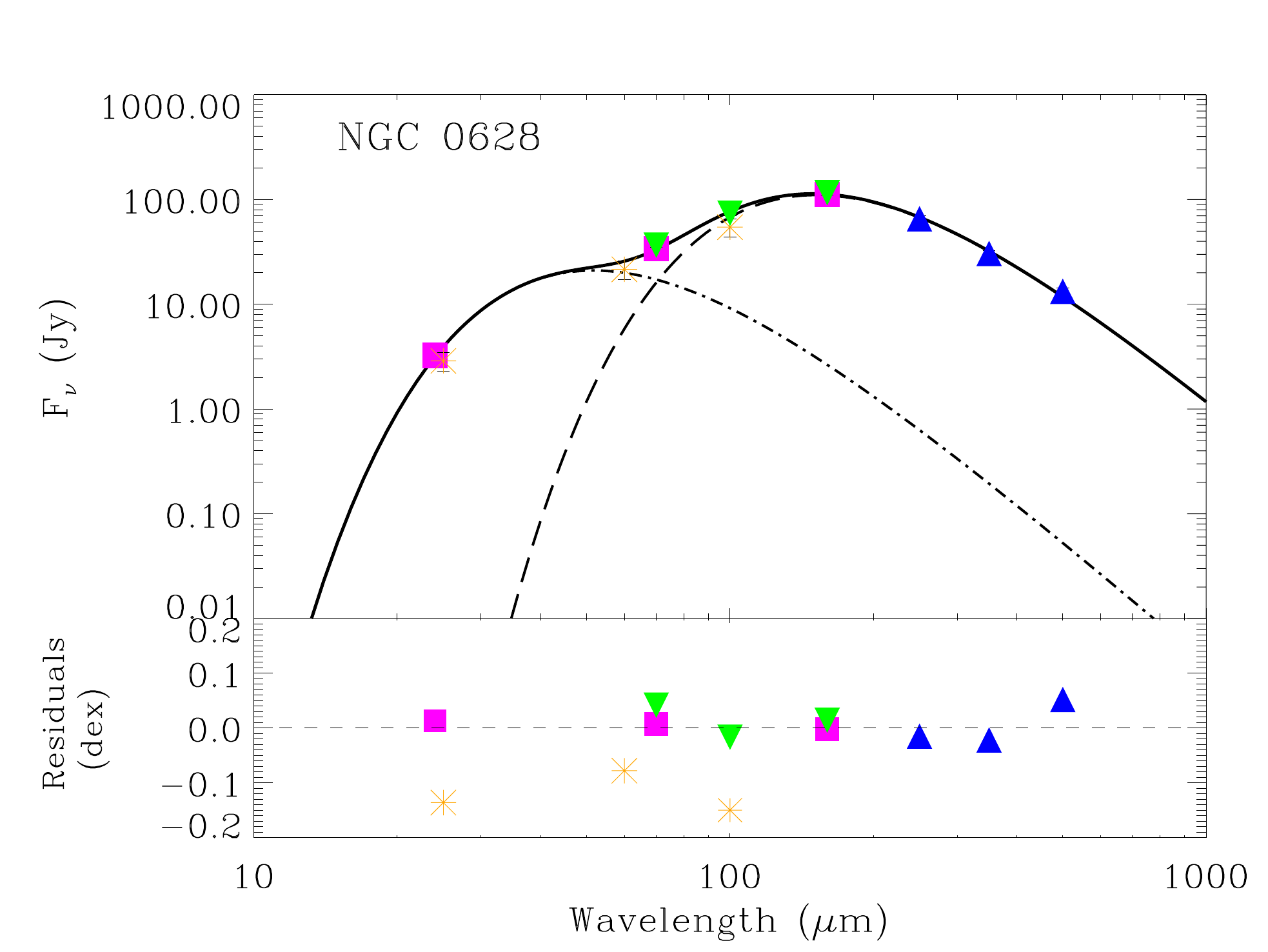} \\
              \includegraphics[width=7.5cm ,height=5.8cm]{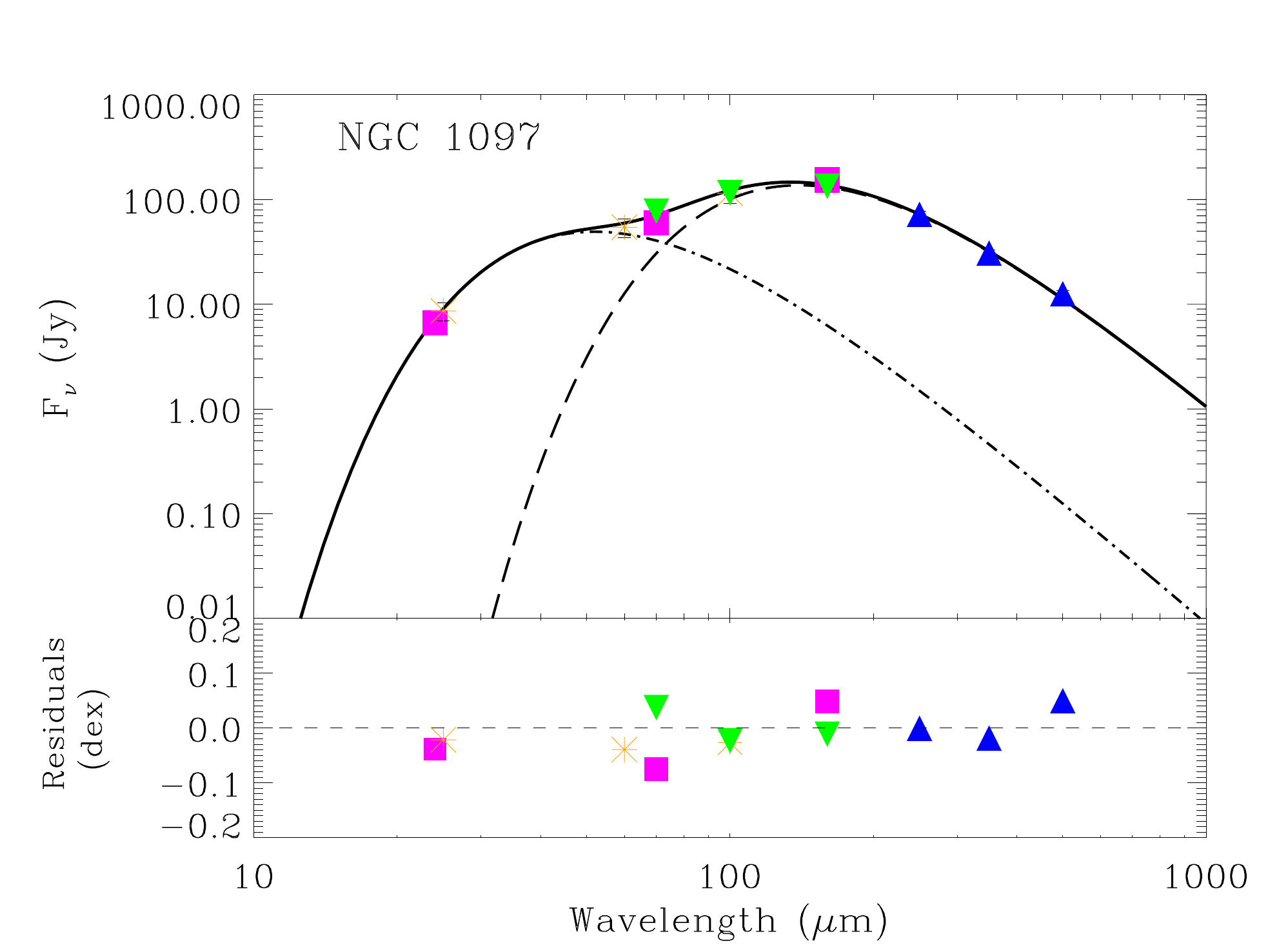}  \\
          \end{tabular}
      \caption{Global SED of our galaxies performed using a two-temperature fit. The emissivity index $\beta$$_w$ of the warm component 
      (dashed dotted line) is fixed at 2.0 and the emissivity index $\beta$$_c$ of the cold component (long dashed line) is a free parameter. IRAS (asterisks), MIPS (squares), PACS (downward triangles), SPIRE (upward triangles) data are overlaid. SCUBA 850 \mic\ flux densities are indicated with empty circles but not used in the fit.
       If the error bars are not shown, errors are smaller than the symbols. The fitted $\beta$$_c$, and temperatures are given in Table~\ref{BB_results}.
       The bottom panel of each plot indicate the residuals from the fit. See $\S$3.1 for details on the modelling technique.}
      \label{2BB_SEDs}          
        \end{figure}
        
\addtocounter {figure}{-1}
        
\begin{figure*}[h!]
    \centering   
        \begin{tabular}{ p{8cm}p{8cm}  }
         \includegraphics[width=7.5cm ,height=5.8cm]{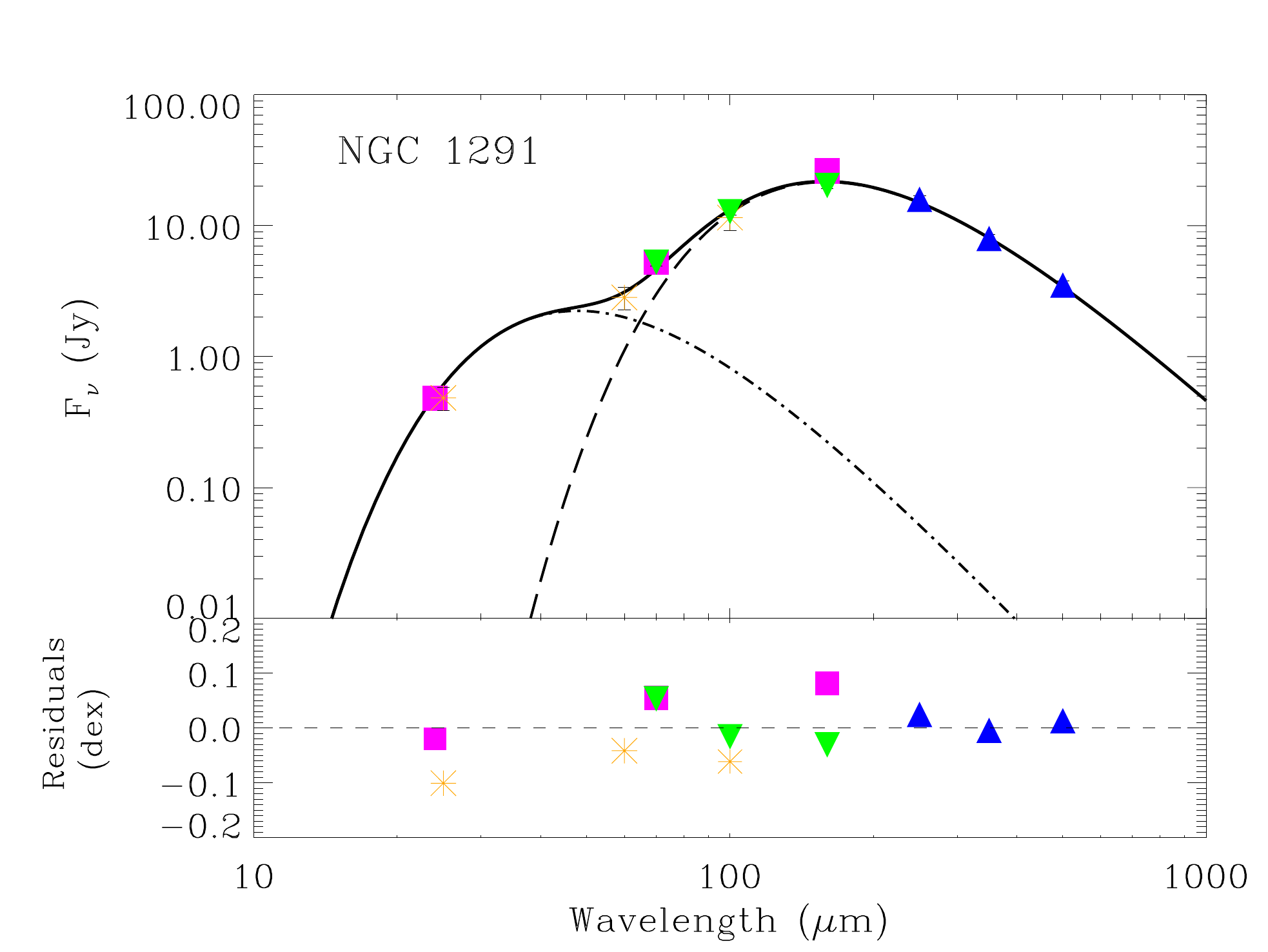}&      
        \includegraphics[width=7.5cm ,height=5.8cm]{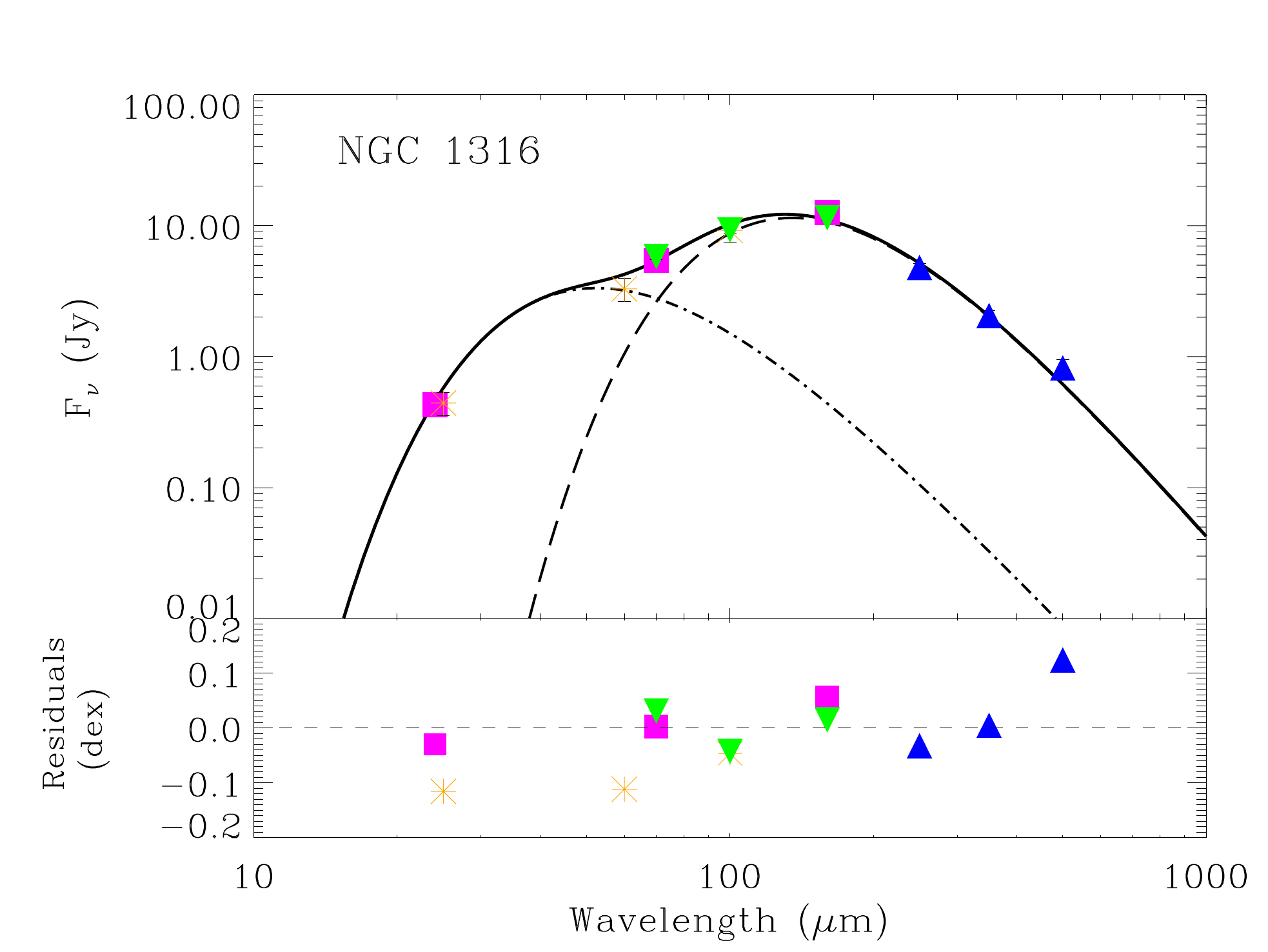}\\
        \includegraphics[width=7.5cm ,height=5.8cm]{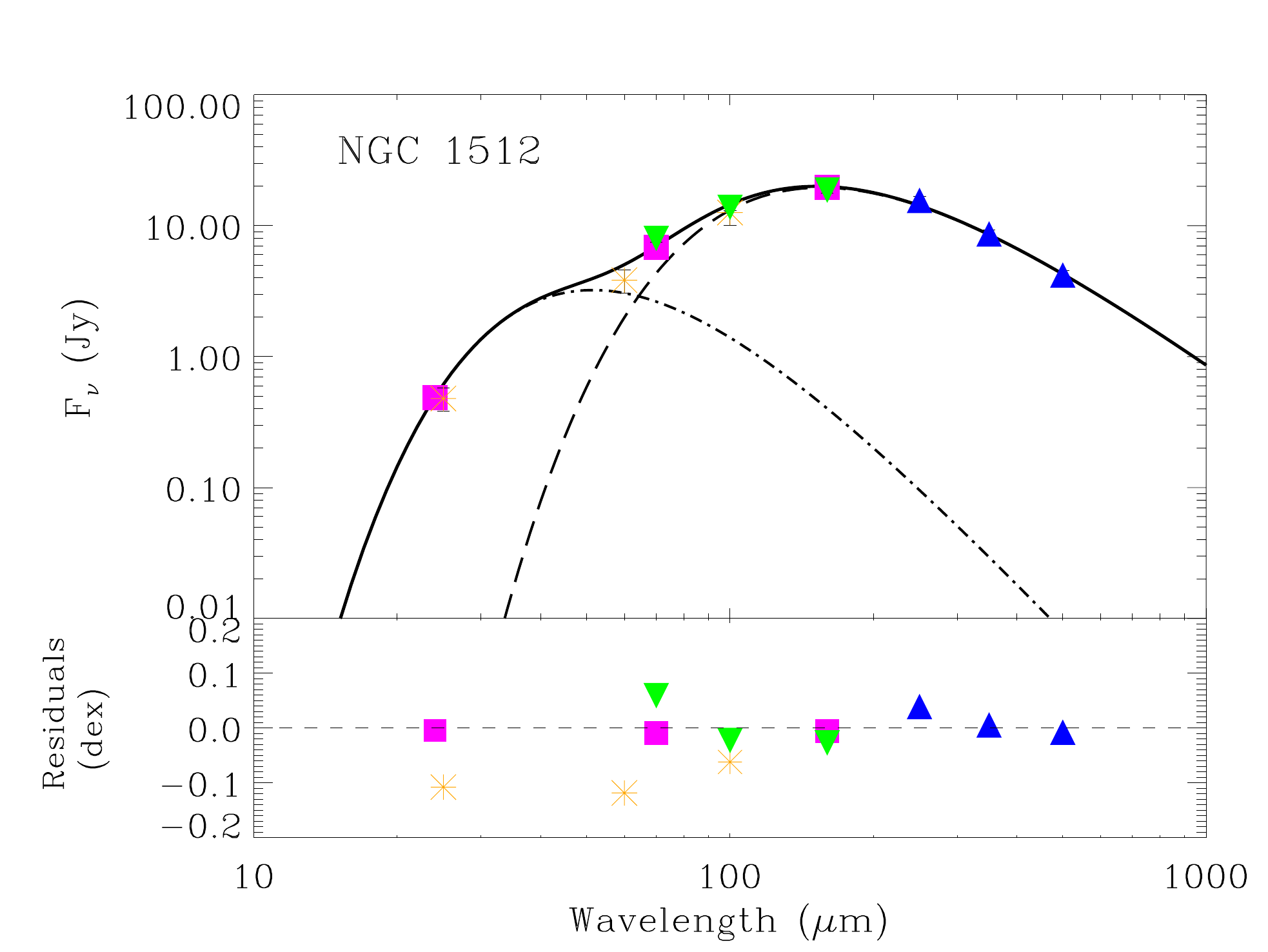}&  
           \includegraphics[width=7.5cm ,height=5.8cm]{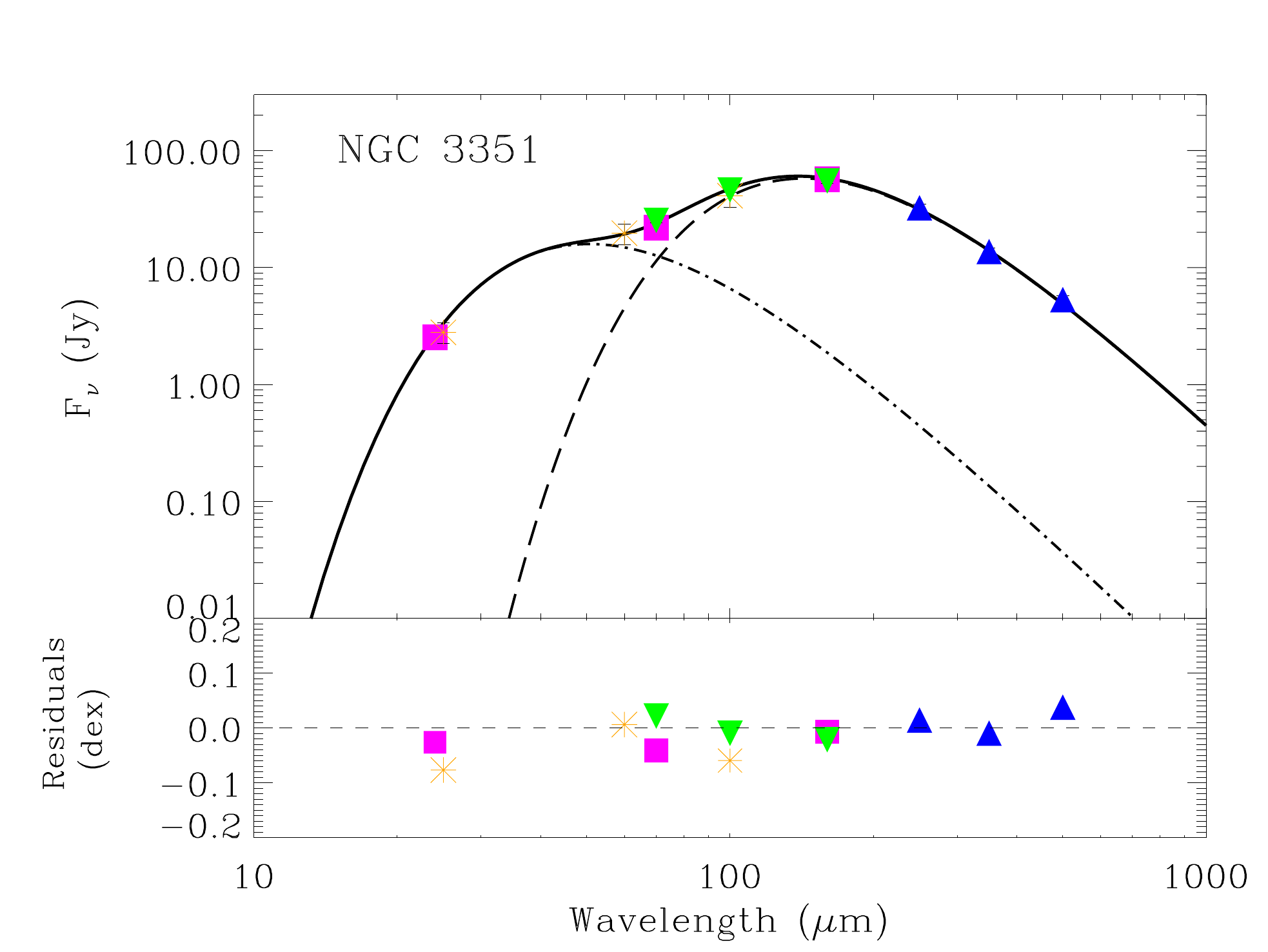} \\
           \includegraphics[width=7.5cm ,height=5.8cm]{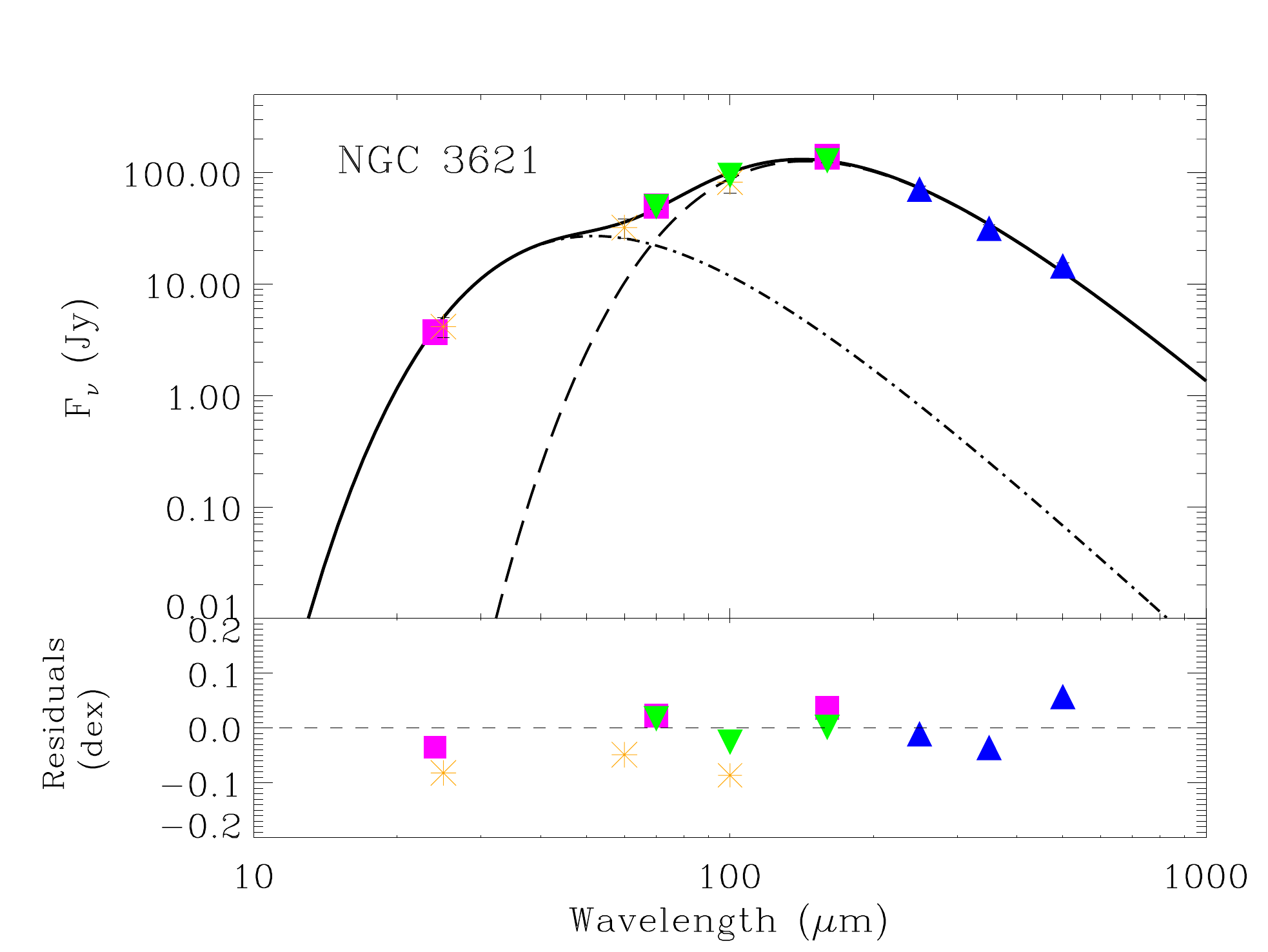} &   
            \includegraphics[width=7.5cm ,height=5.8cm]{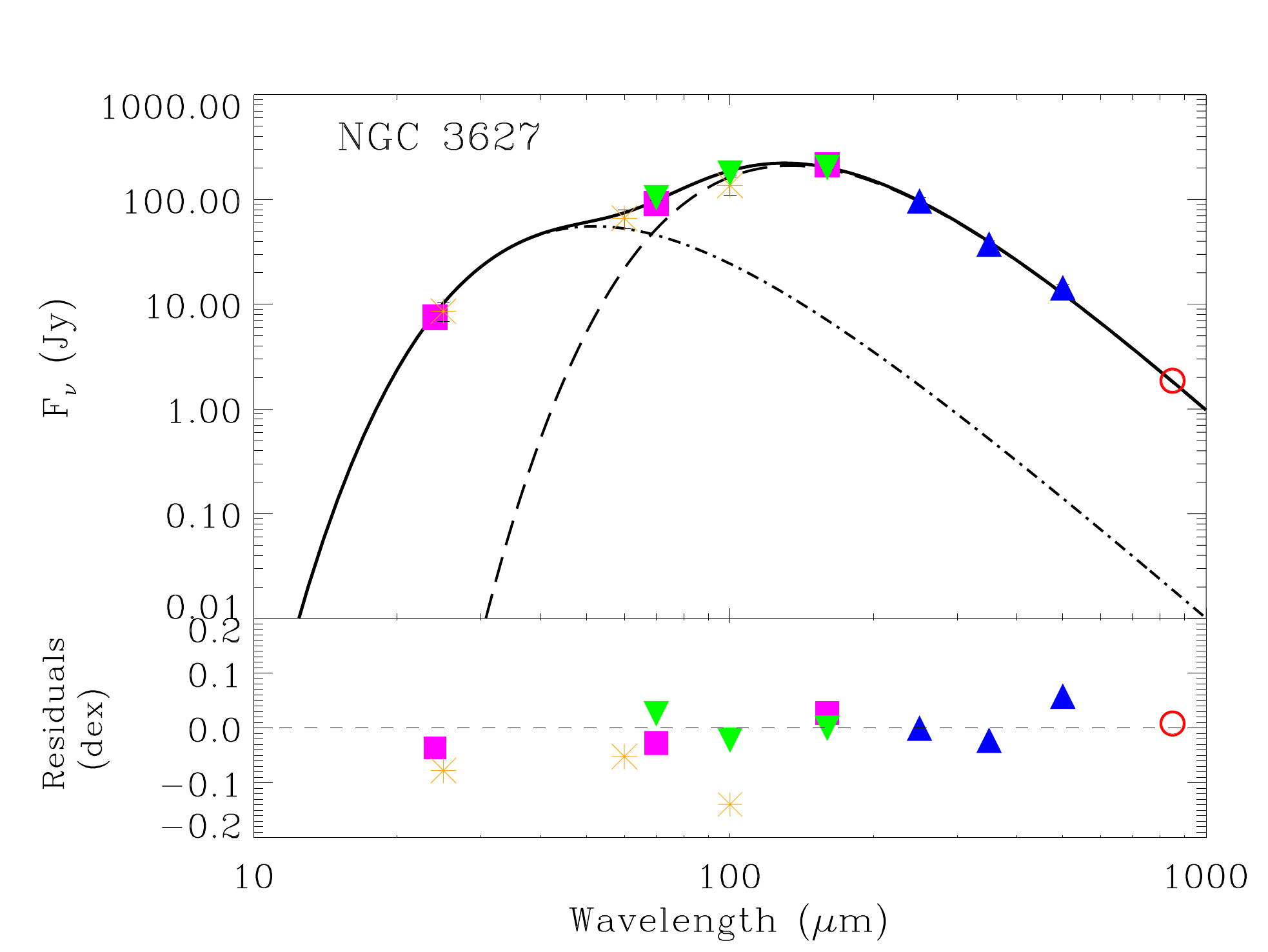}\\
             \includegraphics[width=7.5cm ,height=5.8cm]{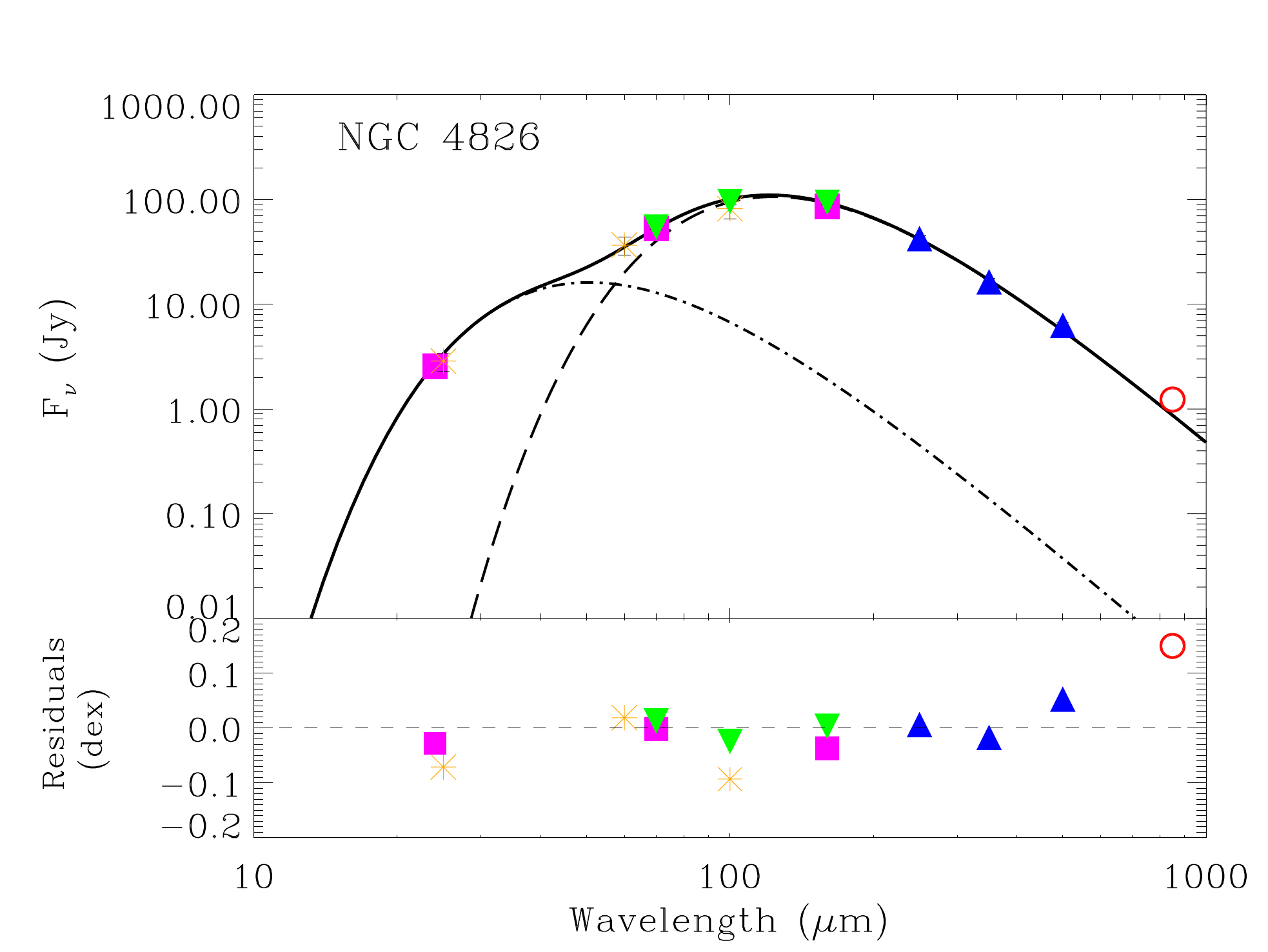} &
              \includegraphics[width=7.5cm ,height=5.8cm]{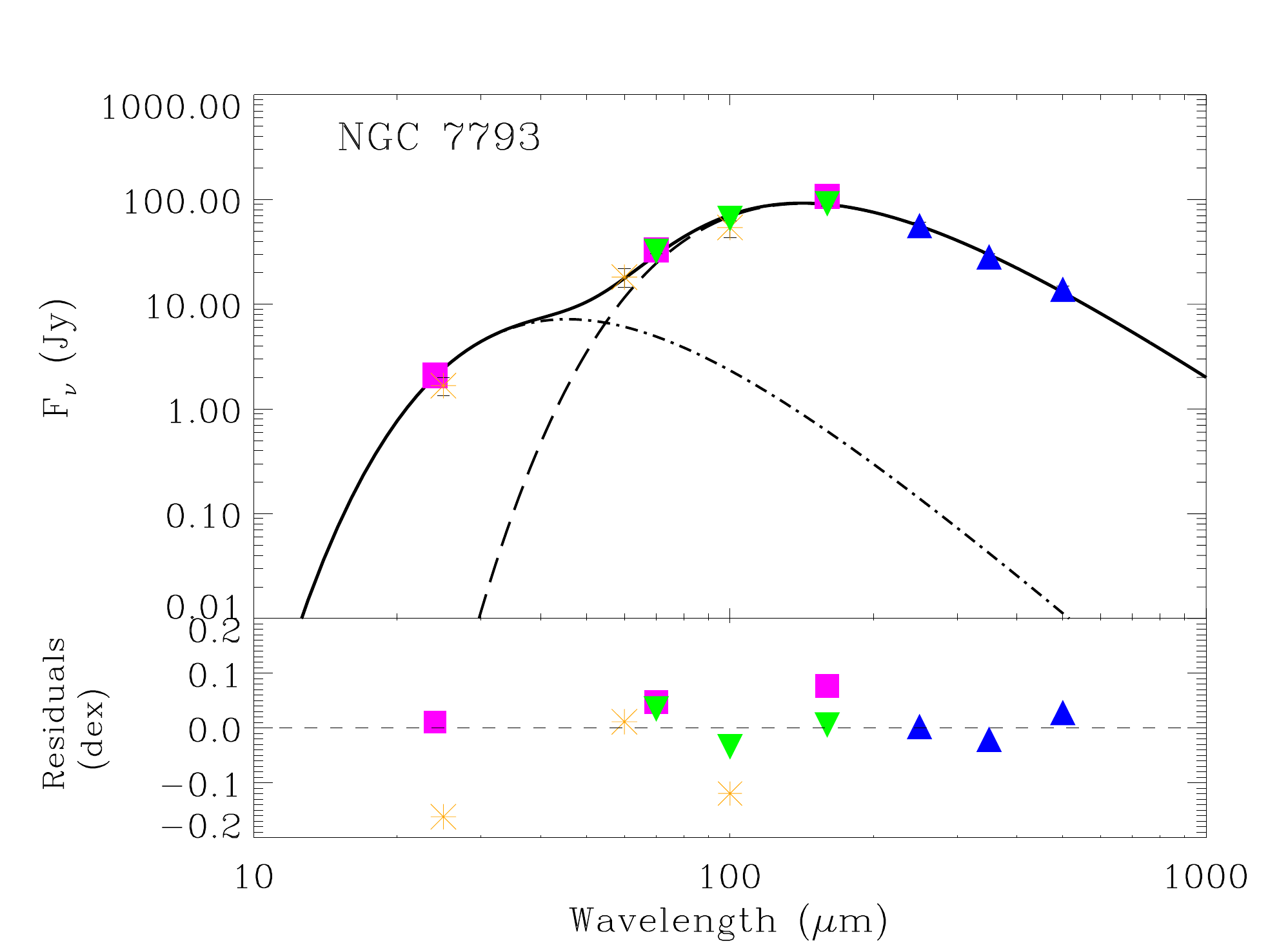}  \\  
          \end{tabular}
      \caption{Continued.}
\end{figure*}

\begin{table*}
\caption{\large Best fit parameters}
\label{BB_results}
 \centering
\begin{tabular}{m{20mm}|ccc|cccc|}
\hline
\hline
   &&&&&&&\\
	  & \multicolumn{3}{c|}{$\beta$$_{c}$ = 2} & \multicolumn{4}{c|}{$\beta$$_{c}$ free}\\
	  & L$_{70\mu m,}$$_{warm}$/L$_{70\mu m,}$$_{tot}$ &T$_{w}~(K)$ & T$_{c}~(K)$ & L$_{70\mu m,}$$_{warm}$/L$_{70\mu m,}$$_{tot}$ & T$_{w}~(K)$ & $\beta$$_{c}$ & T$_{c}~(K)$   \\ 
 &&&&&&&\\
 \hline
 &&&&&&&\\
 NGC~337   		& 0.58 	&51.4 $\pm$ 1.7 & 21.3 $\pm$ 0.8 &0.35		& 55.7 $\pm$ 14.3      & 1.6 $\pm$ 0.2 		& 25.2 $\pm$ 2.4 \\
 NGC~628 	   	& 0.53	&55.9 $\pm$ 1.8 & 19.0 $\pm$ 0.4 &0.52		& 56.0 $\pm$ 5.3 	& 2.0 $\pm$ 0.2 		& 19.1 $\pm$ 1.3 \\
 NGC~1097 		& 0.57	&55.8 $\pm$ 1.6 & 20.7 $\pm$ 0.6 &0.57		& 55.9 $\pm$ 2.4 	& 2.0 $\pm$ 0.2 		& 20.7 $\pm$ 1.4 \\
 NGC~1291	 	& 0.57	&56.2 $\pm$ 2.0 & 17.6 $\pm$ 0.4 &0.36		& 60.8 $\pm$ 10.2 	& 1.5 $\pm$ 0.2 		& 20.4 $\pm$ 1.4 \\
 NGC~1316 	 	& 0.32	&58.9 $\pm$ 8.1 & 22.4 $\pm$ 0.8 &0.52		& 55.2 $\pm$ 11.2 	& 2.5 $\pm$ 0.4 		&19.4 $\pm$ 2.4 \\
 NGC~1512 		& 0.81	&49.9 $\pm$ 1.1 & 16.8 $\pm$ 0.5 &0.38		& 56.4 $\pm$ 16.8	& 0.8 $\pm$ 0.2 		& 25.1 $\pm$ 2.9 \\
 NGC~3351 	 	& 0.51	&57.6 $\pm$ 2.2 & 20.4 $\pm$ 0.5 &0.53		& 57.4 $\pm$ 3.3 	& 2.1 $\pm$ 0.2 		& 20.0 $\pm$ 1.2 \\
 NGC~3621 		& 0.54	&54.9 $\pm$ 1.6 & 19.7 $\pm$ 0.5 &0.47		& 55.9 $\pm$ 5.3 	& 1.8 $\pm$ 0.2		& 20.7 $\pm$ 1.4 \\
 NGC~3627 		& 0.32	&59.2 $\pm$ 6.1 & 22.4 $\pm$ 0.6 &0.47		& 55.8 $\pm$ 5.6 	& 2.3 $\pm$ 0.2 		& 20.2 $\pm$ 1.4 \\
 NGC~4826 		& 0.21	&59.1 $\pm$ 13.5 & 23.7 $\pm$ 0.7 &0.24	& 57.0 $\pm$ 16.4    & 2.1 $\pm$ 0.2 		& 23.1 $\pm$ 1.6 \\
 NGC~7793 		& 0.59	&51.9 $\pm$ 1.7 & 18.7 $\pm$ 0.5 &0.17		& 63.4 $\pm$ 19.5 	& 1.2 $\pm$ 0.2 		& 24.4 $\pm$ 1.9 \\
 &&&&&&&\\
  \hline
\end{tabular}
\begin{list}{}{}
\item ${Note: }$ {\small The temperatures and emissivity indexes given in this table are the median values of the Monte-Carlo distributions (see $\S$3.2).}
\end{list}
 \end{table*} 

We use the global PACS and SPIRE fluxes of KINGFISH galaxies derived by \citet{Dale2012}. Their elliptical apertures (Table~\ref{Fluxes}) are similar to those used to perform the \spitz\ photometry (for SINGS in particular) and encompass the total emission of the galaxy at each wavelength. Global fluxes are measured on original resolution images. We complement the global SED coverage with data from \spitz/MIPS and IRAS. We use the global flux densities of our galaxies for \spitz/MIPS bands derived by \citet{Dale2007}. We obtain the IRAS global flux densities at 25, 60 and 100 \mic\ (accurate to $\sim$20$\%$, when available) through the SCANPI\footnote{http://scanpiops.ipac.caltech.edu:9000/applications/Scanpi/index.html} tool (IRAS Scan Processing and Integration tool) and the HIRES \citep[IRAS High Resolution Image Restoration,][]{Surace2004} atlas. We provide the MIPS, PACS and SPIRE flux densities in Table~\ref{Fluxes}. MIPS and PACS fluxes are in very good agreement. We refer to \citet{Dale2012} for quantitative comparisons between MIPS and PACS global fluxes at 70 and 160 \mic. \\

Cold dust temperatures are often derived using single modified blackbody (MBB) fits. Nevertheless, a galaxy contains a range of dust temperatures that a single MBB will not be able to capture. The warm dust component, in particular, spans a narrow range of dust temperatures, from $\sim$40 to 100 K and contributes to a non negligible fraction of the 70 \mic\ emission. As a result, using a single MBB model fit usually leads to an overestimation of the cold dust temperatures. We note that this effect particularly affects objects showing an enhanced 24-70 \mic\ emission linked with stochastically heated grain emission, like the Magellanic Cloud for instance \citep{Bot2004,Bernard2008}. We thus carry out a two-temperature modeling of our galaxies in order to separate the warm and the cold component, using the MIPS, PACS and SPIRE flux densities listed in Table~\ref{Fluxes} along with MIPS 24 \mic\ and IRAS 25, 60 and 100 \mic\ flux densities. We fit the data {\it d$_{\nu}$} with a model of the form:
\begin{eqnarray}
d_{\nu}  = A_w~\left(\lambda^{-2} B_{\nu}(T_w)\right) + A_c~\left(\lambda^{- \beta_c} B_{\nu}(T_c)\right),
    \label{BBequation}
\end{eqnarray}

\noindent with {\it B$_{\nu}$} the Planck function and the 5 free parameters of our model: {\it $\beta$$_{c}$} is called the emissivity index of the cold component, {\it T$_{w}$} and {\it T$_{c}$} the temperature of the warm and cold component respectively and {\it A$_{w}$} and {\it A$_{c}$} the overall amplitudes of each dust component (see discussion on the naming convention in the next section). 
This equation assumes optically thin emission over our wavelength range. As can be observed in Eq.\ref{BBequation}, we choose to fix the emissivity index of the warm dust component in order to limit the number of free parameters in our model. {\it We decide to fix $\beta$$_w$ to 2.0}. This is a good approximation of the opacity of the graphite/silicate dust models of \citet{Li_Draine_2001}. We note that our choice of $\beta$$_w$ does not influence the cold dust temperatures. Fixing $\beta$$_w$ to 1.5 leads to a small decrease of the cold dust temperatures of less than 0.6$\%$ if $\beta$$_c$=2.0, and less than 1.6$\%$ if $\beta$$_c$ is free. \citet{Tabatabaei2011} (see also Tabatabaei et al., in prep) have synthesized the two MBB approach for M33, with $\beta$$_c$ free and $\beta$$_w$ fixed to 2.0, 1.5 and 1.0 and found that $\beta$$_w$=2.0 best reproduces the observed fluxes, reinforcing our choice.\\
 
We compute the expected photometry for the fitted model by convolving this model with the instrumental spectral responses of the different cameras. We refer to the individual userÕs manuals of each instrument for the usual conventions used for each instrument. We use the IDL function MPCURVEFIT (Markwardt 2006) to perform a Levenberg-Marquardt least-squares fit of the data. We weight the data during the $\chi$$^2$ minimization in order to take the uncertainties on flux measurements into account. \\

\subsection{Uncertainties on the parameters}
 
\citet{Shetty2009} among others show that fits performed solely in the Rayleigh-Jeans regime are very sensitive to instrumental errors. Fits using the full FIR-to-submm coverage usually lead to a significantly better determination of dust properties than fits only performed near the SED peak. In order to quantify the uncertainties on the parameters driven by the errors of our individual fluxes, we perform Monte Carlo simulations for each galaxy. We generate 500 sets of modified constraints. Each flux randomly varies within its error bar, following a normal distribution around its nominal value. The absolute calibration is correlated across the three SPIRE bands (see the SPIRE's observers's manual), namely flux densities in the three bands are expected to move in the same direction if the absolute calibration would be revised. To account for this correlation, we vary the three SPIRE measurements consistently; i.e., they are not allowed to vary relative to each other. We finally apply our two-temperature fit to these 500 modified datasets. \\

  \begin{figure*}
   \centering
   \begin{tabular}{c}
  \includegraphics[width=18cm]{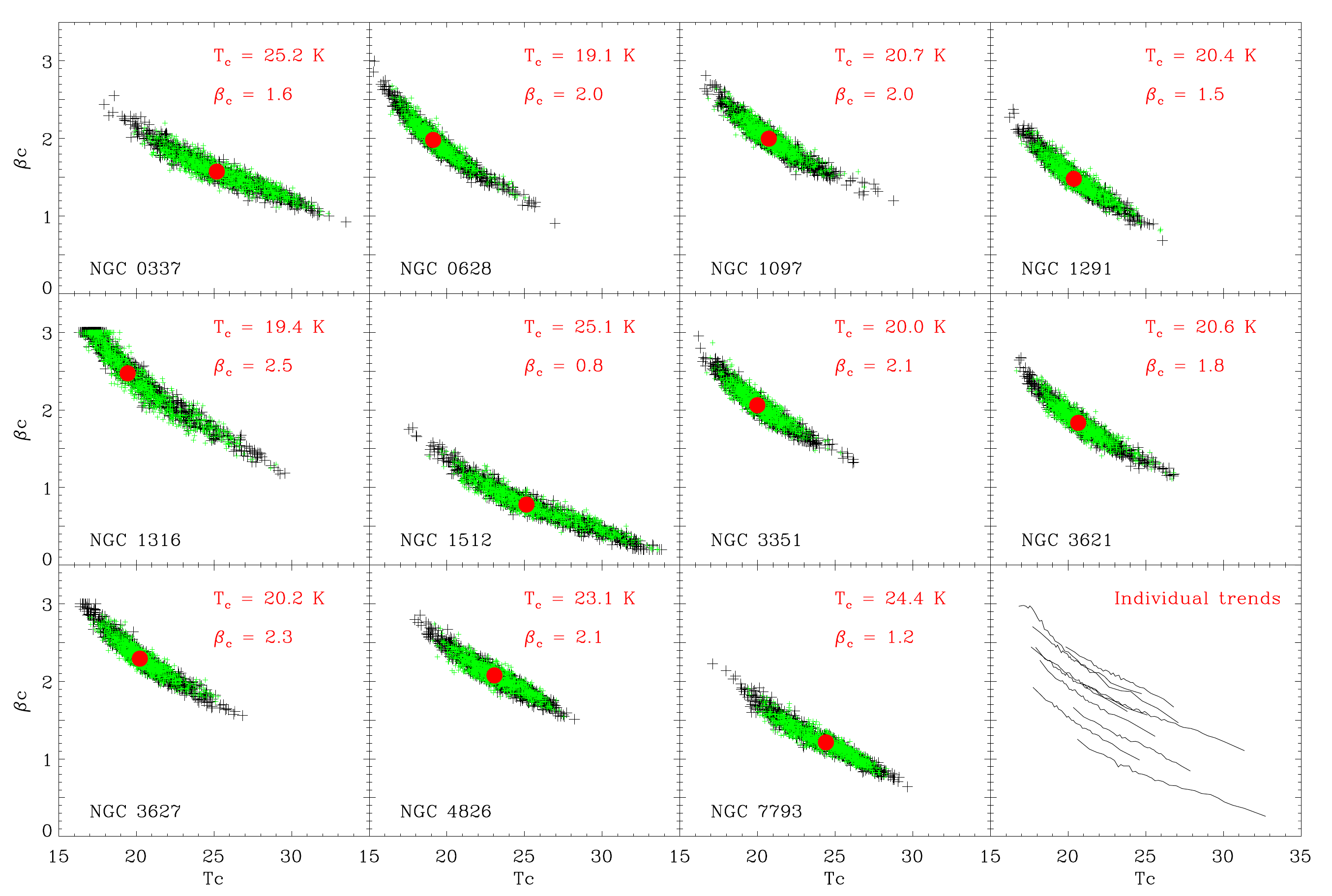} \\
   \end{tabular}
\caption{Distribution of the cold dust parameters in the temperature-emissivity index plane. 500 fits of randomly modified datasets were performed. 
The green crosses indicate the distribution when the correlation between SPIRE calibration errors is applied, the black crosses when no correlation was taken into account. 
The red dot indicates the median values on T$_{c}$ and $\beta$$_{c}$. Those values are reminded in each panel. The last panel summarizes the trends of all the galaxies. }
  \label{T_B_errors} 
\end{figure*}

\subsection{Global dust temperature and emissivity index}

Figure~\ref{2BB_SEDs} shows the global SEDs we obtained. $\beta$$_{c}$ is a free parameter in those models. We overlay the SCUBA data on our SEDs when available (indicated with open circles in Figure~\ref{2BB_SEDs}). Some of the SCUBA maps do not fully cover the outer part of the galaxies or show a poor detection at the rim of the disks due to a possible removal of diffuse emission during the data reduction. Those data are not used in the modelling. Bottom panels for each galaxy indicate the residuals from the fit in logarithm scale.
The differences in shape of the SEDs we obtain highlight the wide range of local ISM environments covered by the sample. Discrepancies particularly appear on the Wien side of the SEDs. We however note that the warm component does not significantly affect the fit to the long wavelength cold component. The 100-to-24 \mic\ flux density ratios vary from $\sim$17 for NGC~1097 to $\sim$37 for NGC~4826, thus from actively to quiescent star forming galaxies. Our SEDs mostly peak at similar wavelengths, residing between $\lambda$=117 \mic\ for NGC~337 and $\lambda$=165 \mic\ for NGC~1512, in their F$_{\nu}$ SEDs. 
We indicate the contribution of the warm component to the 70 \mic\ fluxes in Table~\ref{BB_results}. This contribution can be higher than 50$\%$ in some objects, legitimizing the necessity of two dust components in our simple approach. We also note that the relative fluxes at 60 and 70 \mic\ influences the fits as it samples the SEDs where the warm and cold dust overlap. We try to fit our data excluding the IRAS 60 \mic\ data. This leads to a decrease in our cold dust temperature by less than 1.8$\%$ if $\beta$$_c$ is fixed to 2.0 and less than 8.7$\%$ if $\beta$$_c$ is free.

We summarize the median values of T$_{c}$ and $\beta$$_{c}$ in Table~\ref{BB_results}, with their errors, namely the standard deviation of the parameter distribution obtained using the Monte-Carlo simulations. We also indicate the cold dust temperatures obtained with an emissivity index $\beta$$_c$ fixed to a standard value of 2.0 for comparison. The median reduced ch-squared value is $\sim$1.4 when $\beta$$_c$ is fixed to 2.0 and $\sim$1.1 when $\beta$$_c$ is free, which means that both models lead to satisfying fits. When $\beta$$_{c}$ is free, we observe variations of the cold dust emissivity within our sample, with values ranging from 0.8 to 2.5 (Table~\ref{BB_results}). Those values are consistent, within the uncertainties, with the emissivity index ranges quoted in the literature \citep[][among others]{Hildebrand1983,Dupac2003}. We note that using a free emissivity index also increases the uncertainties on the temperature estimates.

   
   \begin{figure*}
    \centering   
                \vspace{10pt}
    \begin{tabular}{cc}
       {\large MIPS 24 \mic} &
       {\large Temperature map ($\beta$$_c$=2)} \\
	\includegraphics[height=5.5cm]{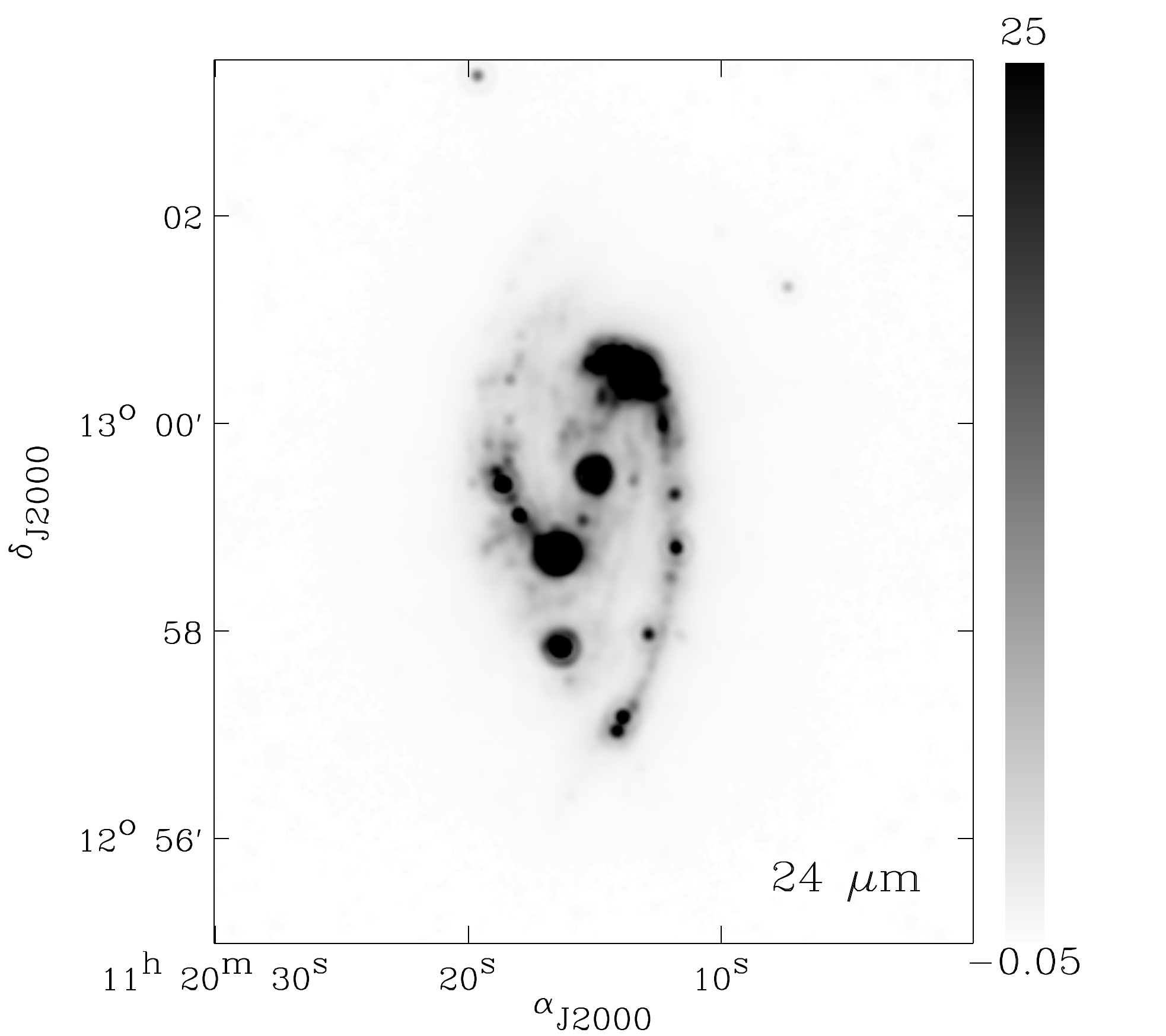}  &
	\includegraphics[height=5.5cm]{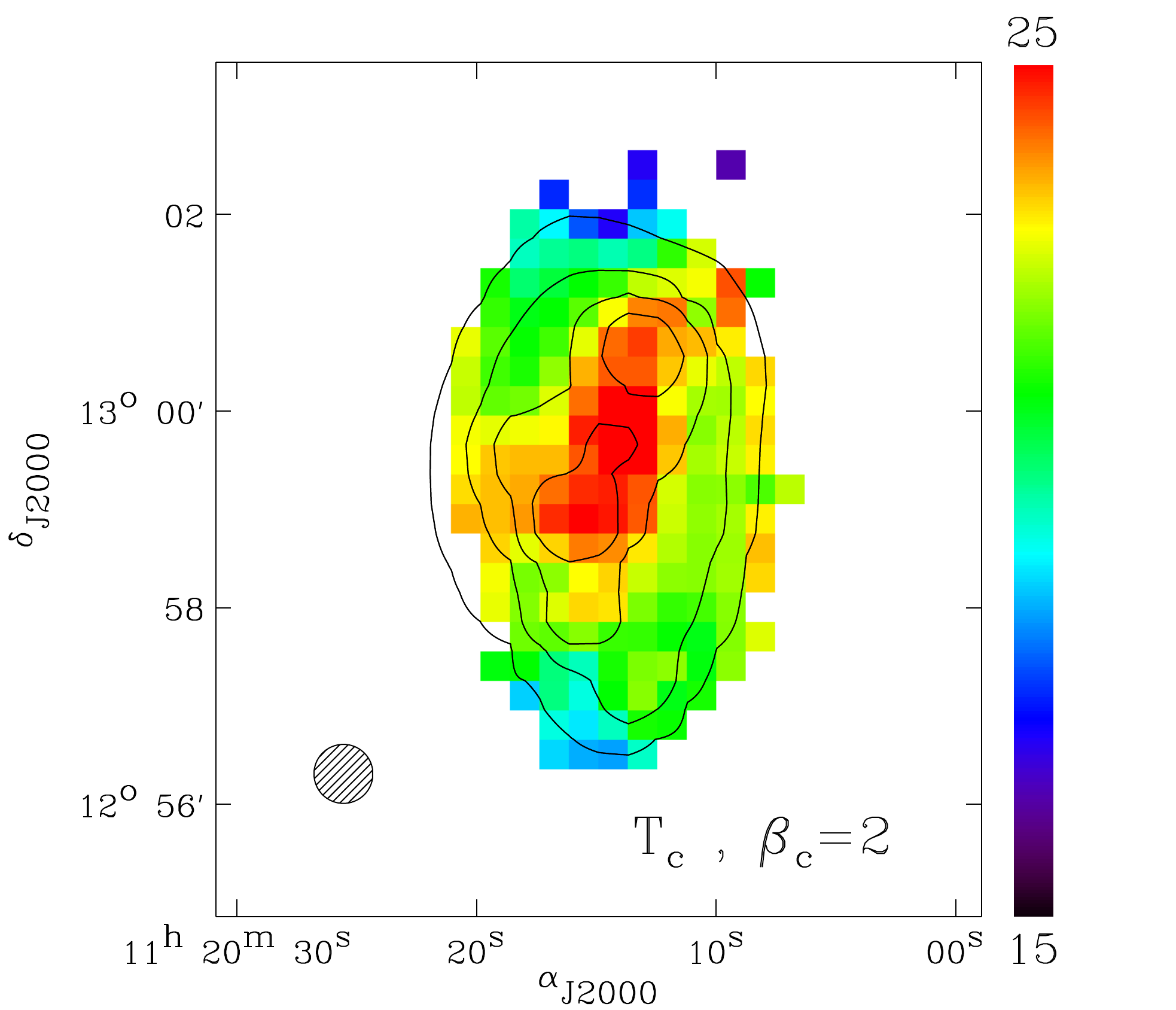}  \\
	&\\
          {\large Emissivity index map ($\beta$$_c$ free)} &
         {\large Temperature map ($\beta$$_c$ free)} \\    
	\includegraphics[height=5.5cm]{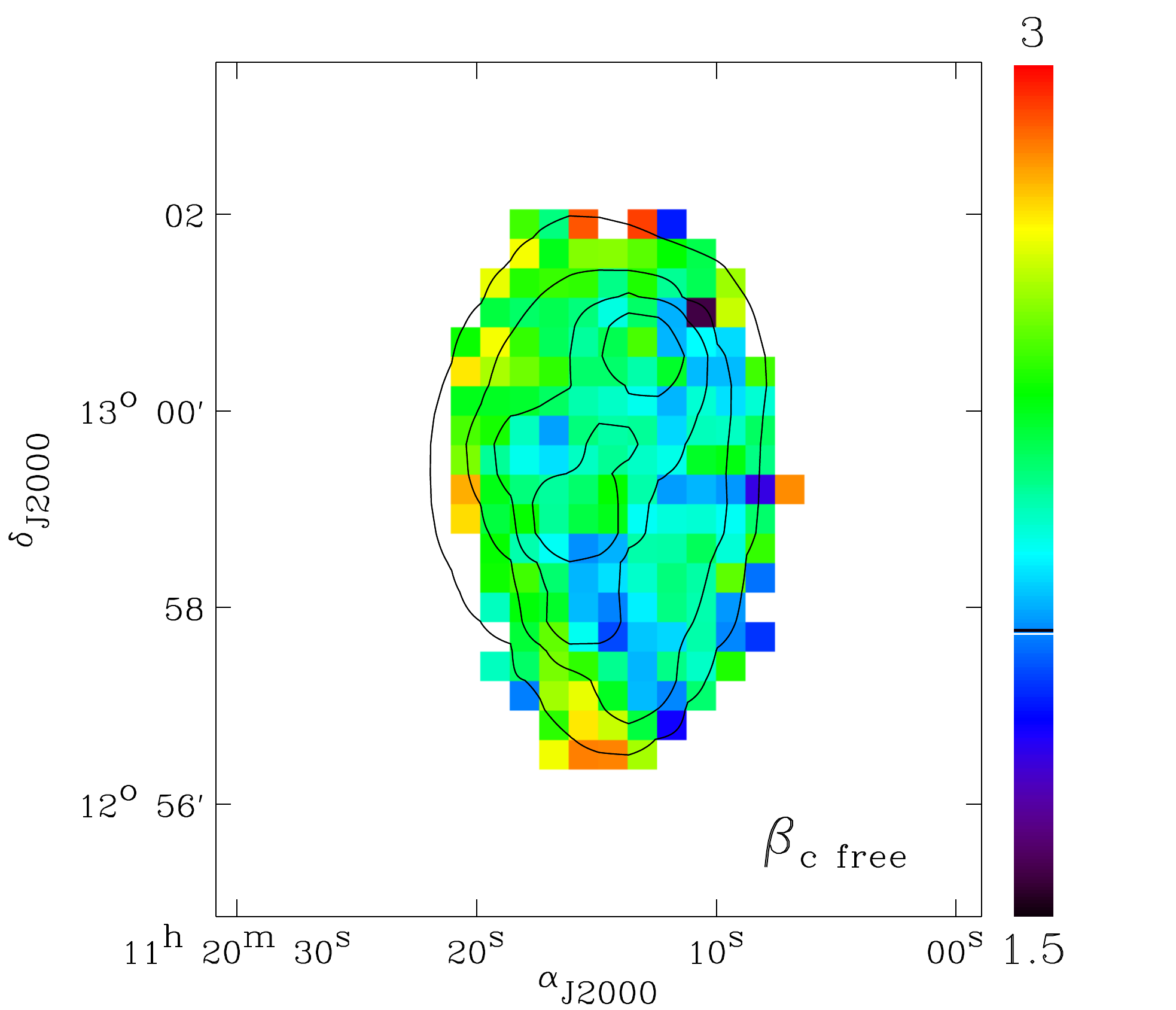}  &
	 \includegraphics[height=5.5cm]{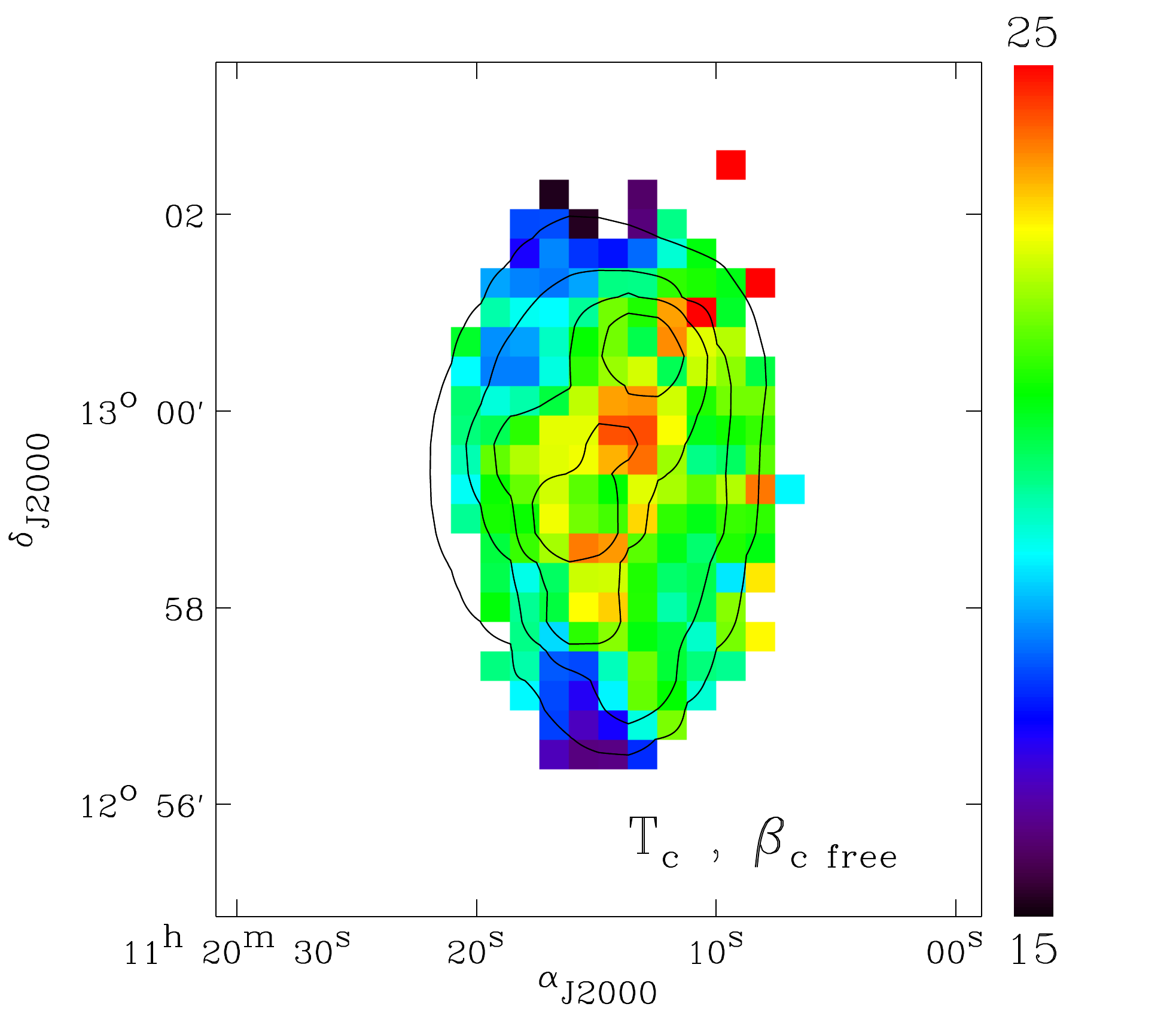} \\
   \end{tabular}
    \caption{Upper left: MIPS 24 \mic\ image (in MJy~sr$^{-1}$) of NGC~3627. Upper right: Cold dust temperature map of NGC~3627 (in K) derived using $\beta$$_c$=2.0. We overlay the MIPS 24 \mic\ contours (smoothed to the resolution of the temperature and beta maps) for comparison. The FWHM of the PSF of SPIRE 500 \mic\ is shown by the dashed circle. Lower left: Emissivity index map of NGC~3627 derived with T$_c$ and $\beta$$_c$ as free parameters. The value $\beta$$_c$=2.0 is indicated on the color scale. Lower right: Cold dust temperature map of NGC~3627 (in K) derived with T$_c$ and $\beta$$_c$ as free parameters. For each map: North is up, east is left. Temperature and emissivity maps for the rest of the sample are available in the appendix of this paper. }
    \label{NGC3627_Temp}
\end{figure*}

Figure~\ref{T_B_errors} enables us to visualize the distribution of the 500 runs, results of the Monte-Carlo simulations, for each galaxy in a T$_{c}$ - $\beta$$_{c}$ parameter space. In all panels, the green crosses are (T$_{c}$, $\beta$$_{c}$) combinations derived from 500 fits taking the correlation between SPIRE calibration uncertainties into account. For comparison, we also plot with black crosses the (T$_{c}$, $\beta$$_{c}$) combinations obtained if SPIRE data are allowed to vary independently from each other. The median values of the temperatures and emissivity indexes are annotated in red. We remind the reader that those are the median values reported in Table~\ref{BB_results}. The last panel gathers the individual trends. Those trends are obtained by taking the average every 30 runs.
We note that, as expected, taking into account the correlations between SPIRE calibration uncertainties narrows the ranges of values reached by the parameters. Nevertheless, this assumption does not change the median values of T$_{c}$ and $\beta$$_{c}$ obtained (less than 2$\%$ difference). We observe a clear T$_{c}$-$\beta$$_{c}$ anti-correlation in all panels of Fig.~\ref{T_B_errors}. \citet{Shetty2009} caution against the interpretation of the inverse T$_{c}$-$\beta$$_{c}$ correlation as a physical property of dust in the ISM since it could be created by measurement uncertainties. Our results clearly confirm that fitting SEDs to noisy fluxes, even with data sampling the thermal dust emission from 24 to 500 \mic, leads to T$_{c}$-$\beta$$_{c}$ anti-correlation. We further discuss this effect in the following section.
 
\begin{figure*} 
   \centering
  \includegraphics[width=13cm ]{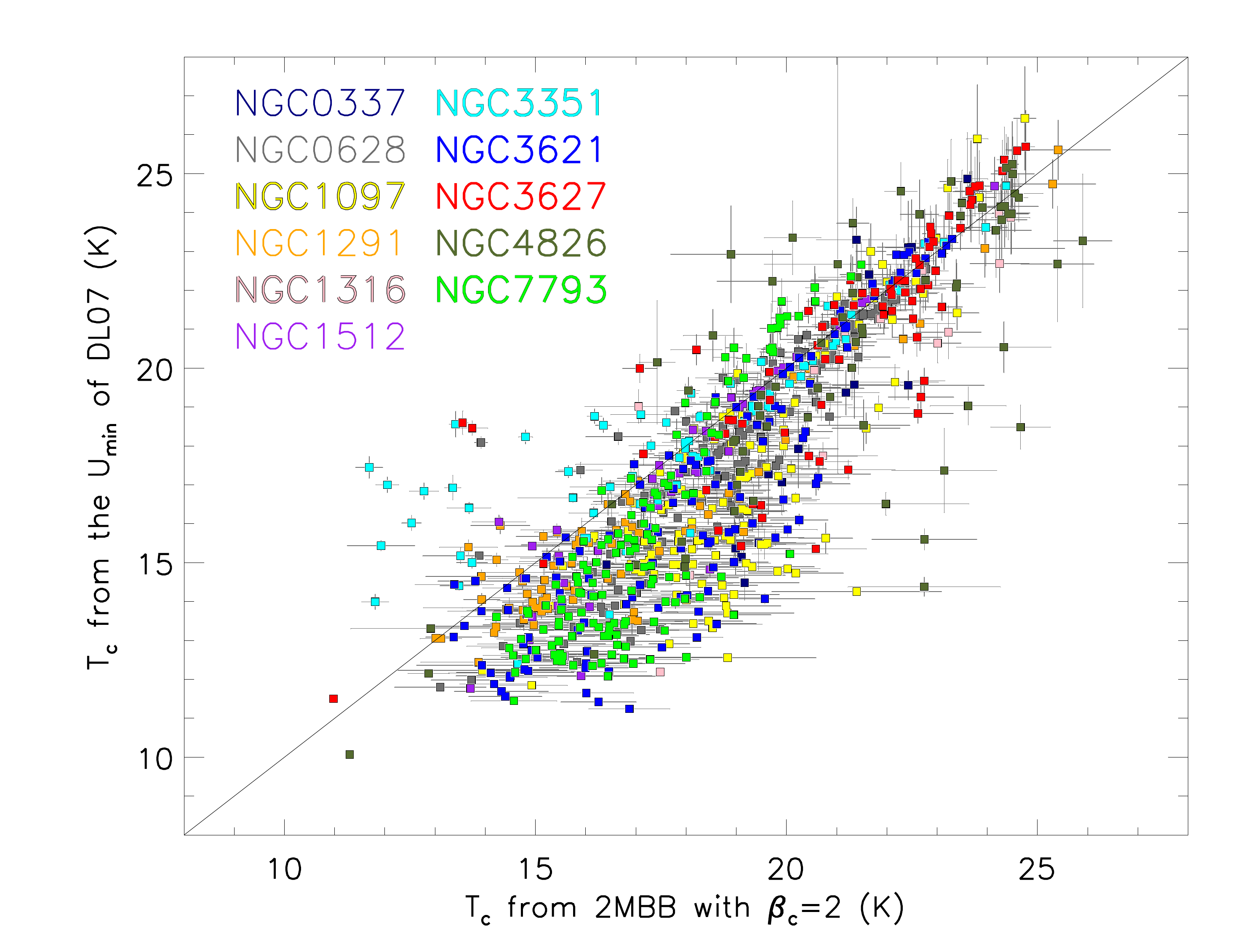}  \\
\caption{Comparison between the cold dust temperature obtained with our two-MBB models and the cold dust temperature derived from the radiation field of the diffuse ISM component (U$_{min}$) from the \citet{Draine_Li_2007} models [DL07]. Squares are pixel-by-pixel values (pixel size: 18\arcsec) averaged 4-by-4 for clarity. The line indicates a ratio of unity.}
  \label{Plot_T_Umin} 
\end{figure*}

\section{ Mapping the cold dust temperature and emissivity }  

\subsection{Dust temperature distribution - $\beta$$_c$ fixed}

The spatial resolution of the \hersc\ instruments now enables us to reproduce the previous study on a local basis in order to resolve the dust temperature and emissivity within galaxies and directly link them with other ISM tracers. We use our two-temperature fitting approach to derive maps of dust temperature and dust emissivity in our objects. In order to constrain the warm dust emission, we use \spitz/MIPS images at 24, 70 and 160 \mic. These maps were obtained as part of the SINGS program \citep[Spitzer Infrared Nearby Galaxies Survey;][]{Kennicutt2003} and reduced using the SINGS Fifth Data Delivery pipeline. MIPS, PACS and SPIRE maps are then convolved to the SPIRE 500 \mic\ resolution (see $\S$2.4 for the method) and projected to a common sample grid with a pixel size of 18\arcsec, thus half of the FWHM of SPIRE at 500 \mic. This pixel size corresponds in linear scale to ISM elements ranging from $\sim$ 340 pc for our closest galaxy NGC~7793 to $\sim$ 1.8 kpc for the furthest galaxy of our sample NGC~1316. The emissivity of the warm component is fixed to a standard emissivity of 2.0. We first use a {\it fixed emissivity index for the cold dust component} to obtain a robust determination of the cold dust temperature distribution based on color temperatures. We study the case with a free $\beta$$_c$ in the following section. 

We exclude pixels with flux densities below a 3-$\sigma$ detection limit in all bandpasses. Uncertainties on the fit parameters are derived on a local basis using Monte Carlo simulations as described in $\S$3.2. We generate 100 sets of modified constraints per pixel and run our SED modelling technique on those sets. We average (median) the 100 dust temperature values obtained in each pixel and derive the final cold dust temperature map. We show the cold temperature map of NGC~3627 obtained with $\beta$$_{c}$ fixed to a standard value of 2.0 in Fig.~\ref{NGC3627_Temp} (upper right panel). We overlay the MIPS 24 \mic\ contours on the maps to compare the temperature distribution with the star forming regions of the galaxies whose 24 \mic\ emission is a usual good tracer. We show the cold dust temperature maps of the whole sample in an appendix in Fig.~\ref{Prop_maps} (top middle panel for each galaxy). We observe a clear radial trend of dust temperatures in our objects. The temperature maxima coincide with the center or the star forming regions. The dust grain temperatures decrease in the outer parts of galaxies. This radial trend is consistent with results obtained with \hersc\ observations in nearby galaxies \citep[][or Hinz et al. in prep]{Bendo2011,Boquien2011,Engelbracht2010}. 

The source of heating of the cold dust is still poorly understood. The emission of old stars, traced by the IRAC 3.6 \mic\ observations, smoothly decreases with radius, at our resolution, in our objects. A radial trend of dust temperatures may thus suggest that the old stellar populations could contribute to the heating of the cold dust while the youngest star-forming population is closely associated with the warm dust \citep{Calzetti2007}. However, the star formation activity of our galaxies tends also to decrease radially in some of our objects as suggested by the 24 \mic\ emission, and the resolution of our dust temperature maps may not allow us to properly disentangle between the different possible heating sources. We observe nonetheless that the cold dust temperature distribution follows the bars in the galaxies NGC~1097 and NGC~3627 or strongly peaks in the nucleus of NGC~1316 where the 24 \mic\ emission peaks as well (see Fig.~\ref{Prop_maps}, top left panels), suggesting a more significant role of the star forming regions than the ISRF in heating the cold dust in those objects. \\

[A12b] perform a pixel-by-pixel SED modeling of the whole KINGFISH sample using the \citet{Draine_Li_2007} dust models on \spitz\ + \hersc\ data up to 500 \mic. These models couple a ``PDR component'' reproducing the emission of photodissociation regions with a range of radiation fields and a ``diffuse ISM component'' exposed to a single radiation field intensity U$_{min}$. We note that in this model, most of the cold dust is expected to reside in the diffuse component. 

The energy absorbed by a cold grain in this diffuse phase is:
\begin{equation}
\Gamma_{abs}\propto \int_{0}^{\infty} Q_{abs}(\lambda)U_{\lambda}(\lambda)d\lambda \propto U_{min},
 \end{equation}
 \noindent with {\it Q$_{abs}$} the absorption efficiency of the grains and {\it U} the intensity of the radiation field of the diffuse ISM, thus U$_{min}$.
 
 The energy emitted by a cold grain in this diffuse phase is:
 \begin{equation}
\Gamma_{em}\propto \int_{0}^{\infty} \nu^{\beta}B_{\nu}(T_{eq})d\nu \propto T_c^{4+\beta},
 \end{equation}
  \noindent with {\it B$_{\nu}$} the Planck function and {\it T$_{eq}$} the equilibrium temperature in the phase, namely the temperature of the cold dust T$_c$.
  
Assuming thermal equilibrium conditions, we can write that $\Gamma$$_{abs}$=$\Gamma$$_{em}$. We can thus associate the radiation intensity U$_{min}$ and the cold dust temperature T$_c$. If we assume that dust grains possess an emissivity index $\beta$=2.0 and normalize the temperature to the equilibrium dust temperature of the Galaxy \citep[17.5K, ][]{Boulanger1996}, we obtain: 
\begin{equation}
T_c (K) = 17.5 \times U^{1/(4+\beta)}_{min} = 17.5 \times U^{1/6}_{min},
 \end{equation}

\begin{figure*}
   \centering
  \includegraphics[width=17cm ]{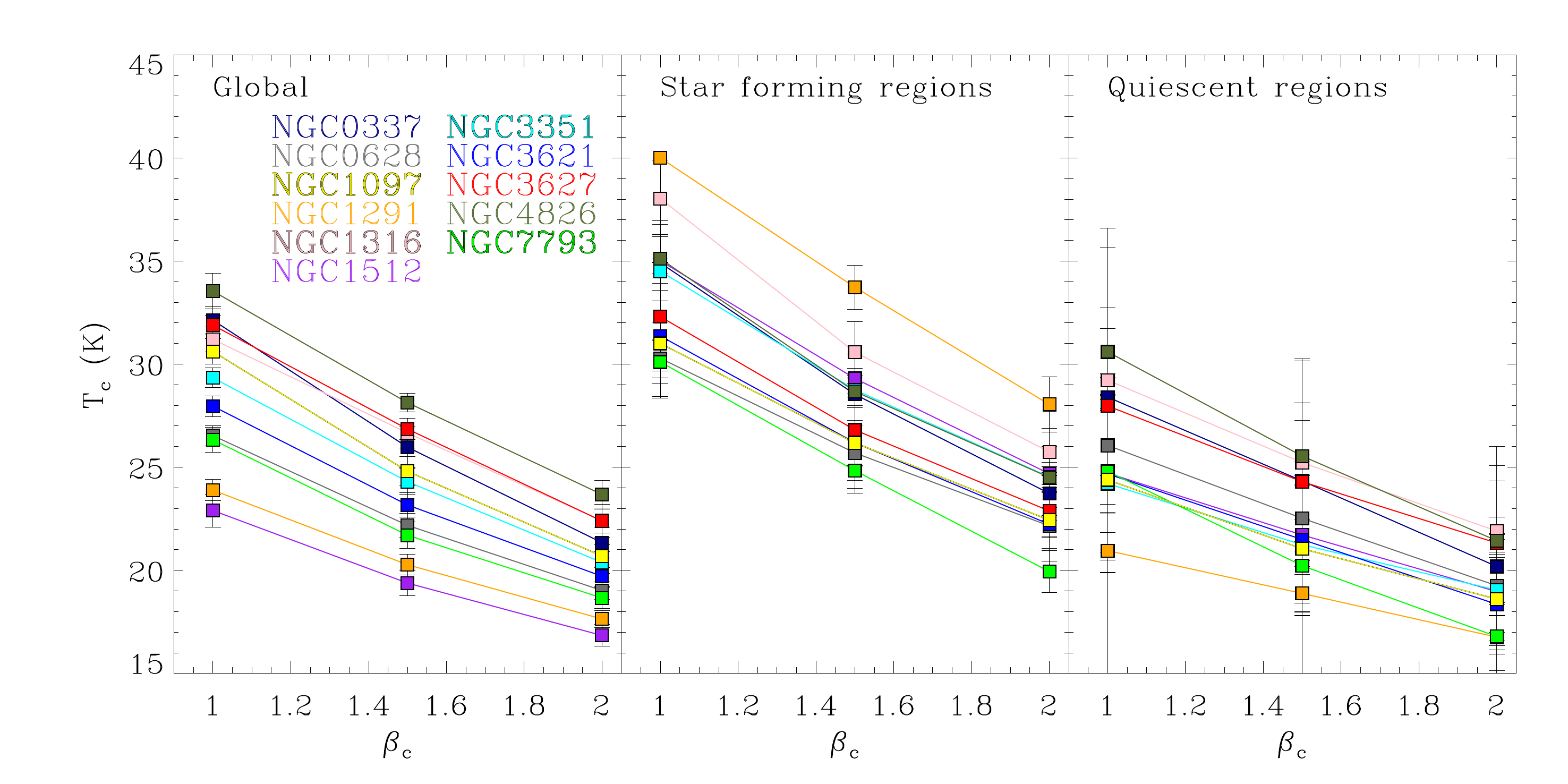} \\
\caption{Dependence of the dust temperatures with the chosen cold emissivity index. Left: Global T$_{c}$ obtained using $\beta$$_{c}$ = 1, 1.5 and a standard emissivity index of 2.0. Galaxies are color-coded. Middle: Median of the pixel-by-pixel T$_{c}$ obtained for star forming regions. Right: Median of the pixel-by-pixel T$_{c}$ obtained for less active regions.}
  \label{Evol_Temp}
\end{figure*}

We can thus compare our derived cold temperature maps obtained with a fixed emissivity $\beta$$_c$=2.0 with the U$_{min}$ map obtained by [A12b] using this formula. Dust temperatures have very similar distributions in both cases. We plot the pixel-by-pixel comparison in Fig~\ref{Plot_T_Umin}. We observe that in regions showing a ``hotter" cold dust temperature (co-spatial with star forming regions or high starlight intensities), the temperatures obtained in both cases are very similar. The points diverge in regions which show a lower dust temperature, with colder temperatures deduced using the U$_{min}$ parameter obtained from the \citet{Draine_Li_2007} and the conversion of Eq.~6. This could be explained by the fact that the \citet{Draine_Li_2007} model assumes a range of dust temperature and thus allows to reach colder dust than a MBB fit with $\beta$$_c$ fixed to 2.0. Our hypothesis of an emissivity index of $\beta$=2.0 to translate U$_{min}$ into temperatures slightly influences our results but not enough to explain the trend we observe. We refer to [A12b] for a complete description of the U$_{min}$ maps for the whole KINGFISH sample. \\
 
We also derive temperature maps with an emissivity index of the cold dust component fixed to $\beta$$_c$=1.5 and $\beta$$_c$=1.0. This choice does not modify the relative distribution of cold dust temperatures in our objects but influences the temperature ranges obtained, scaling temperatures to higher values than in the ``$\beta$$_{c}$ = 2.0" case. 
We want to compare this effect in bright star-forming regions as well as in more quiescent regions. We separate our pixels into two different categories based on their star formation rate, using the MIPS 24 \mic\ band as a qualitative indicator for the SFR and the relation from \citet{Calzetti2007} :
{\begin{eqnarray}
SFR(M_\odot~yr^{-1}) = 1.24 \times 10^{-38}[L(24 \mu m)(erg~s^{-1})]^{0.88} 
 \end{eqnarray}}
We fix the separation between active and less active regions at SFR $=$ 4 $\times$ 10$^{-3}$ \msun~yr$^{-1}$~kpc$^{-2}$ and plot the variations of dust temperatures as a function of the emissivity index in Fig.~\ref{Evol_Temp}. The left panel indicates the dust temperatures obtained from integrated fluxes using $\beta$$_c$ = 1.0, 1.5 and 2.0. Galaxies are color-coded. The errors are derived from our Monte Carlo simulations performed on global fluxes ($\S$3.2). The middle panel shows the median values of the pixel-by-pixel dust temperatures if we restrict the study to the bright star forming regions of the galaxies. The right panel shows the median values of the pixel-by-pixel dust temperatures if we restrict the study to the more quiescent regions of the galaxies. For the middle and right panels, errors are the standard deviations of the distributions of dust temperatures. 

As mentioned earlier, we observe that using a lower emissivity index $\beta$$_c$ shifts the temperature range to higher values. The median temperatures of bright star forming regions are systematically higher than the median temperatures of more quiescent regions, as intuitively expected. With $\beta$$_c$=1.5, the temperatures increase by up to 25$\%$ for the bright regions and by up to 21 $\%$ for quiescent regions. With $\beta$$_c$=1.0, the temperatures increase by up to 51$\%$ for the bright regions and by up to 48 $\%$ for less active regions. The choice of $\beta$$_c$ has a major impact on the dust temperatures obtained and on the dust masses usually derived from those blackbody fits (see $\S$4.3). Our comparison between quiescent and active regions seems to indicate that the choice of $\beta$$_c$ has a non-linear influence on the dust temperatures derived, with a slightly larger impact in bright regions.

\begin{figure*} 
   \centering
  \includegraphics[width=13cm ]{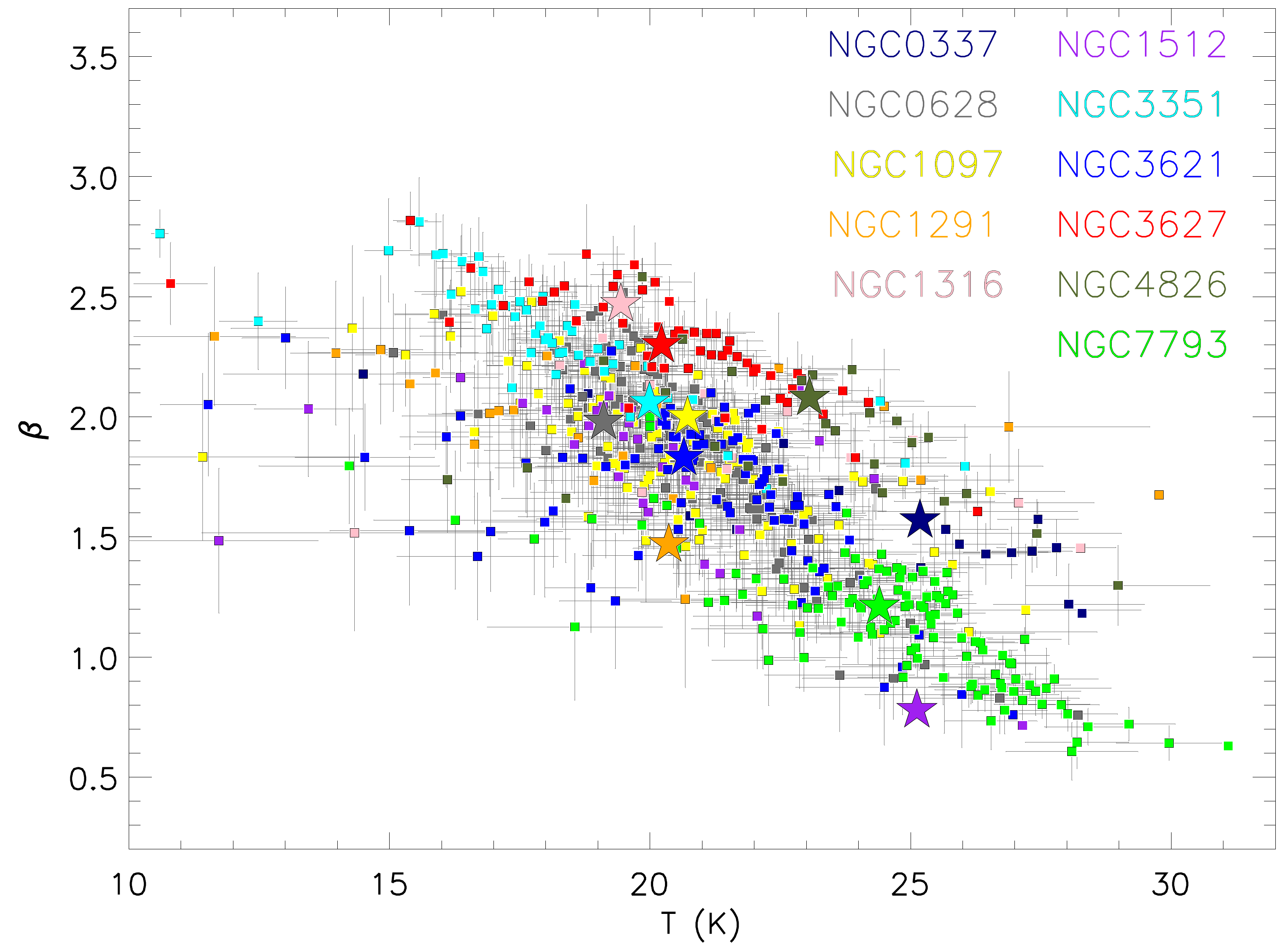}  \\
\caption{Temperature dependence of the spectral index in our sample. Global estimates are shown with stars. Squares are pixel-by-pixel values (pixel size: 18\arcsec) averaged 4-by-4 for clarity. }
  \label{Plot_T_B} 
\end{figure*}

\subsection{Investigations on emissivity index variations}

The coverage and resolution of \hersc\ enable us for the first time to properly constrain the submillimeter slope of resolved SEDs, allowing for more complex analysis of potential variations in the dust emissivity at local scales. To investigate these variations, we apply the same local MBB fitting technique {\it with both T$_c$ and $\beta$$_{c}$ as free parameters}. We show the emissivity index map and the cold temperature map of the galaxy NGC~3627 obtained with those assumptions in Fig.~\ref{NGC3627_Temp} (lower panels). We gather the cold dust temperature maps of the whole sample in Appendix Fig.~\ref{Prop_maps} (bottom left and middle middle panel for each galaxy). We overlay the MIPS 24 \mic\ contours as a comparison. The same color scale is used for the two cold dust temperature maps. We note that the choice of $\beta$$_c$ (2.0 or free) strongly influences the cold dust temperature ranges obtained for NGC~337 and NGC~7793. We used different color scales for these galaxies.\\

{\it Temperatures - }Choosing $\beta$$_c$ as a free parameter modifies the temperature distribution in our galaxies. The dust temperature distribution is less homogenous in the outer parts of galaxies (in NGC~628 or NGC~1097 for instance). This is probably linked to the fact that our model becomes very sensitive to noise at low-surface brightnesses and was previously observed in \citet{Planck_collabo_2011_MagellanicClouds} for the Large Magellanic Cloud and in \citet{Foyle2012} for M83. When $\beta$ is used as a free parameter in our model, we observe that the distribution of cold dust temperatures follows a systematic behavior with the morphology of the galaxy. Barred spiral galaxies such as NGC~1097, NGC~1291, NGC1512, NGC~3351 and NGC~3627 show distributions of their dust temperature map similar to those obtained with a fixed $\beta$$_{c}$, i.e. a distribution that follows the star forming regions or that is decreasing in the outer part of the galaxies. NGC~3627 presents an enhancement of the dust temperature at the end of its bar. Bars alter the abundance distribution of the gaseous component in galaxies \citep{Friedli1994} and affect the star formation activity \citep[][]{Ellison2011}. Strong concentrations of molecular gas are detected at the end of the bar of NGC~3627 \citep{Warren2010} which is probably responsible for the high star formation activity and thus for the high dust temperature observed. On the contrary, the grand-design spiral NGC~628 or bulge-less objects like NGC~3621 or NGC~7793 show cold dust temperatures that are homogenous throughout the structure of the galaxies. The fact that the cold dust temperature distribution does not seem to correlate with any dust heating source (neither star forming regions, nor old stellar populations or radius) within the galaxies is worrying and questions the use of a free $\beta$$_c$ in the model, at least for this type of galaxy.\\

{\it Emissivities - }Here again the emissivity index maps seem to show a distribution that depends on the morphology of the galaxy. Barred spiral galaxies show an homogeneous distribution of emissivity index, with median values similar to the $\beta$$_{c}$ values determined globally. On the contrary, we observe a radial decrease of the emissivity index in the non-barred galaxies NGC~628, NGC~3621 or NGC~7793 when $\beta$$_c$ is free. This radial decrease of $\beta$$_c$ was also observed in M33 by \citet{Tabatabaei2011} or in Andromeda by \citet{Smith2012}. We nevertheless note a clear inverse relation between temperature and emissivity index in some of these objects. We discuss this further in the following section. We do not observe a clear trend in NGC~337 and NGC~4826. Finally, we obtain low local emissivity values for the galaxies NGC~337 and NGC~7793 compared to the standard value of 2.0 as shown by the emissivity index ranges of each maps. This low emissivity indices are consistent with the integrated values obtained at global scales. Indeed, we find global emissivity indices $\beta$$_{c}$ of 1.6 and 1.2 for NGC~337 and NGC~7793 respectively). \\


\begin{figure*}
    \centering   
    \begin{tabular}{cc}
	\includegraphics[height=6cm]{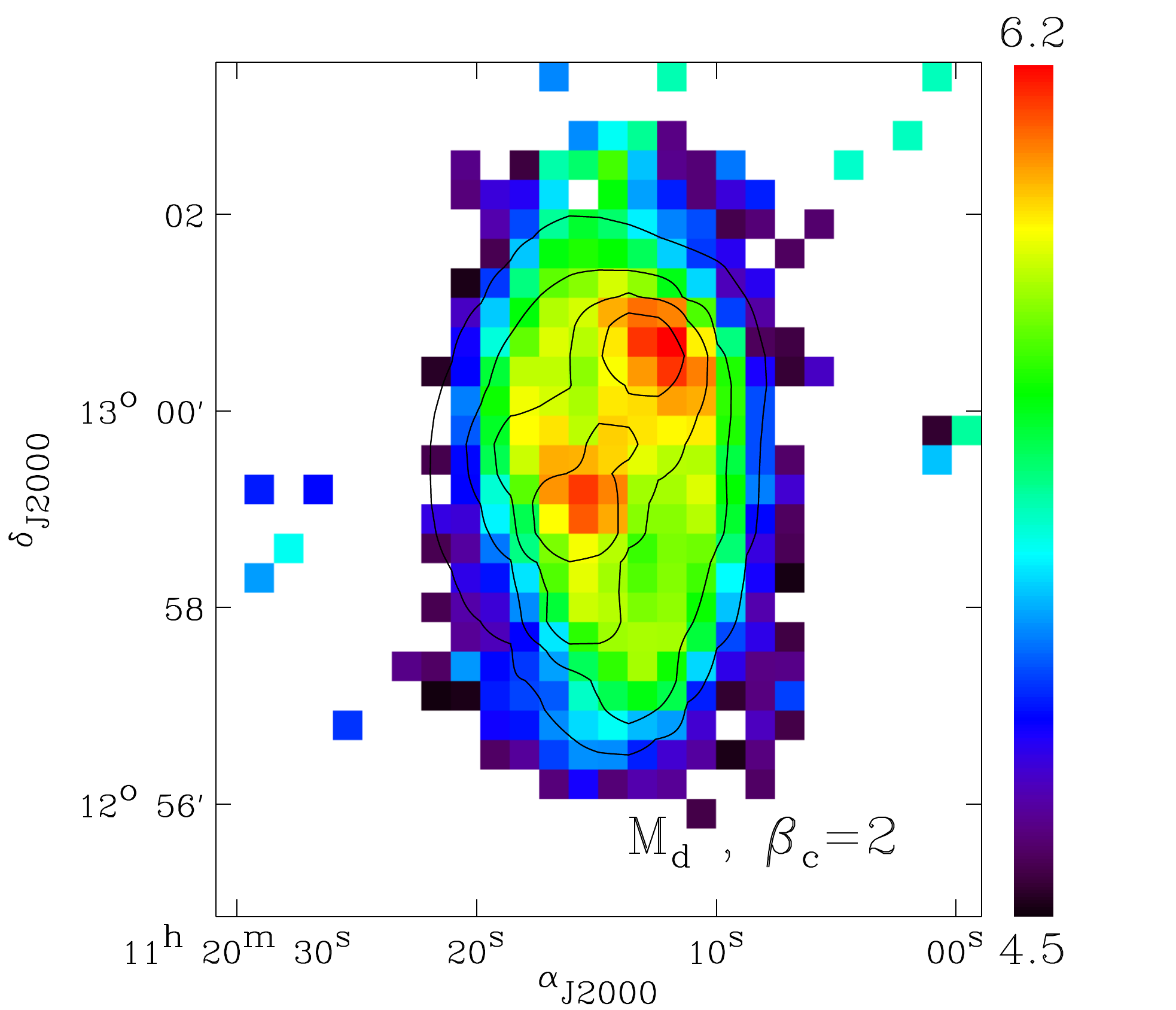} & 
		\includegraphics[height=6cm]{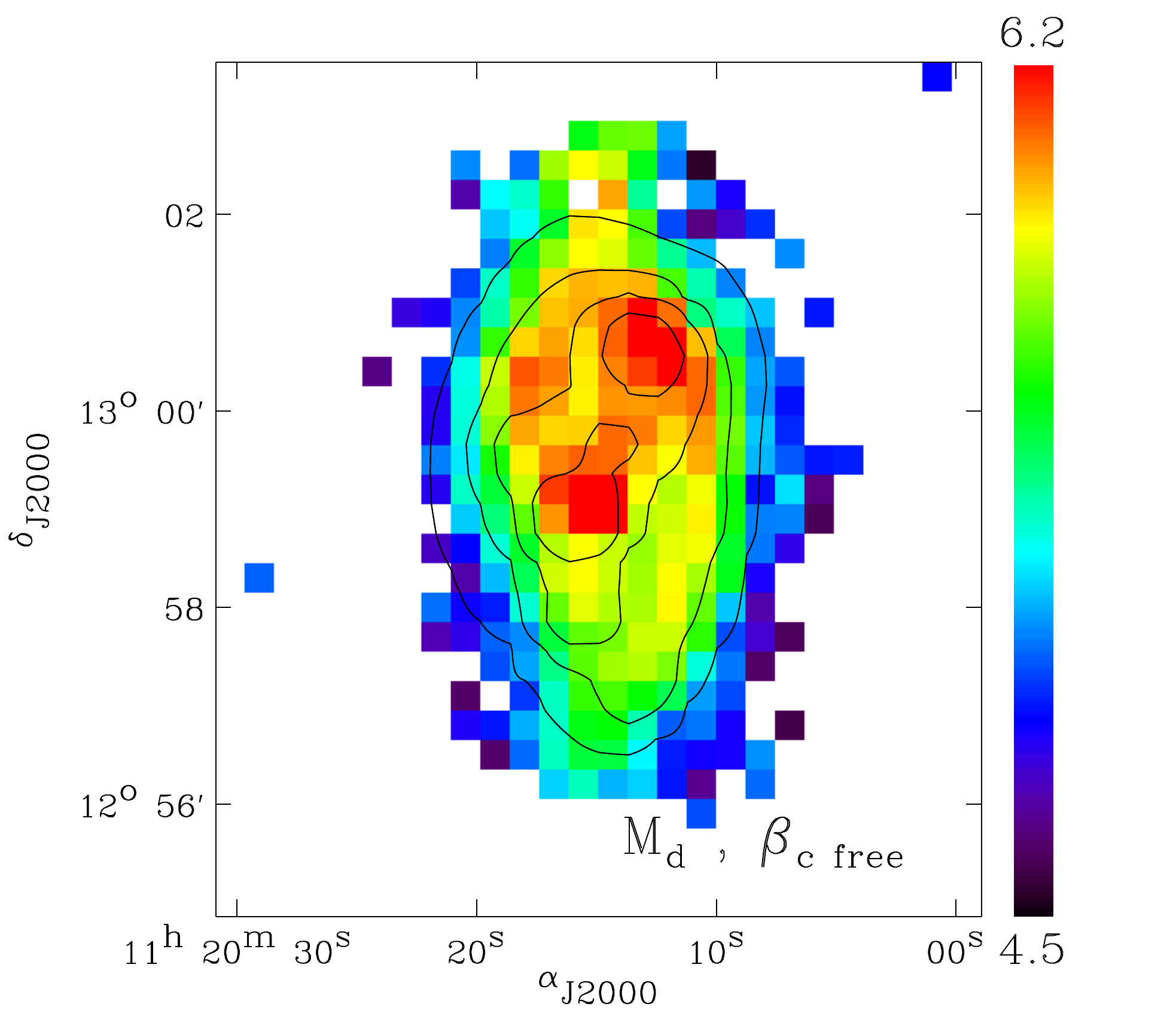} \\
   \end{tabular}
    \caption{Dust mass surface density map of NGC~3627 (in \msun~kpc$^{-2}$, log scale) derived with $\beta$$_c$ = 2.0 (left) and $\beta$$_c$ free (right). We overlay the MIPS 24 \mic\ for comparison. The dust mass surface density maps for the rest of the sample are available in the appendix of this paper.}
    \label{NGC3627_Dustmass}
\end{figure*}

{\it T$_{c}$ - $\beta$$_{c}$ anti-correlation - }Figure~\ref{Plot_T_B} gathers pixels that show a detection above our signal-to-noise threshold in each \hersc\ band for our 11 galaxies in a T$_{c}$ - $\beta$$_{c}$ diagram. Global values of Table~\ref{BB_results} are also overlaid with stars. We estimate the errors using the standard deviation of the Monte Carlo simulations of uncertainties performed for each pixel. They include calibration and sky background uncertainties. An anti-correlation is clearly observed between the temperature and the emissivity, with higher emissivity values in colder ISM elements. This relation is coherent with laboratory measurements \citep[][and references therein]{Mennella1998, Meny2007} that have reported a strong dependence between the temperature and the spectral index of the absorption coefficients, with higher $\beta$ values found for dust grains in cold cores.
The anti-correlation between T$_c$ and $\beta$$_c$ does not disappear if we restrict ourselves to the pixels showing the smallest uncertainties. This might indicate that Eddington bias (i.e. uncertainties on measurements) could not be sufficient to explain the anti-correlation we observe, even if this assumption is difficult to assess.
The discrepancies between the temperature maps derived with a fixed and free $\beta$$_c$, with temperature distribution not correlated with potential dust heating source in non-barred galaxies and the well-known degeneracy between temperature and emissivity index of course question the physical sense of letting both temperature and emissivity vary at the same time. We also caution the fact that even with the good resolution of \hersc\ instruments, the ISM elements studied in a 18\arcsec\ $\times$ 18\arcsec\ pixel are large for the galaxies of the sample, even for our closest object ($\sim$0.3 kpc for NGC~7793). We thus probe a large mixture of dust temperatures for each ISM element. It can therefore be challenging to disentangle a real relation or an anti-correlation created by the addition of several dust temperatures, with an ``artificial" flattening of the slope when cold dust is present along the line of sight. We refer to \citet{Juvela2012} for a detailed analysis of the possible effect of temperature mixing on the T - $\beta$ relation. A compromise to avoid these degeneracies for our non-barred galaxies could be to use $\beta$ as a free parameter to derive an average emissivity index across the galaxy and then fix $\beta$ to this value to derive temperature and mass surface density maps \citep[as done in][for instance]{Planck_collabo_2011_NearbyMolecularClouds}.
Studies on resolved structures in the Milky Way \citep[e.g.][]{Paradis2009} or in closer galaxies like the Magellanic Clouds are necessary to restrict the dust temperature range probed per resolved elements and contribute to resolve these issues.   \\

\begin{table*}
\caption{\large Global dust masses in units of 10$^5$ solar masses}
\label{Dust_masses}
 \centering
 \begin{tabular}{m{12mm}|ccc|ccccc|}
\hline
\hline
&&&&&&&&\\
	  & \multicolumn{3}{c|}{Global fit} & \multicolumn{5}{c|}{Sum of the dust masses obtained per pixel} \\
&&&&&&&&\\
	  	    & $\beta$$_{c}$ = 2.0   & $\beta$$_{c}$ free &            [D12] $^{1}$					& $\beta$$_{c}$ = 2.0     & $\beta$$_{c}$ = 1.5  & $\beta$$_{c}$ = 1.0      & $\beta$$_{c}$ free    & [A12b] $^{2}$\\
&&&&&&&&\\
 \hline
&&&&&&&&\\

NGC~337    & 361$^{+25}_{-22}$  & 
			205$^{+38}_{-28}$&
			225$^{+73}_{-73}$ 	&   
			
			347$^{+46}_{-37}$   &   
			180$^{+18}_{-16}$     &   
			94$^{+12}_{-9}$  &   
			214$^{+40}_{-32}$ &   
			240$^{+62}_{-62}$ (306) \rule[-7pt]{0pt}{5pt}\\
			
NGC~628    & 441$^{+20}_{-18}$ & 
			437$^{+66}_{-52}$& 
			291$^{+37}_{-37}$ 	&   
			
			381$^{+46}_{-36}$   &   
			207$^{+21}_{-17}$     &   
			113$^{+11}_{-9}$  &   
			371$^{+55}_{-46}$ &   
			329$^{+53}_{-53}$ (355) \rule[-7pt]{0pt}{5pt}\\ 
			
NGC~1097  & 
			1390$^{+73}_{-66}$ & 
			1380$^{+190}_{-152}$ & 
			904$^{+229}_{-229}$ 	&  
			 
			1309$^{+121}_{-109}$ &   
			749$^{+69}_{-67}$ &  
			415$^{+38}_{-37}$ &   
			1420$^{+231}_{-203}$ &  
			1241$^{+226}_{226}$ (1338) \rule[-7pt]{0pt}{5pt}\\
			
NGC~1291  & 
			283$^{+14}_{-13}$ & 
			151$^{+22}_{-17}$& 
			293$^{+79}_{-79}$ &   
			
			214$^{+45}_{-39}$   &   
			127$^{+23}_{-21}$     &   
			73$^{+12}_{-10}$  &   
			241$^{+44}_{-55}$ &   
			194$^{+106}_{-106}$  (422) \rule[-7pt]{0pt}{5pt} \\
			
NGC~1316  & 
			170$^{+11}_{-10}$ & 
			312$^{+95}_{-61}$  & 
			126$^{+13}_{-13}$ 	&   
			
			215$^{+22}_{-42}$   &   
			125$^{+14}_{-26}$     &   
			72$^{+11}_{-17}$  &   
			186$^{+30}_{-33}$ &  
			129$^{+21}_{-21}$  (183) \rule[-7pt]{0pt}{5pt}\\ 
			
NGC~1512  & 
			462$^{+32}_{-29}$  & 
			95$^{+22}_{-15}$ & 
			164$^{+19}_{-19}$ &   
			
			250$^{+37}_{-38}$   &  
			143$^{+20}_{-18}$      &   
			81$^{+9}_{-10}$  &   
			195$^{+34}_{-35}$ &   
			188$^{+35}_{-35}$  (231) \rule[-7pt]{0pt}{5pt}\\
			
NGC~3351  & 
			258$^{+13}_{-12}$ & 
			287$^{+37}_{-30}$&
			196$^{+20}_{-20}$ 	&   
			
			279$^{+28}_{-22}$   &   
			160$^{+14}_{-11}$     &  
			91$^{+9}_{-6}$  &   
			468$^{+75}_{-64}$ &   
			238$^{+34}_{-34}$  (283) \rule[-7pt]{0pt}{5pt}\\ 
			
NGC~3621  & 
			369$^{+19}_{-17}$ & 
			294$^{+40}_{-32}$& 
			242$^{+84}_{-84}$ &   
			
			358$^{+44}_{-36}$   &   
			191$^{+22}_{-18}$     &   
			103$^{+16}_{-12}$  &   
			283$^{+36}_{-31}$ &   
			333$^{+58}_{-58}$ (409) \rule[-7pt]{0pt}{5pt}\\ 
			
NGC~3627  & 
			601$^{+29}_{-27}$  & 
			892$^{+129}_{-109}$& 
			439$^{+97}_{-97}$ &   
			
			571$^{+24}_{-30}$   &   
			317$^{+19}_{-19}$     &   
			170$^{+19}_{-16}$   &  
			782$^{+80}_{-66}$ &   
			475$^{+99}_{-99}$  (543) \rule[-7pt]{0pt}{5pt}\\
			
NGC~4826  & 
			76$^{+4}_{-4}$  & 
			85$^{+12}_{-9}$ & 
			52$^{+6}_{-6}$ 	&   
			
			75$^{+5}_{-5}$     &   
			40$^{+5}_{-4}$        &   
			21$^{+4}_{-3}$   &  
			86$^{+8}_{-7}$  &   
			59$^{+9}_{-9}$  (61) \rule[-7pt]{0pt}{5pt}\\
			
NGC~7793  & 
			140$^{+8}_{-7}$ & 
			49$^{+7}_{-6}$ & 
			131$^{+28}_{-28}$ &   
			
			153$^{+15}_{-12}$   &   
			75$^{+7}_{-6}$        &   
			37$^{+3}_{-3}$   &  
			48$^{+5}_{-5}$  &   
			131$^{+42}_{-42}$ (145) \rule[-7pt]{0pt}{5pt}\\

&&&&&&&&\\
  \hline
\end{tabular}
\begin{list}{}{}
\item[$^{1}$] {\small Total dust masses derived by \citet{Dale2012} ([D12]) using the \citet{Draine_Li_2007} model to fit the global SED. }
\item[$^{2}$] {\small The dust mass maps are derived by [A12b] using the \citet{Draine_Li_2007} model applied pixel-by-pixel. The dust masses are derived adding the dust masses of pixels that are above our signal-to-noise threshold. The total dust masses they derive (whole galaxy) are given in parenthesis.}
\end{list}
 \end{table*} 
 
\section{Dust masses}
   
\subsection {Dust mass maps} 

Since most of the dust mass of a galaxy resides in its cold phase, we can directly derive dust mass maps from the cold temperature maps previously generated. To do so, we use the relation:

\begin{equation}
M_{dust}=\frac{S_{\nu_o} D^2}{\kappa_{\nu_o} B_{\nu_o}(T_c)}
\end{equation}

\noindent with {\it S$_{\nu}$} the flux density, {\it $\nu_o$} the reference frequency, {\it $\kappa$$_{\nu_o}$} the absorption opacity of dust grains at that reference frequency, {\it D} the distance to the galaxy and {\it B(T$_c$)} the Planck function for a given temperature {\it T$_c$}. \citet{Dale2012} derive global dust masses of the KINGFISH sample from MBBs using the reference frequency $\nu_o$ = c / 250 \mic\ and the SPIRE 250 \mic\ flux densities. Similar techniques have been used to derive dust mass surface density maps in very nearby galaxies by \citet{Foyle2012} for M83, \citet{Parkin2012} in Centaurus A or Mentuch et al. (submitted to ApJ) for the Whirlpool system. Since the 500 \mic\ band is less sensitive to temperature variations than the 250 \mic\ band, we prefer to use $\nu_o$ = c / 500 \mic\  as the reference frequency. We note that using 500 \mic\ as a reference wavelength for dust calculation could make our masses dependent on submm excess if present in our objects. However, we do not expect submm excess to be significant at 500 \mic\ for our selection of mostly metal-rich galaxies \citep{Draine2007,Gordon2010,Galametz2011}. We choose a dust opacity $\kappa_{abs}$(250 \mic)=3.98 cm$^2$~g$^{-1}$ (taken from www.astro.princeton.edu/$\sim$draine/dust/dustmix.html and a Milky Way R$_V$=3.1 model) and for each case ($\beta$ fixed or free) derive the $\kappa_{abs}$(500 \mic) such as:

\begin{equation}
\kappa_{abs}(500~\mu m) = \kappa_{abs}(250~\mu m) \times \left(\frac{500}{250}\right)^{-\beta}
\end{equation}

In the Milky Way R$_V$=3.1 model we use, $\beta$ is $\sim$2 in the 100-1000 \mic\ range. Technically, our choice of fixing the scaling term of our dust masses from the reference wavelength 250 \mic\ depend on where the intrinsic emissivity deviates from this model. \citet{Li_Draine_2001} made small modifications to their amorphous silicate opacity to better match the average high Galactic latitude spectrum measured by COBE-FIRAS from $\lambda$$>$250 \mic. We thus choose to modify our emissivity from the same wavelength.
We show the dust mass surface density map obtained for the galaxy NGC~3627 with $\beta$$_c$=2.0 and $\beta$$_c$ free (in \msun~kpc$^{-2}$, log scale) in Fig.~\ref{NGC3627_Dustmass}. We overlay MIPS 24 \mic\ contours for comparison. We gather the dust mass surface density maps (with $\beta$$_c$=2.0 and free) for the whole sample in appendix in Fig.~\ref{Prop_maps} (right panels for each galaxy). We observe that peaks of this distribution are usually located in the center of the galaxies (NGC~337, NGC~3351 for example). Nevertheless, peaks are also found at the bar ends in NGC~3627, coinciding with peaks in the molecular hydrogen detected through CO observations \citep{Leroy2009}, linking the dust concentrations with the cold gas reservoirs. This result is similar to that of \citet{Foyle2012} who find that the highest dust concentration in M83 is located in the nucleus and at the ends of the bar of the galaxy. We observe peaks of the dust mass surface density in the northeastern and southwestern arms of NGC~1512, here again at the end of the bar where the dust emission peaks from \spitz/IRAC to \hersc/SPIRE wavelengths. Unfortunately, no CO observations are available for this galaxy that would allow comparisons with the molecular gas distribution.  We observe high concentrations of dust at the northwest and southeast sides of the nucleus in NGC~1316, which correspond to peaks of dust emission at PACS and SPIRE wavelengths (see the resolved blobs in Fig.~\ref{Images}). These dusty structures coincide with reservoirs of molecular hydrogen detected in CO \citep{Lanz2010}. \citet{Horellou2001} have suggested that this molecular gas could have an external origin (accretion of material during merging processes). We however remind the reader that NGC~1316 possesses a low-luminosity X-ray active galactic nucleus (AGN). Our submillimeter wavelengths thus potentially contain non-thermal emission from the AGN that could affect the SEDs and the dust mass map we obtain, as discussed in \cite{Bendo2011} for M81 or in \citet{Parkin2012} for Centaurus A. A refinement of our resolved fit combining non-thermal emission from the AGN and dust associated with the X-ray bubbles would be necessary to understand how this influences the distribution of the dust mass surface density in this object.

We note that a non-negligible amount of dust is detected in the ring of NGC~1291, with higher dust mass surface densities than in the nucleus in the west arm in particular. \citet{VanDriel1988} observations of H{\sc i} in this galaxy have revealed a concentration toward the ring and a central hole in the distribution. Dust thus seems to trace the atomic gas in this galaxy. A proper comparison of the ring/nucleus temperature and masses in NGC~1291 is detailed in Hinz et al. (submitted to ApJ). We finally note that the choice of $\beta$$_c$ (fixed or free) does not affect the locations of the dust mass surface density peaks in our galaxies but directly influences the total dust masses derived for each galaxy, as shown in the following section.

\subsection {Dependencies of the derived total dust masses}

In this final section, we aim to quantitatively compare the total dust mass derived for our objects with the different modelling methods used in this paper. Table~\ref{Dust_masses} summarizes the total dust masses obtained by adding the individual dust masses determined pixel-by-pixel with our two-MBB modelling technique in the four different cases $\beta$$_c$=2.0, 1.5, 1.0 or free. The global fit section indicates the total dust masses calculated from temperatures derived with a two-MBB fit of the integrated fluxes. The errors on the dust mass are directly deduced from the temperature errors, the main source of uncertainties for our the dust estimates. The non-linearity of the Planck function at IR wavelengths is translated into the asymmetric errors of our dust mass estimates. For the local MBB fits, the final errors are obtained by summing the individual uncertainties. [A12b] perform a pixel-by-pixel SED modeling of the KINGFISH galaxies using the \citet{Draine_Li_2007} model and produce dust mass maps for the whole sample. We sum up the local dust masses they obtain in the regions we selected (3-$\sigma$ detections in all \hersc\ bands) in our galaxies and add these values to Table~\ref{Dust_masses} for comparison.

\begin{figure}
   \centering
   \begin{tabular}{c}
      \includegraphics[width=8.5cm]{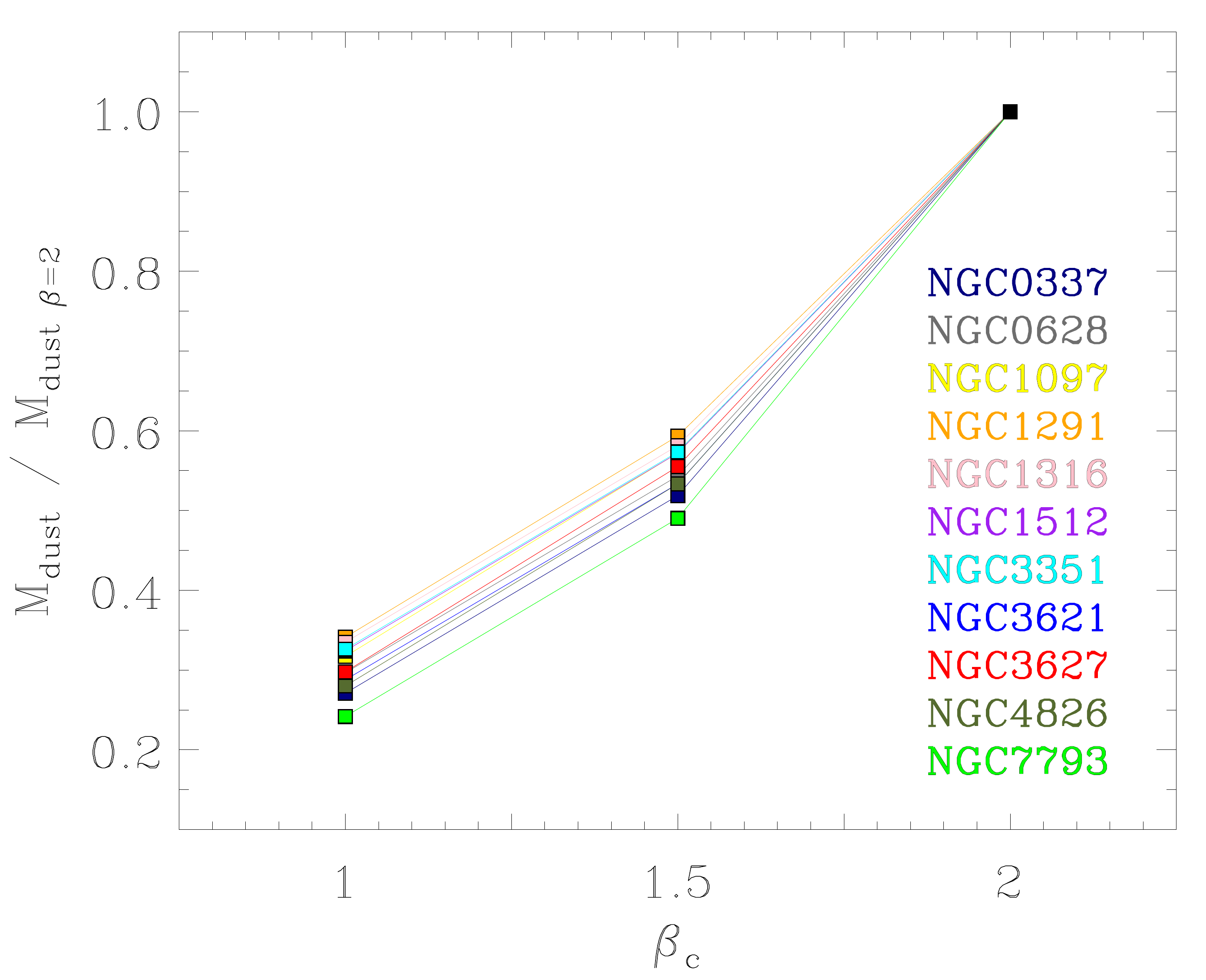} \\
   \end{tabular}
\caption{Dependence of the total dust mass with the cold dust emissivity index. The x-axis indicates the emissivity index in the modelling. Dust masses are normalized to those deduced with a standard emissivity index of 2.0. }
  \label{Mdust_beta_dep}
\end{figure}

\subsubsection {Influence of the emissivity index} 

Emissivity spectral indices ranging between 1.0 and 2.0 are commonly used in the literature. Table~\ref{Dust_masses} highlights the strong dependence of the total dust mass estimate with the choice of $\beta$$_c$. Figure~\ref{Mdust_beta_dep} directly illustrates the decrease of the dust mass with the emissivity index. We note for instance that MBB fits using a cold dust emissivity index of 1.5 lead to a decrease of the total dust mass of $\sim$50 $\%$ compared to those performed with a standard emissivity index of 2.0 and that MBB fits using an emissivity index as low as 1.0 will lead to a decrease of $\sim$70 $\%$. For NGC~337 and NGC~7793, the distributions of dust temperatures and emissivity indices derived from the Monte Carlo simulations on integrated fluxes (see Fig.~\ref{T_B_errors}) seem to rule out the use of a single temperature dust model with $\beta$$_c$=2.0 to model the cold dust emission at global scale. At local scales, the cold dust emissivity maps of NGC~337 and NGC~7793 show that $\beta$$_c$ is always lower than 2.0, leading to the same conclusion. We note that a MBB fit with $\beta$$_c$=2.0 is not a valid assumption at global scale for NGC~1512 either. The emissivity map does not, however, show such a trend at local scales. We conclude that the choice in the emissivity index value has a direct and significant impact on the dust mass deduced and thus on the interpretation made from these values (gas-to-dust mass ratios for instance). It also complicates direct comparisons between studies that call for different models and assumptions.

\begin{figure}
   \centering
   \begin{tabular}{m{9cm}}
  \includegraphics[width=8.5cm ]{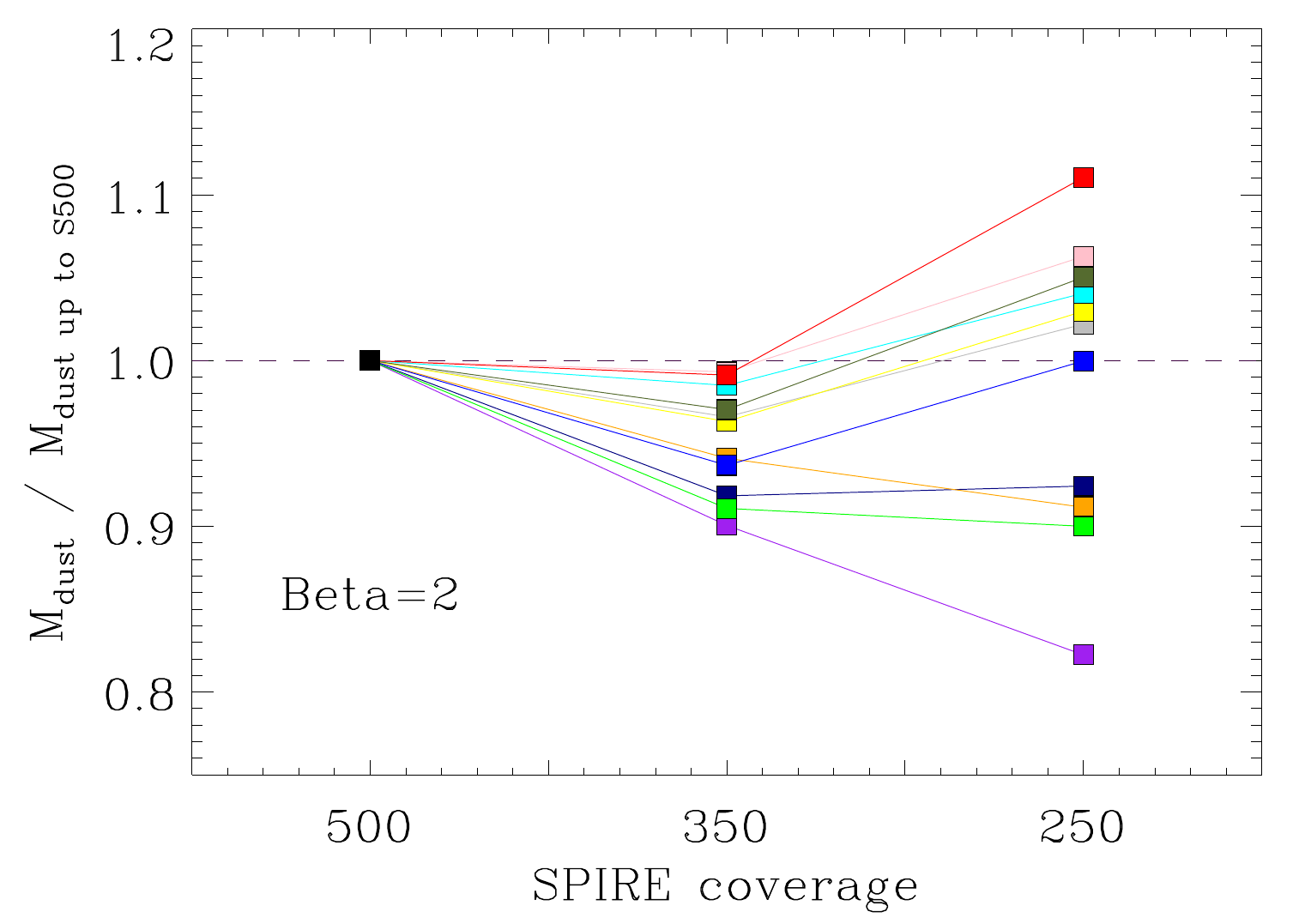} \\
    \includegraphics[width=8.5cm ]{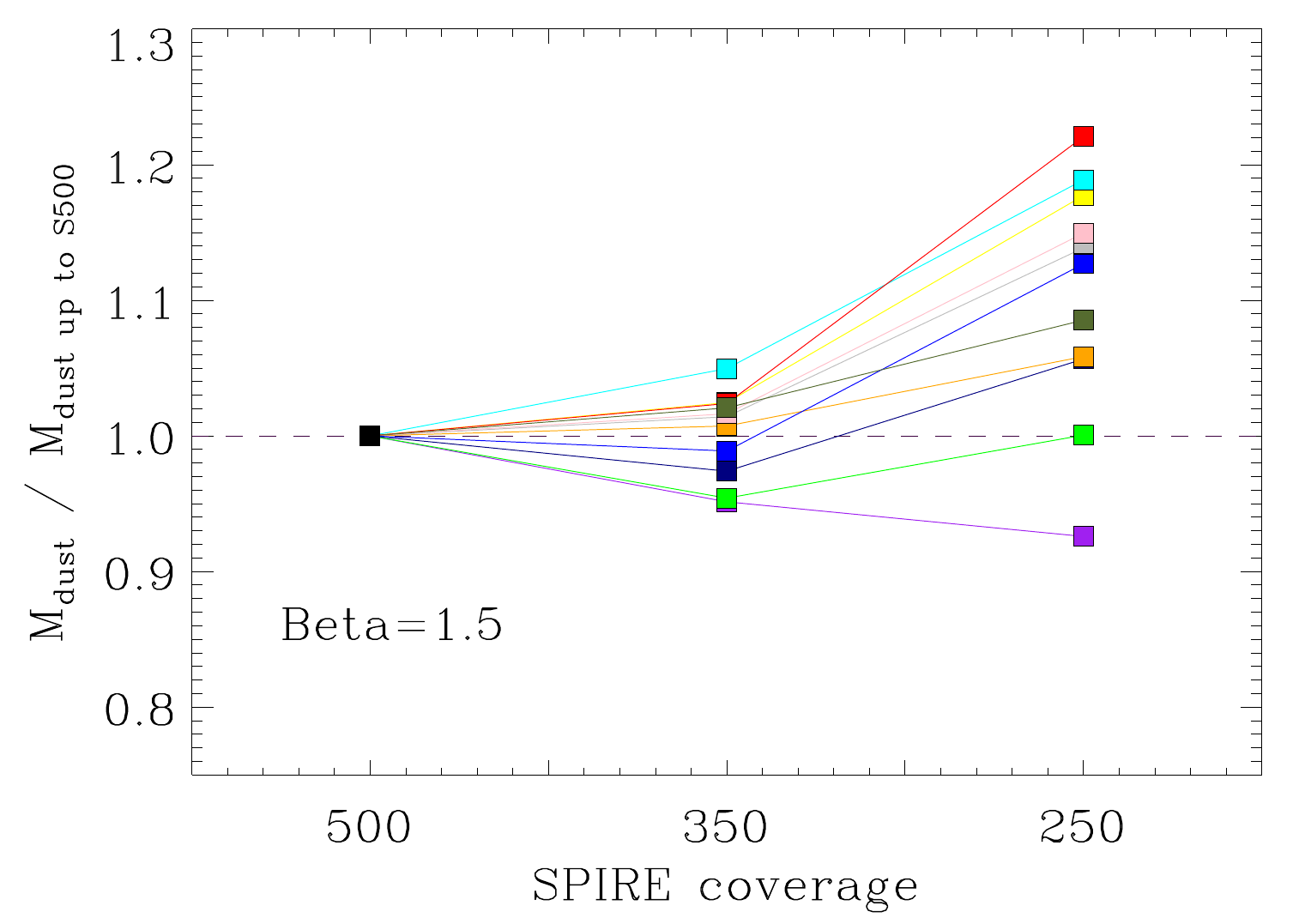} \\
      \includegraphics[width=8.5cm ]{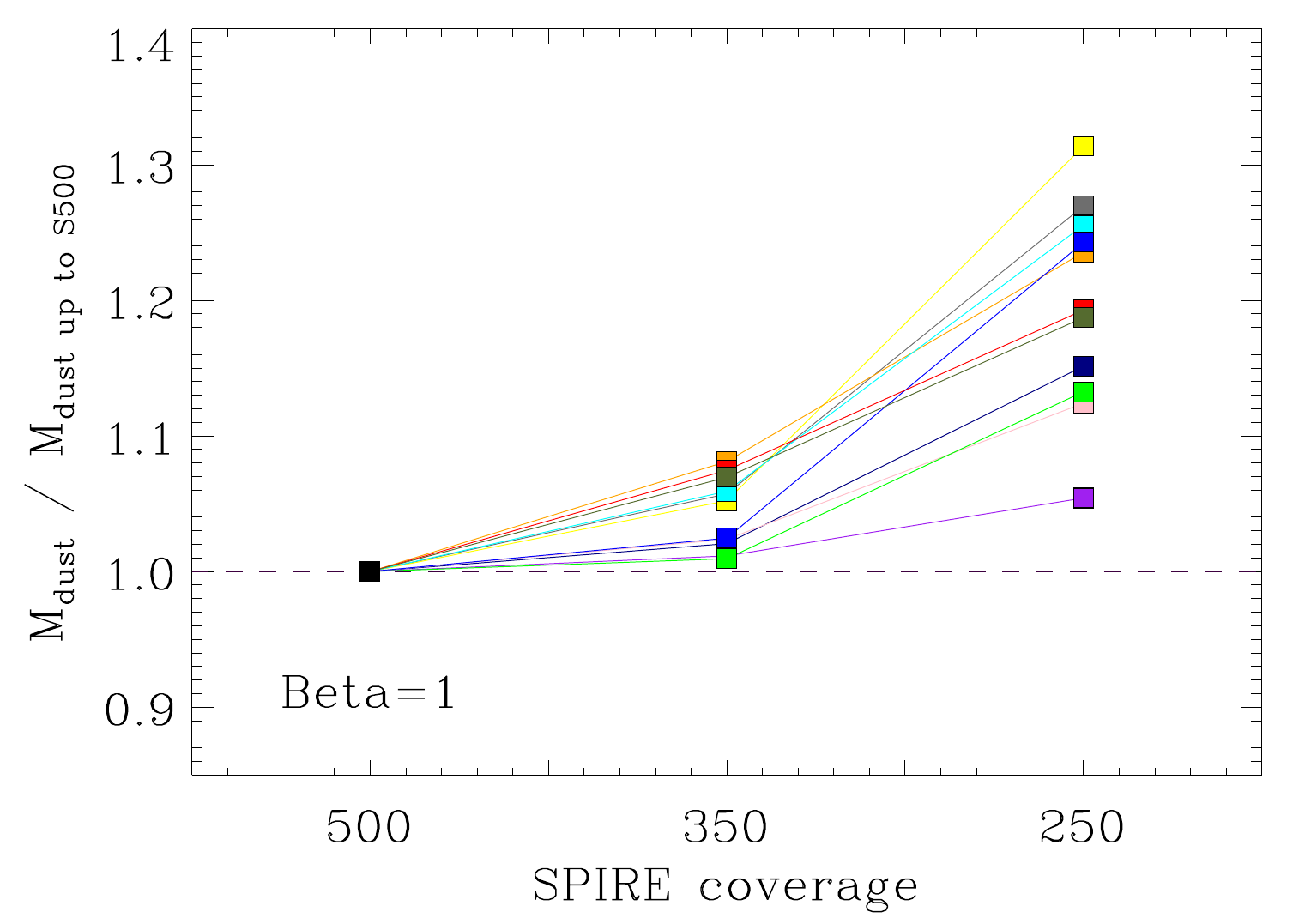} \\
   \end{tabular}
\caption{Dependence of the total dust mass with submm data included in the fits (two-MBB technique). The x-axis indicates the longest wavelength used. For instance, `350' means that data from 24 to 350 \mic\ are used. Dust masses are normalized to dust masses obtained using a complete coverage from 24 to 500 \mic\ (called ``M$_{dust~up~to~500}$"). From top to bottom, $\beta$$_c$ was fixed to 2.0, 1.5 and 1.0. For convention on galaxy colors, see Fig.~\ref{Mdust_beta_dep}. }
  \label{Mdust_submm_dep}
\end{figure}

\subsubsection {Influence of the SPIRE data} 

Several studies have stressed the importance of observations longward of 160 \mic\ to properly determine dust temperature and masses \citep[][]{Gordon2010}. \citet{Galametz2011} show that metal-rich galaxies usually have dust SEDs that peak at long wavelengths and require constraints above 160 \mic\ to position the peak and sample the submm slope of their SEDs and thus correctly probe the dust temperature distribution. On the other hand, metal-poor galaxies modeled with submm data could show higher dust masses than when modeled without submm data, due to the flattening of the submm slope or submm excess often detected in those objects. We can directly compare our dust masses derived with submm data and those derived with data up to the 160 \mic\ constraint of \spitz. We observe that dust masses obtained using \hersc\ data can differ from those derived with \spitz-based fits available in \citet{Draine2007}, especially for the galaxies NGC~337, NGC~628, NGC~1097 and NGC~1316. Do we see any trend with metallicity? Most of the galaxies of the sample show a decrease in the total dust mass. NGC~1316, the most metal-rich galaxy of the sample, shows the second highest decrease in our dust masses. The total dust masses of NGC~3621 or NGC 7793 - galaxies that have lower metallicities - are, on the contrary, rather unaffected by the addition of \hersc\ data. Of course, any statistical interpretation of metallicity effects is rather hazardous for this small sample. Nevertheless, these trends with metallicity are consistent with the dependence expected from \citet{Galametz2011} and are already mentioned in the study of \citet{Dale2012} performed for the whole KINGFISH sample. To test the influence of SPIRE data on the dust mass we obtain at global scale, we apply our two-MBB fitting procedure on data up to 250 \mic\ or data up to 350 \mic. The emissivity index is fixed to 2.0, 1.5 or 1.0. Figure~\ref{Mdust_submm_dep} shows how the total dust mass estimates evolve with the number of SPIRE submm data included in the fits. We observe that two MBB fits with $\beta$$_c$=2.0 performed with data up to 250 \mic\ could differ from those derived using the complete SPIRE coverage by $\pm$10$\%$ (-17$\%$ for NGC~1512 due to the very flat submm slope of this object not sampled when we restrict the fit to data below 250 \mic).  [A12a]  performed comparisons between dust masses derived with data up to 160 \mic\ or data including the full SPIRE coverage for NGC~628 and NGC~6946 and found similar values (difference of $\sim$ 16$\%$ and 10$\%$ respectively). This difference is even more pronounced with $\beta$$_c$=1.5 or 1, reaching up to 30 $\%$ in the second case. We finally note that using a two-MBB fit and data up to 350 \mic\ lead to a systematic underestimation of the dust mass (by up to 10$\%$) if $\beta$$_c$ is fixed to 2 for the galaxies of our sample.

\subsubsection {Comparisons with \citet{Skibba2011} } 

\citet{Skibba2011} have estimated dust masses for KINGFISH galaxies using single modified blackbody fits with an emissivity index $\beta$ of 1.5 and, as in this study, SPIRE 500 \mic\ flux densities as a reference wavelength in their dust mass calculation. From our global two-MBB fits with $\beta$$_c$ fixed to 1.5, we obtain dust masses that are larger by a factor of 1.1 to 2.6 than those obtained with the same cold dust emissivity index by \citet{Skibba2011}. This is probably linked with their use of a unique modified blackbody to fit the 70-to-500 \mic\ range that leads to systematically higher temperature estimates for the cold dust (warm dust contributing to the 70 \mic\ flux as mentioned in $\S$3.4) and thus to lower dust masses.

\subsubsection {Comparisons between MBB fits and the \citet{Draine_Li_2007} model} 

We observe that, as already noticed in \citet{Dale2012}, global dust masses derived with the \citet{Draine_Li_2007} model only differ from models using MBBs and a standard emissivity index of 2.0 by less than 35$\%$. A local estimation of the dust masses leads to similar results, namely an agreement within 30$\%$. It can be explained by the fact that in the wavelength range covered by SPIRE bands, the SED of the diffuse ISM component predicted by the \citet{Draine_Li_2007} model (corresponding to the bulk of the dust mass in these models) is quite similar to a single temperature modified blackbody with $\beta$$_c$ = 2. We nevertheless note a significant difference between the dust mass obtained with a global fit for NGC~1512 when $\beta$$_c$ = 2.0 and that obtained by \citet{Dale2012}. As mentioned previously, this galaxy shows a flat submm slope (global $\beta$$_c$ $<$ 1.0, Table~\ref{BB_results}). By fixing $\beta$$_c$ to 2.0, we implicitly let our two-MBB model invoke colder dust to account for the SPIRE 500 \mic\ emission. We observe that the \citet{Draine_Li_2007} model used to fit the global SED of this galaxy in \citet{Dale2012} tends to under-predict the SPIRE 350 and 500 \mic\ fluxes for this galaxy, which probably leads to a underestimation of the cold dust mass in this particular object. It is thus not surprising that we derive high dust masses for this galaxy. This could imply that cold dust could be present in this galaxy (and thus more dust mass), especially in the outer regions of the galaxy (not modeled in this study if the detection is below our chosen signal-to-noise threshold) or that the global emissivity index is not well reproduced by standard graphite grains used to model carbon dust in the \citet{Draine_Li_2007} model.\\

\subsubsection {Resolution effects} 

We want to compare our global dust masses with our resolved dust masses. In our local study, we restrict ourselves to regions above our signal-to-noise threshold in all \hersc\ bands in our galaxies. Unfortunately, we are consequently missing dust mass in the faint outer parts of the galaxies. Making a direct comparison with the total dust mass derived from integrated fluxes and the sum of the individual dust masses obtained in the different pixels is thus not directly possible from this study. Nevertheless, we observe that the dust masses obtained globally using the \citet{Draine_Li_2007} model by \citet{Dale2012} are systematically lower that those derived by summing the local dust masses (Table~\ref{Dust_masses}, last column, in parenthesis), by 11 $\%$ for NGC~7793 to 69$\%$ for NGC~3621, one of the lowest metallicity of the sample. Since dust masses derived using a MBB technique and an emissivity index of 2.0 are very close to those obtained using the \citet{Draine_Li_2007} model, we can easily predict the same effect for studies using MBBs, models that are intensively used by the scientific community. 
Those results agree with \citet{Galliano2011} who observe similar resolution effects while deriving dust masses in a strip of the LMC. The dust masses they derive from integrated fluxes are significantly lower than dust masses obtained from a local scale study. The dilution of cold massive regions in hotter regions is invoked as a possible explanation of this effect. If this hypothesis is valid, we should expect an increase of our dust mass estimates until the resolution of our submm maps reach scales where cold dust dominates (size of giant molecular clouds?). Observations at higher resolution, with ALMA for instance, will help to address this issue. These differences of total dust mass estimates with resolution highlight the necessity of a resolved study to properly quantify the total dust mass budget of galaxies, when it is possible. Global dust mass of high redshift galaxies might be underestimated because of the lack of resolution. Moreover, our sample does not cover the full diversity of the SEDs of distant galaxies, especially the very luminous IR galaxies detected at high-redshift. In some cases, resolution effects may be even more significant. The choice of the fitting method therefore becomes decisive.  \\


\section{Conclusions}

We combine \spitz/MIPS and \hersc\ data to derive global properties of the cold dust in a sample of 11 nearby galaxies using a two-temperature fit to model their SEDs. At global scales,

\begin{itemize}
  \item Our objects show a wide range of SEDs, with a clear evolution of the 100-to-24 \mic\ flux density ratios from actively to quiescent star-forming objects and variations on the global cold dust parameter $\beta$$_c$ we derive, from $\beta$$_c$ = 0.8 to 2.5.
  
  \item We perform a Monte Carlo simulation in order to quantify the effect of uncertainties on the parameters derived from our fitting technique. We show that even with an exhaustive sampling of the global thermal dust emission, an artificial T-$\beta$ anti-correlation can be created by uncertainties in the flux measurements.
  
  \end{itemize}

We use the unprecedented spatial resolution of \hersc\ to derive resolved dust properties within our objects and study the robustness of our result with our SED model assumptions (emissivity index, SED coverage, resolution). Our study at local scales shows that:

\begin{itemize}

  \item  When we fix the cold emissivity index to a standard $\beta$$_c$=2.0, we obtain a smooth distribution of the cold temperatures, with a radial decrease toward the outer parts of galaxies. In some objects, the temperature distribution follows the distribution of star-forming regions (NGC~3627 for instance). Fixing $\beta$$_c$ to 1.5 or 1.0 influences the temperature range derived but does not strongly modify the temperature relative distribution.
  
  \item When $\beta$$_c$ is free, barred spirals (e.g. NGC~1097, NGC~1316, NGC~1512, NGC~3351 or NGC~3627) show temperature distributions that are similar to those obtained with a fixed $\beta$$_c$ and homogeneous emissivity index values across the galaxies. For non-barred galaxies, $\beta$$_c$ seems to decrease with radius. We nevertheless obtain homogeneous cold temperature distributions for those objects. The fact that these temperature maps do not seem to relate to any dust heating sources questions the use of a free $\beta$ in those objects. Applying a $\beta$-free model to deduce an average emissivity index across the galaxy and derive the temperature map using this median value could help to avoid potential degeneracies between T and $\beta$ in those objects.
 
 \item Gathering our local results in a T$_c$-$\beta$$_c$ diagram clearly leads to a correlation between the two parameters. This correlation could be linked with {\it 1)} a real emissivity variation of dust grains with temperature {\it 2)} noise effects and uncertainties on flux measurements {\it 3)} temperature mixing along the line of sight.

  \item The dust mass surface density usually peaks in the center of galaxies but also seems to coincide with extra-nuclear peaks of star formation at the end of the bars (usually corresponding to molecular gas reservoirs). 
    
  \item The dust mass estimates seem to be affected by resolution, with a systematic increase of total dust masses in the ``resolved" case compared to dust masses globally determined. 
  
 \item The submm coverage of our SEDs with \hersc\ observations modifies the dust masses compared to those derived from \spitz-based fits. Two MBB fits performed with data up to 250 \mic\ differ from those derived using the complete SPIRE coverage (data up to 500 \mic) by up to 17$\%$ if $\beta$$_c$=2.0, up to 30$\%$ if $\beta$$_c$=1.0.
 
  \item Dust masses derived globally or on a local basis using the \citet{Draine_Li_2007} model only differ from models using modified blackbodies and $\beta$$_c$=2.0 by less than 30$\%$. Dust masses derived from modified blackbody fits using a shallower parameter $\beta$$_c$ are systematically lower that those derived with $\beta$$_c$=2.0 (lower by 25 to 46 $\%$ if $\beta$=1.5 for instance in our sample). Our results highlight the necessity to better investigate the variations of the cold dust properties to properly quantify the total dust mass if we aim for instance, to use it as a gas tracer. 

\end{itemize}

This study aimed to investigate the properties of cold dust unveiled by \hersc\ observations of nearby galaxies. Questions remain due to possible degeneracies between the grain emissivities and temperatures. Moreover, this paper uses isothermal fits, which implies that the intrinsic emissivity index of the dust grains and the emissivity index directly obtained from our fits are one and the same. The presence of several dust grain temperatures in our objects both locally and at global scales also makes a direct extrapolation of dust properties difficult. Nevertheless, combining those results with other derived quantities such as the elemental abundances in the ISM or reliable gas-to-dust mass ratios as well as further investigations on closer objects will help us to disentangle between the different hypotheses.


\section*{Acknowledgments}

We would first like to thank the referee for his/her helpful comments that helped to improve the clarity of the paper. The research of C.D.W. is supported by grants from the Natural Sciences and Engineering Research Council of Canada.
PACS has been developed by MPE (Germany); UVIE (Austria); KU Leuven, CSL, IMEC (Belgium); CEA, LAM (France); MPIA (Germany); INAF-IFSI/OAA/OAP/OAT, LENS, SISSA (Italy); IAC (Spain). This development has been supported by BMVIT (Austria), ESA-PRODEX (Belgium), CEA/CNES (France), DLR (Germany), ASI/INAF (Italy), and CICYT/MCYT (Spain). 
SPIRE has been developed by a consortium of institutes led by Cardiff Univ. (UK) and including: Univ. Lethbridge (Canada); NAOC (China); CEA, LAM (France); IFSI, Univ. Padua (Italy);IAC (Spain); Stockholm Observatory (Sweden); Imperial College London, RAL, UCL-MSSL, UKATC, Univ. Sussex (UK); and Caltech, JPL, NHSC, Univ. Colorado (USA). This development has been supported by national funding agencies: CSA (Canada); NAOC (China); CEA, CNES, CNRS (France); ASI (Italy); MCINN (Spain); SNSB (Sweden); STFC, UKSA (UK); and NASA (USA).


\bibliographystyle{mn2e}
\bibliography{mybiblio.bib}

\appendix
\section{Dust temperature, emissivity index and dust mass surface density maps}

\begin{figure*}
    \centering
    \begin{tabular}{m{5.4cm} m{5.4cm} m{5.4cm}}
   {\large NGC~337} &&  \\
   \centering{\large MIPS 24 \mic} & 
   \centering{\large Temperature map ($\beta$$_c$=2)} & 
   \centering{\large Dust mass surface density ($\beta$$_c$=2)} \\
   	\tabularnewline
	\includegraphics[height=4.3cm]{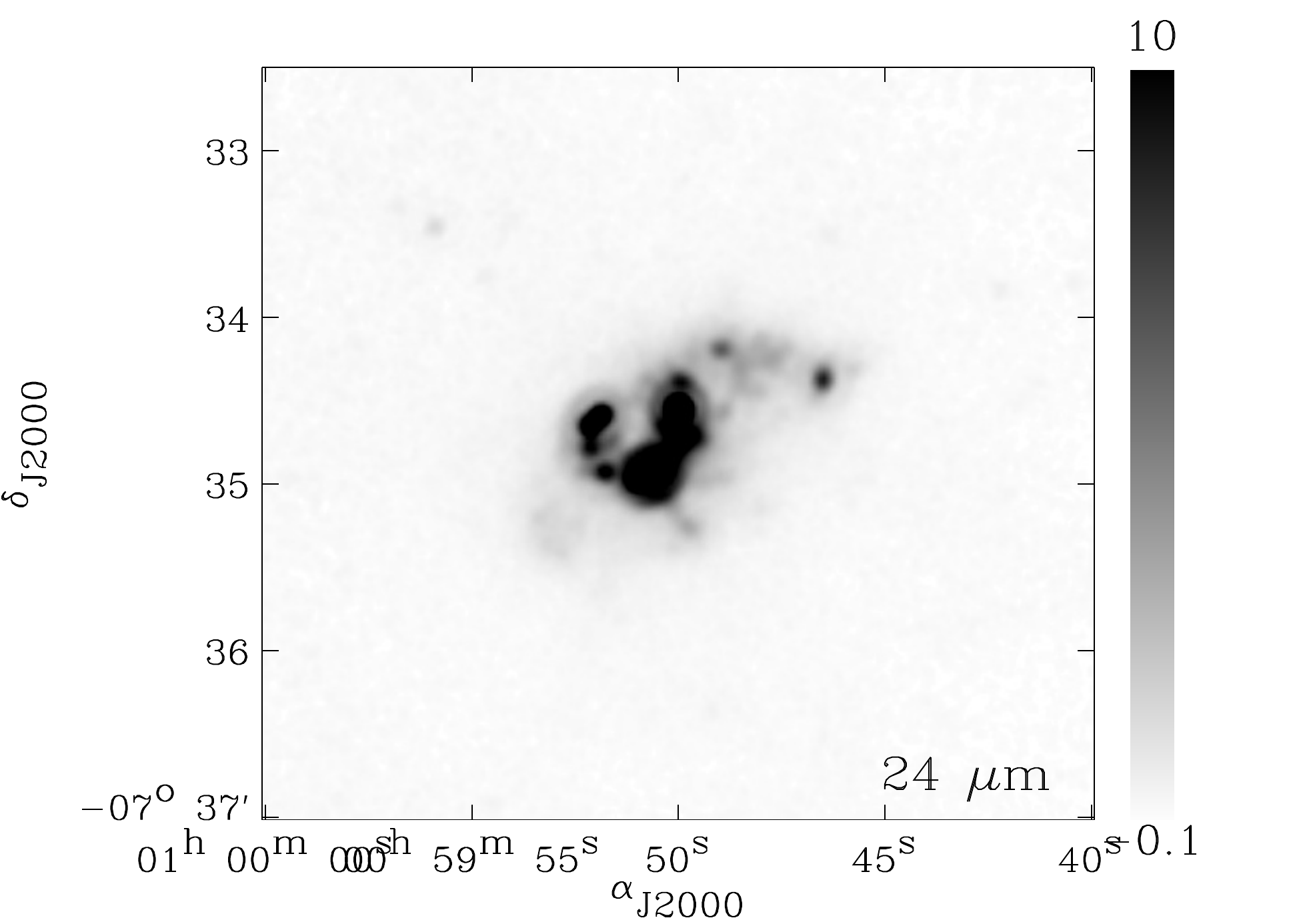} &
	\includegraphics[height=4.3cm]{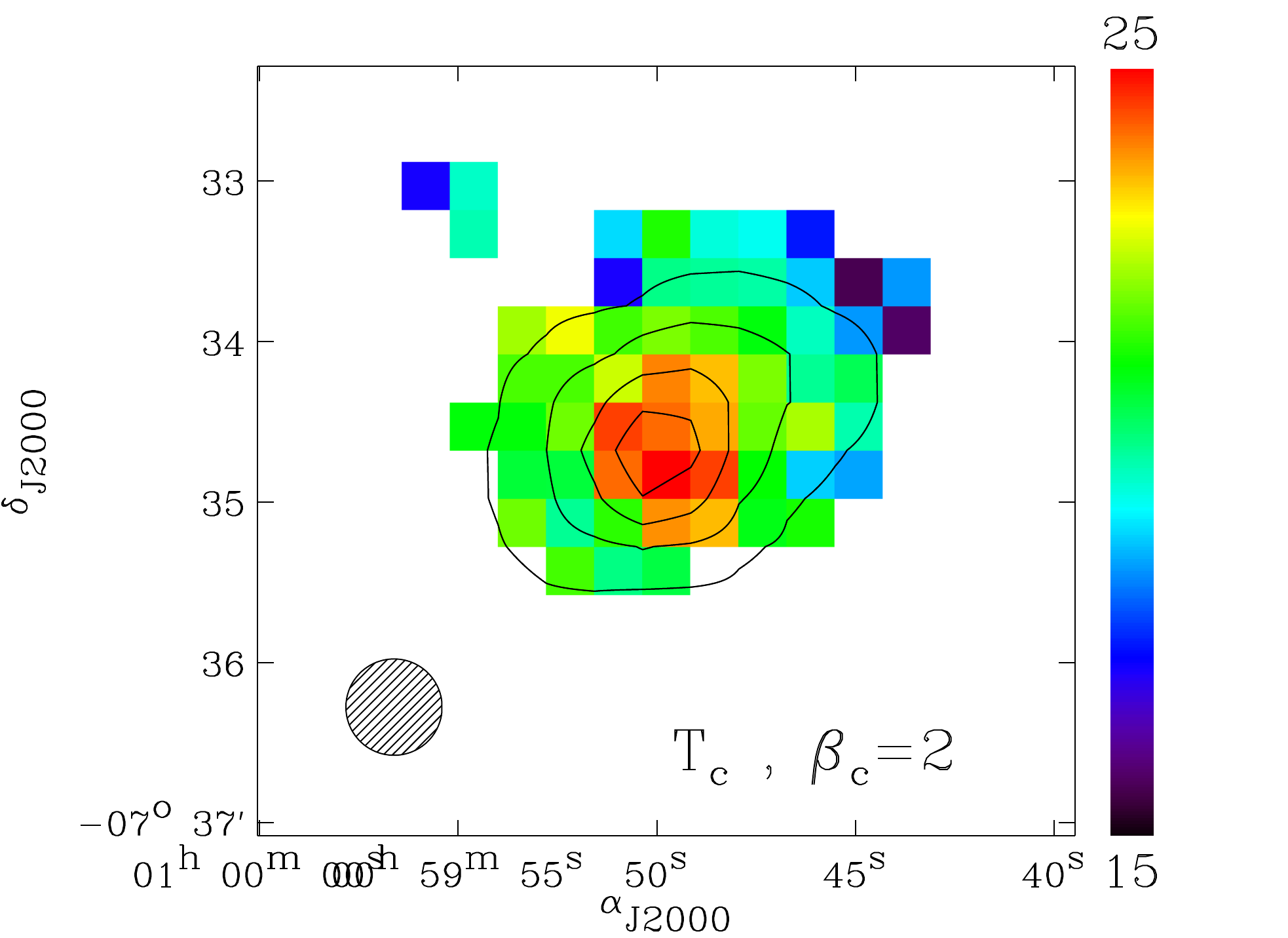} &
	\includegraphics[height=4.3cm]{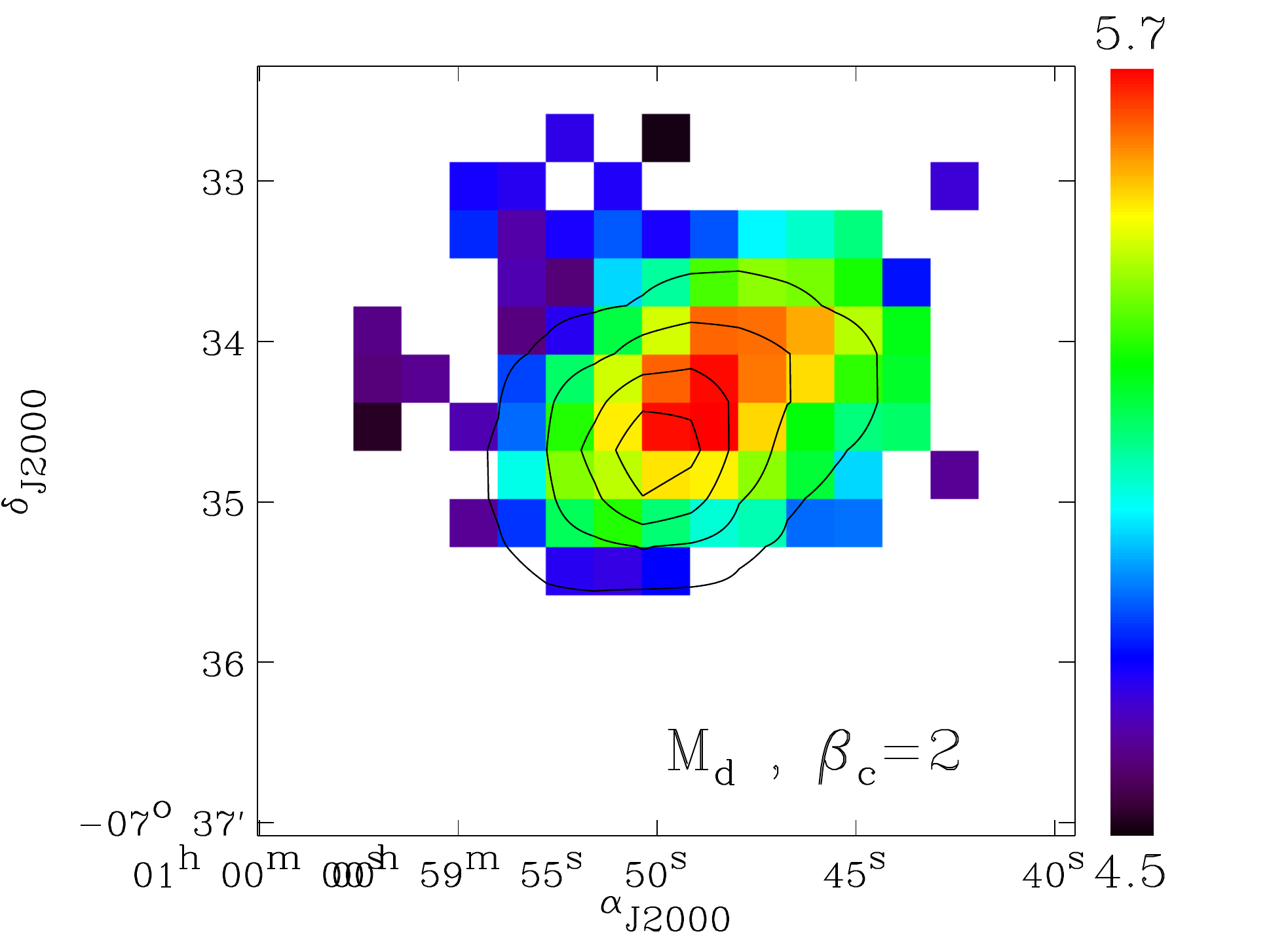} \\
     \centering{\large Emissivity index map} & 
     \centering{\large Temperature map ($\beta$$_c$ free)} & 
     \centering{\large Dust mass surface density ($\beta$$_c$ free)} \\
   	\tabularnewline
	\includegraphics[height=4.3cm]{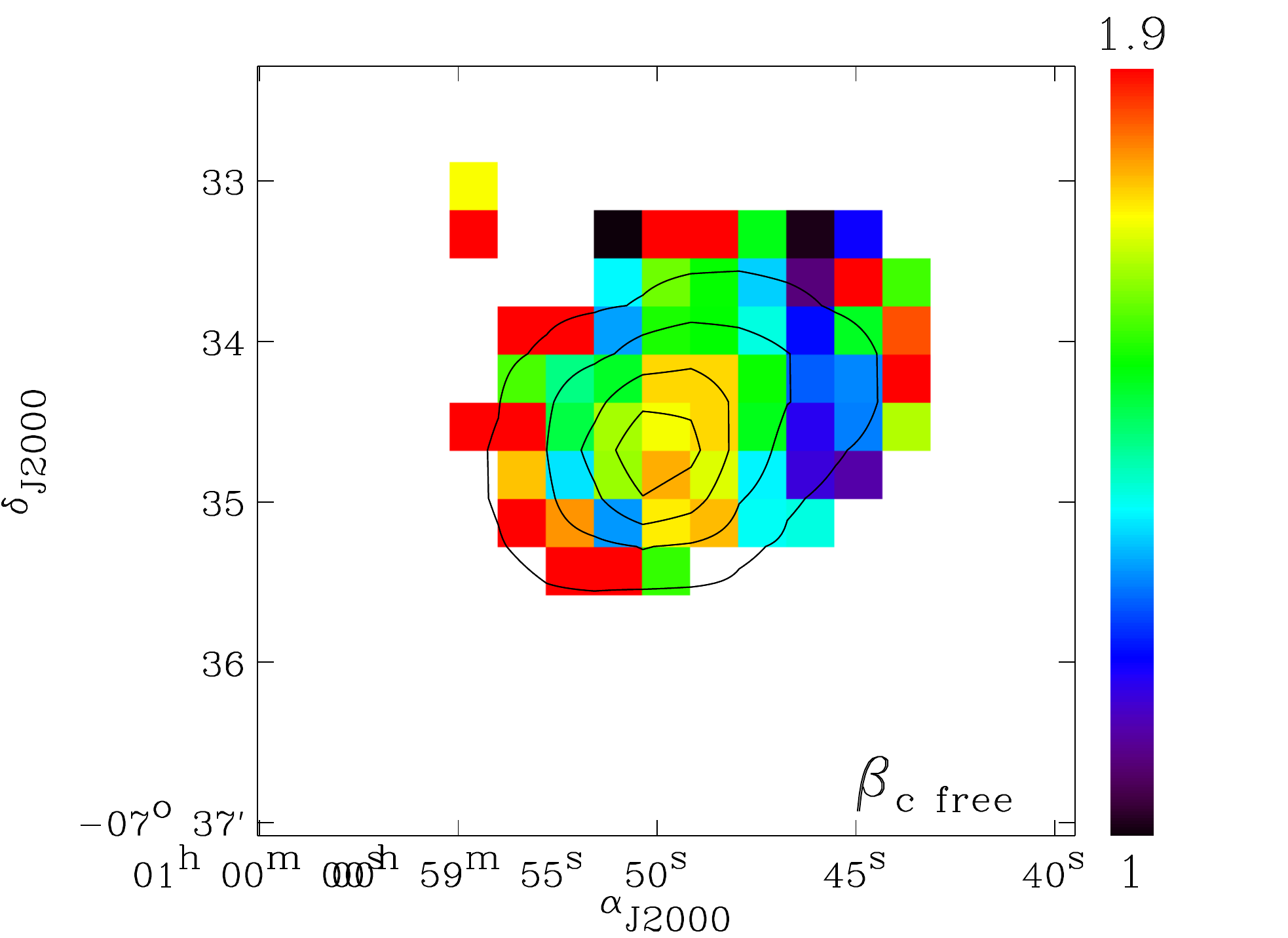}  &
	 \includegraphics[height=4.3cm]{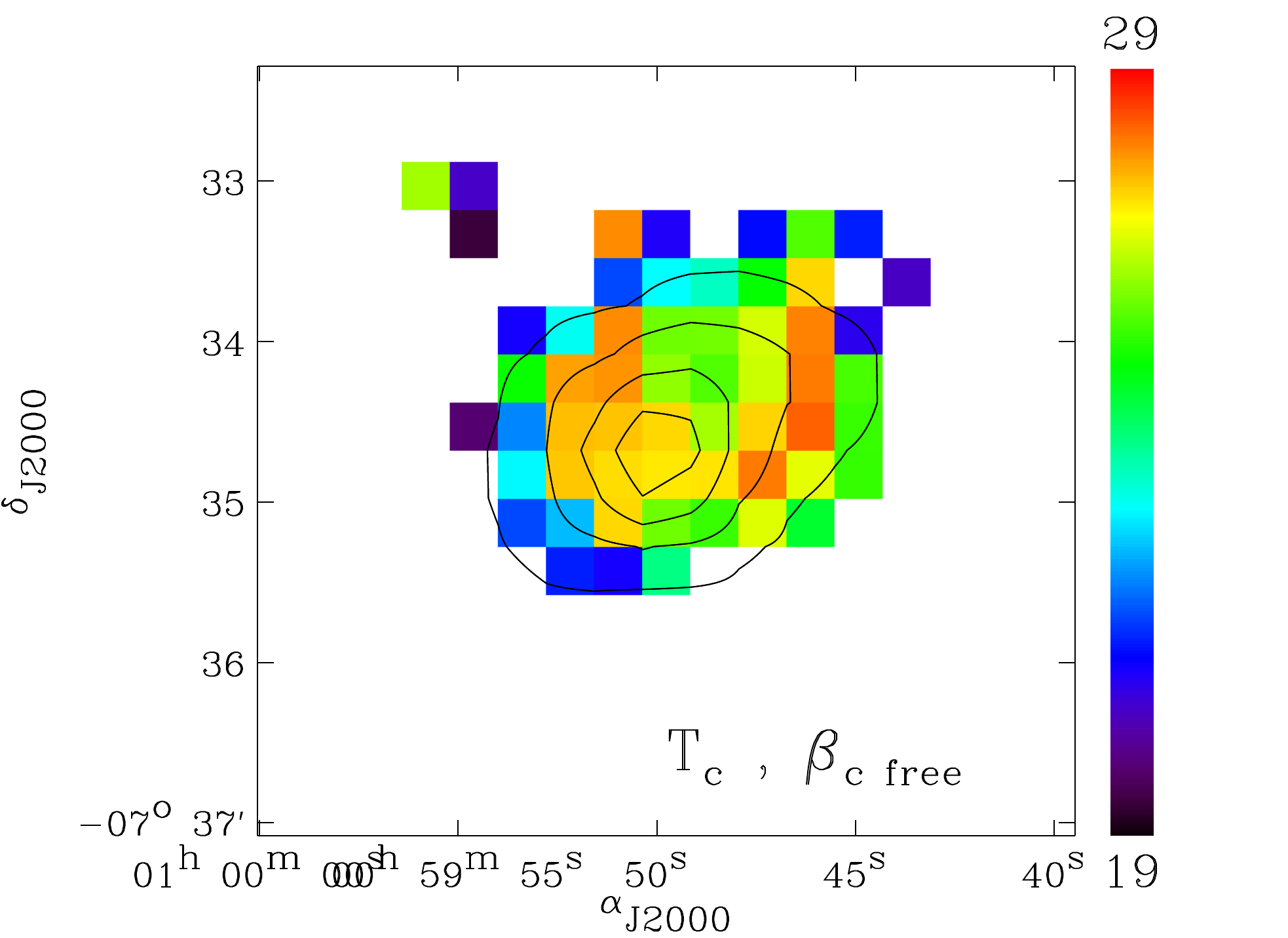} &
	 \includegraphics[height=4.3cm]{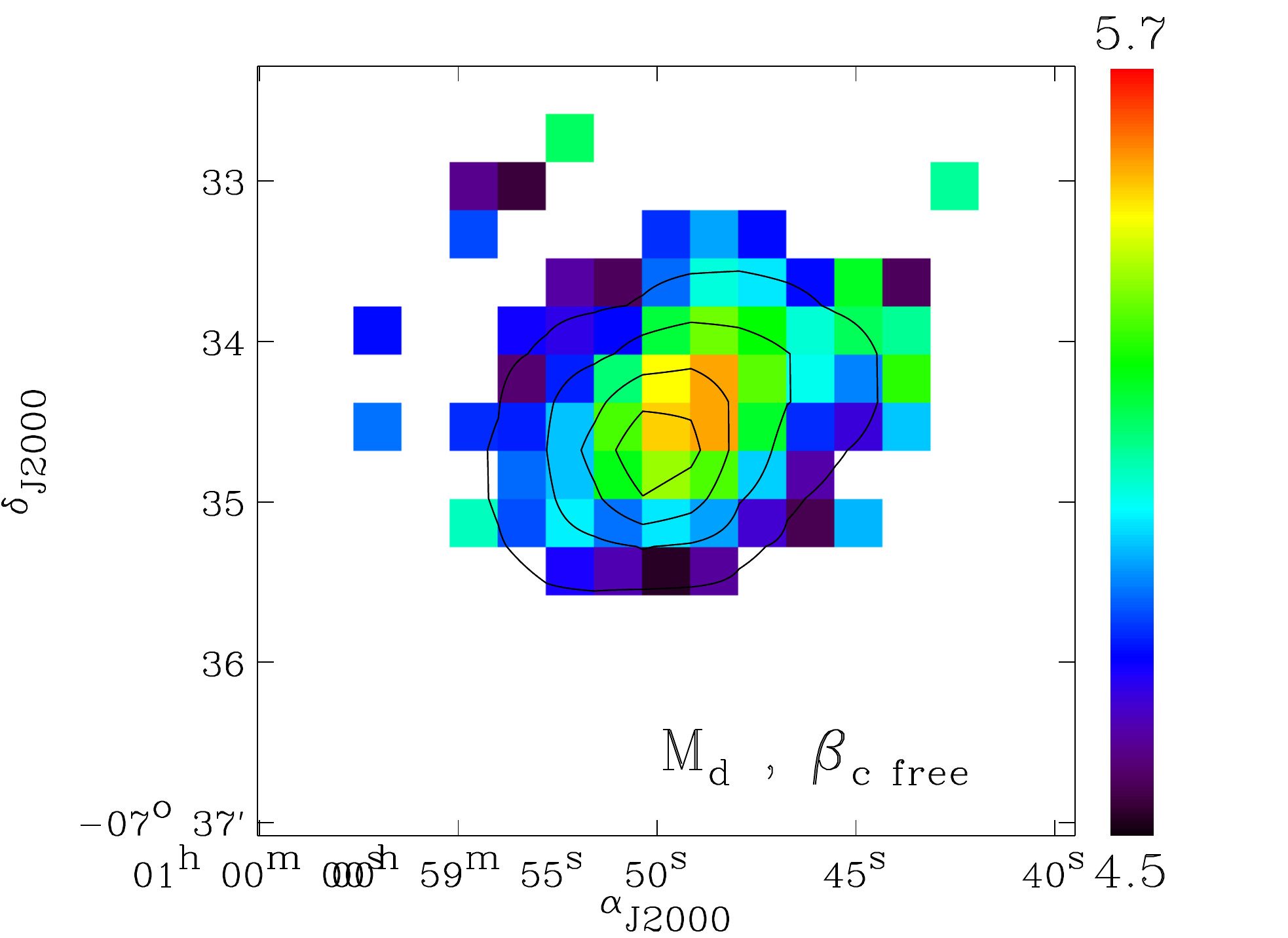} \\
          &&    \\
     \end{tabular}
    \begin{tabular}{m{5.4cm} m{5.4cm} m{5.4cm}}	 
  {\large NGC~628} && \\
   \centering{\large MIPS 24 \mic} & 
   \centering{\large Temperature map ($\beta$$_c$=2)} & 
   \centering{\large Dust mass surface density ($\beta$$_c$=2)} \\
   	\tabularnewline
	\includegraphics[height=4.3cm]{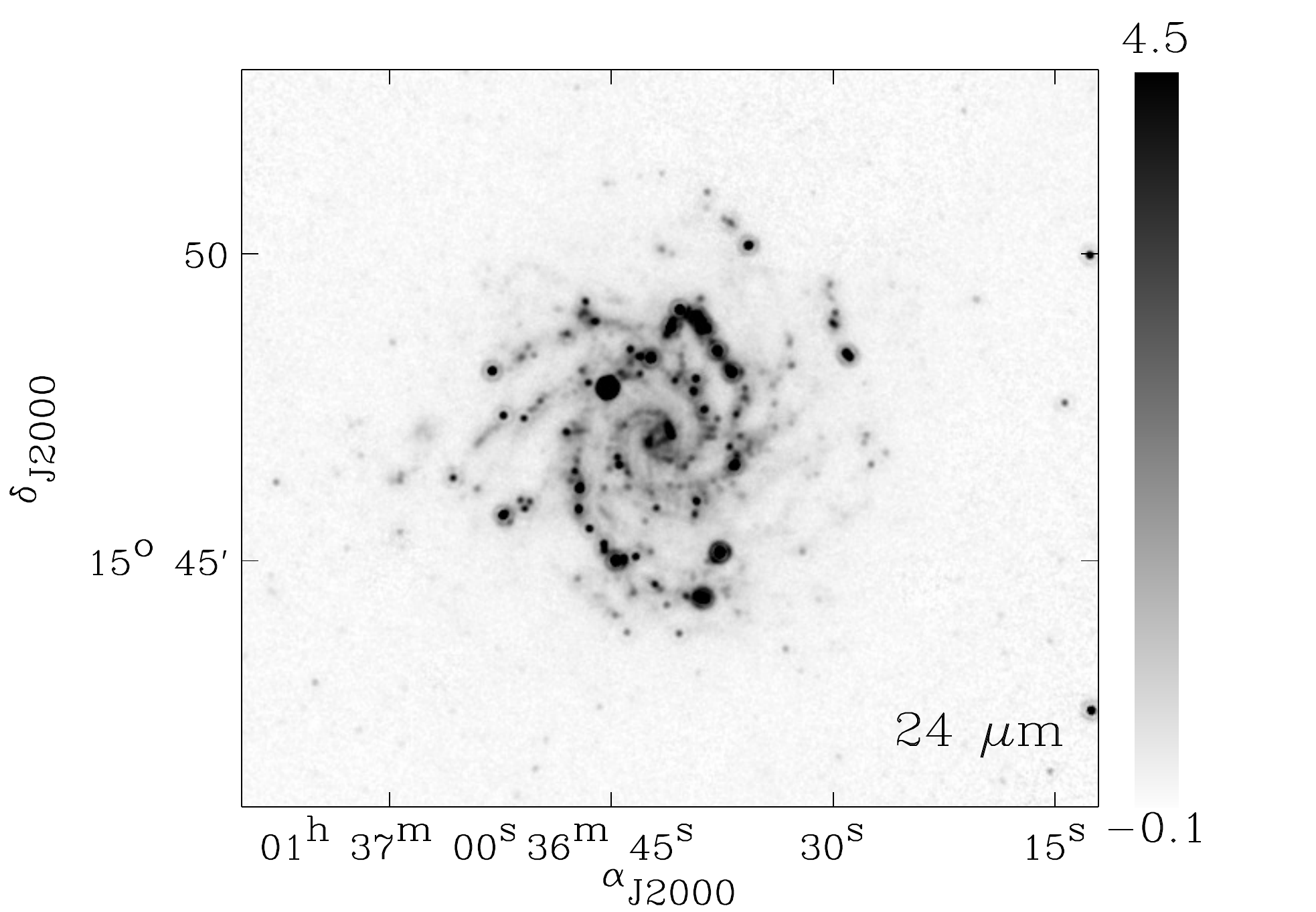} &
	\includegraphics[height=4.3cm]{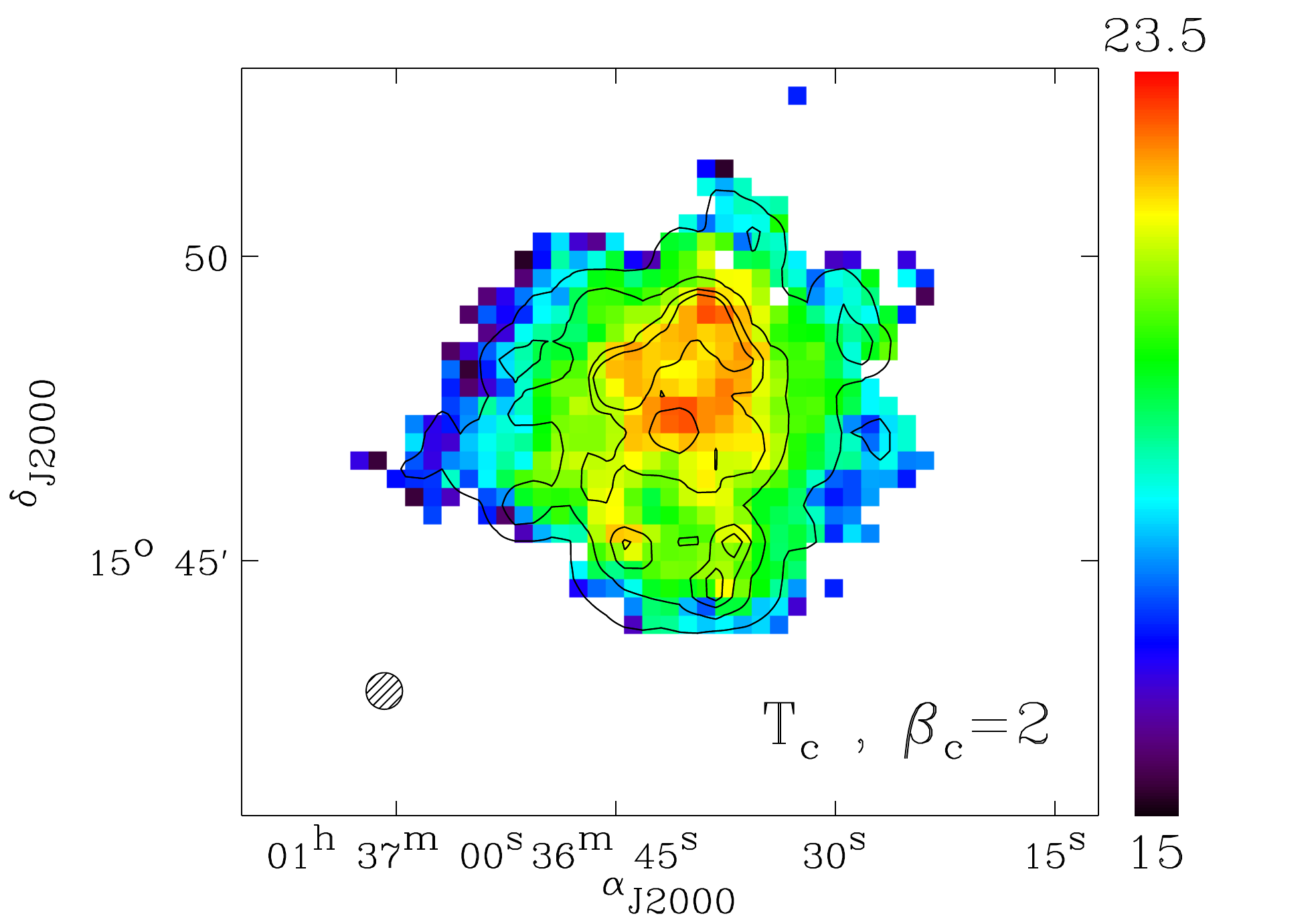} &
	\includegraphics[height=4.3cm]{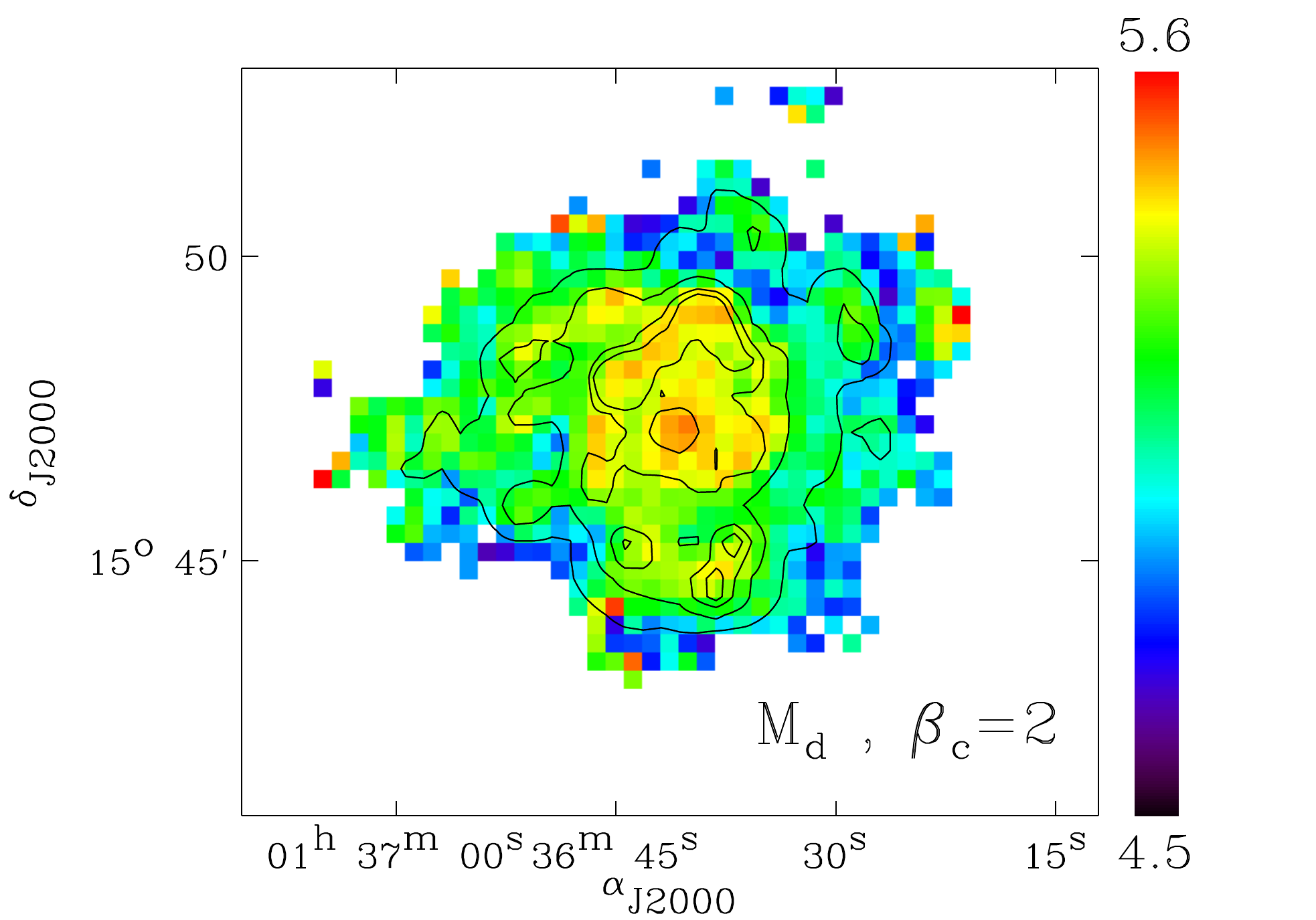} \\
     \centering{\large Emissivity index map} & 
     \centering{\large Temperature map ($\beta$$_c$ free)} & 
     \centering{\large Dust mass surface density ($\beta$$_c$ free)} \\
   	\tabularnewline
	\includegraphics[height=4.3cm]{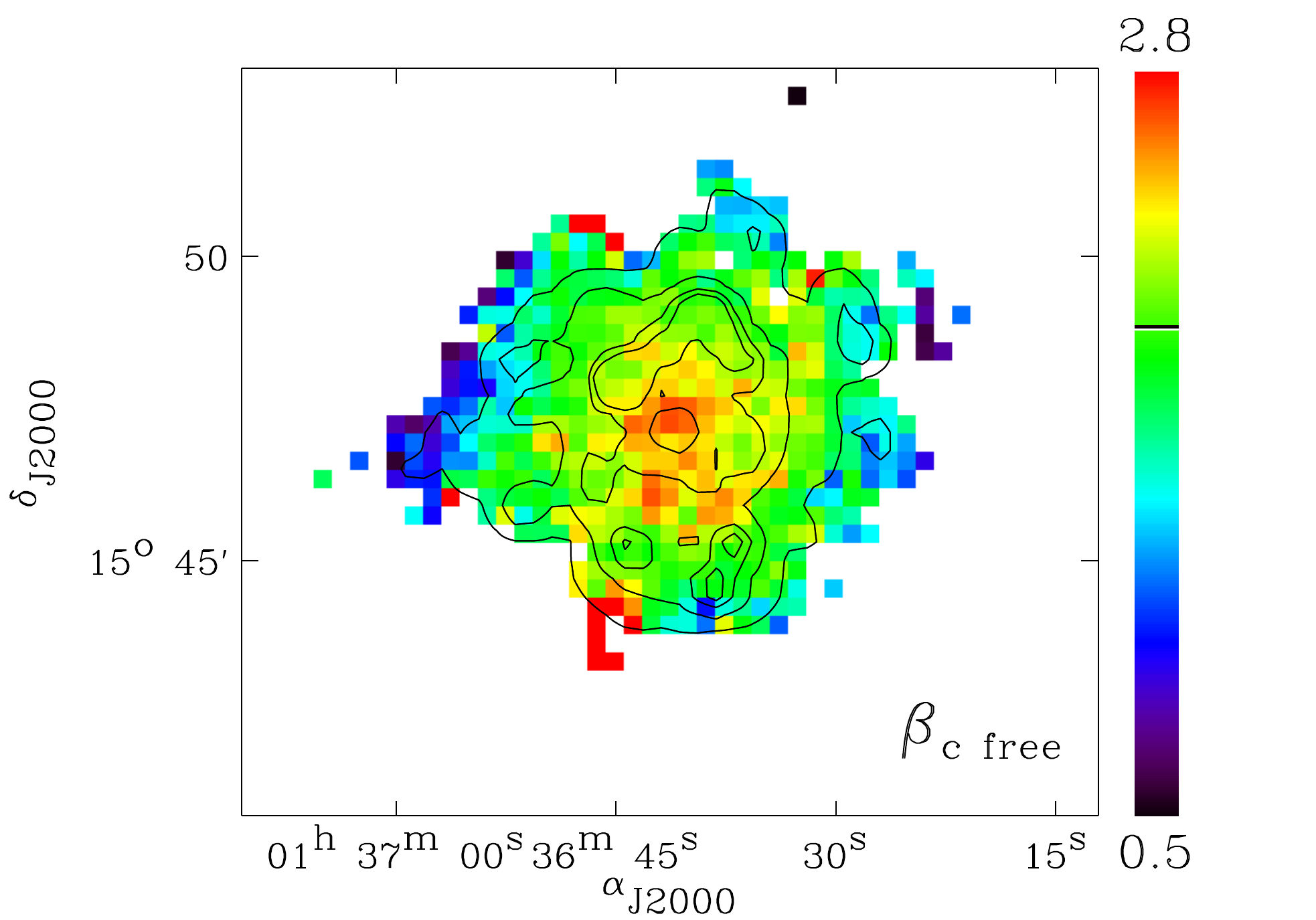}  &
	 \includegraphics[height=4.3cm]{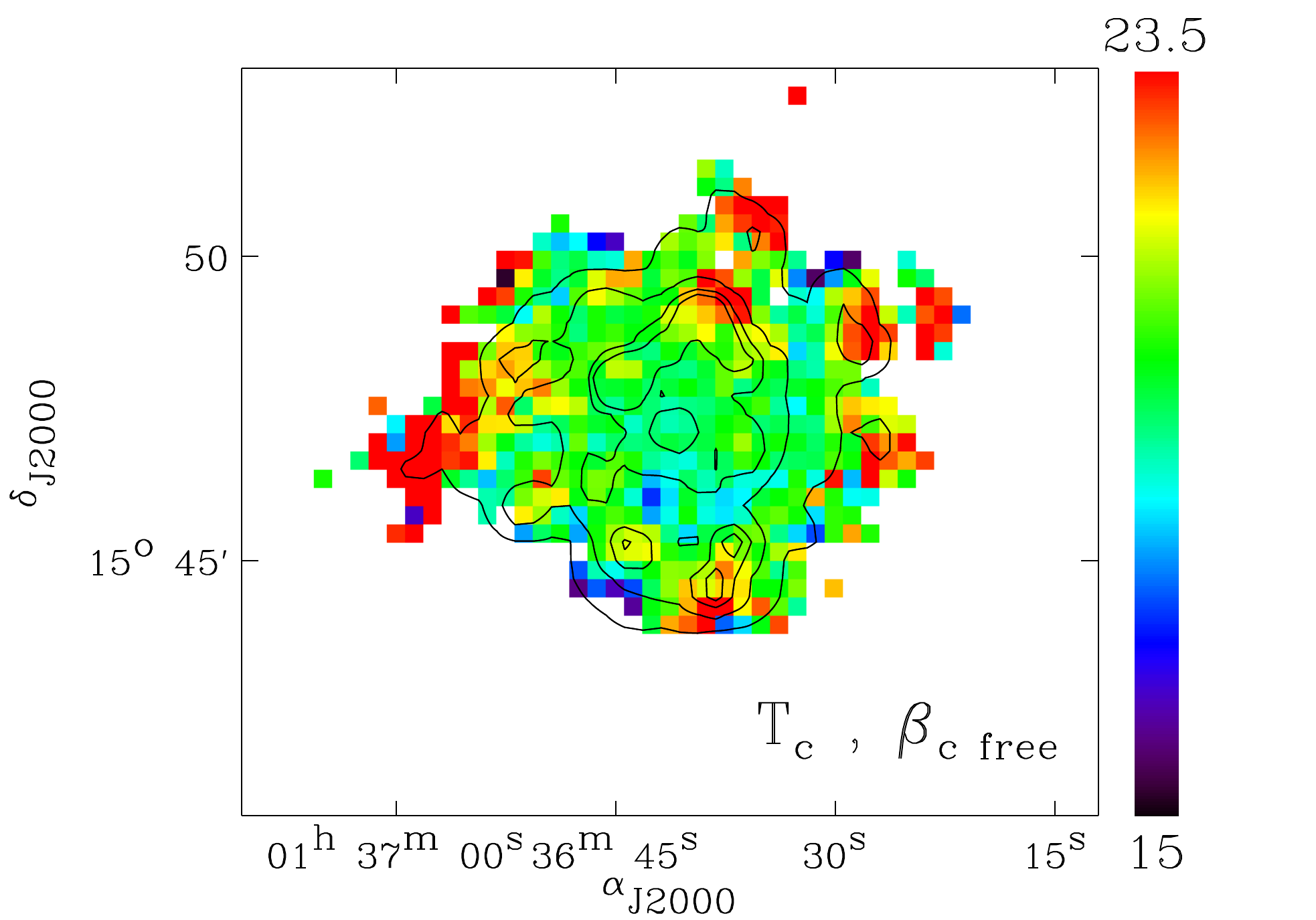} &
	 \includegraphics[height=4.3cm]{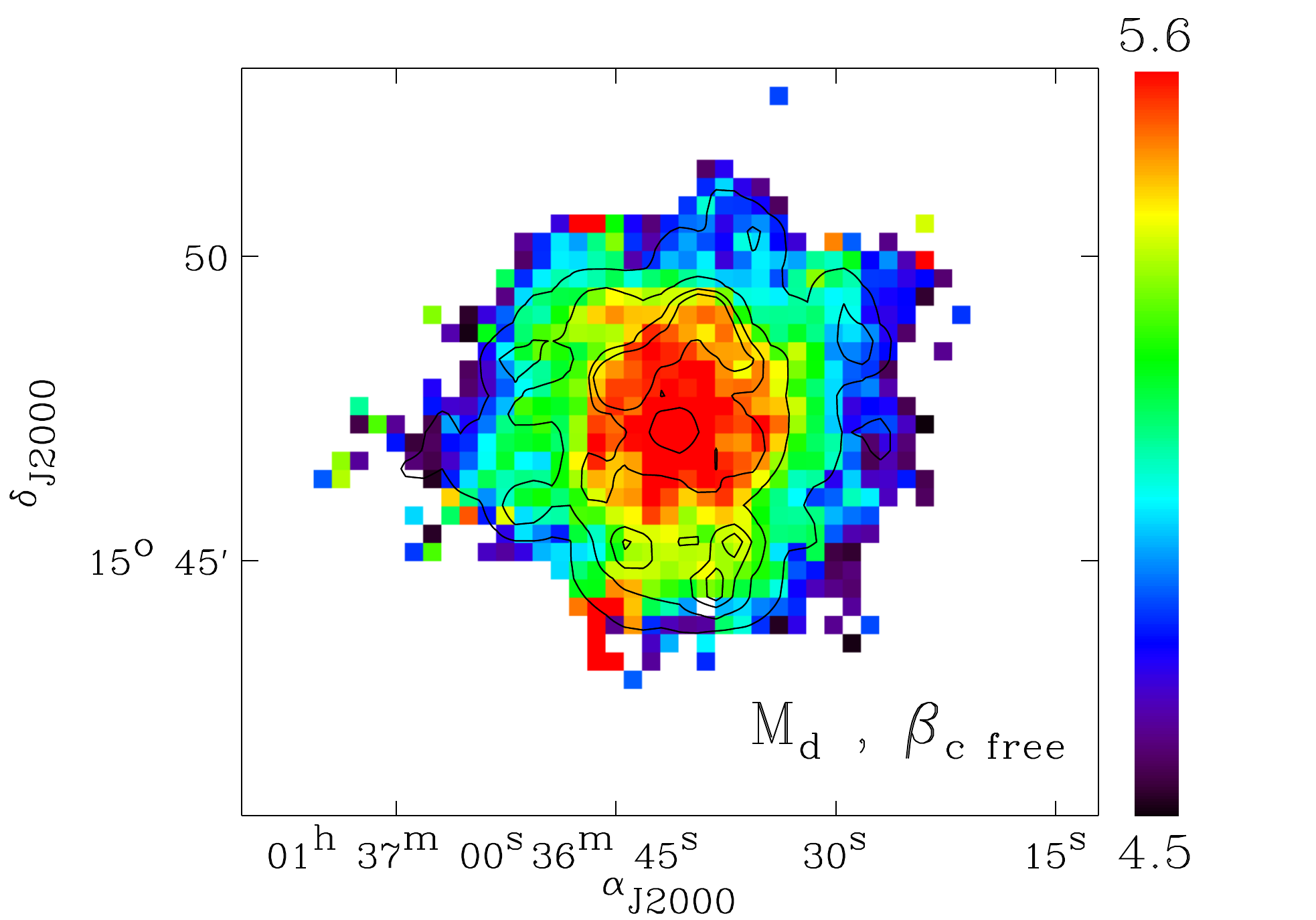} \\
&&  \\	 
      \end{tabular}
    \caption{For each galaxy, North is up, East is left. Top: MIPS 24 \mic\ image (MJy~sr$^{-1}$), cold dust temperature map (in K) and dust mass surface density map (in \msun~kpc$^{-2}$, log scale) derived with $\beta$$_c$ = 2.0. All images are convolved to the SPIRE500 PSF before deriving the parameter maps. The FWHM of the PSF of SPIRE 500 \mic\ is shown by the dashed circle in the top middle panel for each galaxy. Bottom: Emissivity index map, cold dust temperature map (in K) and dust mass surface density map (in \msun~kpc$^{-2}$, log scale) derived with T$_c$ and $\beta$$_c$ as free parameters. The value $\beta$$_c$=2.0 is indicated on the color scale of the emissivity index map. We overlay the MIPS 24 \mic\ contours (smoothed to the resolution of the temperature and beta maps) for comparison.}
    \label{Prop_maps}
\end{figure*}

\addtocounter {figure}{-1}

\begin{figure*}
    \centering
    
    \begin{tabular}{m{5.7cm} m{5.7cm} m{5.7cm}}	   
  {\large NGC~1097} && \\
   \centering{\large MIPS 24 \mic} & 
   \centering{\large Temperature map ($\beta$$_c$=2)} &
    \centering{\large Dust mass surface density ($\beta$$_c$=2)} \\
   	\tabularnewline
	\includegraphics[height=4.3cm]{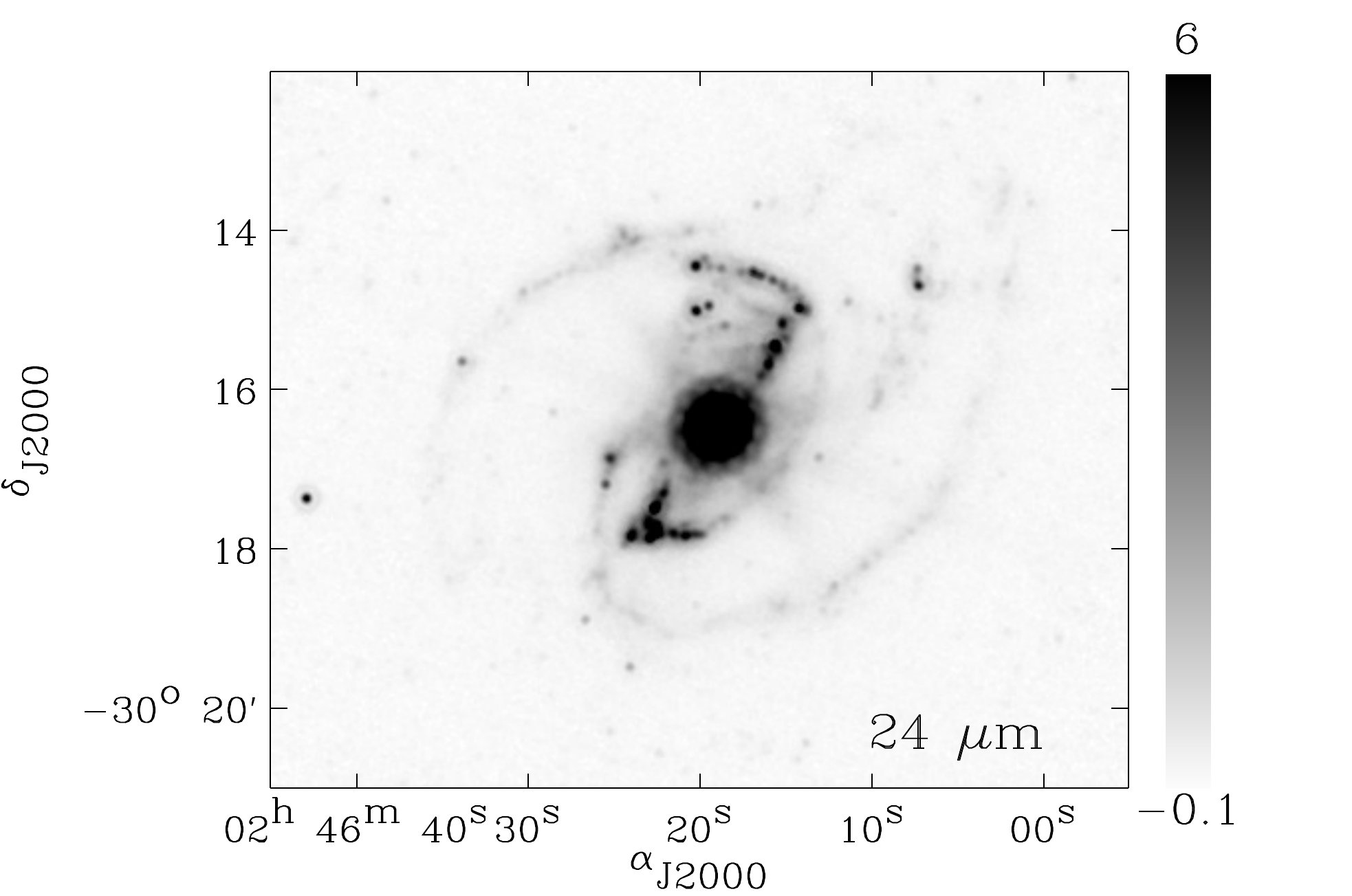} &
	\includegraphics[height=4.3cm]{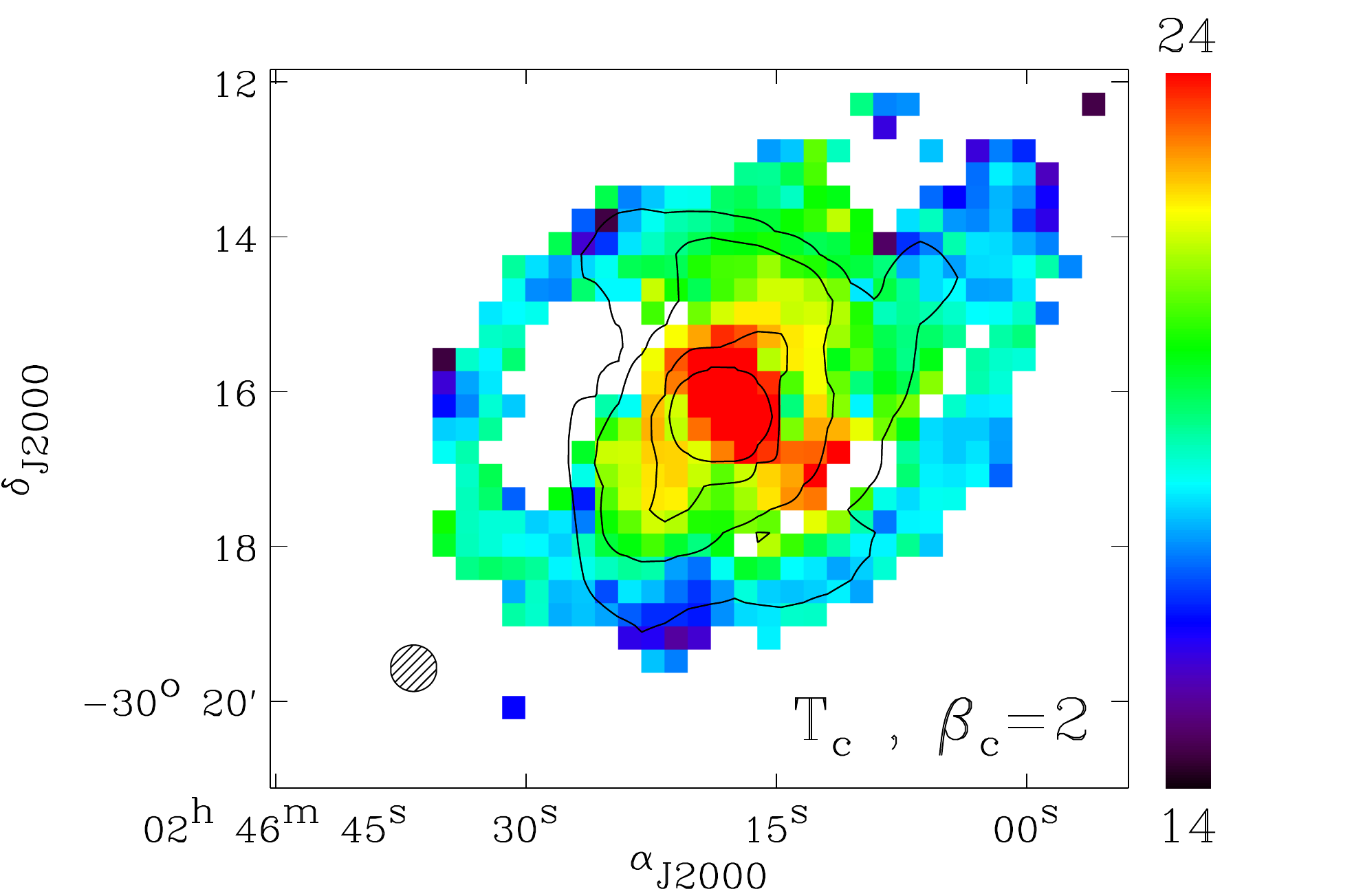} &
	\includegraphics[height=4.3cm]{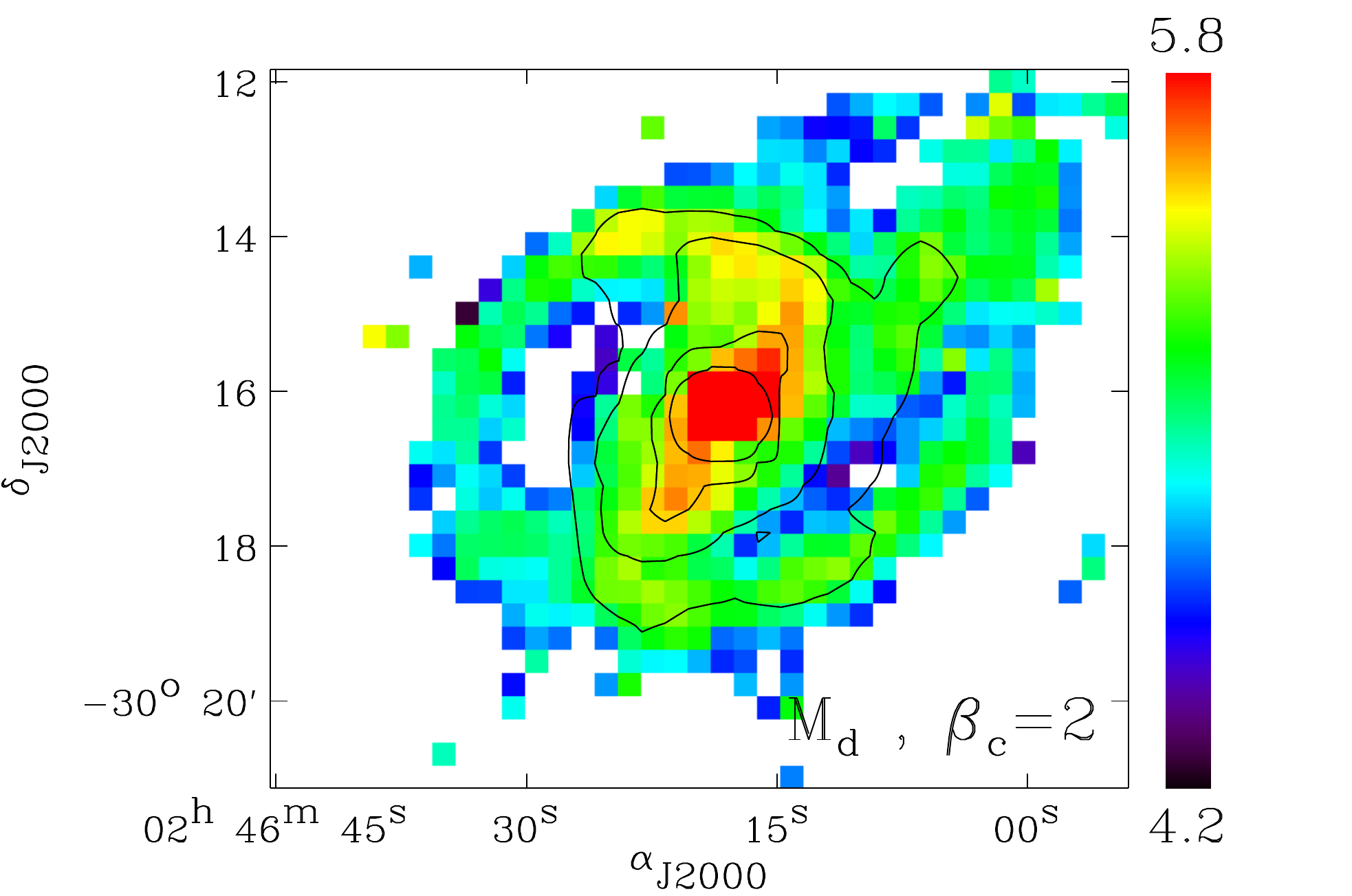} \\
     \centering{\large Emissivity index map} & 
     \centering{\large Temperature map ($\beta$$_c$ free)} &
      \centering{\large Dust mass surface density ($\beta$$_c$ free)} \\
   	\tabularnewline
	\includegraphics[height=4.3cm]{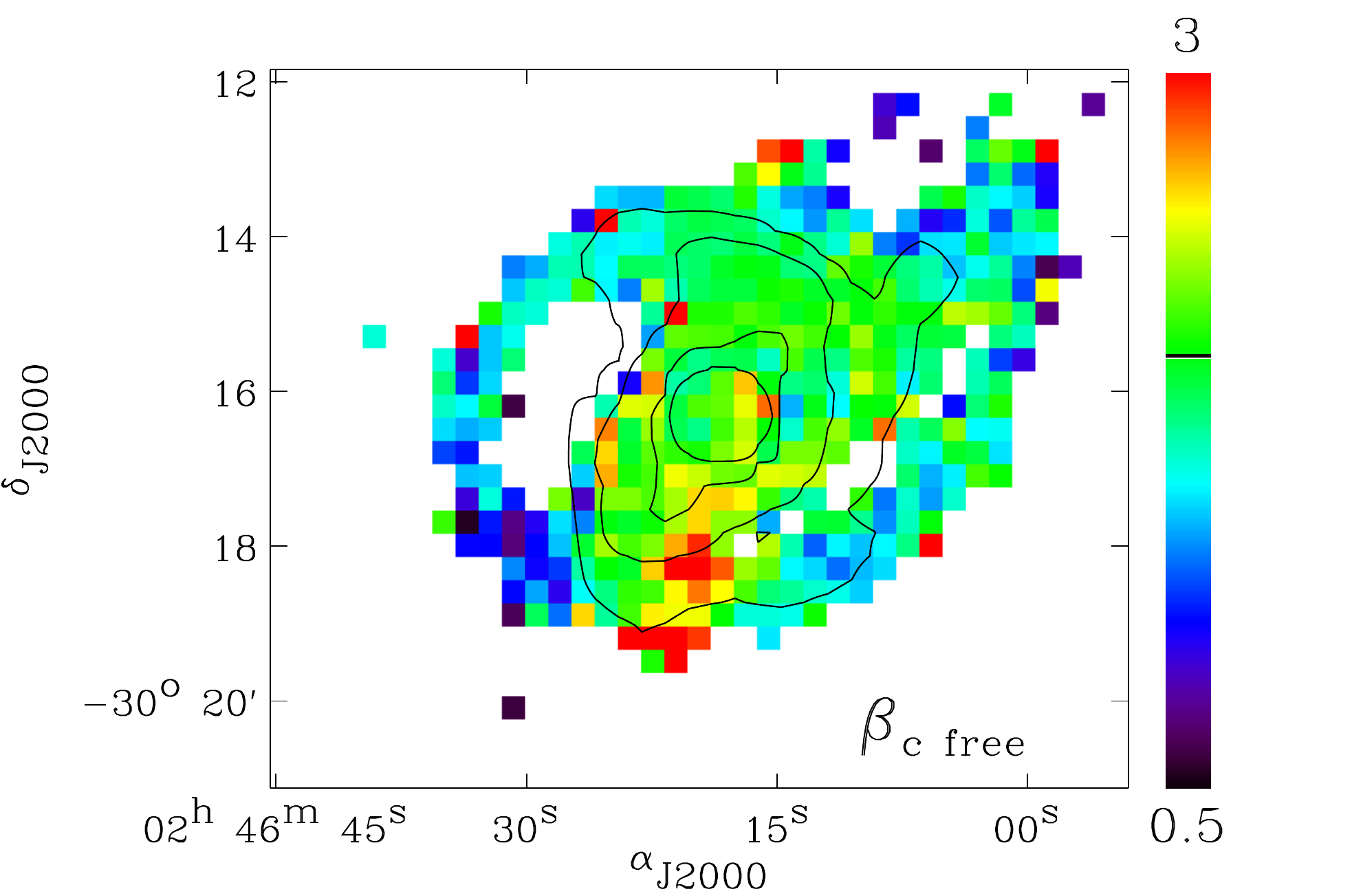}  &
	 \includegraphics[height=4.3cm]{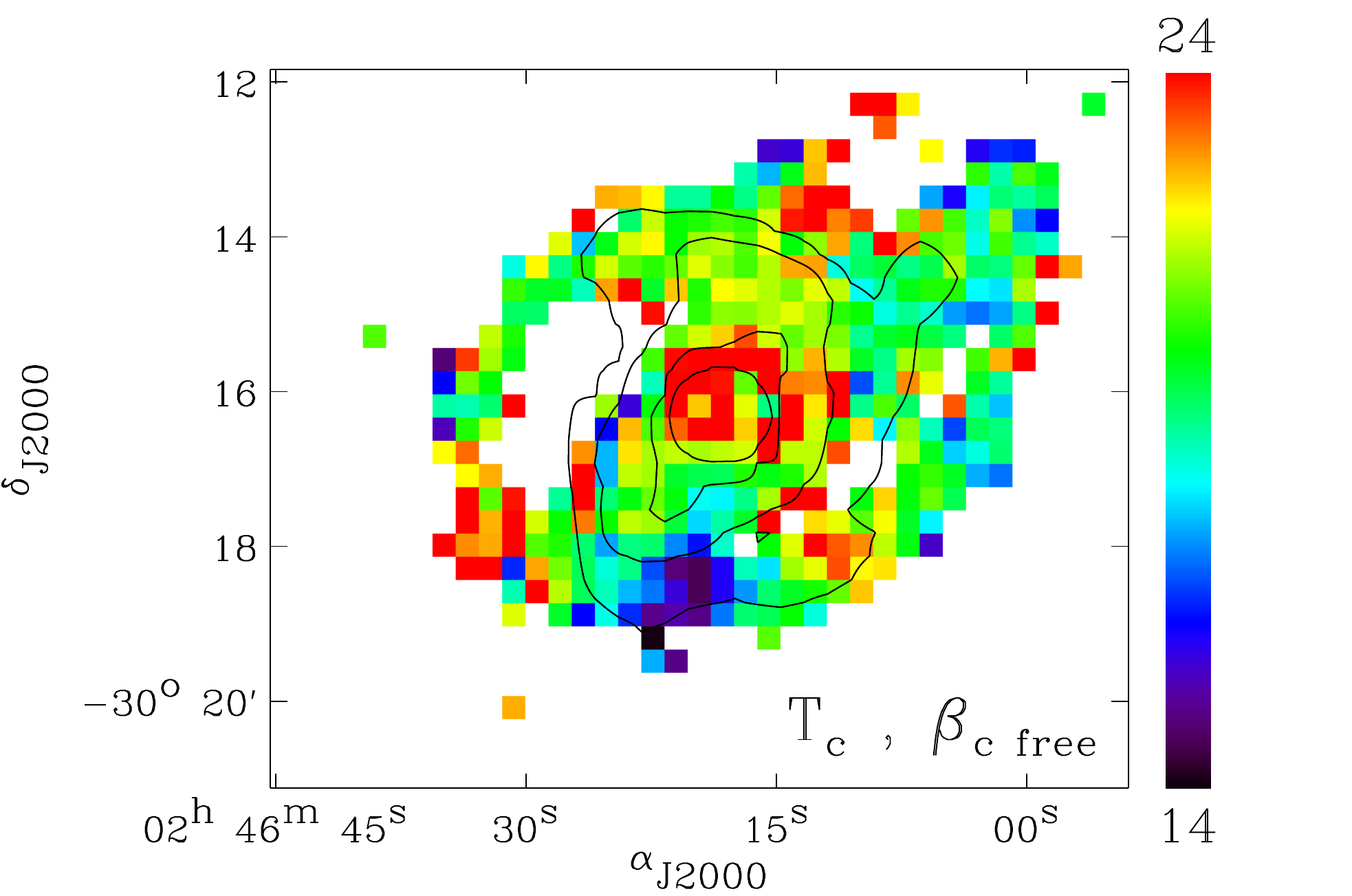} &
	 \includegraphics[height=4.3cm]{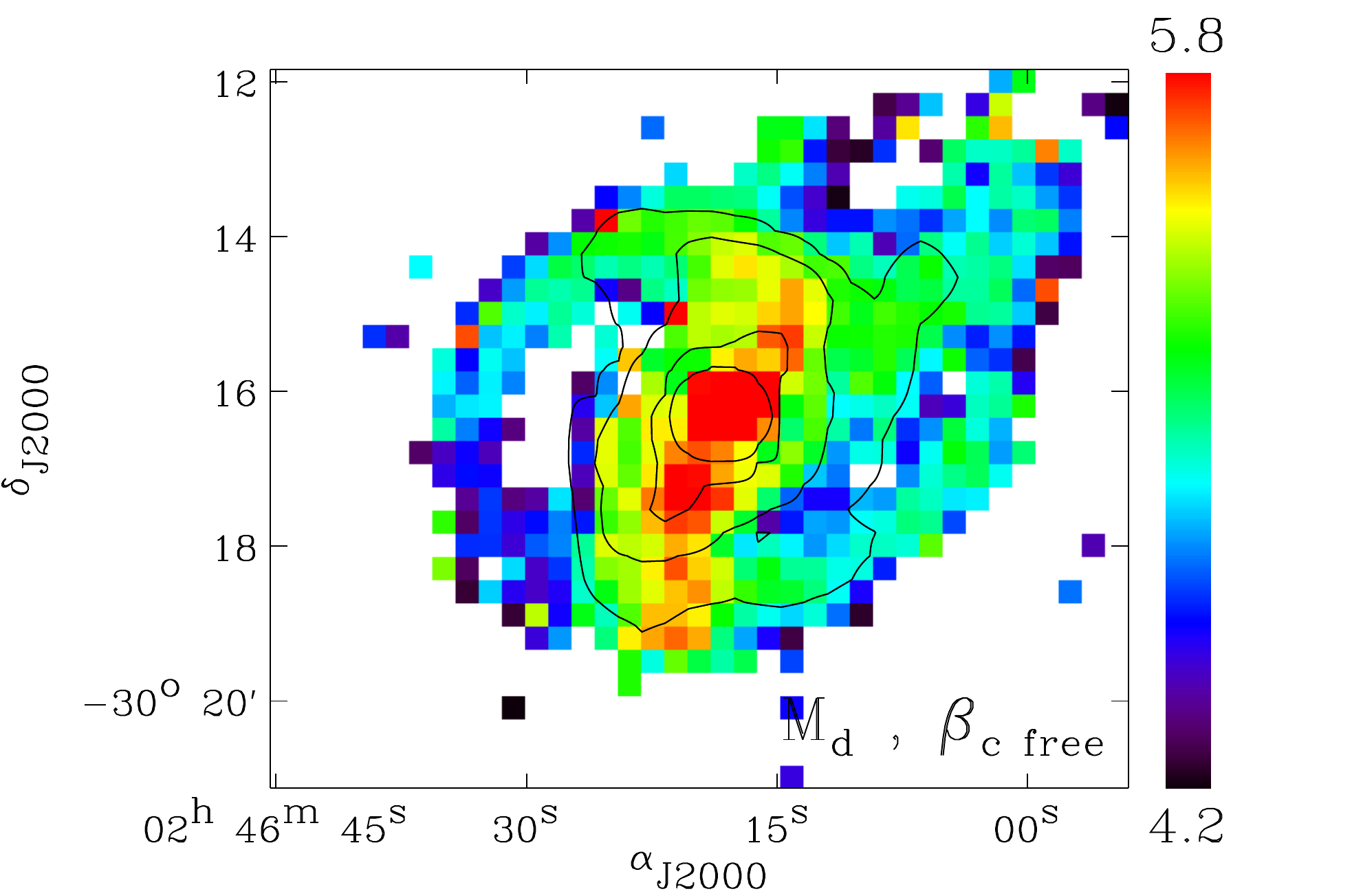} \\
&&  \\     
&&    \\
  {\large NGC~1291} && \\
   \centering{\large MIPS 24 \mic} & 
   \centering{\large Temperature map ($\beta$$_c$=2)} & 
   \centering{\large Dust mass surface density ($\beta$$_c$=2)} \\
   	\tabularnewline
	\includegraphics[height=4.3cm]{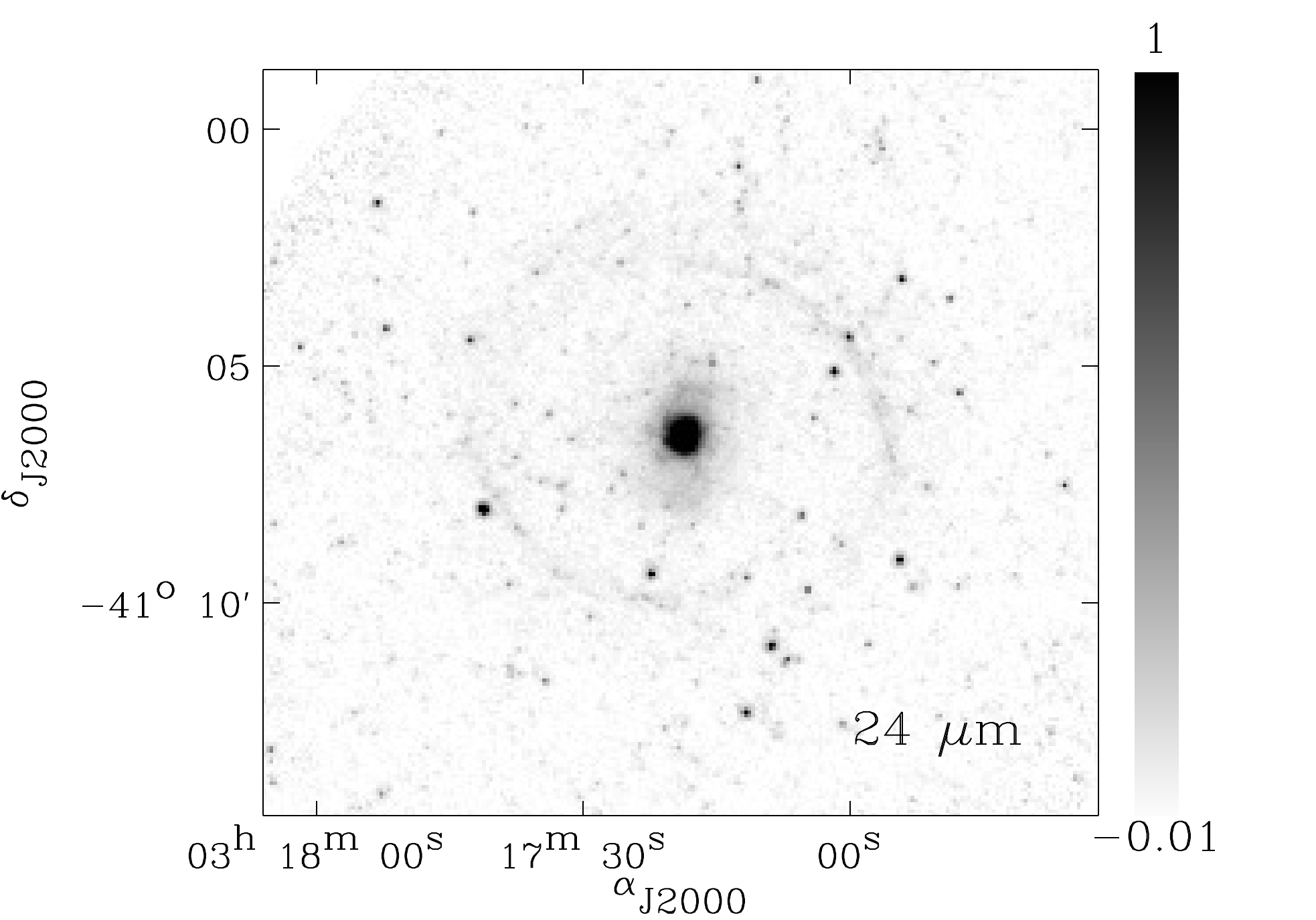} &
	\includegraphics[height=4.3cm]{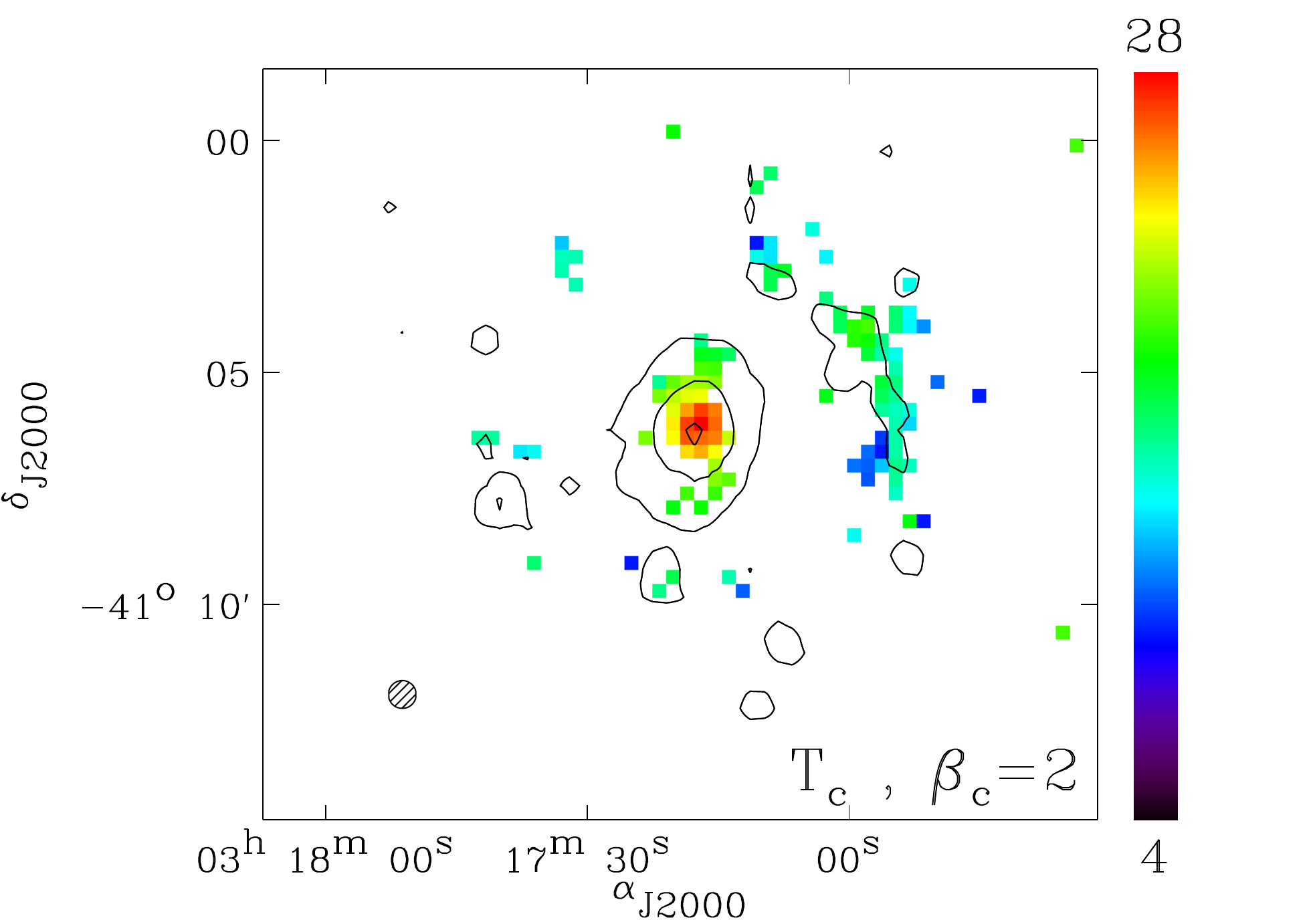} &
	\includegraphics[height=4.3cm]{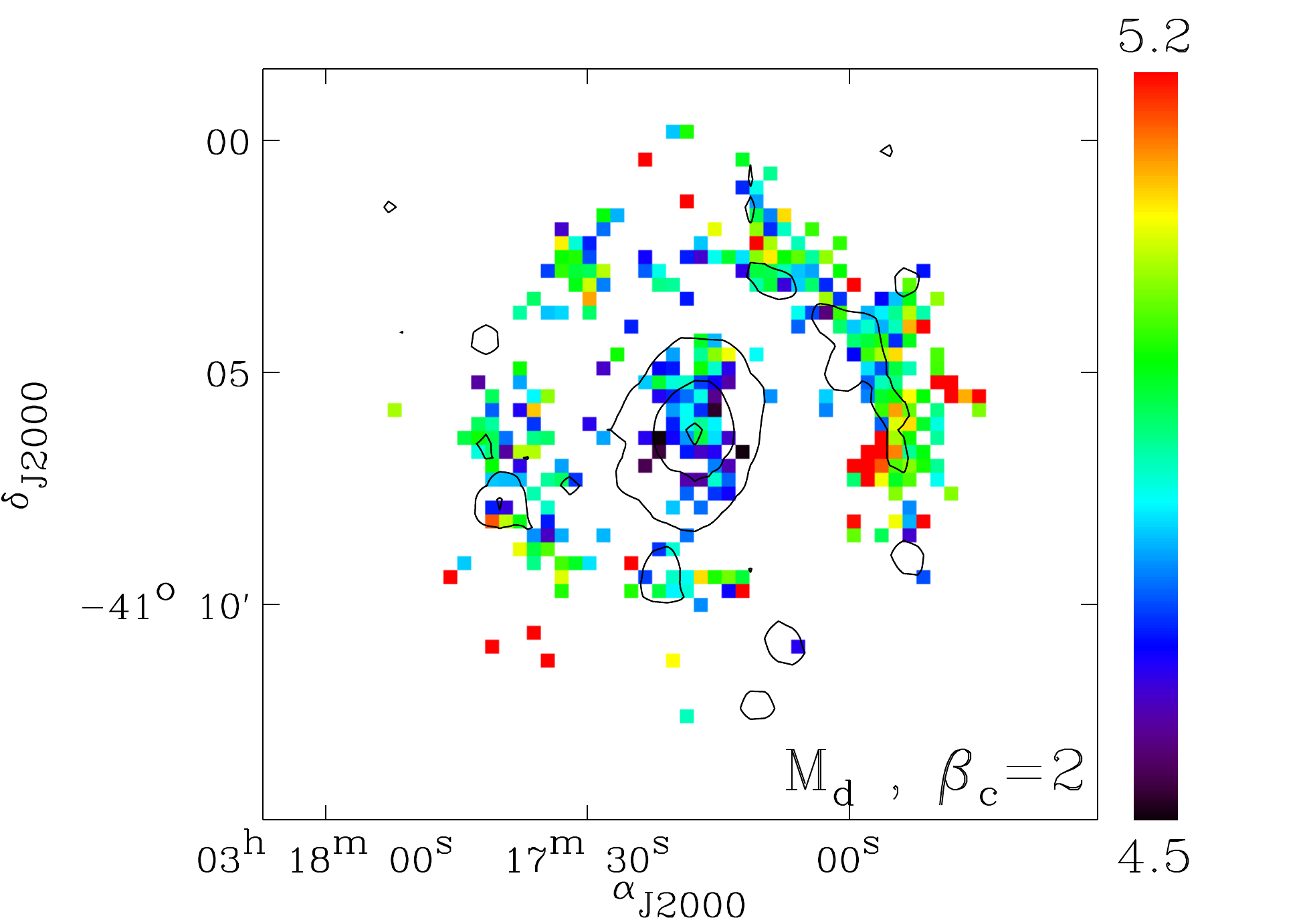} \\
     \centering{\large Emissivity index map} & 
     \centering{\large Temperature map ($\beta$$_c$ free)} & 
     \centering{\large Dust mass surface density ($\beta$$_c$ free)} \\
   	\tabularnewline
	\includegraphics[height=4.3cm]{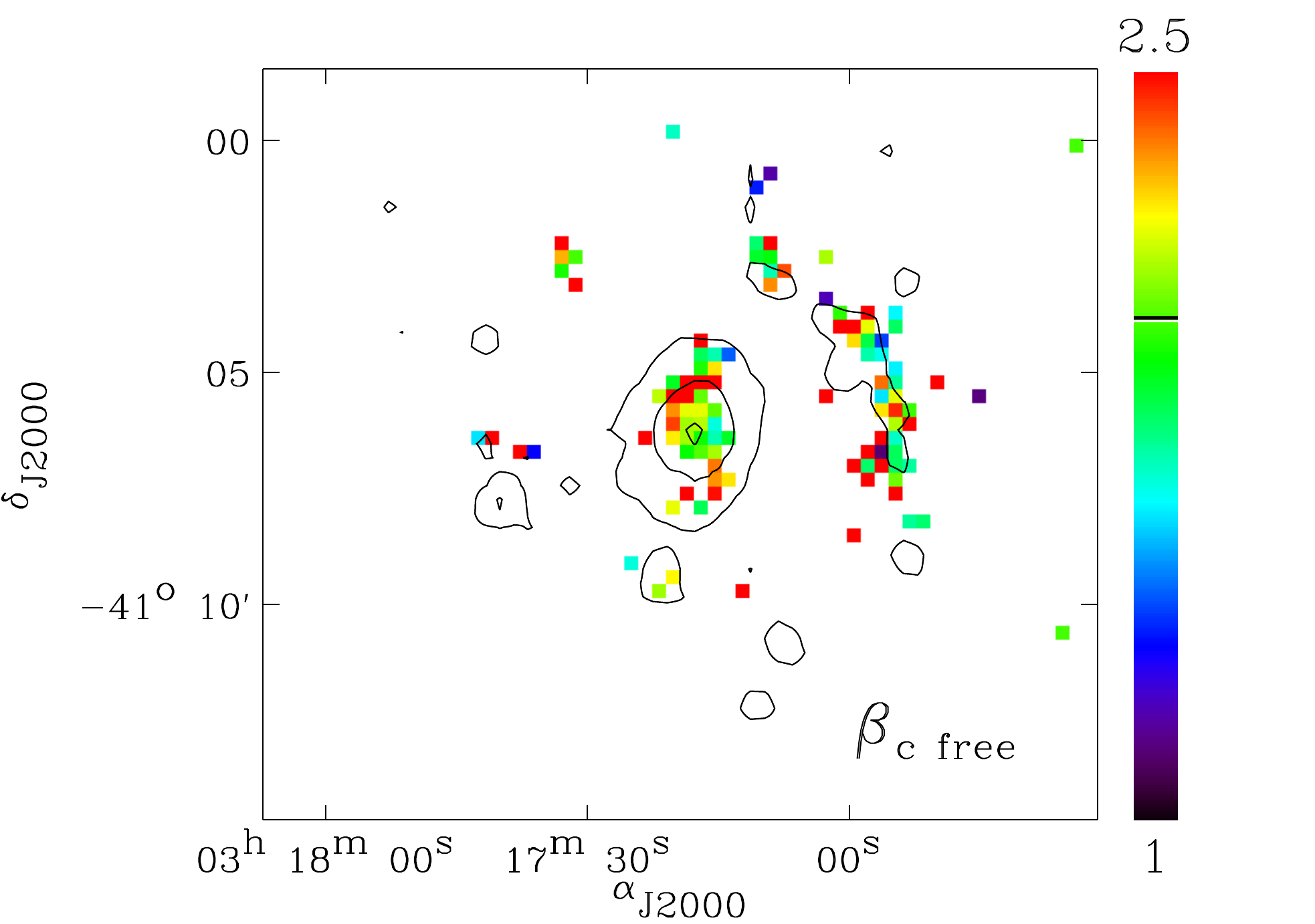}  &
	 \includegraphics[height=4.3cm]{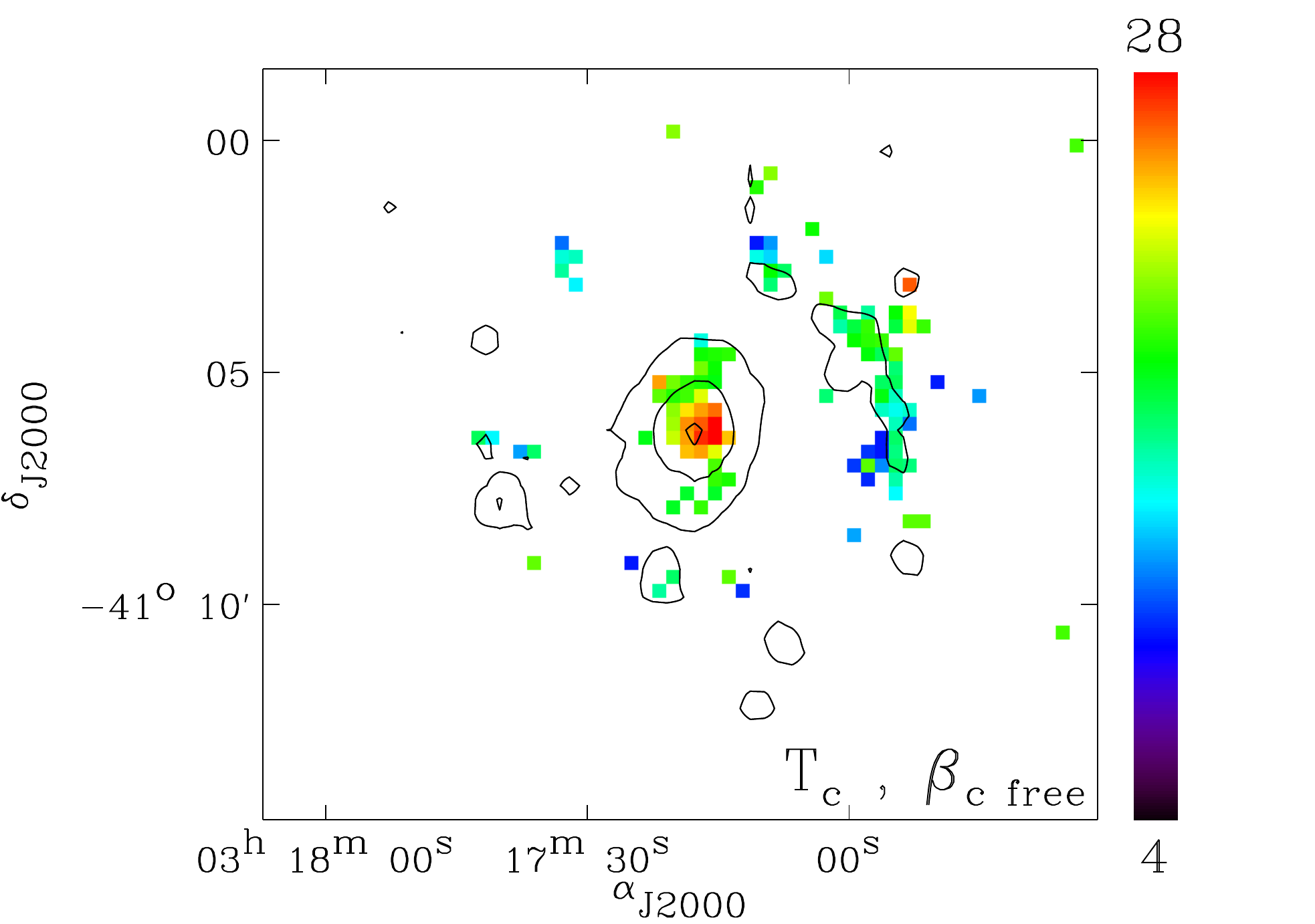} &
	 \includegraphics[height=4.3cm]{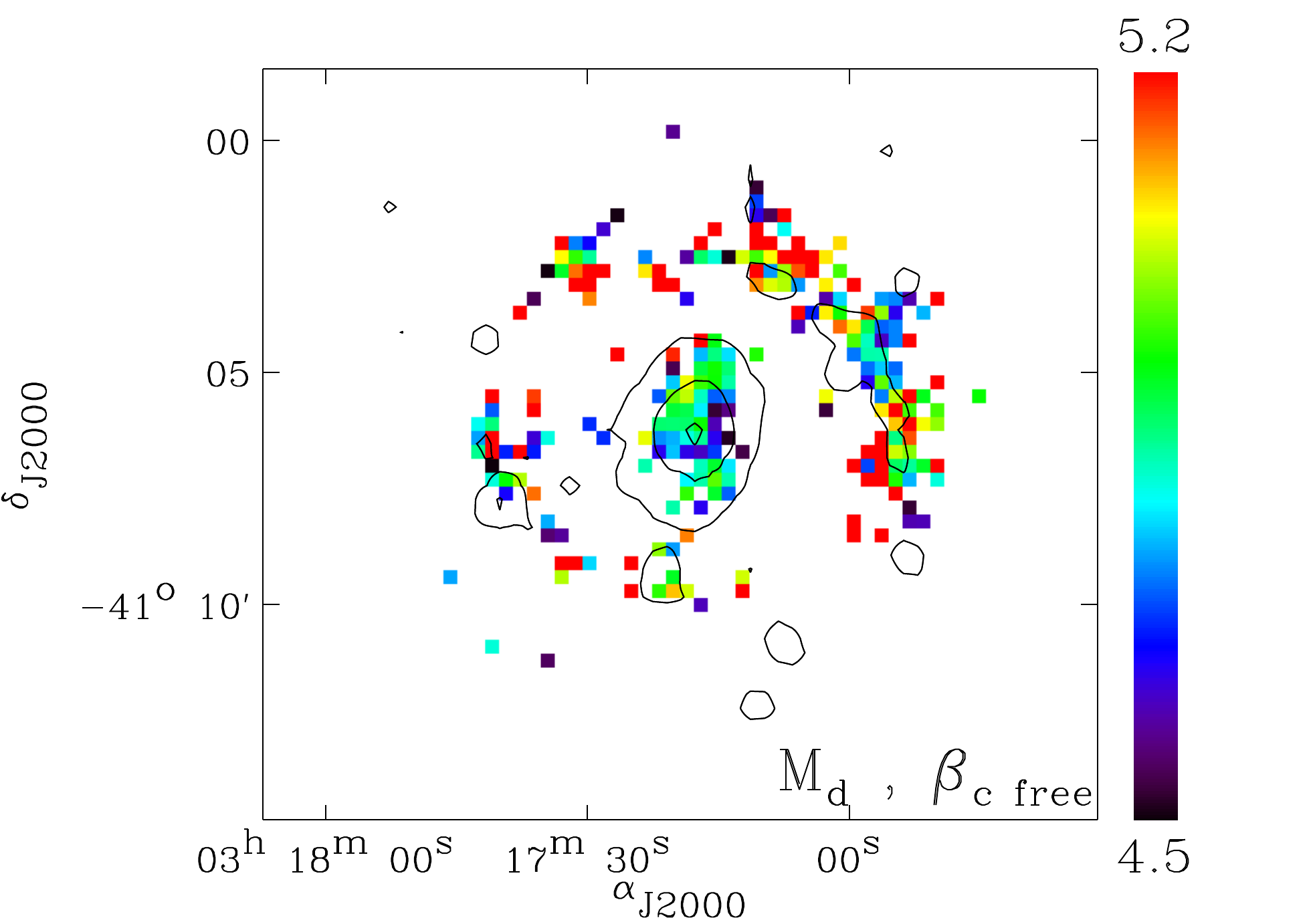} \\          
                            \end{tabular}
\caption{Continued. }
\end{figure*}

\addtocounter {figure}{-1}
        
\begin{figure*}
    \centering   

    \begin{tabular}{m{5.5cm} m{5.5cm} m{5.5cm}}
  {\large NGC~1316} &&\\
   \centering{\large MIPS 24 \mic} & 
   \centering{\large Temperature map ($\beta$$_c$=2)} &
    \centering{\large Dust mass surface density ($\beta$$_c$=2)} \\
   	\tabularnewline
	\includegraphics[height=4.4cm]{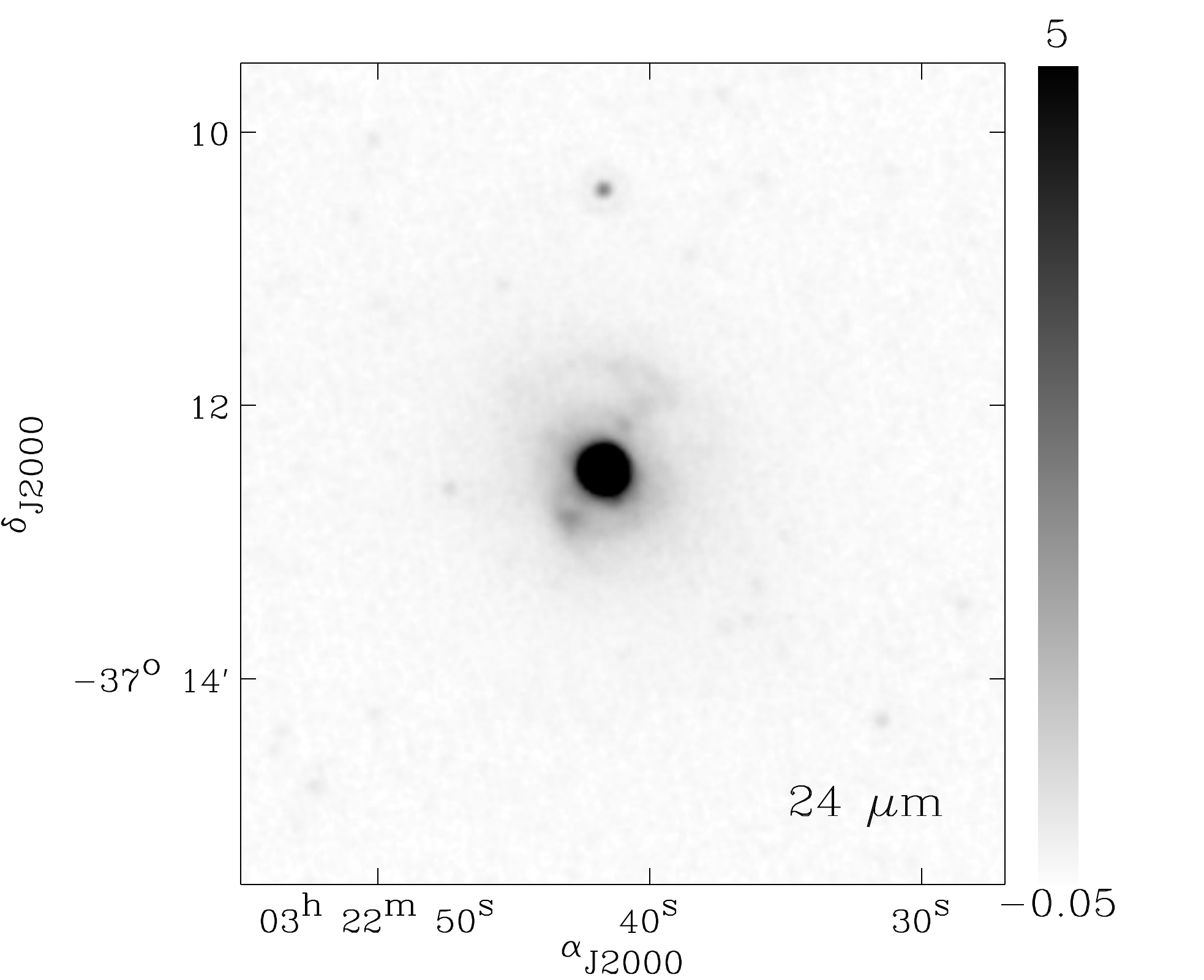} &
	\includegraphics[height=4.4cm]{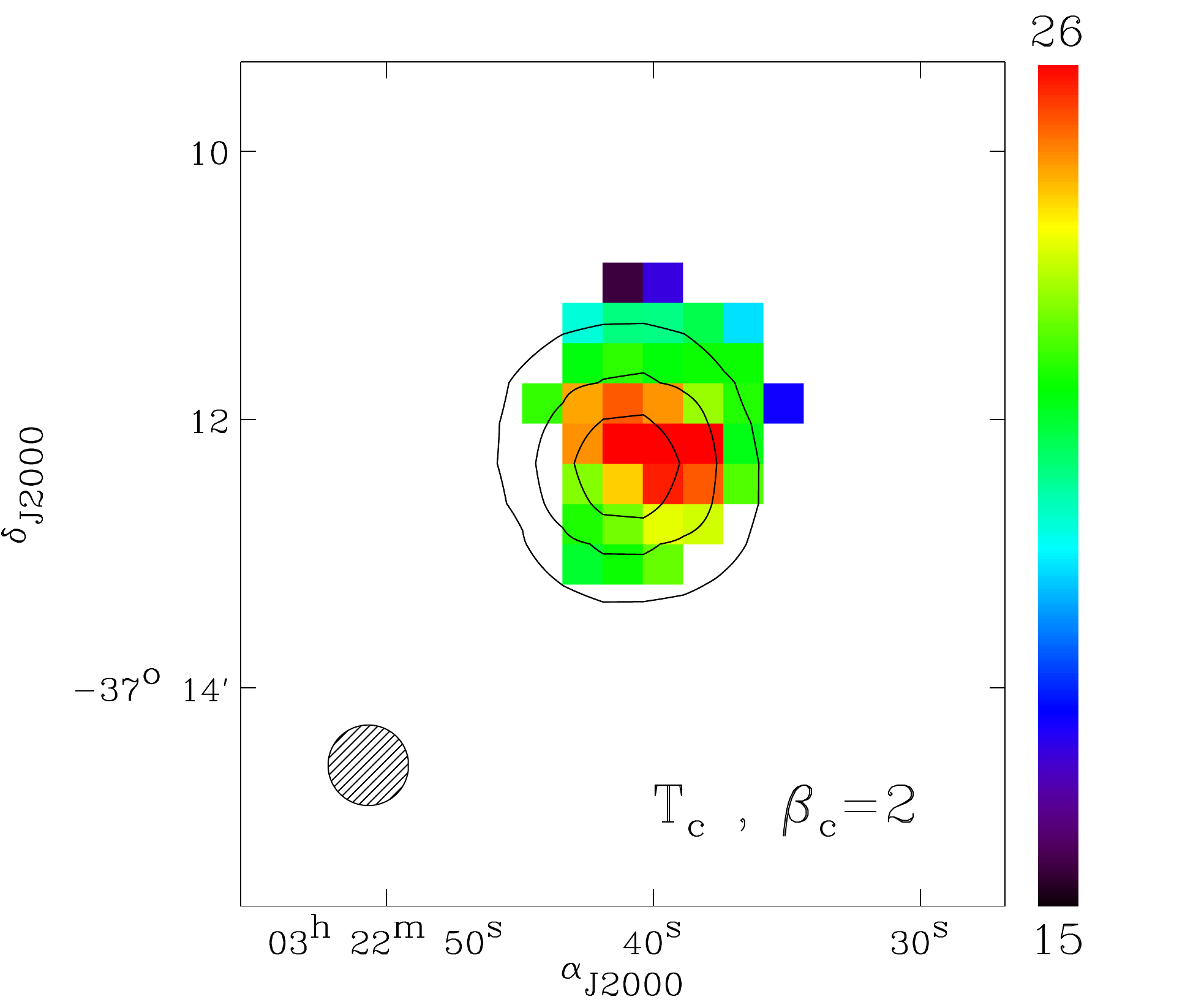} &
	\includegraphics[height=4.4cm]{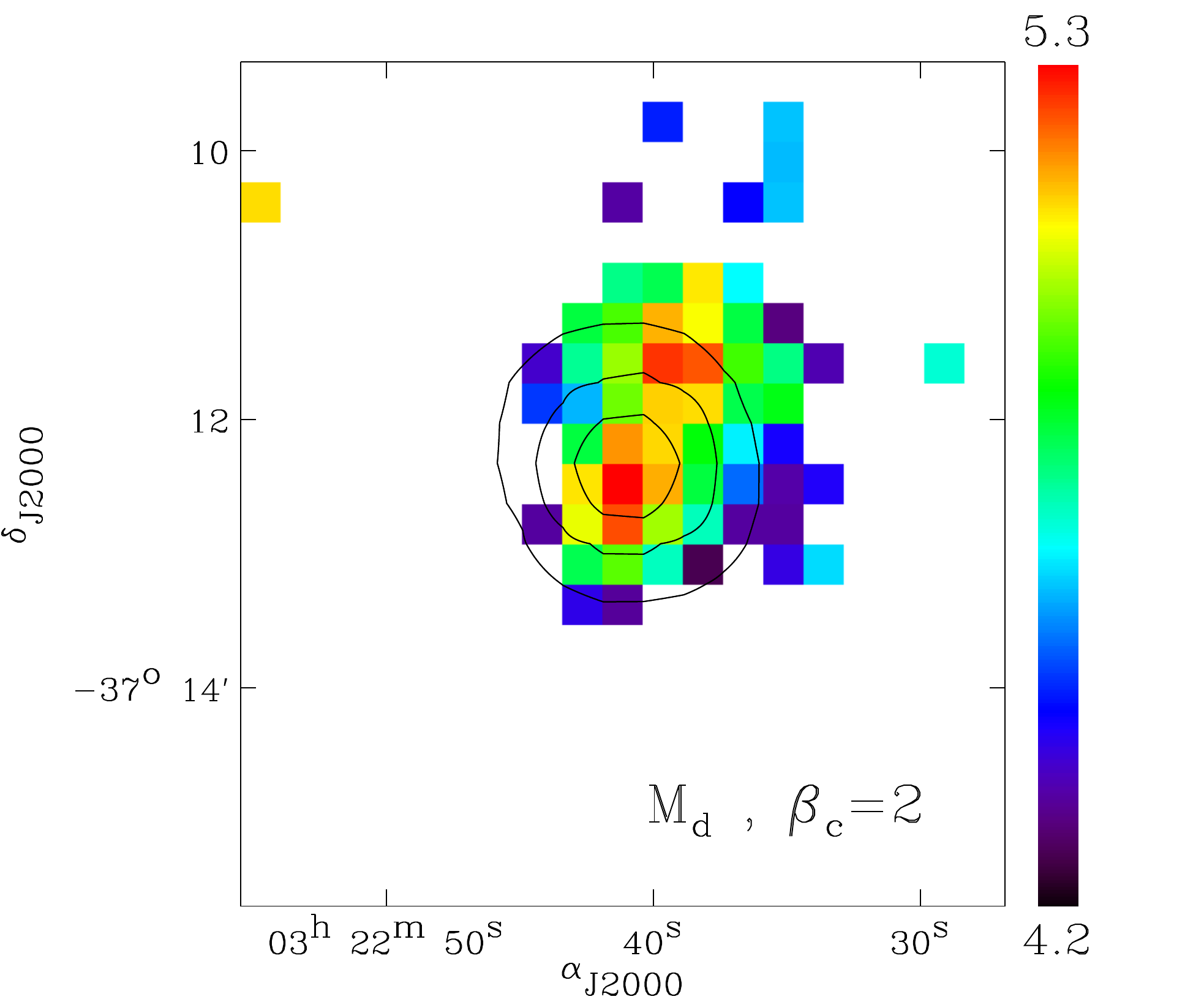} \\
     \centering{\large Emissivity index map} & 
     \centering{\large Temperature map ($\beta$$_c$ free)} & 
     \centering{\large Dust mass surface density ($\beta$$_c$ free)} \\
   	\tabularnewline
	\includegraphics[height=4.4cm]{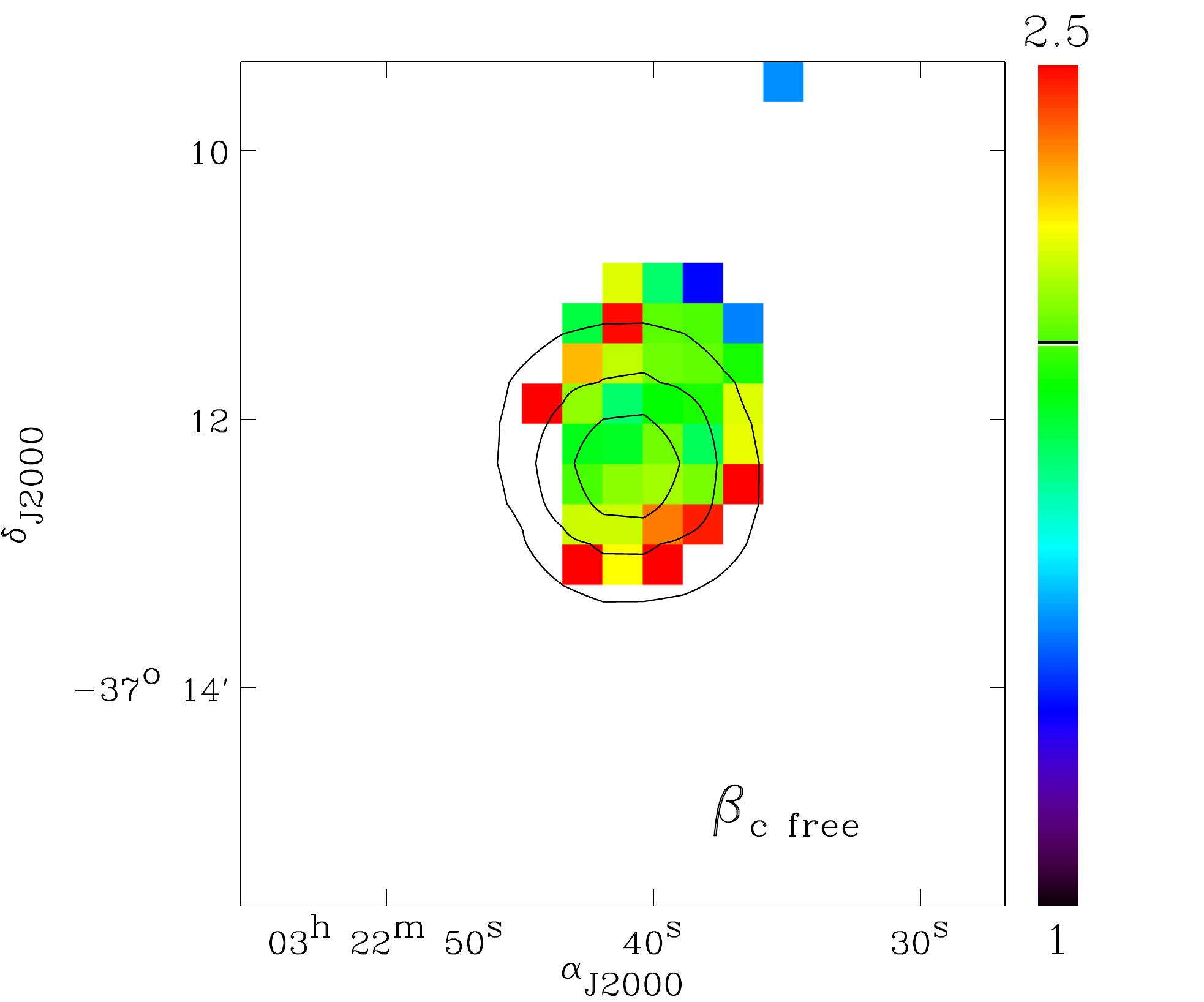}  &
	 \includegraphics[height=4.4cm]{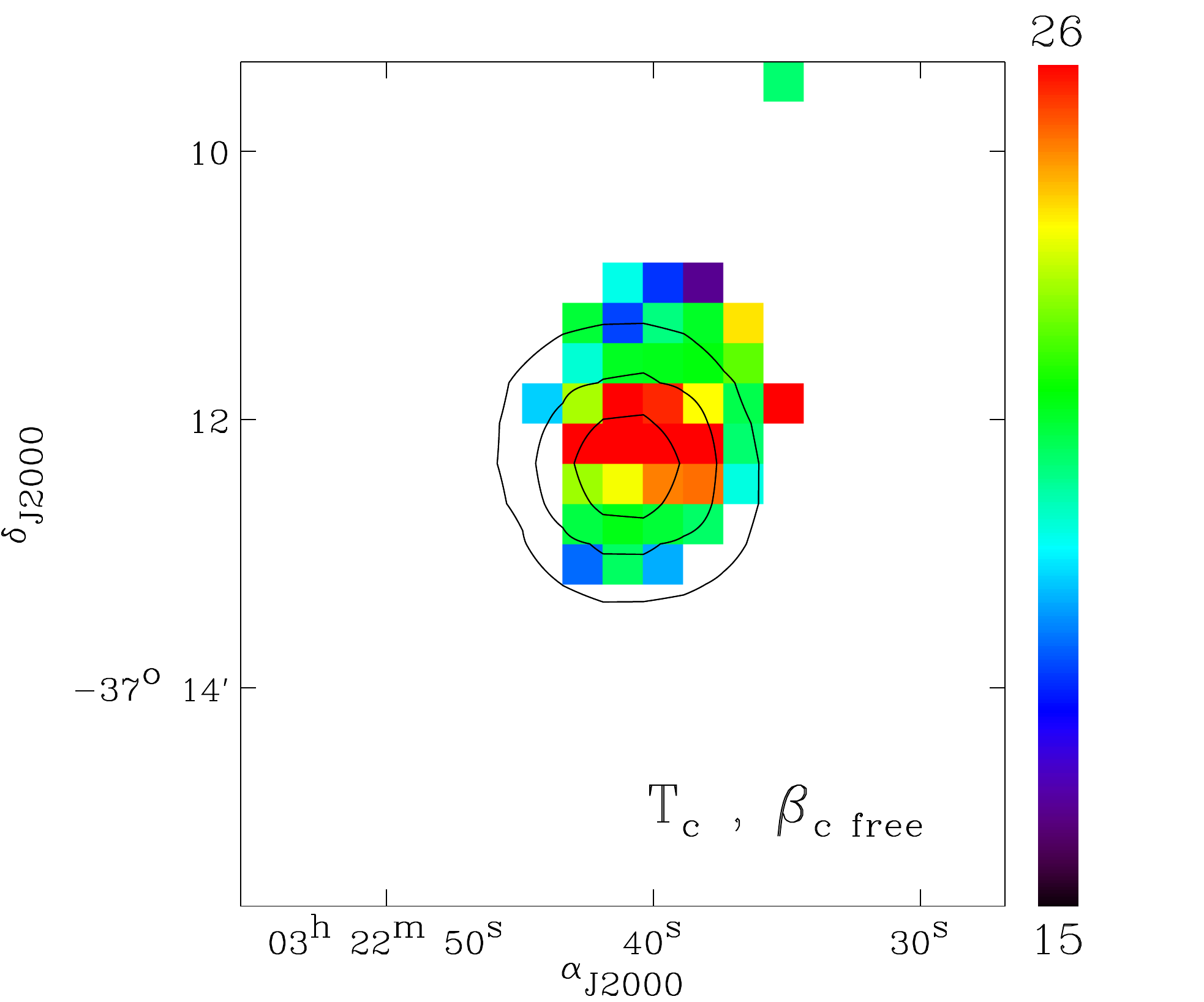} &
	 \includegraphics[height=4.4cm]{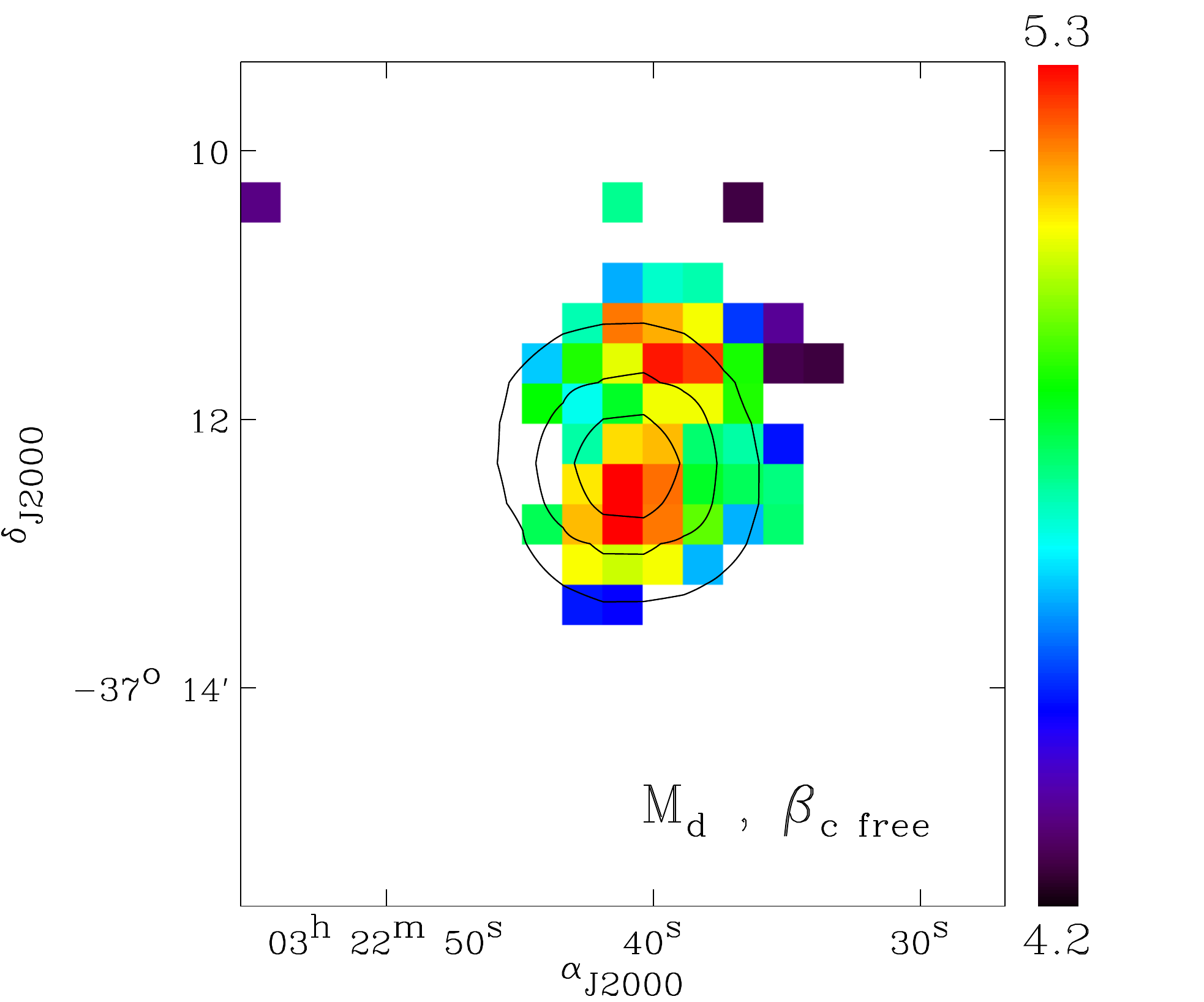} \\	
&&  \\     
&&    \\
     \end{tabular}
     
    \begin{tabular}{m{5.7cm} m{5.7cm} m{5.7cm}}
  {\large NGC1512} && \\
   \centering{\large MIPS 24 \mic} & 
   \centering{\large Temperature map ($\beta$$_c$=2)} & 
   \centering{\large Dust mass surface density ($\beta$$_c$=2)} \\
   	\tabularnewline
	\includegraphics[height=4.7cm]{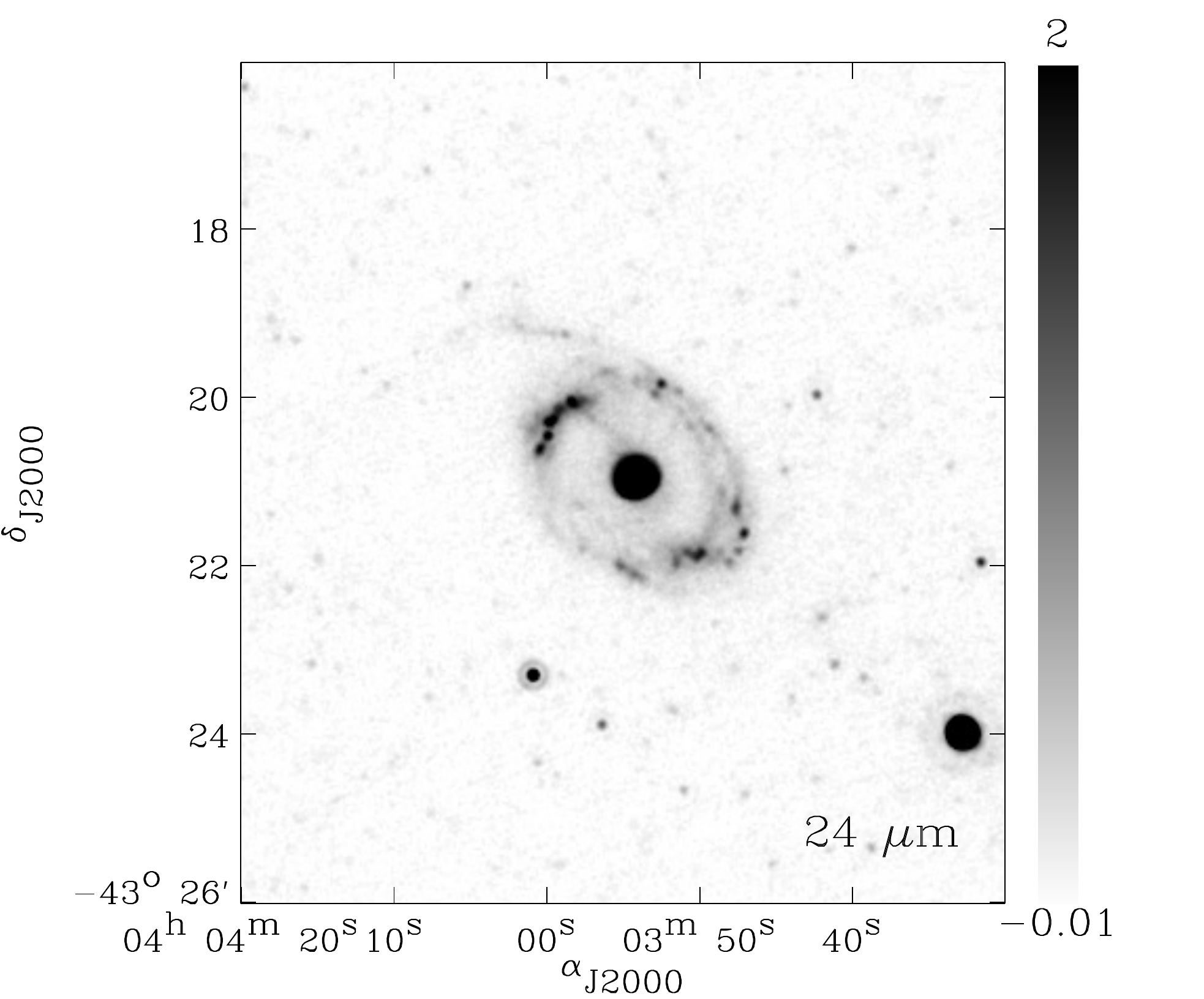} &
	\includegraphics[height=4.7cm]{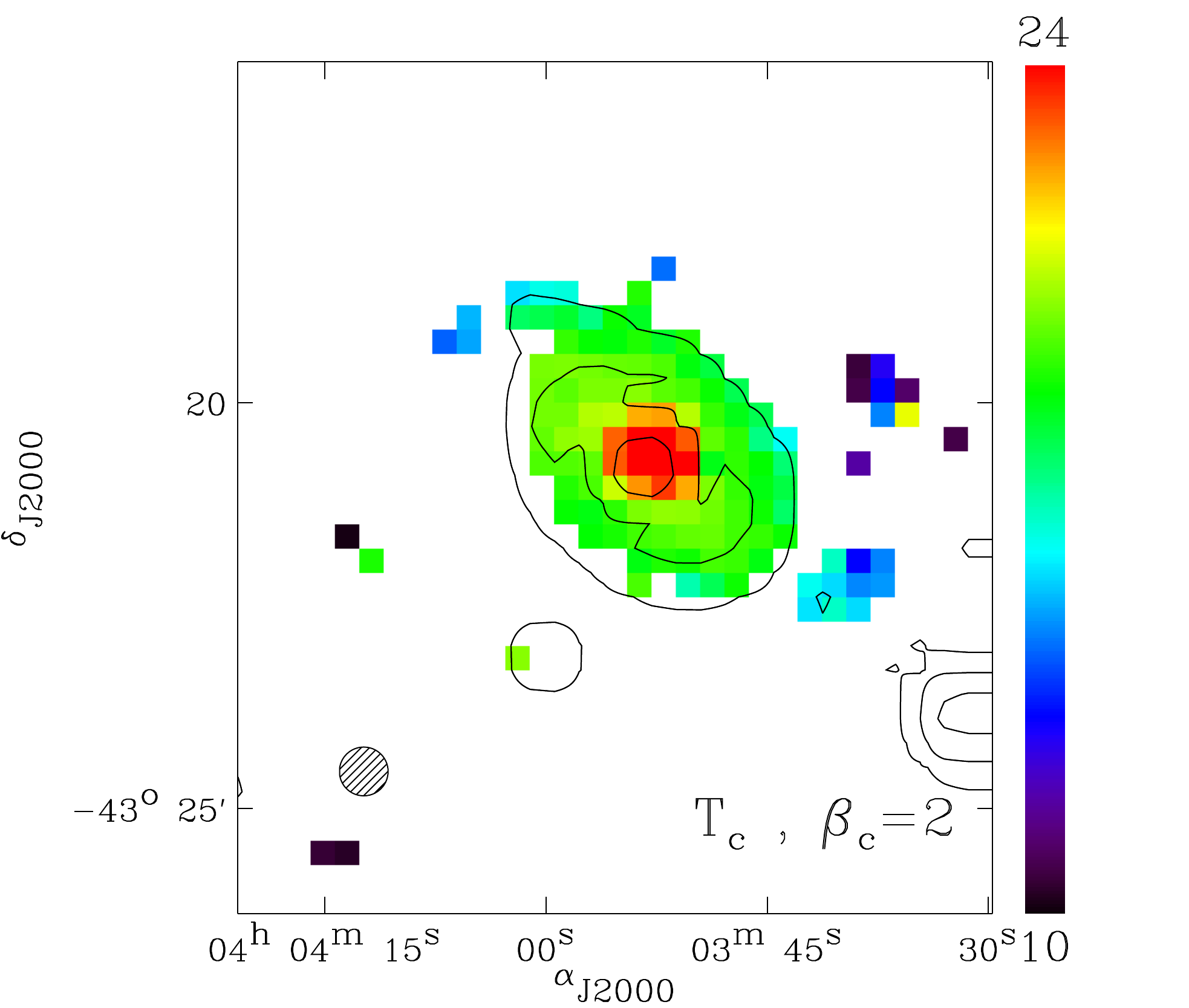} &
	\includegraphics[height=4.7cm]{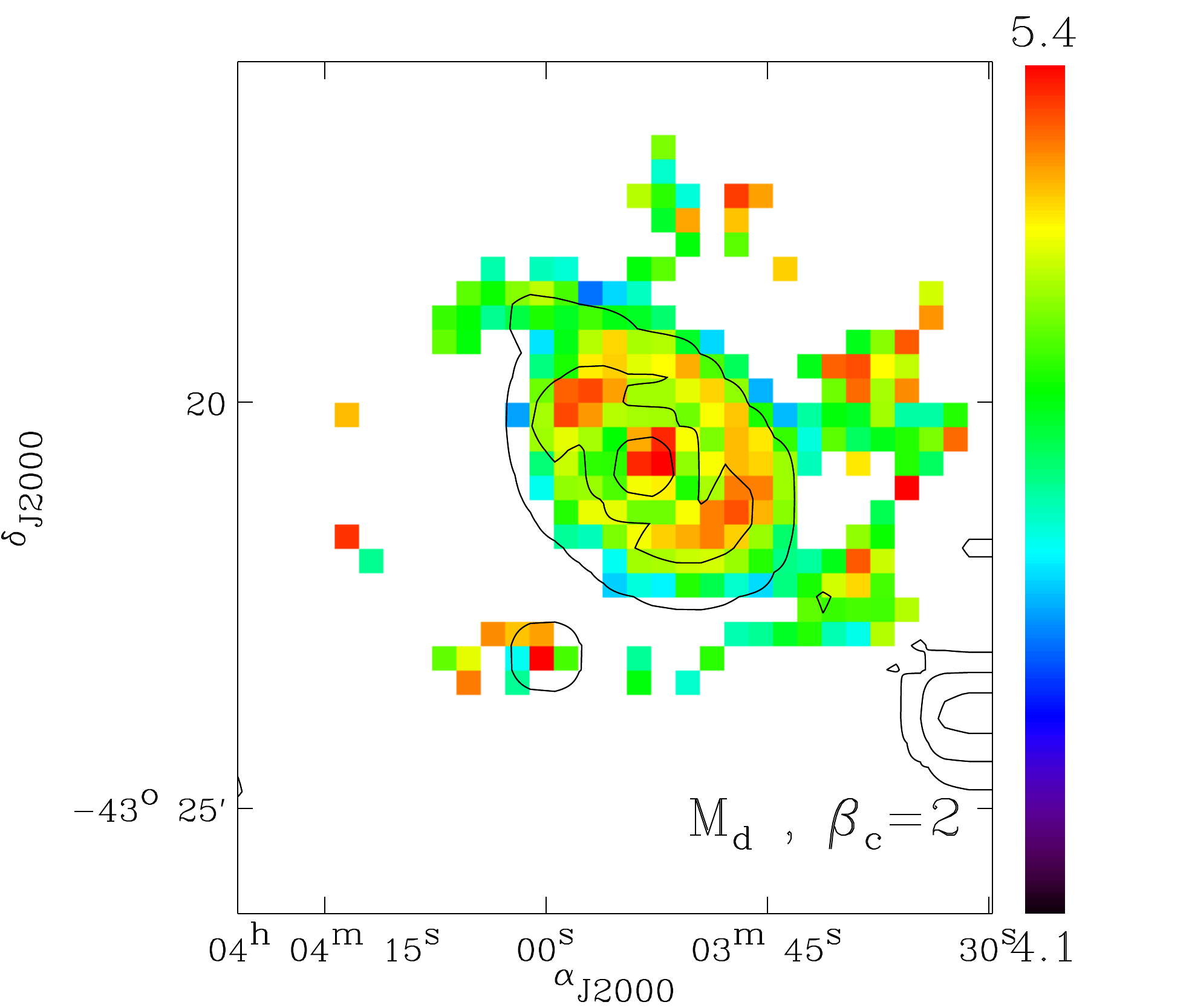} \\
     \centering{\large Emissivity index map} & 
     \centering{\large Temperature map ($\beta$$_c$ free)} & 
     \centering{\large Dust mass surface density ($\beta$$_c$ free)} \\
   	\tabularnewline
	\includegraphics[height=4.7cm]{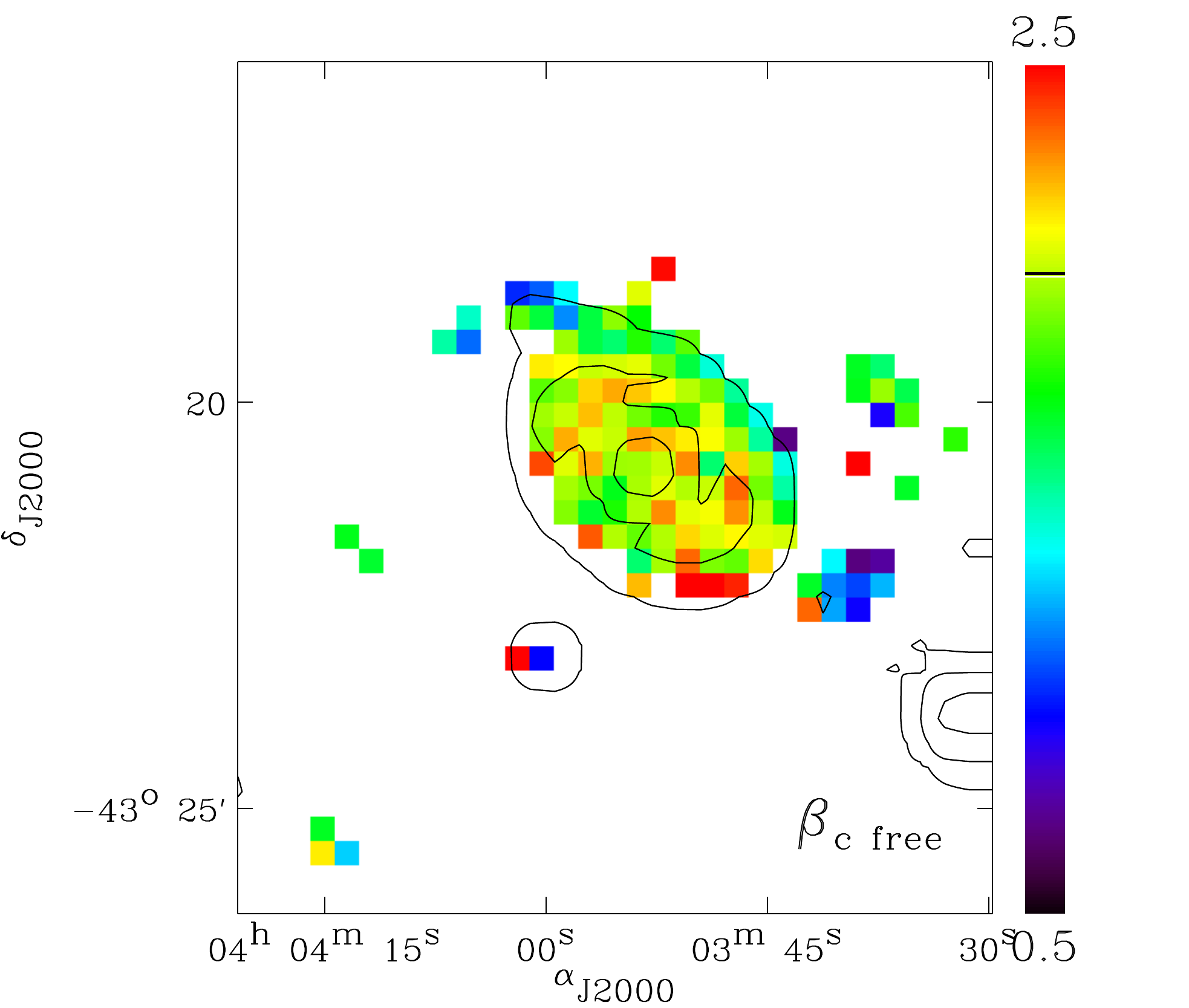}  &
	 \includegraphics[height=4.7cm]{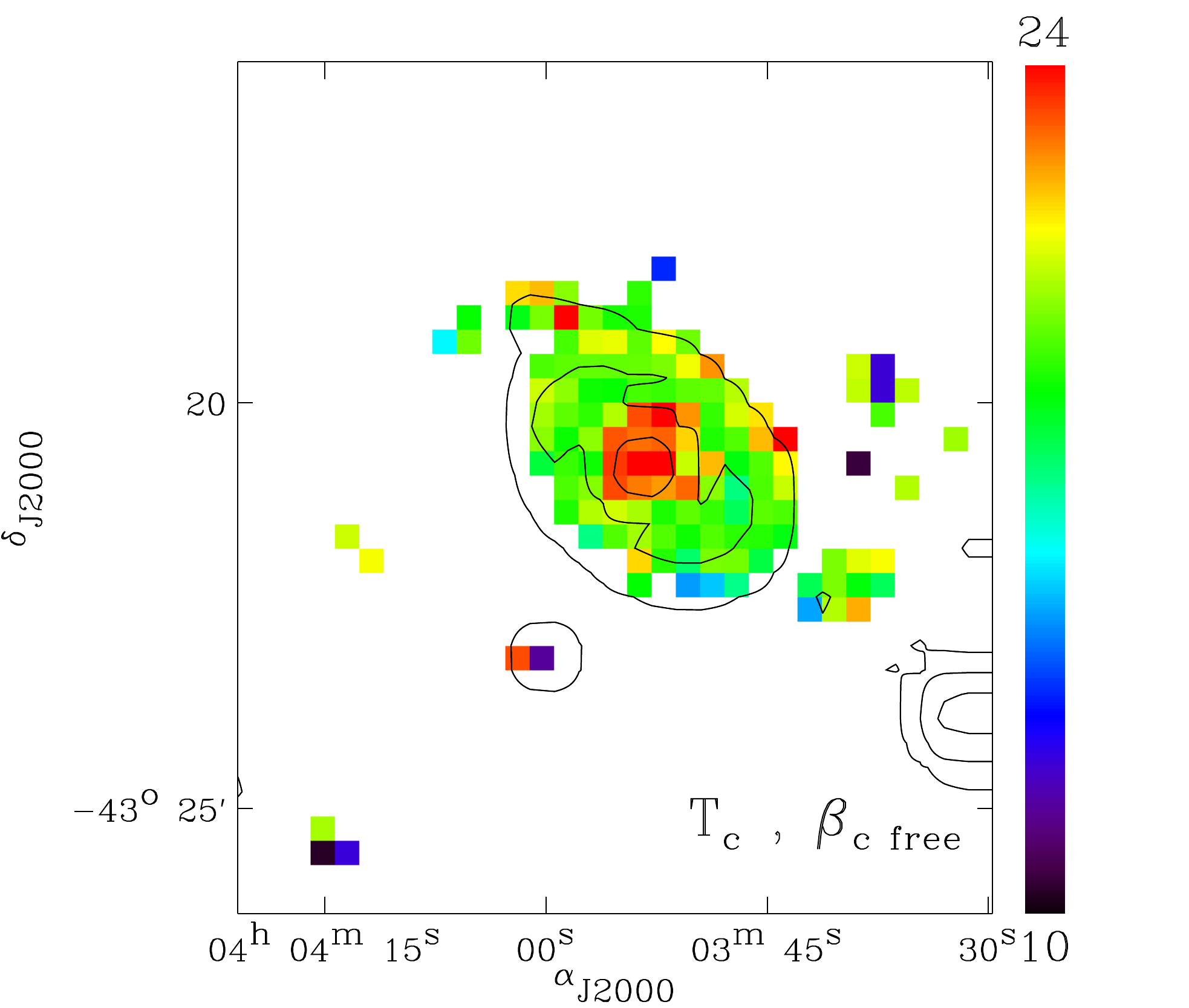} &
	 \includegraphics[height=4.7cm]{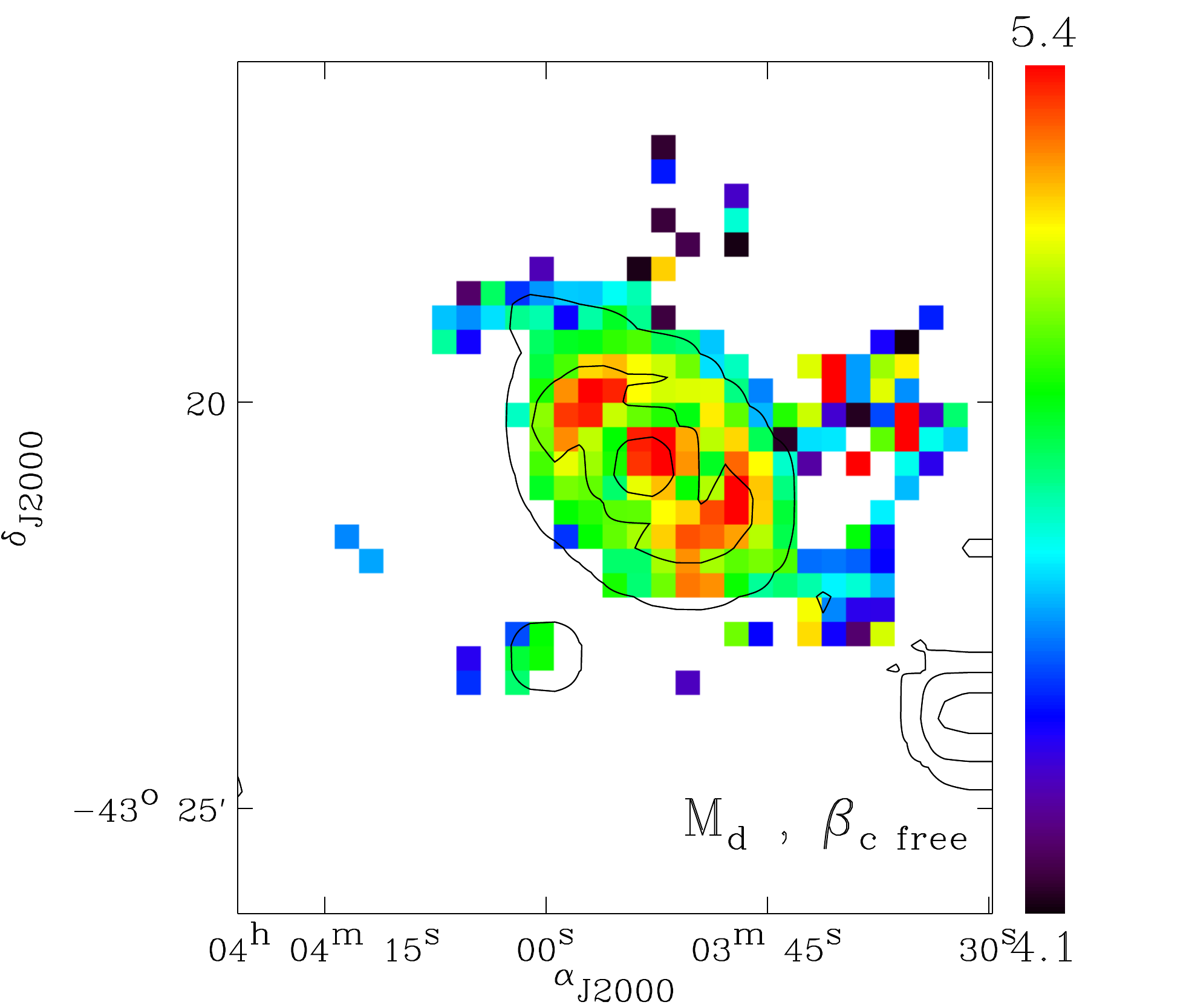} \\	 
   &&    \\                     
                            \end{tabular}
\caption{Continued. }
\end{figure*}

\addtocounter {figure}{-1}
        
\begin{figure*}
    \centering   

    \begin{tabular}{m{5.7cm} m{5.7cm} m{5.7cm}}

 {\large NGC~3351}&&\\
   \centering{\large MIPS 24 \mic} & 
   \centering{\large Temperature map ($\beta$$_c$=2)} &
    \centering{\large Dust mass surface density ($\beta$$_c$=2)} \\
   	\tabularnewline
	\includegraphics[height=4.7cm]{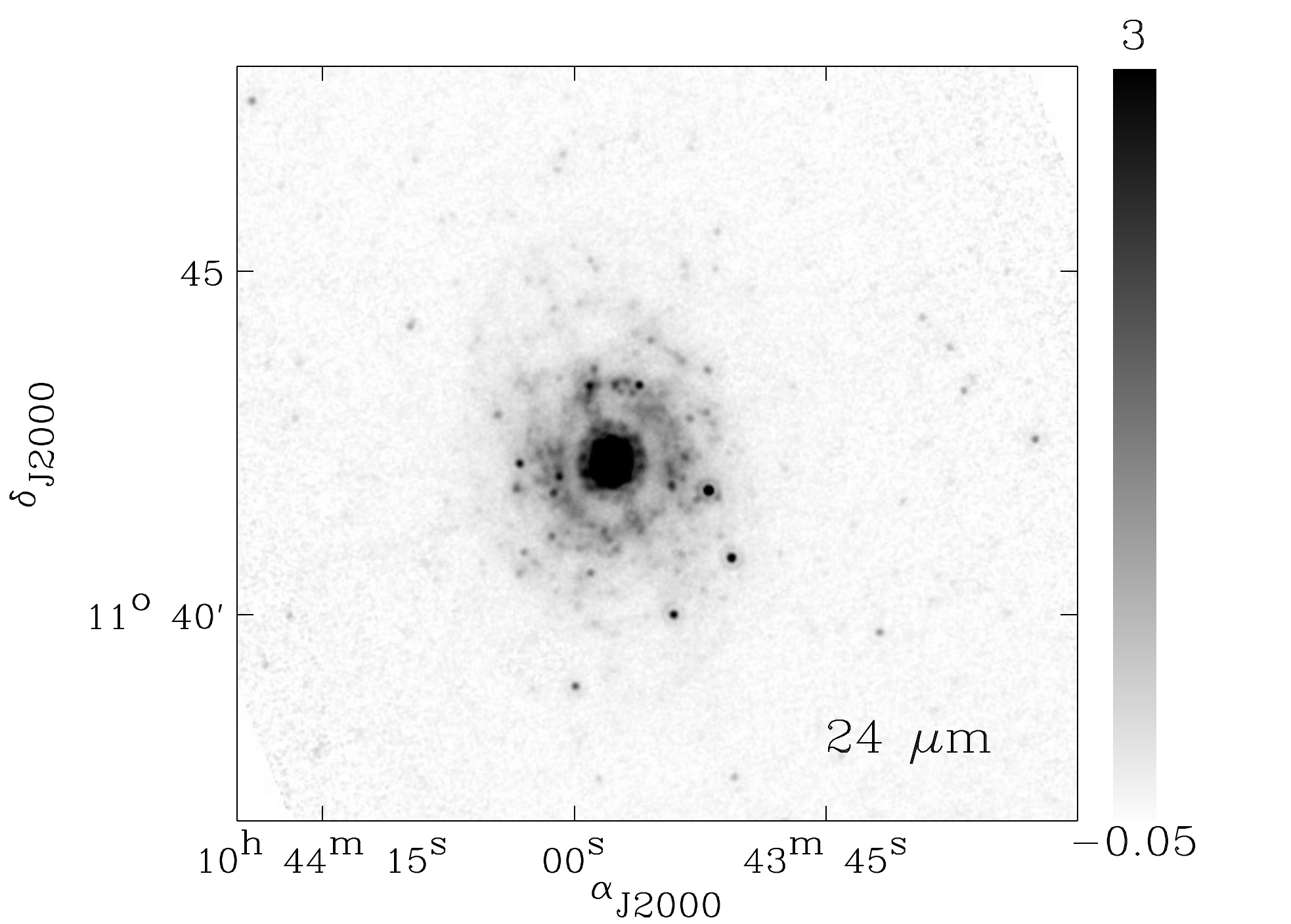} &
	\includegraphics[height=4.7cm]{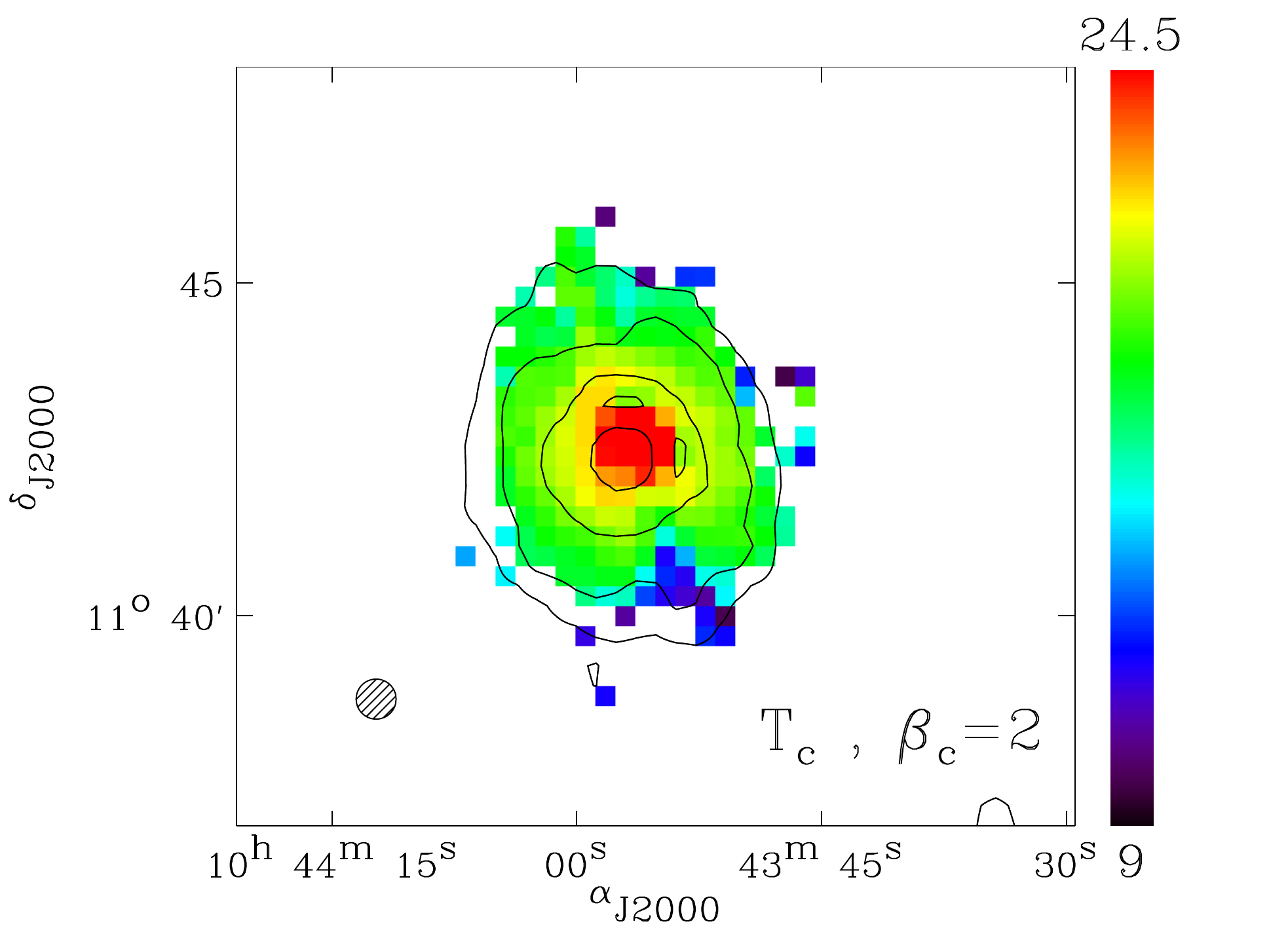} &
	\includegraphics[height=4.7cm]{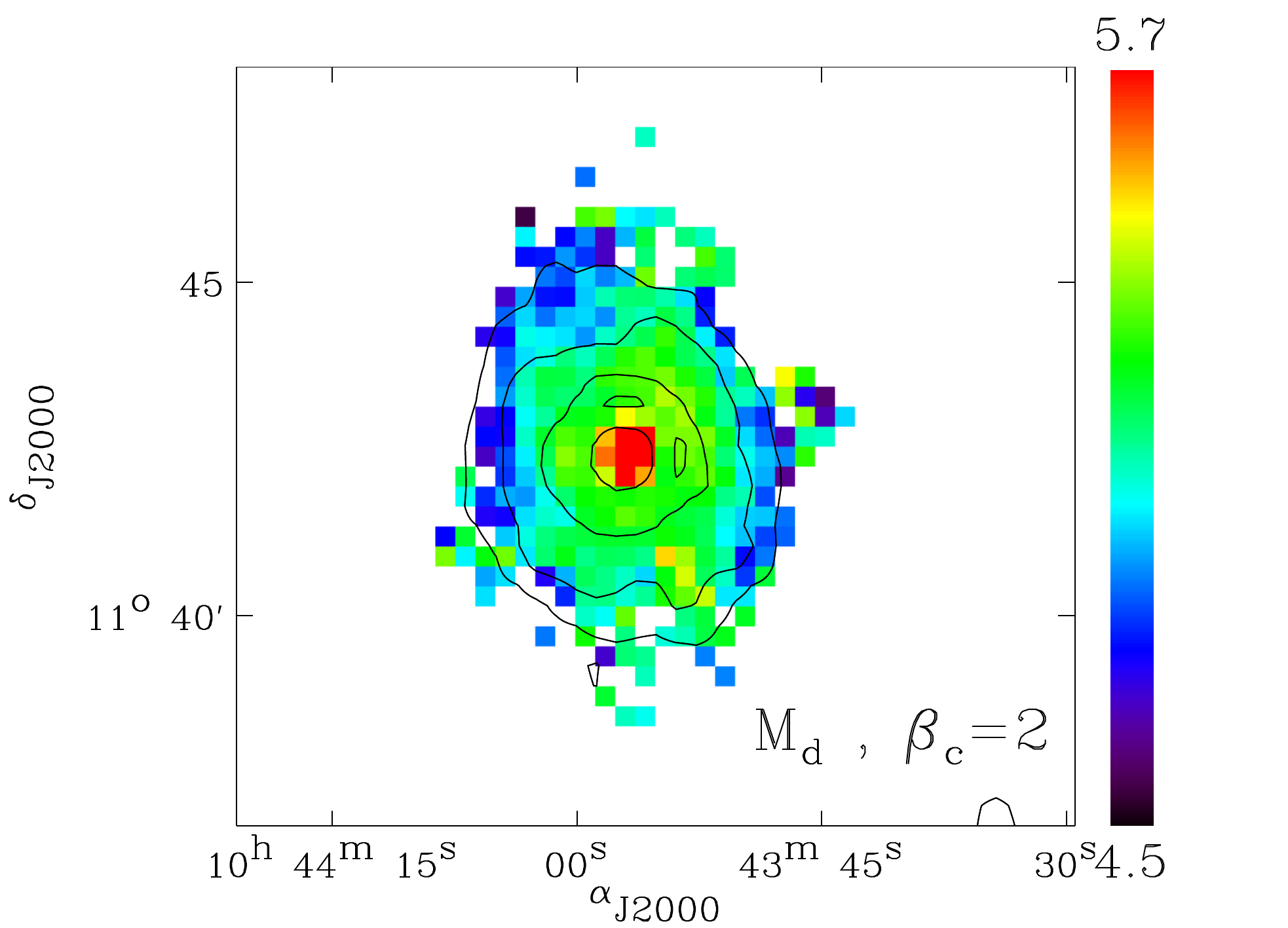} \\
     \centering{\large Emissivity index map} & 
     \centering{\large Temperature map ($\beta$$_c$ free)} & 
     \centering{\large Dust mass surface density ($\beta$$_c$ free)} \\
   	\tabularnewline
	\includegraphics[height=4.7cm]{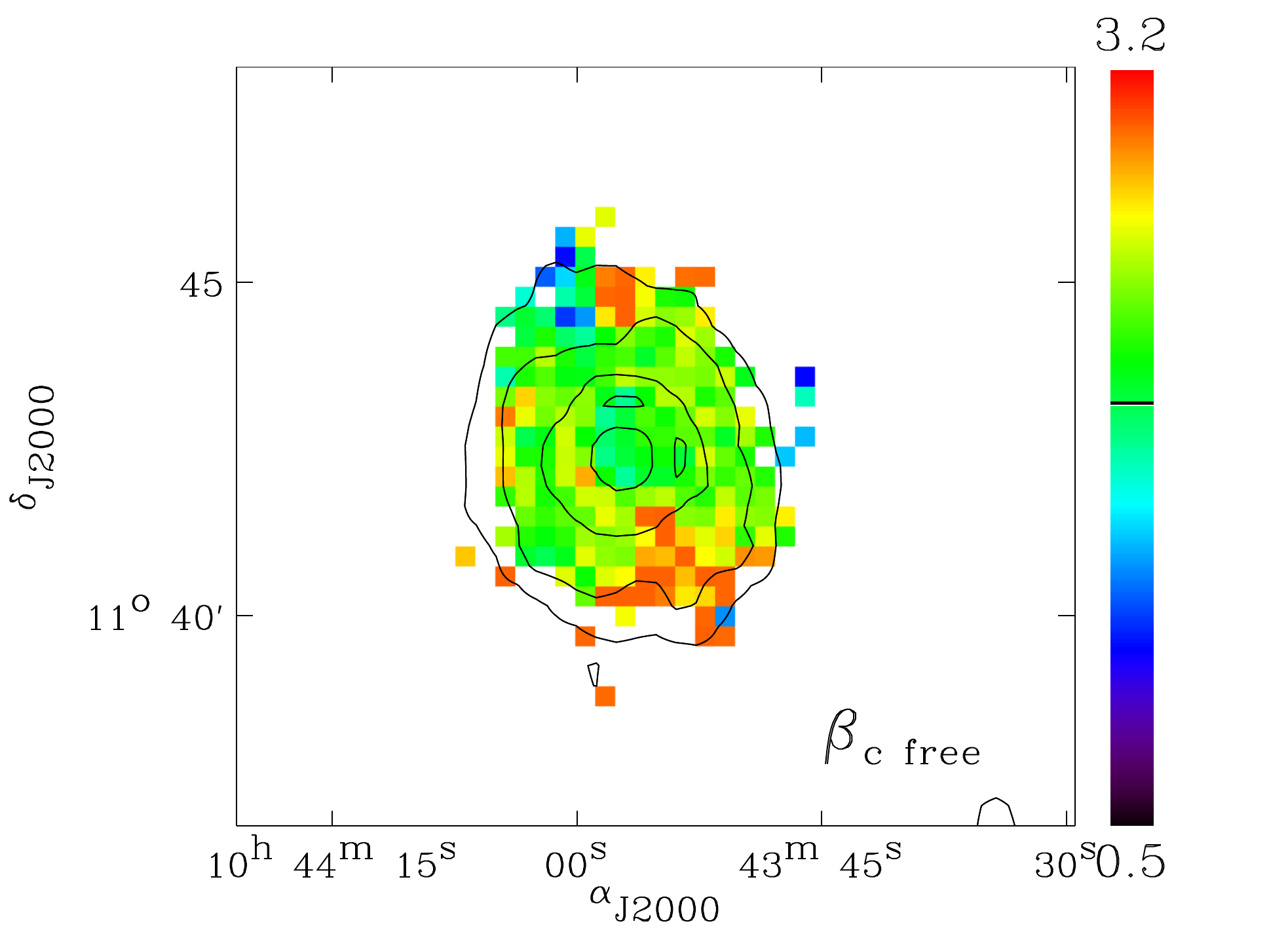}  &
	 \includegraphics[height=4.7cm]{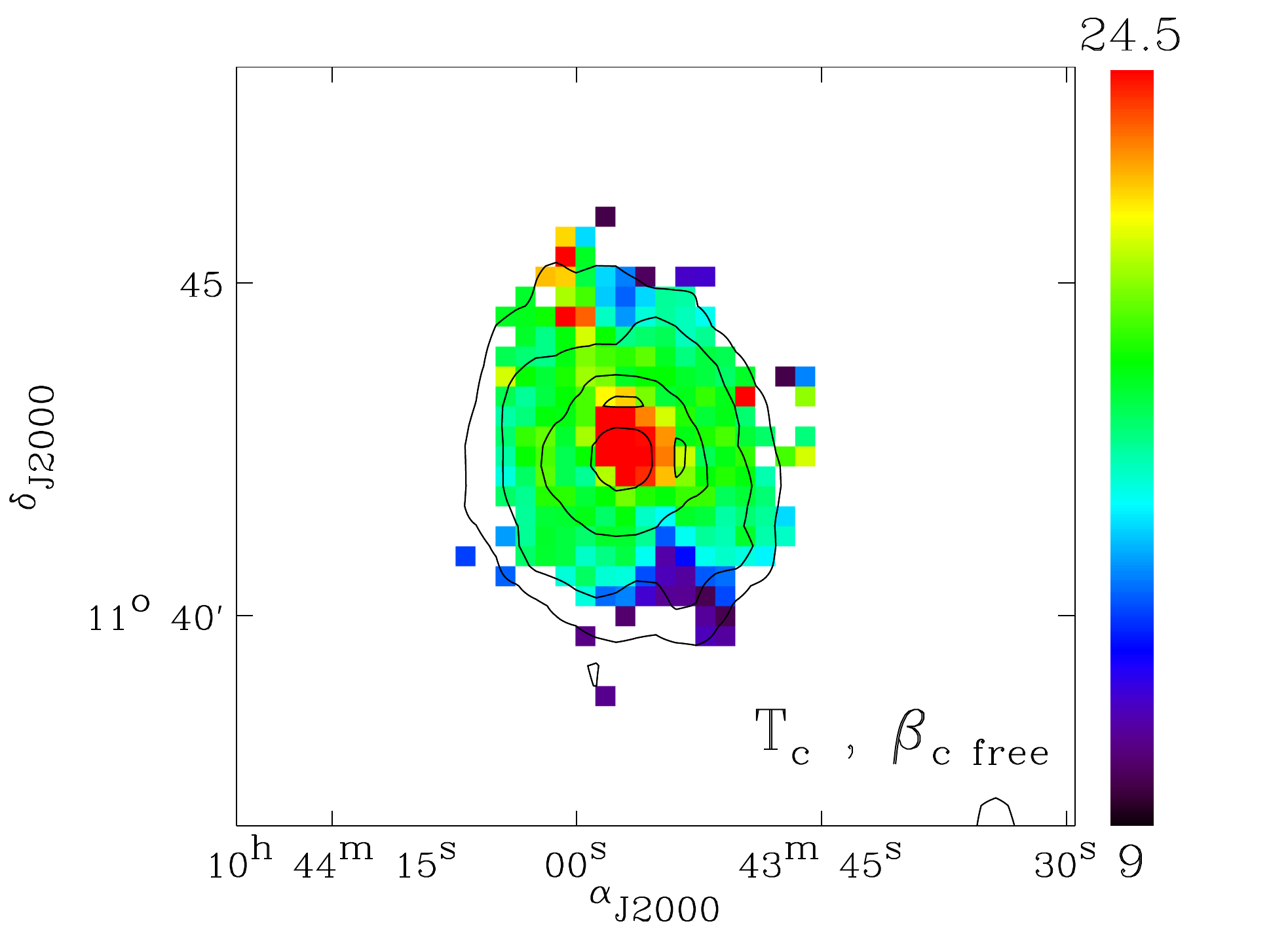} &
	 \includegraphics[height=4.7cm]{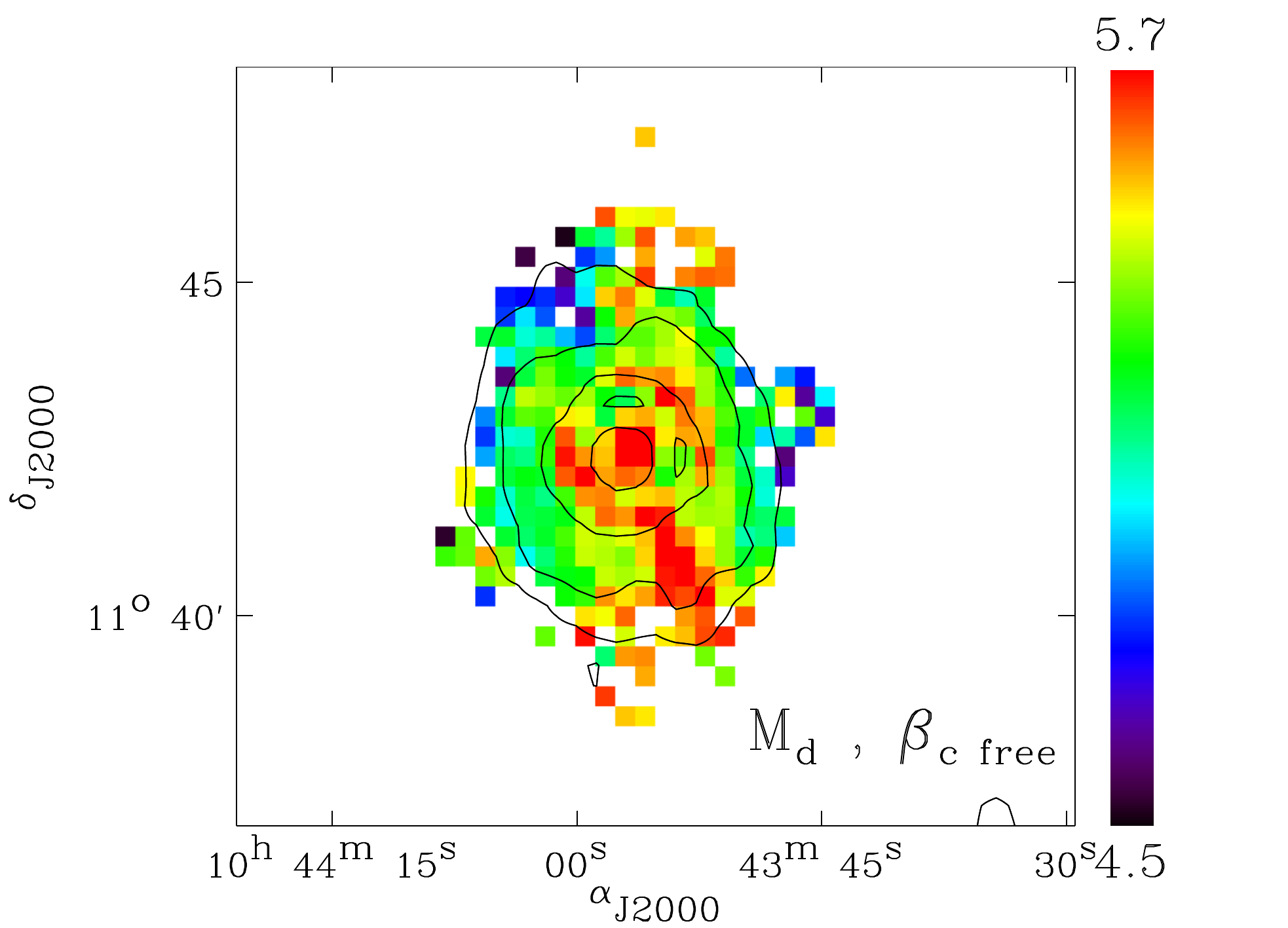} \\	
&&  \\     
&&    \\
 {\large NGC~3621}&&\\
   \centering{\large MIPS 24 \mic} & 
   \centering{\large Temperature map ($\beta$$_c$=2)} & 
   \centering{\large Dust mass surface density ($\beta$$_c$=2)} \\
   	\tabularnewline
	\includegraphics[height=4.3cm]{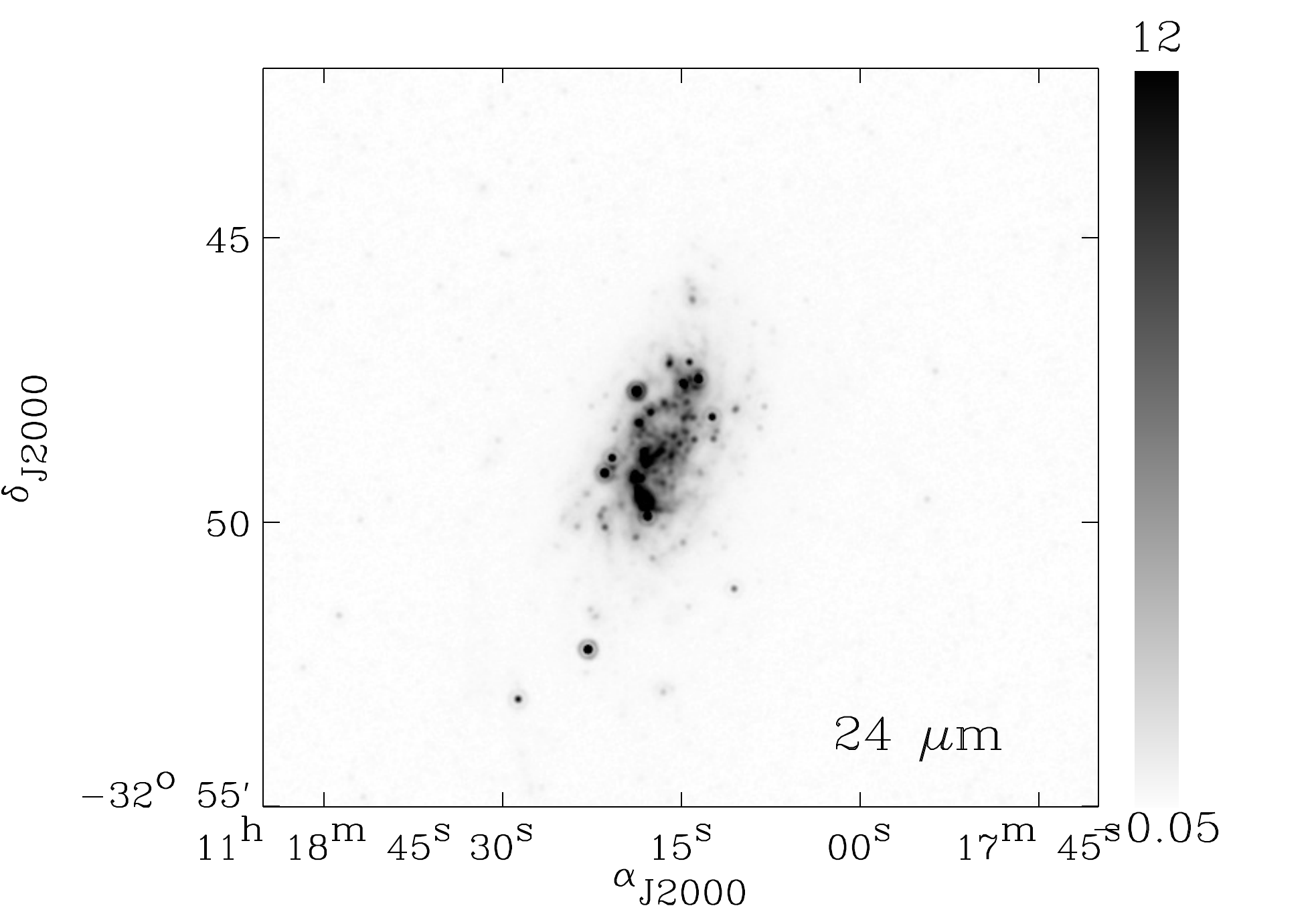} &
	\includegraphics[height=4.3cm]{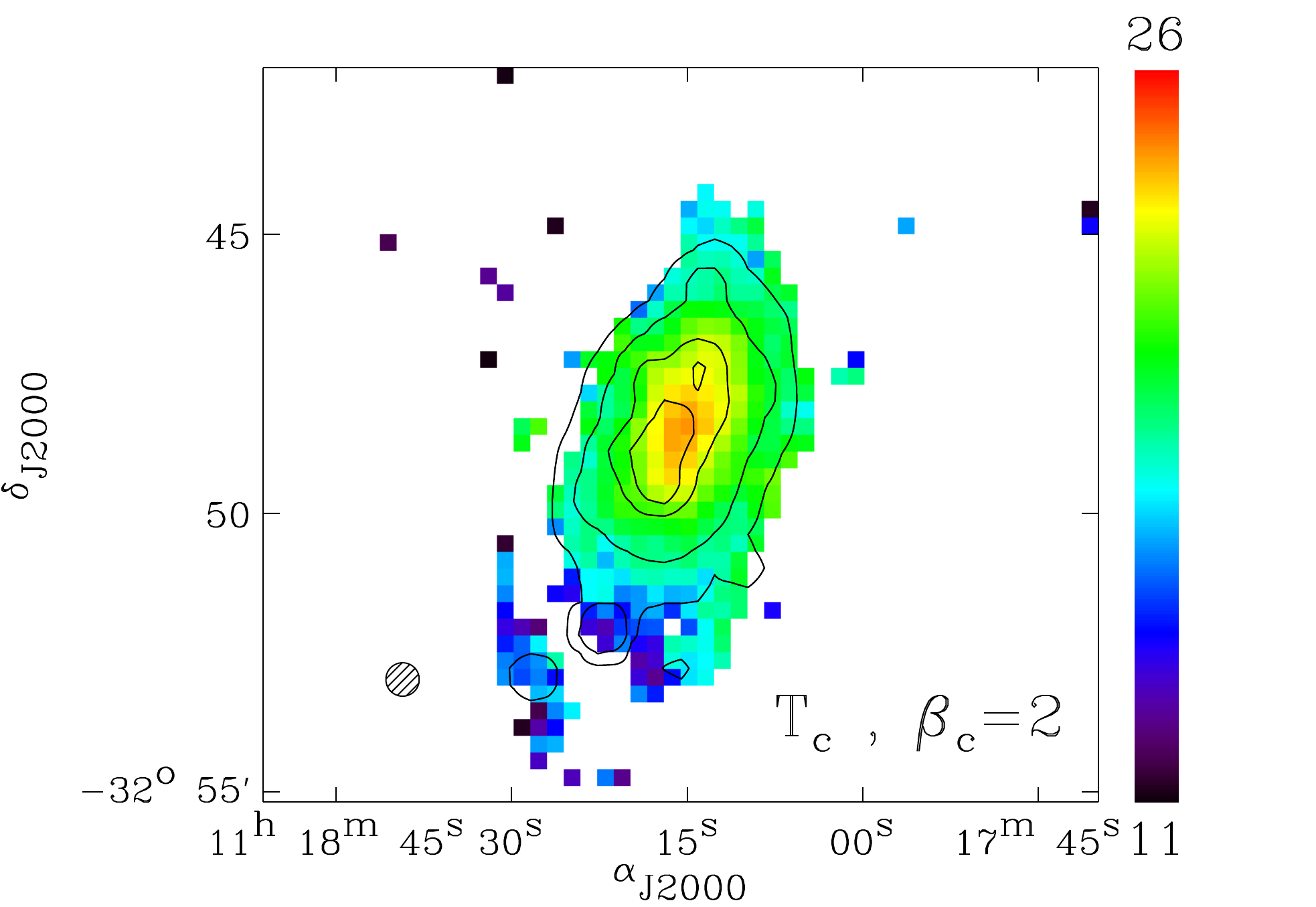} &
	\includegraphics[height=4.3cm]{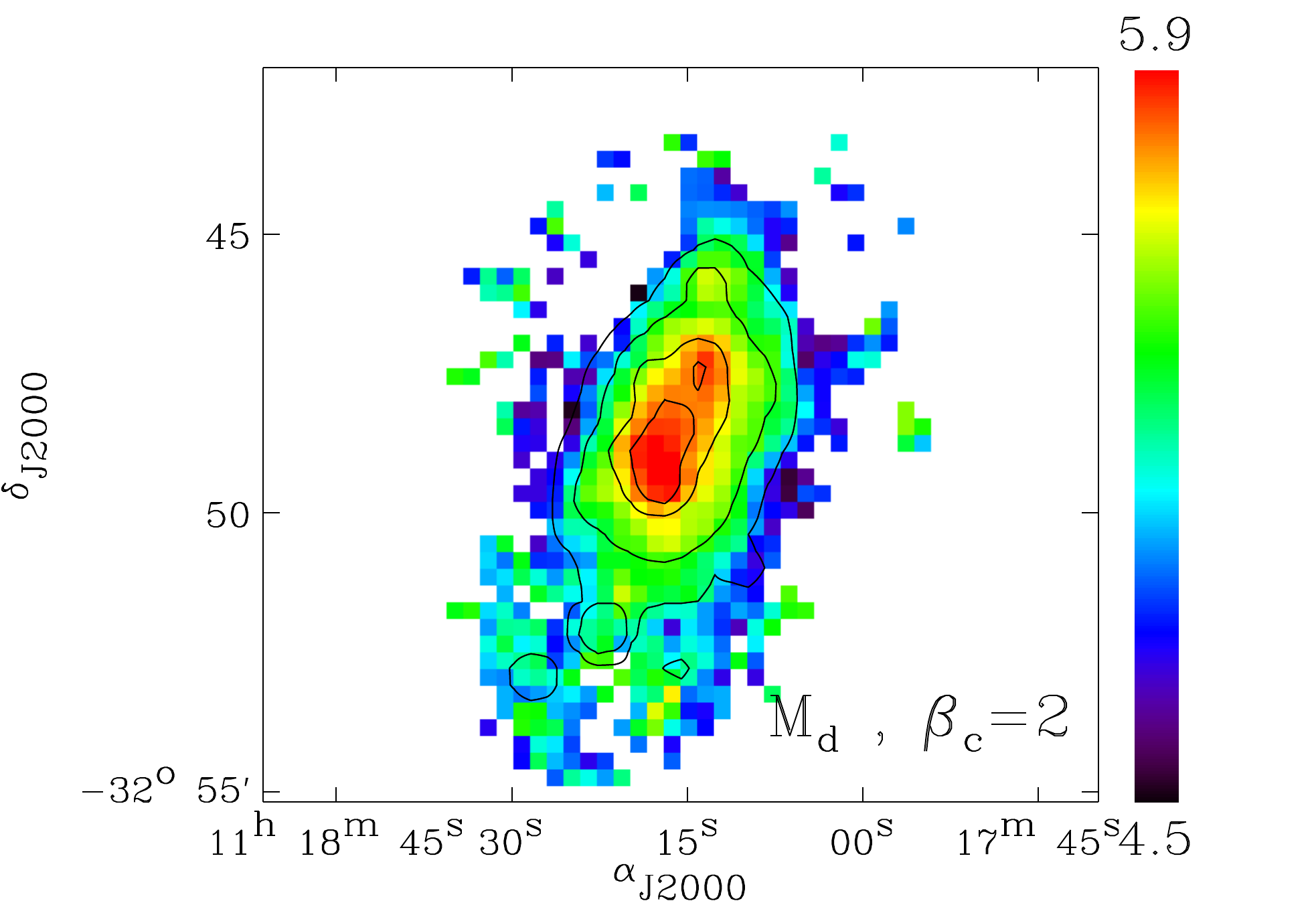} \\
     \centering{\large Emissivity index map} &
      \centering{\large Temperature map ($\beta$$_c$ free)} & 
      \centering{\large Dust mass surface density ($\beta$$_c$ free)} \\
   	\tabularnewline
	\includegraphics[height=4.3cm]{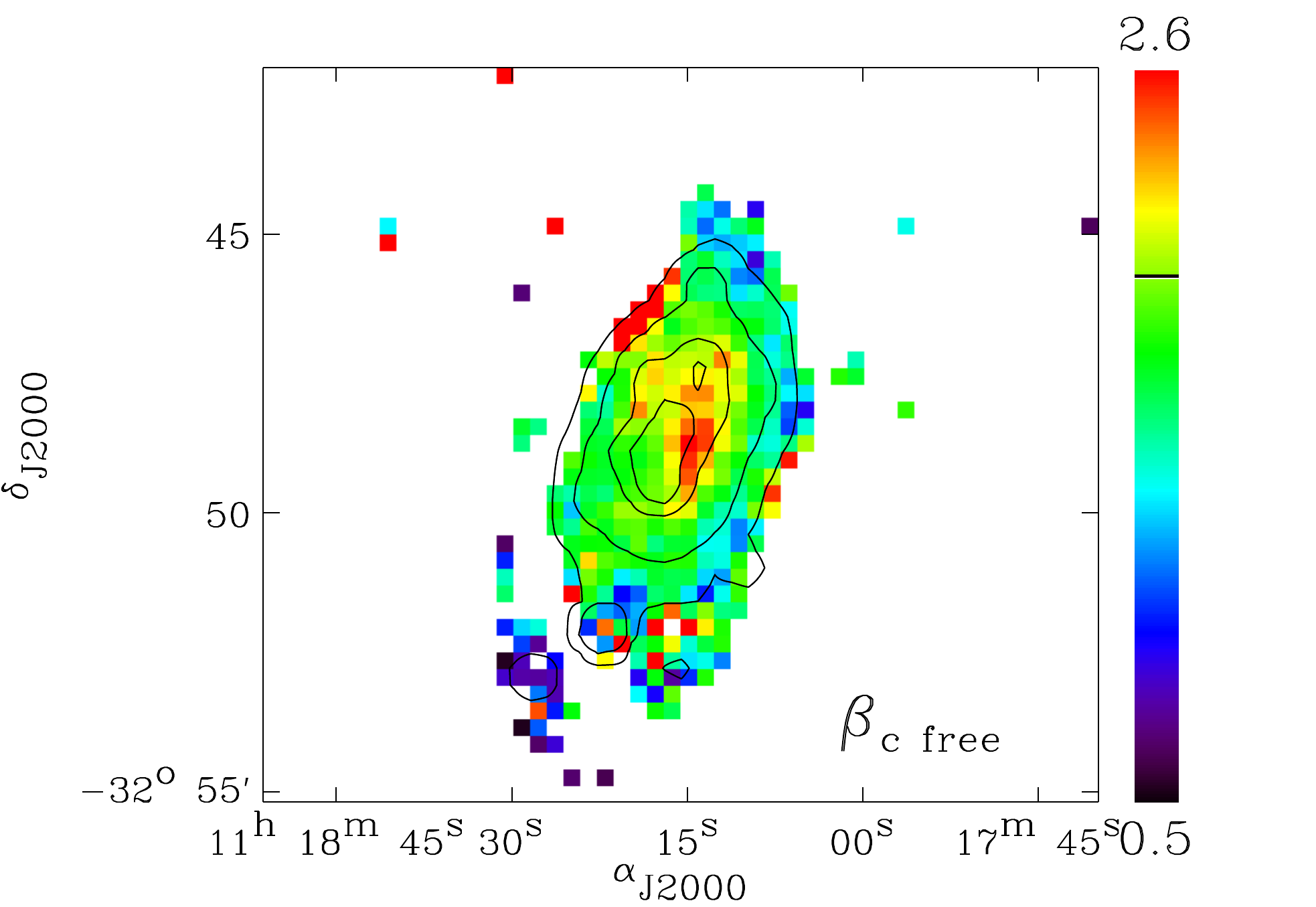}  &
	 \includegraphics[height=4.3cm]{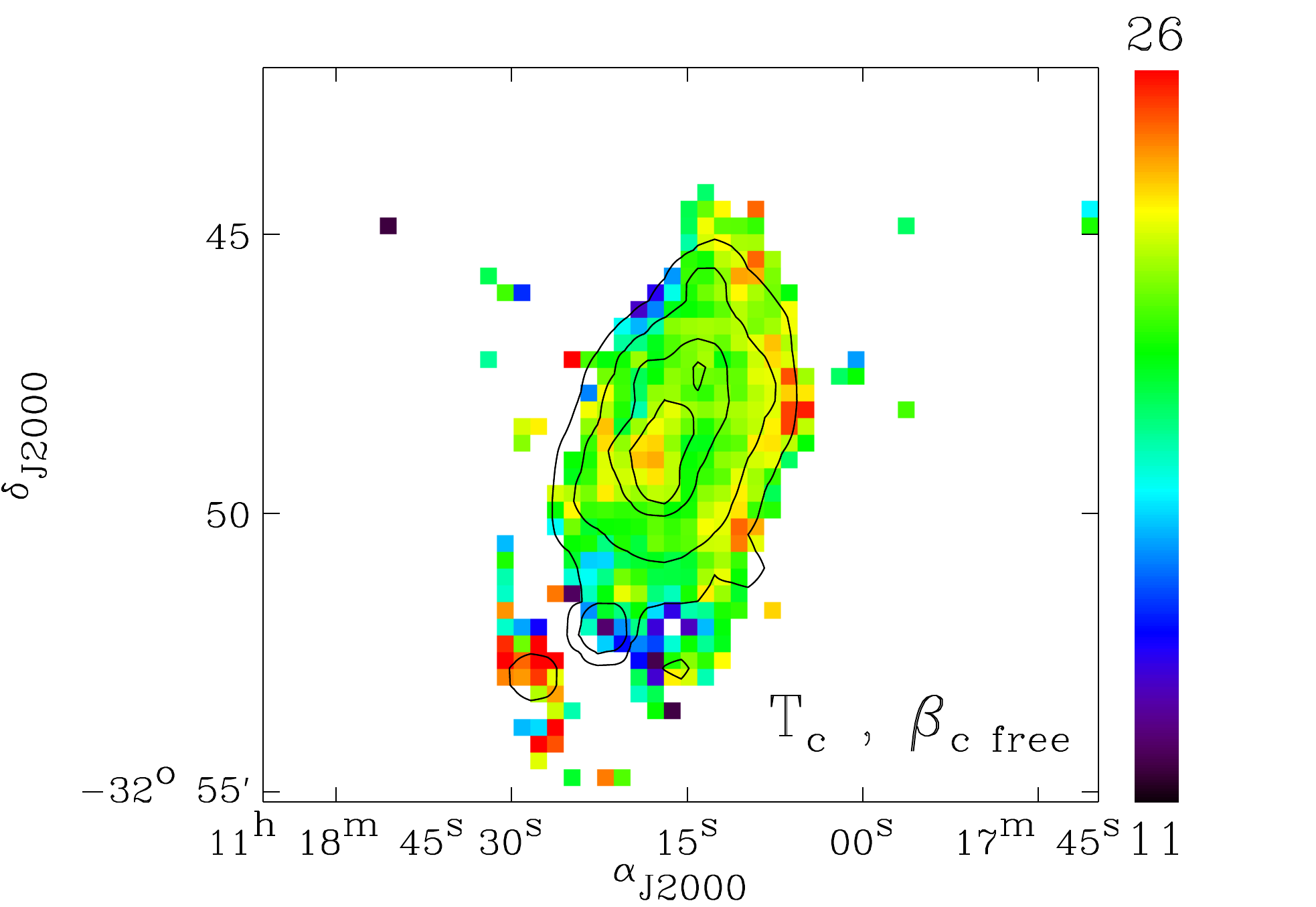} &
	 \includegraphics[height=4.3cm]{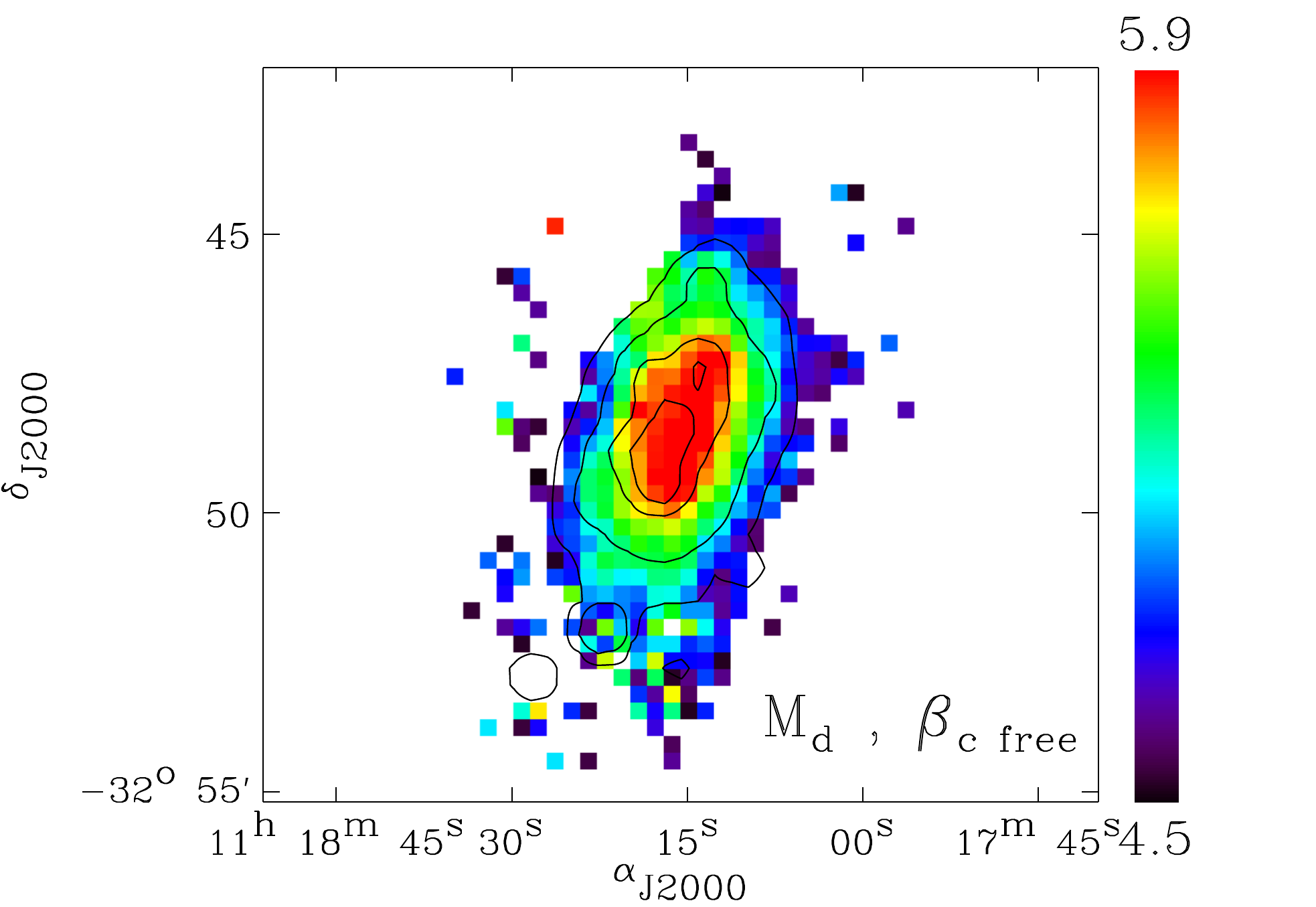} \\	        
                            \end{tabular}
\caption{Continued. }
\end{figure*}

\addtocounter {figure}{-1}
        
\begin{figure*}
    \centering   
    \begin{tabular}{m{5.7cm} m{5.7cm} m{5.7cm}}

 {\large NGC~3627} &&   \\
   \centering{\large MIPS 24 \mic} &
    \centering{\large Temperature map ($\beta$$_c$=2)} &
     \centering{\large Dust mass surface density ($\beta$$_c$=2)} \\
   	\tabularnewline
	\includegraphics[height=4.7cm]{NGC3627_MIPS24} &
	\includegraphics[height=4.7cm]{NGC3627_Temp_Median_beta2_24um} &
	\includegraphics[height=4.7cm]{NGC3627_Mdust_beta2_24um} \\
     \centering{\large Emissivity index map} &
      \centering{\large Temperature map ($\beta$$_c$ free)} &
       \centering{\large Dust mass surface density ($\beta$$_c$ free)} \\
   	\tabularnewline
	\includegraphics[height=4.7cm]{NGC3627_Beta_Median_24um}  &
	 \includegraphics[height=4.7cm]{NGC3627_Temp_Median_betafree_24um} &
	 \includegraphics[height=4.7cm]{NGC3627_Mdust_betafree_24um} \\
&&  \\     
&&    \\
   \end{tabular}
   
    \begin{tabular}{m{5.5cm} m{5.5cm} m{5.5cm}}

    {\large NGC~4826} && \\
   \centering{\large MIPS 24 \mic} &
    \centering{\large Temperature map ($\beta$$_c$=2)} &
     \centering{\large Dust mass surface density ($\beta$$_c$=2)} \\
   	\tabularnewline
	\includegraphics[height=4.1cm]{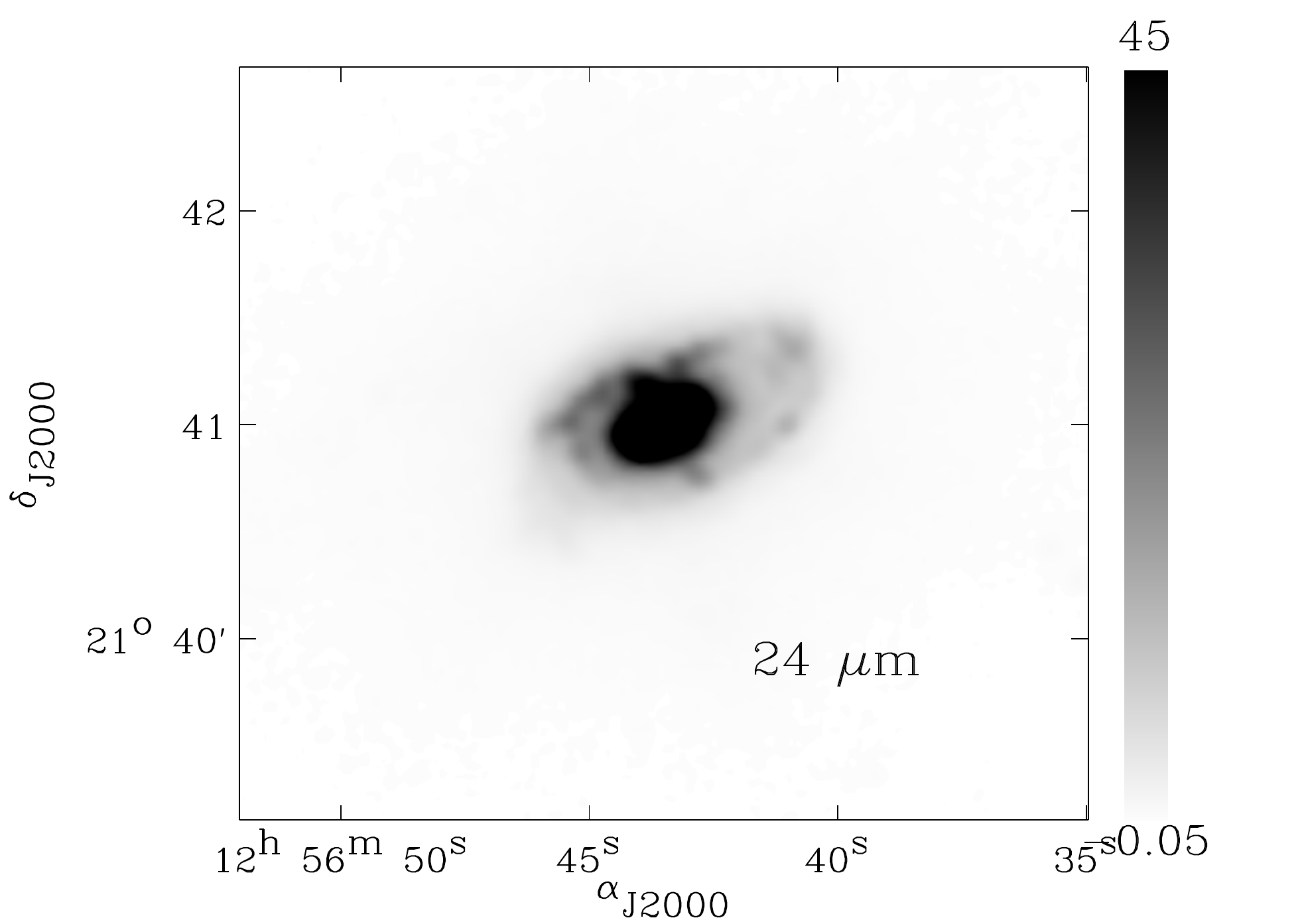} &
	\includegraphics[height=4.1cm]{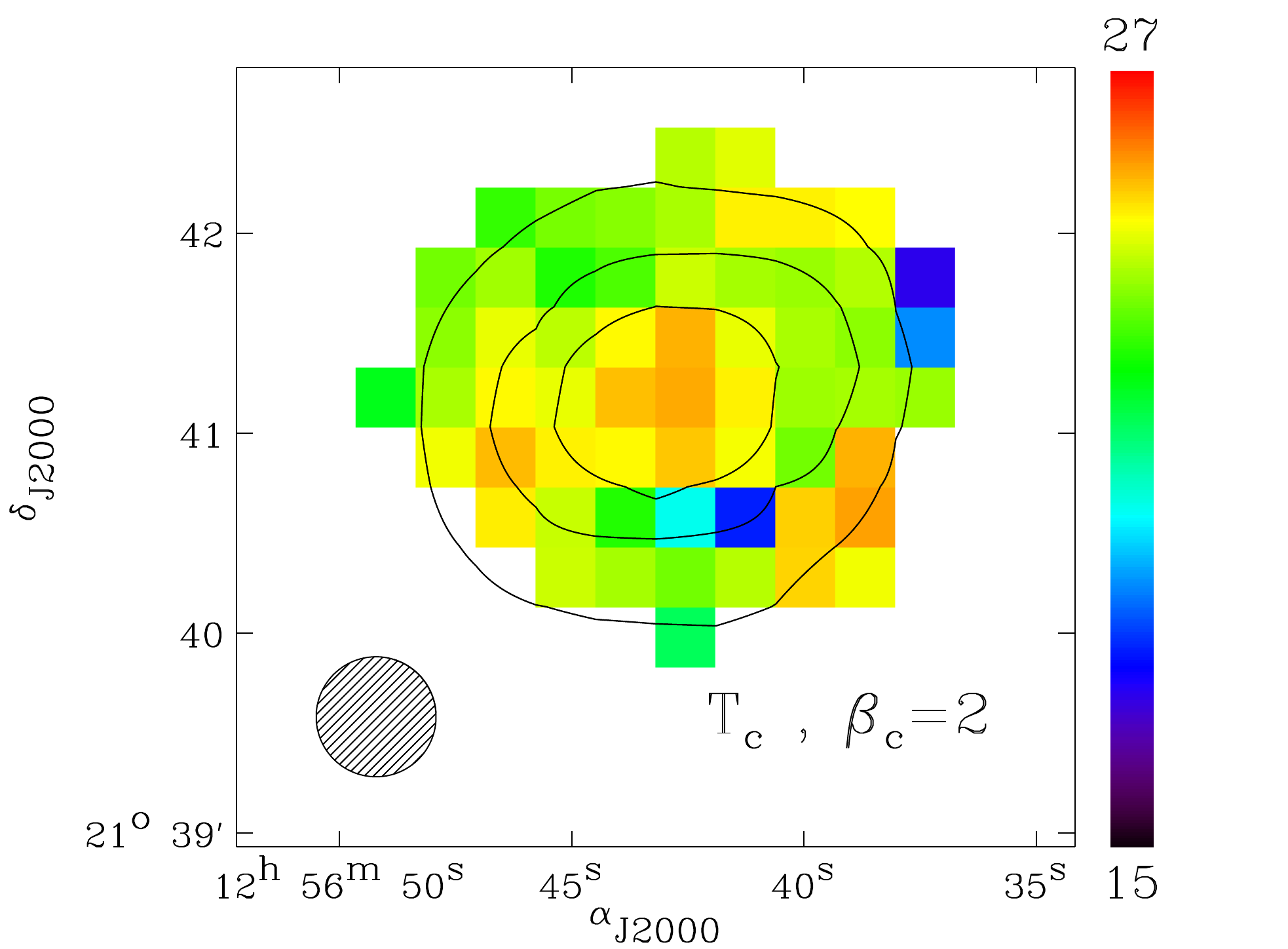} &
	\includegraphics[height=4.1cm]{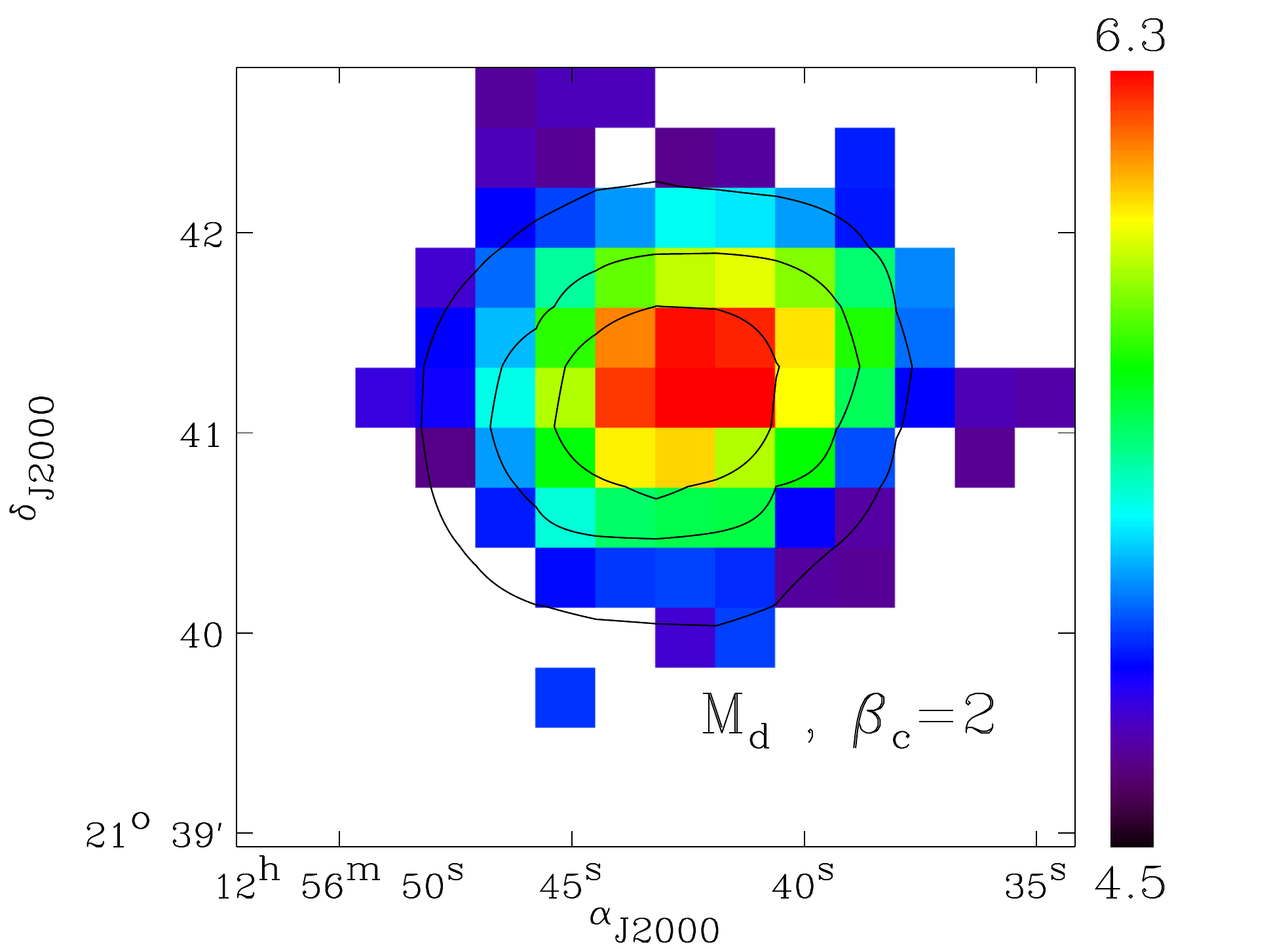} \\
     \centering{\large Emissivity index map} &
      \centering{\large Temperature map ($\beta$$_c$ free)} &
       \centering{\large Dust mass surface density ($\beta$$_c$ free)} \\
   	\tabularnewline
	\includegraphics[height=4.1cm]{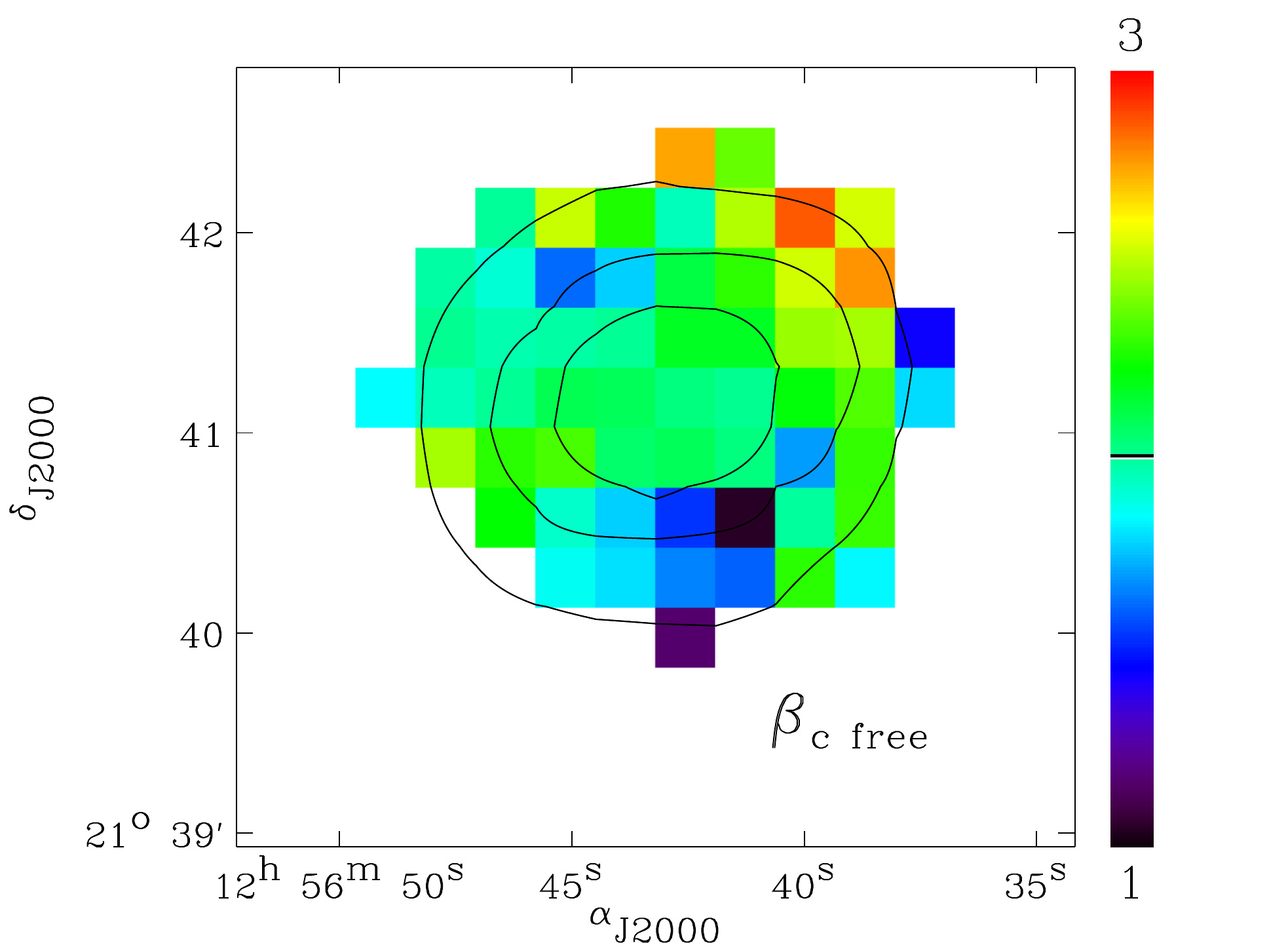}  &
	 \includegraphics[height=4.1cm]{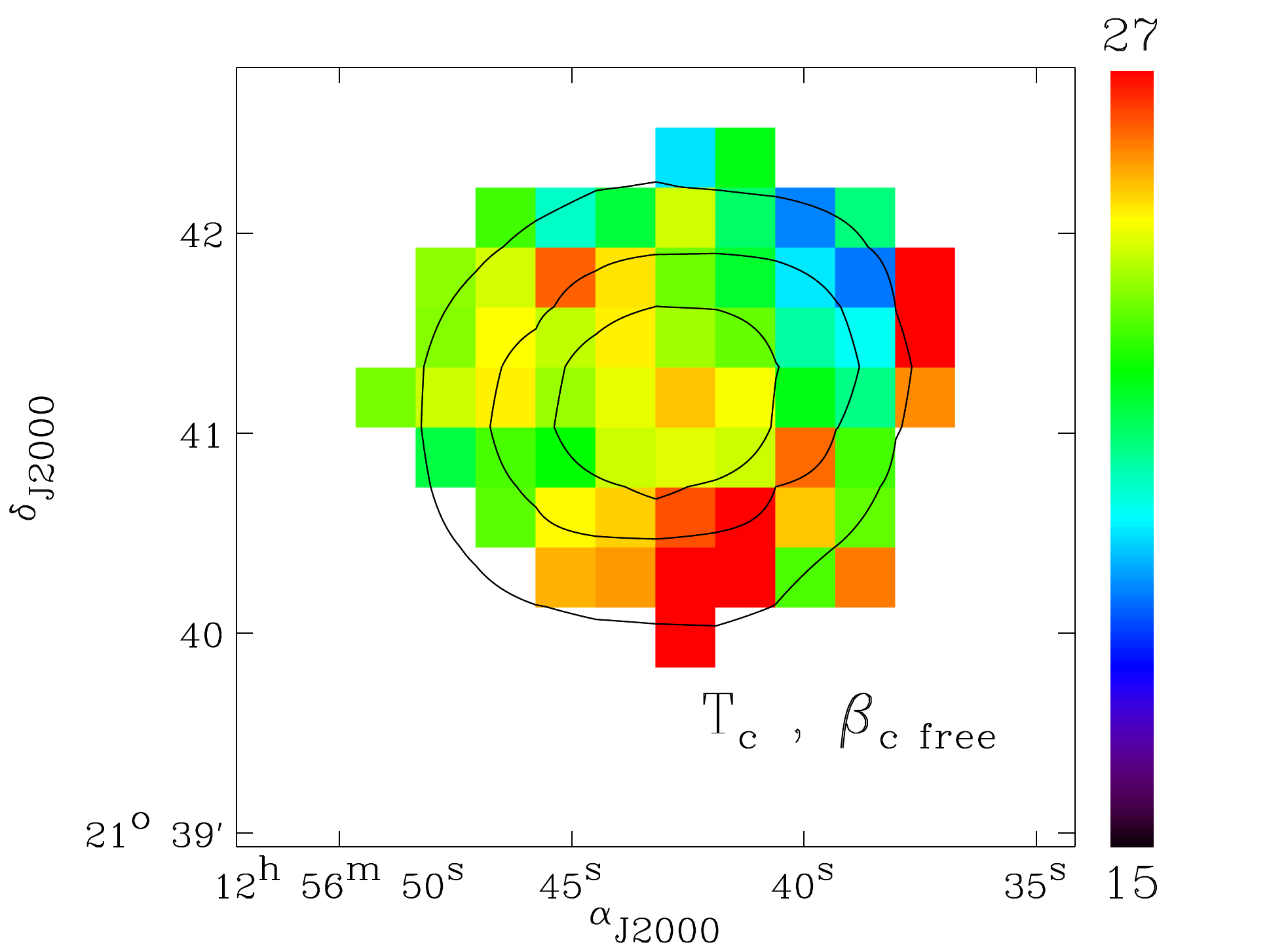} &
	 \includegraphics[height=4.1cm]{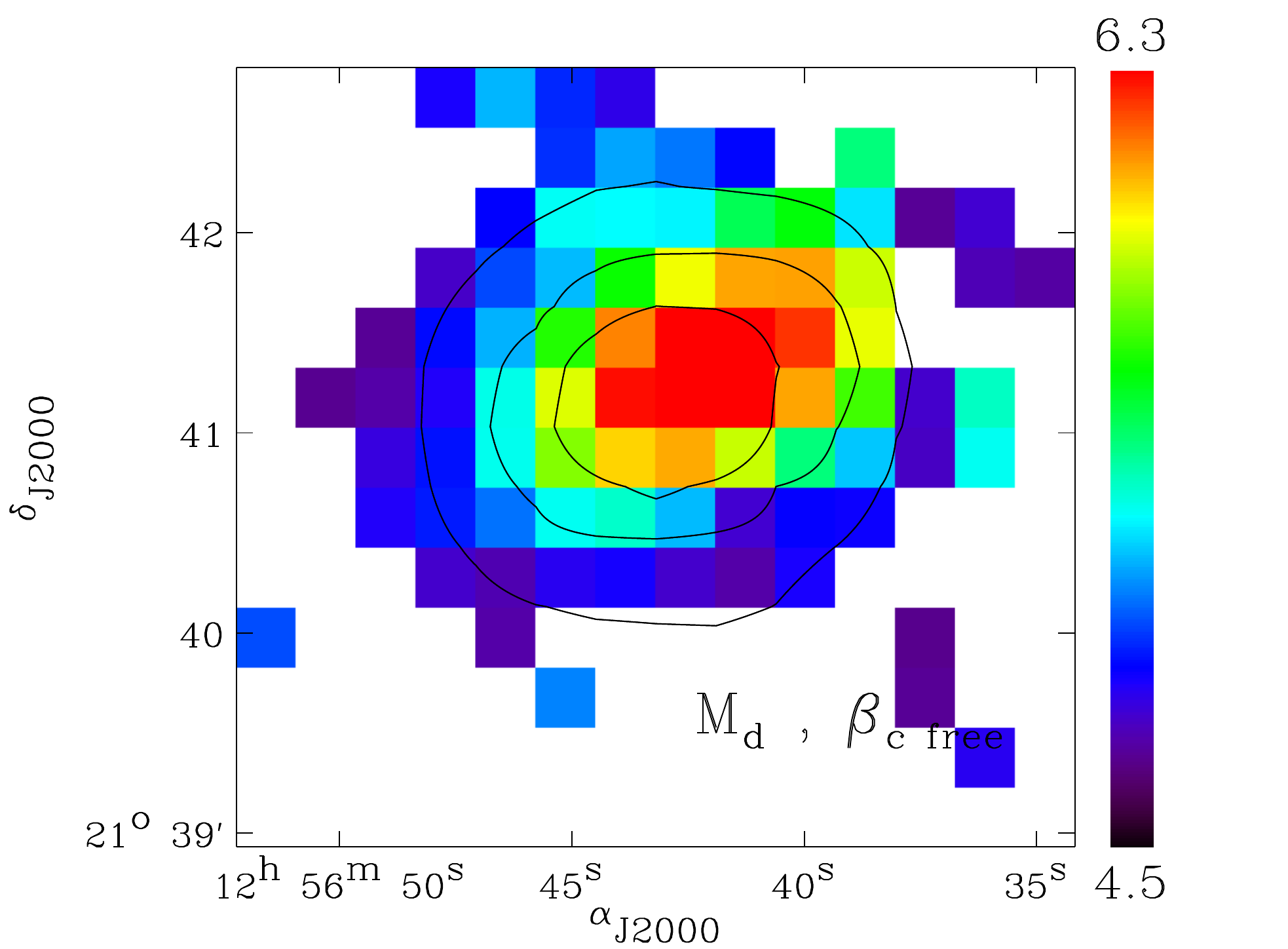} \\
       &&    \\
                            \end{tabular}
\caption{Continued. }
\end{figure*}

\addtocounter {figure}{-1}
        
\begin{figure*}
    \centering   

    \begin{tabular}{m{5.8cm} m{5.8cm} m{5.8cm}}

    {\large NGC~7793} &&\\		
   \centering{\large MIPS 24 \mic} &
    \centering{\large Temperature map ($\beta$$_c$=2)} &
     \centering{\large Dust mass surface density ($\beta$$_c$=2)} \\
   	\tabularnewline
	\includegraphics[height=4.3cm]{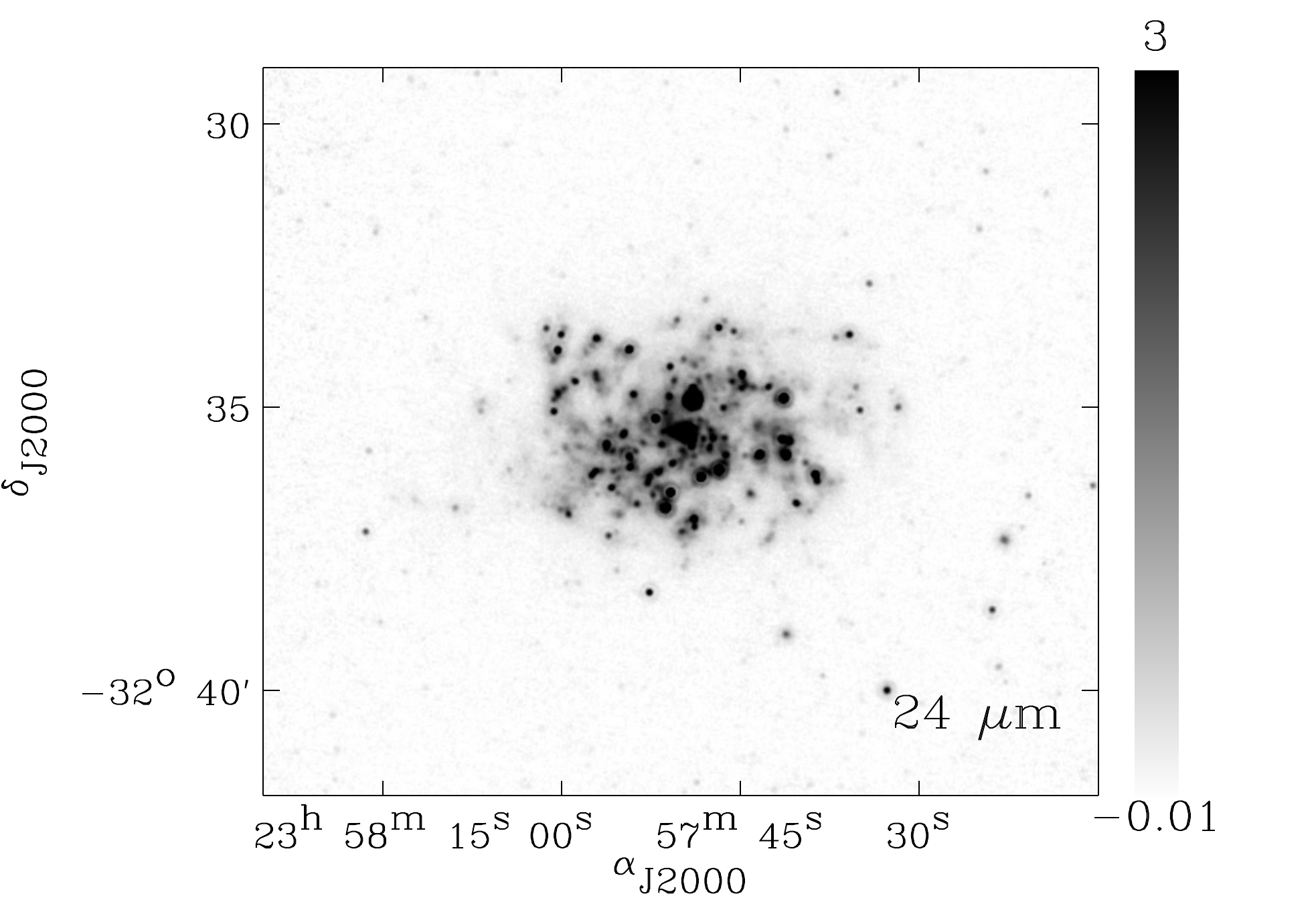} &
	\includegraphics[height=4.3cm]{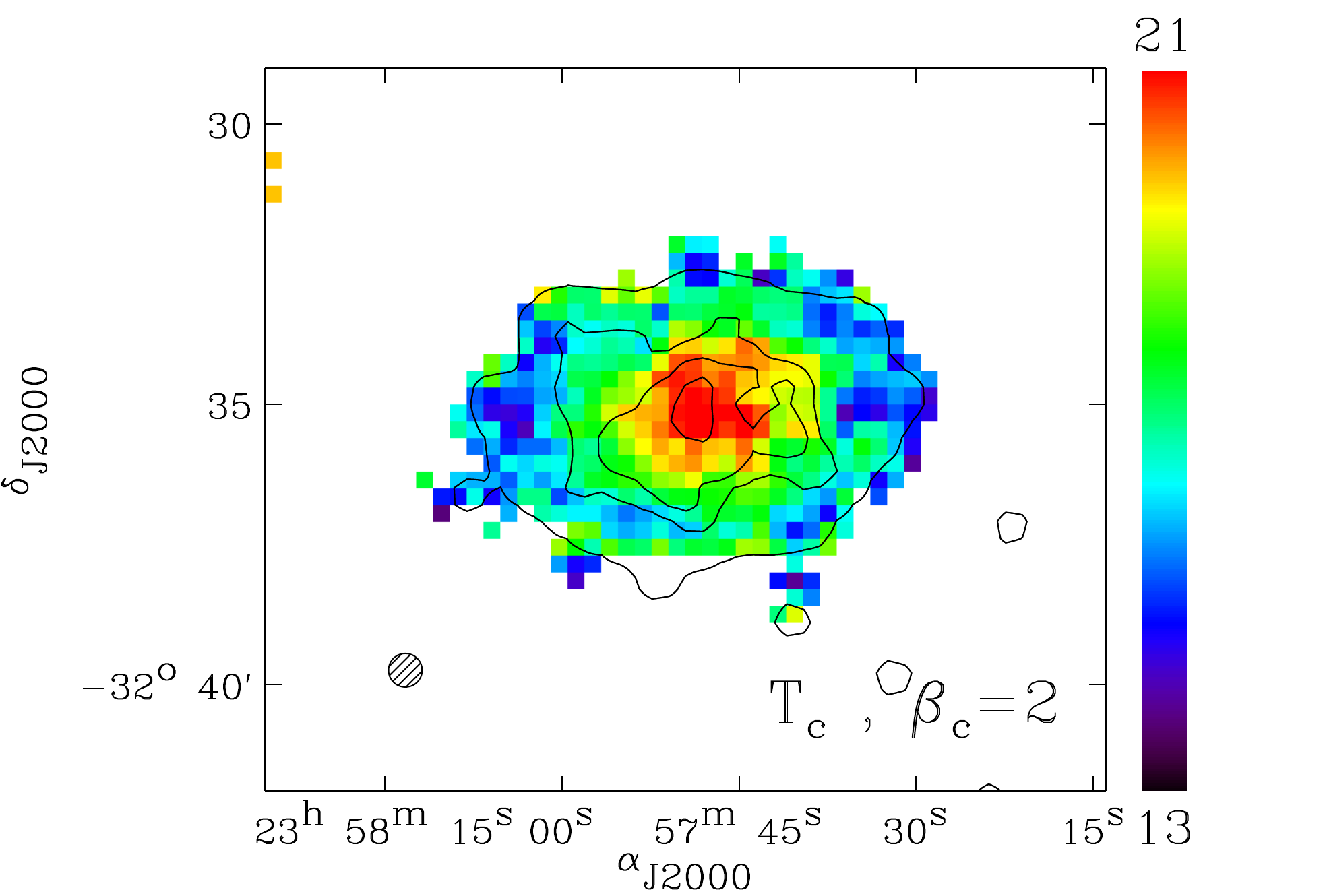} &
	\includegraphics[height=4.3cm]{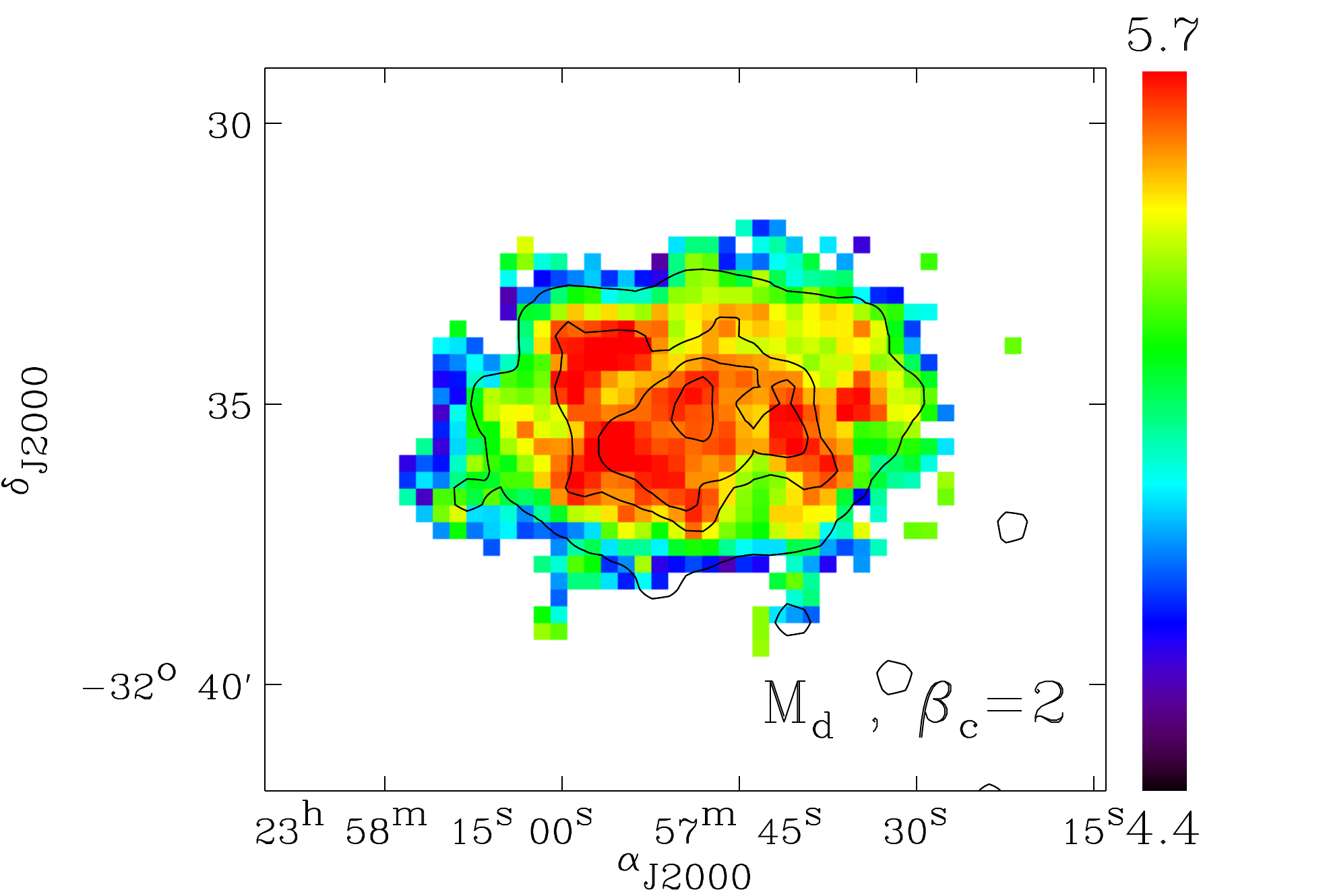} \\
     \centering{\large Emissivity index map} &
      \centering{\large Temperature map ($\beta$$_c$ free)} &
       \centering{\large Dust mass surface density ($\beta$$_c$ free)} \\
   	\tabularnewline
	\includegraphics[height=4.3cm]{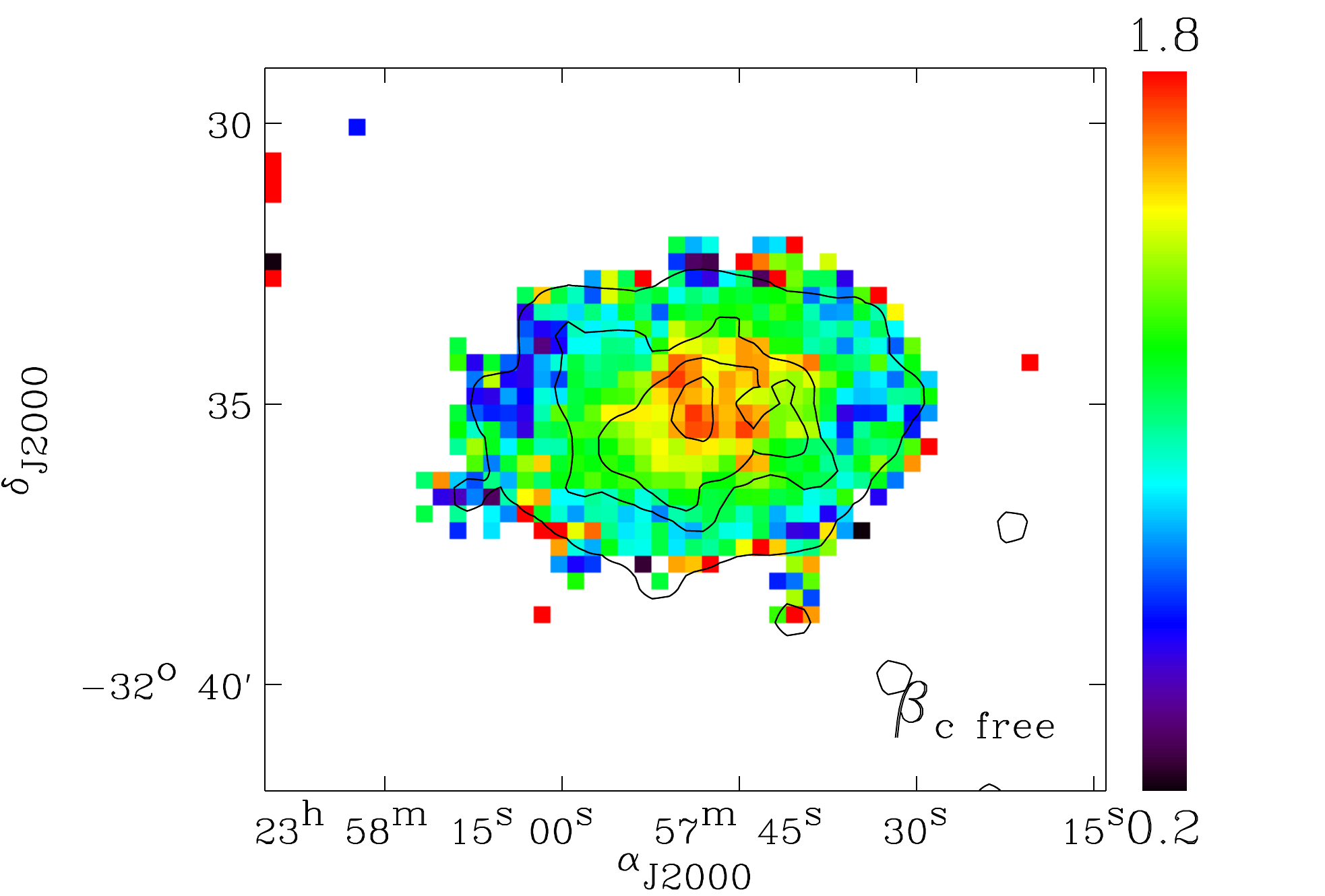}  &
	 \includegraphics[height=4.3cm]{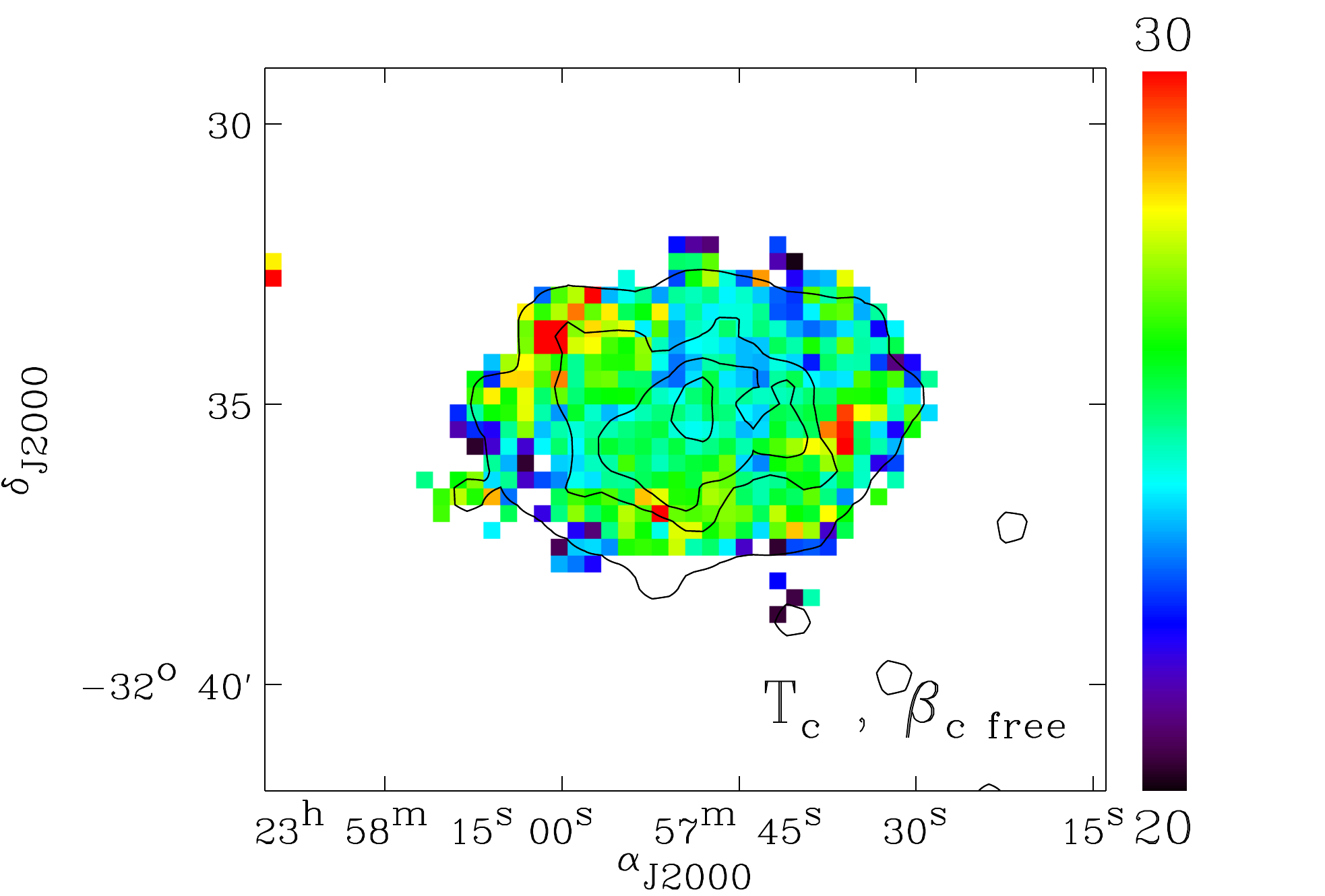} &
	 \includegraphics[height=4.3cm]{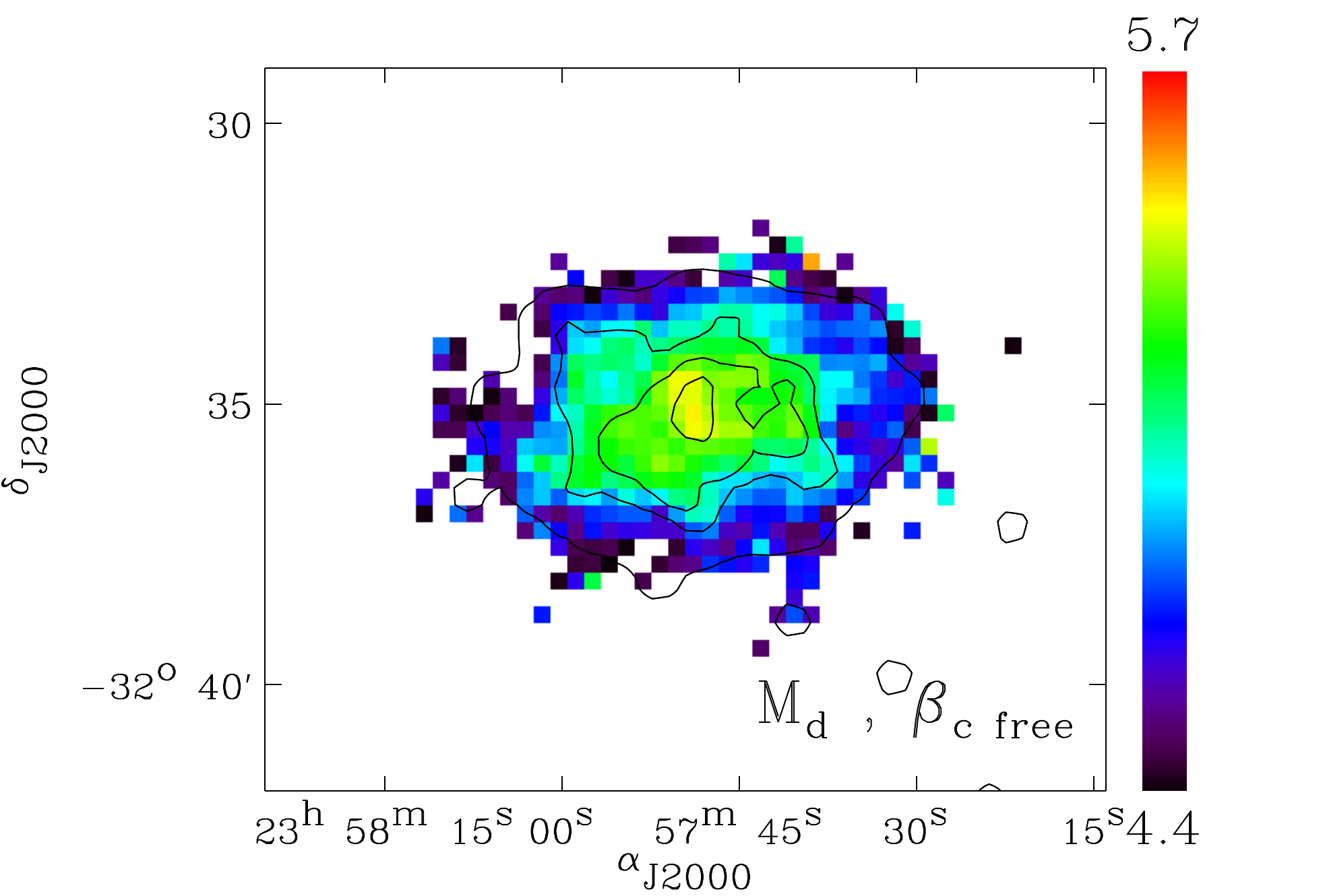} \\
          \end{tabular}
    \caption{Continued. }
\end{figure*}


\end{document}